\documentclass[fleqn,usenatbib]{mnras}

\usepackage{newtxtext,newtxmath}

\usepackage[T1]{fontenc}

\DeclareRobustCommand{\VAN}[3]{#2}
\let\VANthebibliography\thebibliography
\def\thebibliography{\DeclareRobustCommand{\VAN}[3]{##3}\VANthebibliography}

\usepackage{graphicx}	
\graphicspath{{./}{figures/}{figuresapdx/}}
\usepackage{comment}    
\usepackage{color}
\usepackage{amsmath}	
\usepackage{deluxetable}
\usepackage{subcaption} 
\usepackage{gensymb}
\usepackage{ulem}
\usepackage{longtable}

\usepackage{academicons}
\usepackage{xcolor}
\usepackage{scalerel}
\usepackage{tikz}
\usetikzlibrary{svg.path}

\definecolor{orcidlogocol}{HTML}{A6CE39}
\tikzset{
  orcidlogo/.pic={
    \fill[orcidlogocol] svg{M256,128c0,70.7-57.3,128-128,128C57.3,256,0,198.7,0,128C0,57.3,57.3,0,128,0C198.7,0,256,57.3,256,128z};
    \fill[white] svg{M86.3,186.2H70.9V79.1h15.4v48.4V186.2z}
                 svg{M108.9,79.1h41.6c39.6,0,57,28.3,57,53.6c0,27.5-21.5,53.6-56.8,53.6h-41.8V79.1z M124.3,172.4h24.5c34.9,0,42.9-26.5,42.9-39.7c0-21.5-13.7-39.7-43.7-39.7h-23.7V172.4z}
                 svg{M88.7,56.8c0,5.5-4.5,10.1-10.1,10.1c-5.6,0-10.1-4.6-10.1-10.1c0-5.6,4.5-10.1,10.1-10.1C84.2,46.7,88.7,51.3,88.7,56.8z};
  }
}

\newcommand\orcidicon[1]{\href{https://orcid.org/#1}{\mbox{\scalerel*{
\begin{tikzpicture}[yscale=-1,transform shape]
\pic{orcidlogo};
\end{tikzpicture}
}{|}}}}

\usepackage{subfiles}
\usepackage{hyperref}

\newcommand{\redmar}[1]{\textcolor{black}{#1}} 
 
\newcommand{\redsep}[1]{\textcolor{black}{#1}}

\title[Luhman 16AB IGRINS atlas]{A 1.48--2.48 µm $R$=\redsep{28\,000} spectroscopic atlas of the L7.5 and T0.5 components of the nearest pair of brown dwarfs: Luhman 16AB} 

\author[Ishikawa et al.]{Hiroyuki Tako Ishikawa$^{1}$\thanks{E-mail: hishikaw@uwo.ca}\orcidicon{0000-0001-6309-4380}, 
Stanimir Metchev$^{1,2}$\thanks{E-mail: smetchev@uwo.ca}\orcidicon{0000-0003-3050-8203}, 
Megan E. Tannock$^{1,3}$ \orcidicon{0000-0002-9445-2870}, 
Gregory N. Mace$^{4}$ \orcidicon{0000-0001-7875-6391}, 
\newauthor
Callie E. Hood$^{5}$ \orcidicon{0000-0003-1150-7889}, 
Jonathan J. Fortney$^{5}$ \orcidicon{0000-0002-9843-4354}, 
Sagnick Mukherjee$^{5}$ \orcidicon{0000-0003-1622-1302},
Paulo Miles-P\'aez$^{6}$ \orcidicon{0000-0003-2446-8882}, 
\newauthor
Radostin Kurtev$^{7}$ \orcidicon{0000-0002-9740-9974} 
\\
$^{1}$Department of Physics and Astronomy, The University of Western Ontario, 1151 Richmond St, London, Ontario, N6A~3K7, Canada\\
$^{2}$Institute for Earth and Space Exploration, The University of Western Ontario, 1151 Richmond St, London,\\ Ontario, N6A~3K7, Canada\\
$^{3}$Cosmographic Software, New Haven, CT 06513, USA\\
$^{4}$Department of Astronomy, The University of Texas at Austin, 2515 Speedway, Austin, TX 78712, USA\\
$^{5}$Department of Astronomy \& Astrophysics, University of California, Santa Cruz, CA 95064, USA\\
$^{6}$Centro de Astrobiolog\'ia, CSIC-INTA, Camino Bajo del Castillo s/n, 28692 Villanueva de la Ca\~nada, Madrid, Spain\\
$^{7}$ Departamento de Fisica y Astronomia, Facultad de Ciencias, Universidad de Valparaiso, Av. Gran Bretana 1111, Casilla 5030, Valparaiso, Chile
}

\date{Accepted 2025 March 12. Received 2025 March 10; in original form 2024 August 17}

\pubyear{2025} 

\begin{document}
\label{firstpage}
\pagerange{\pageref{firstpage}--\pageref{lastpage}}
\maketitle

\begin{abstract} 
We present a high signal-to-noise (SNR $\sim$ 450), high-dispersion ($R \equiv \lambda / \Delta \lambda \sim$ \redsep{28\,000}) $H$- and $K$-band spectroscopic atlas of the L7.5 and T0.5 components of the Luhman 16AB binary (WISE J104915.57$-$531906.1AB): the closest pair of brown dwarfs, and one of the best substellar benchmarks.  
The spectra were combined from a \redsep{70-day}
spectroscopic monitoring campaign of the binary with IGRINS on Gemini South. 
We fit model photospheres to the combined high-quality spectra to estimate atmospheric parameters. 
The model is based on the Sonora model atmosphere further incorporating the effects of clouds and disequilibrium. 
\redsep{We detect ammonia (NH$_3$) lines in both binary components, making Luhman 16A the warmest object where individual NH$_3$ lines were identified.}
We discover hydrogen (H$_2$), hydrogen sulfide (H$_2$S), \textcolor{black}{and hydrogen fluoride (HF) lines in both components, following recent reports of these species in either cooler (H$_2$, H$_2$S in a T6 dwarf) or warmer (HF in young late-M or mid-L dwarfs) objects.}
Methane (CH$_4$) shows a small contribution\redsep{,}
with lines sensitive to the slight temperature difference 
spanning the L/T transition.
\redsep{Against model expectations,}
we do not \redsep{detect}
FeH \redsep{lines,}
\redsep{ implying more efficient iron rainout}
than incorporated in the models.
We find \redsep{various}
unidentified features \textcolor{black}{in water-dominated regions, likely the result of residual inaccuracies in the water line lists.}
We searched for planetary-mass companions by periodogram analysis of radial velocities over 70 days \redsep{but detected no}
significant signal. 
\redsep{The}
upper limits of projected planetary mass \redsep{are}
$M\sin{i}=$ 0.2 $M_{\mathrm{J}}$ and 0.3 $M_{\mathrm{J}}$ at P $\sim$ 1 day, and 0.4 $M_{\mathrm{J}}$ and 0.7 $M_{\mathrm{J}}$ at P $\sim$ 10 days for Luhman 16A and B, respectively.
\end{abstract}

\begin{keywords}
(stars:) brown dwarfs -- (stars:) binaries: visual -- infrared: stars -- techniques: spectroscopic -- stars: atmospheres -- line: identification
\end{keywords}

\section{Introduction}
\label{sec:intro}

Reliable atmospheric characterisation of substellar objects depends on the accuracy of the spectral modelling and thus, on the correct understanding of spectral lines. Recent updates \redsep{in}
molecular line lists have offered significant improvements in accuracy, particularly for water and methane in the near-infrared spectra \citep[e.g.,][]{2020JQSRT.25507228T, 2021ApJS..254...34G, 2022JQSRT.27707949G, 2023ApJ...953..170H}. 
The near-infrared is well suited to spectroscopic characterisation at high dispersion because it spans the emission peak of brown dwarfs. 
However, there are still only a handful of high-SNR high-dispersion spectroscopic studies with broad wavelength near-infrared coverage of brown dwarfs. 
\citet{Tannock2022} performed a detailed analysis of $R\sim\redsep{28\,000}$ $H$- and $K$-band spectra of a T6 brown dwarf, 2MASS J08173001--6155158, assessing the accuracy of the latest line lists of H$_2$O, CH$_4$, and NH$_3$. 
They also reported the first detection of H$_2$ and H$_2$S molecular lines in an atmosphere outside the Solar System.
We extend the identification and assessment of these spectral lines to warmer objects near the L/T transition by presenting a comparable high-dispersion analysis on the nearest brown dwarf binary system, Luhman~16AB (\citealt{Luhman2013}; also known as WISE J104915.57$-$531906.1AB or WISE 1049$-$5319). 

At a distance of only 1.9960 $\pm$ 0.0002 pc \citep{2024AN....34530158B}, Luhman 16 is the third nearest stellar or substellar system, after $\alpha$ Cen and Barnard's star, and the nearest star or brown dwarf to be discovered in almost a century. 
Luhman 16AB consists of an L7.5 and a T0.5 dwarf \citep{Burgasser2013b} and has been extensively studied since its discovery, 
with measurements of the photometric amplitudes and periods of the two components \citep{Gillon2013, Burgasser2013a, Biller2013b, 2021ApJ...906...64A, 2024ApJ...965..182F}, 
\redsep{Doppler imaging to map their surface inhomogeneities \citep{Crossfield2014, 2024MNRAS.533.3114C},}
spectroscopic monitoring of the cloud distributions on both objects \citep{Buenzli2015, Karalidi2016, Kellogg2017, 2021ApJ...920..108H},
and Bayesian inference of the temperature and C/O ratio of the primary \citep{2022ApJS..258...31K}.
The proximity of the binary has also enabled precise astrometric monitoring of the system, resulting in the accurate determination of the binary orbit and the individual component masses. 
\citet{2024AN....34530158B} constrained an orbital period of 
\redmar{$29.0^{+4.9}_{-4.2}$ }
years and masses of 
$35.4 \pm 0.2$
Jupiter masses ($M_{\mathrm{J}}$) and 
$29.4 \pm 0.2$ 
$M_{\mathrm{J}}$ for components A and B, respectively. 

Its proximity and apparent brightness enable high-SNR high-dispersion spectroscopy despite the intrinsic faintness of this type of objects. With two coeval components that span the L/T transition, the Luhman 16AB binary can also elucidate the role of CO/$\mathrm{CH_{4}}$ disequilibrium chemistry that characterizes this transition. In addition, the wealth of astrometric monitoring and dynamic mass estimates help narrow down the search for additional components in the system.

The potential presence of planets in this neighbouring system is also of interest because the low masses of the binary components could reveal some of the lowest-mass planets through direct imaging, astrometry, or radial velocity monitoring.
\citet{Melso2015} and \citet{Bedin2017, 2024AN....34530158B} excluded the presence of widely separated massive planets ($\gtrsim$26 $M_{\mathrm{\oplus}}$ with orbital periods of 400--5000 days or $\gtrsim$50 $M_{\mathrm{\oplus}}$ with 2--5000-day periods) through direct imaging and astrometry, respectively.
Such planets are not necessarily expected given the 
\redmar{$3.67^{+0.43}_{-0.40}$ }
au binary semi-major axis \citep{2024AN....34530158B} and the substellar masses of the components. 
Low-mass stars tend to have lower-mass planets on narrower orbits (e.g., \citealt{Batalha2014}, and references therein), and brown dwarf hosts could be expected to follow that trend. 

Transits or radial velocity monitoring are more sensitive to close-in planets. The long-duration TESS monitoring of Luhman 16AB by \citet{2021ApJ...906...64A, 2024ApJ...965..182F} does not reveal any transit signatures. While this excludes transiting orbital geometries, close-in planets on inclined orbits have yet to be surveyed with radial velocities around either binary component. Being spatially resolved at about $1\arcsec$ and having components of similarly high apparent brightness ($K_{s,A}$ = 9.6 mag and $K_{s,B}$ = 9.8 mag; \citealt{Kniazev2013}), the Luhman 16AB binary potentially offers an opportunity for precise radial velocity (RV) monitoring even without a stabilized RV spectrograph: by using one binary component as an RV reference for the other.
We used our Immersion GRating INfrared Spectrometer ~\citep[IGRINS;][]{2010SPIE.7735E..1MY, 2014SPIE.9147E..1DP, 2016SPIE.9908E..0CM, 2018SPIE10702E..0QM} series of observations with Gemini South to search for close-in giant planets.

\section{Observations and Data Reduction} \label{sec:IGRINS}

We observed the Luhman 16AB system with IGRINS on Gemini South under Gemini program ID GS-2018A-Q-114 (PI: S. Metchev). IGRINS is a cross-dispersed spectrograph that simultaneously covers the $H$ and $K$ bands from 1.45 to 2.48 µm with a spectral resolution of $R \sim \redsep{28\,000}$.

\subsection{Observation} \label{sec:observation}

We observed the two components of the binary system simultaneously, with both objects on the IGRINS spectrograph slit. 
Observations took place on 33 nights over a 70-night span between 21 April 2018 and 29 June 2018, and are summarized in Table~\ref{tab:observations}. 

Each observation was composed of an AB dither pair with exposure times of 300s. One or two sets of AB pairs were obtained on each night.
Low-quality data due to poor seeing and bad weather was excluded, and we used 93 exposures for our final analysis.
All 93 spectra were combined into a single spectrum for atlas creation, parameter estimation (Section~\ref{sec:photosphere}, and atmospheric analysis (Section \ref{sec:discussion}). However, the spectrum at each epoch was treated as a separate data point for the periodic analysis of RVs (Section \ref{sec:planetsearch}).

We tabulate the correspondence between the diffraction order ($m$) and the wavelength range in Table \ref{tab:orderswavelengths}.

\begin{figure*}
    \centering
        \includegraphics[width=0.995\textwidth, trim={3.4cm 0.5cm 3.5cm 1.5cm}, clip]{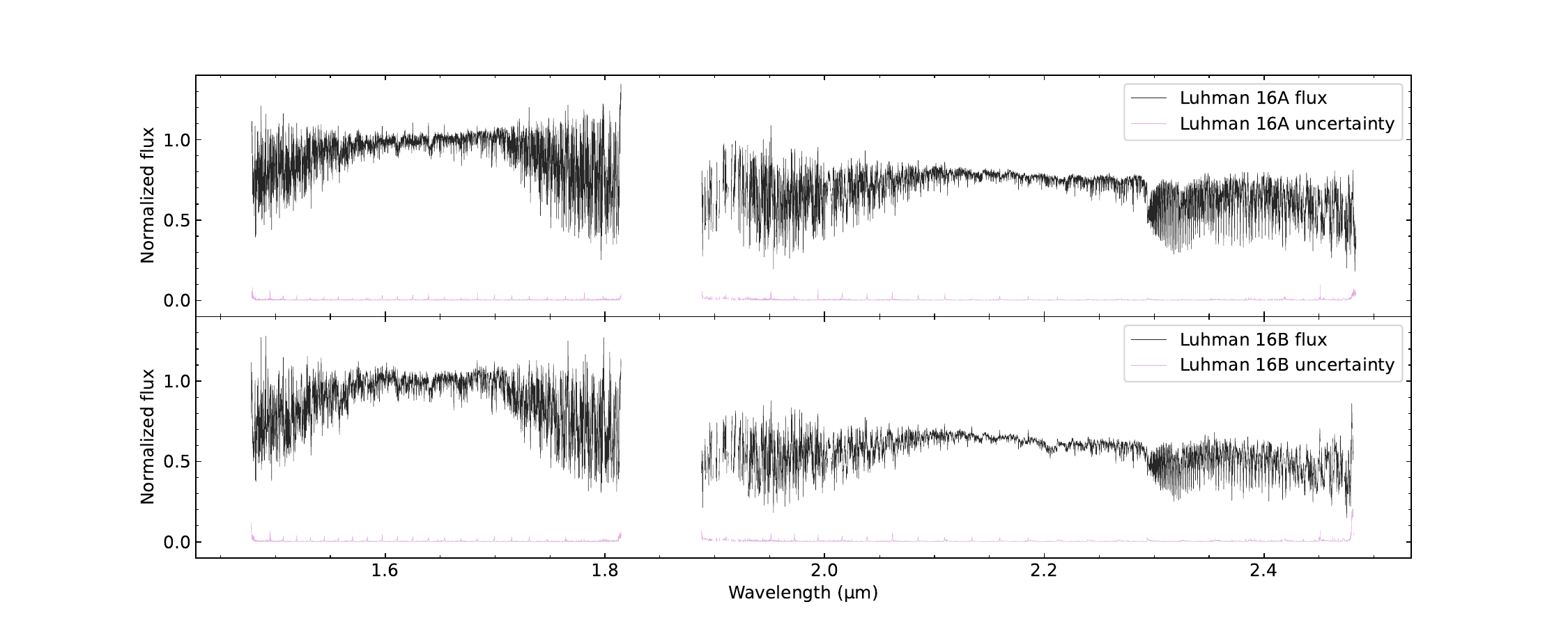} 
    \caption{
    The full $H$- and $K$-band IGRINS spectra of Luhman 16A (upper panel) and Luhman 16B (lower panel) with all 93 epochs combined, and all 46 orders stitched together. 
    The relative fluxes between orders are calibrated based on medium-dispersion spectra from \citet{Kellogg2017}.
    The entire spectra are normalized by the pseudo-continuum over 1.6--1.7~µm at
    the $H$-band peak for each object.  
    These data look noisy but actually have SNR $\simeq$ \redsep{600}
    at the peak of the $H$-band spectra 
    \redsep{with average SNRs of $\sim$250 over both bands for both objects.}
    Most of the conspicuous downward spikes are real absorption features\redsep{.}
    They can be seen in detail in the order-by-order spectral atlas in Appendix \ref{sec:atlas}.
    }\label{fig:stitchedLuhman16AB}
\end{figure*}

\begin{table*}
	\centering
	\caption{The wavelengths of the IGRINS diffraction orders \citep[from][]{Stahl2021} and the known molecular absorbers in each order. 
	}
	\label{tab:orderswavelengths}
	\begin{tabular}{cccl@{\hskip 0.75in}cccl} 
		\hline
		Band & Order & Wavelength & Molecular Line & Band & Order & Wavelength & Molecular Line \\
		& $m$ & Coverage (µm) & Absorbers & & $m$ & Coverage (µm) & Absorbers \\
		\hline
        $H$ & 121 & 1.482--1.494 & H$_2$O & $K$ & 94 & 1.894--1.910 & H$_2$O, H$_2$S \\  
        $H$ & 120 & 1.493--1.506 & H$_2$O & $K$ & 93 & 1.909--1.930 & H$_2$O \\  
        $H$ & 119 & 1.504--1.519 & H$_2$O & $K$ & 92 & 1.929--1.950 & H$_2$O \\  
        $H$ & 118 & 1.517--1.531 & H$_2$O & $K$ & 91 & 1.949--1.972 & H$_2$O \\  
        $H$ & 117 & 1.529--1.543 & H$_2$O & $K$ & 90 & 1.971--1.993 & H$_2$O, H$_2$S \\  
        $H$ & 116 & 1.541--1.556 & H$_2$O & $K$ & 89 & 1.992--2.015 & H$_2$O, H$_2$S, NH$_3$ \\  
        $H$ & 115 & 1.554--1.569 & H$_2$O, CO & $K$ & 88 & 2.014--2.038 & H$_2$O, NH$_3$ \\  
        $H$ & 114 & 1.567--1.583 & H$_2$O, FeH, CO, H$_2$S & $K$ & 87 & 2.037--2.061 & H$_2$O \\  
        $H$ & 113 & 1.581--1.596 & H$_2$O, FeH, CO, H$_2$S & $K$ & 86 & 2.060--2.085 & H$_2$O \\  
        $H$ & 112 & 1.594--1.610 & H$_2$O, FeH, CO, H$_2$S & $K$ & 85 & 2.084--2.109 & H$_2$O \\  
        $H$ & 111 & 1.608--1.624 & H$_2$O, FeH, CO, H$_2$S & $K$ & 84 & 2.108--2.134 & H$_2$O, H$_2$ \\  
        $H$ & 110 & 1.622--1.639 & H$_2$O, FeH, CO, H$_2$S, CH$_4$ & $K$ & 83 & 2.133--2.159 & H$_2$O \\  
        $H$ & 109 & 1.637--1.653 & H$_2$O, FeH, H$_2$S, CH$_4$ & $K$ & 82 & 2.158--2.185 & H$_2$O, NH$_3$ \\  
        $H$ & 108 & 1.651--1.668 & H$_2$O, FeH, CH$_4$ & $K$ & 81 & 2.184--2.212 & H$_2$O, NH$_3$, CH$_4$ \\  
        $H$ & 107 & 1.666--1.683 & H$_2$O, FeH, CH$_4$ & $K$ & 80 & 2.211--2.239 & H$_2$O, NH$_3$, CH$_4$, H$_2$ \\  
        $H$ & 106 & 1.681--1.699 & H$_2$O, FeH, CH$_4$ & $K$ & 79 & 2.238--2.267 & H$_2$O, NH$_3$, CH$_4$ \\  
        $H$ & 105 & 1.697--1.715 & H$_2$O, FeH & $K$ & 78 & 2.266--2.295 & H$_2$O, CO, NH$_3$, CH$_4$ \\  
        $H$ & 104 & 1.713--1.730 & H$_2$O, FeH & $K$ & 77 & 2.294--2.326 & CO, H$_2$O \\  
        $H$ & 103 & 1.728--1.747 & H$_2$O & $K$ & 76 & 2.325--2.355 & CO, H$_2$O, NH$_3$, CH$_4$ \\  
        $H$ & 102 & 1.745--1.764 & H$_2$O & $K$ & 75 & 2.354--2.383 & CO, H$_2$O, NH$_3$, CH$_4$ \\  
        $H$ & 101 & 1.762--1.781 & H$_2$O & $K$ & 74 & 2.389--2.414 & CO, H$_2$O, CH$_4$, H$_2$ \\  
        $H$ & 100 & 1.779--1.798 & H$_2$O & $K$ & 73 & 2.420--2.445 & H$_2$O, CO, CH$_4$ \\  
        $H$ & 99  & 1.797--1.812 & H$_2$O & $K$ & 72 & 2.452--2.478 & H$_2$O, CO, H$_2$S \\  
		\hline
	\end{tabular}
\end{table*}

\subsection{Data Reduction} 
\label{sec:reductions}

\subsubsection{Spectroscopic Extraction and Telluric Correction}

Since the Luhman~16AB system was observed with both objects on the slit, a custom reduction in addition to the standard procedures with IGRINS Pipeline Package (PLP; \citealt{plp}) was required.

The data were first reduced using the PLP.
The PLP performed bad pixel correction, pattern noise removal, and sky background subtraction.
Each order was rectified onto a rectangular grid and the wavelength solution was derived from sky OH emission and telluric absorption features.
Post-processing the two-dimensional IGRINS PLP output was required to separate the simultaneously observed components of Luhman~16.
The custom post-processing IDL software first reduced the background by subtracting the median from each column (spatial direction) of the two-dimensional spectrum.
Each spectral order was then extracted and collapsed along the spectral direction so the median spatial profile was determined.
The A and B spectra were optimally extracted \redsep{by fitting each column's peak flux}
to the normalized spatial profile and then taking the sum under the profile.
We masked some parts of spectra that were affected by two-dimensional patterns of bad pixels too large to be corrected in the reduction process. 

The same extraction was performed on the A0V 
standard \redsep{star observed for the telluric correction \citep{2003PASP..115..389V}.}
Uncertainties \redsep{for each order}
were derived from the same extraction of two-dimensional variance spectra from the PLP.
The telluric standard spectra \redsep{were}
divided by a Vega model to pre-reduce the effect of 
A0V intrinsic features.  
The target spectrum and the Vega-corrected telluric standard were then cross-correlated to align the telluric features in the target frame.
The target spectrum was divided by the telluric standard and the barycentric velocity correction was applied to the wavelength solution.
The telluric-corrected output files were one-dimensional spectra of the flux, variance, and signal-to-noise for both the A and B components for each observation.

\redsep{
Some strong telluric features still left residuals, which may impact the model-fitting analyses. 
To identify and mask the regions with prominent telluric features, 
we simulated transmission spectra for the Earth's atmosphere using the Planetary Spectrum Generator ~\citep[PSG;][]{2018JQSRT.217...86V} for
the longitude, latitude, and altitude of the Gemini South Observatory. 
We masked regions where telluric absorption reduces atmospheric transmission to $<$35\%
and regions with OH emission lines \citep{2000A&A...354.1134R}. 
We slight increased the transmission threshold to 40\% for the four orders ($m=$~86--89) where strong CO$_2$ telluric lines also left non-negligible residuals.
The PSG spectra and the transmission thresholds are displayed in the spectral atlas of Appendix \ref{sec:atlas}.}

We obtain a final high signal-to-noise (SNR) spectrum for each component in each order by combining all spectra of individual exposures via a weighted average. 
\redsep{By combining data from different epochs, the wavelength range 
lost in the telluric masking is partly recovered because the 
relative velocity shifts of the target and telluric spectra vary
at different epochs. 
For each wavelength point, we calculate the weighted standard errors from 93 spectra from different epochs to estimate the uncertainty of the combined spectrum. This 
includes not only the statistical uncertainty based on the derivation in the reduction phase but also systematic errors potentially raised from telluric residuals or changes in instrumental or observing conditions.}

The combined spectra have SNRs with an average of $\sim$\redsep{250}
and a maximum of $\sim$\redsep{600.}
We show in Figure \ref{fig:stitchedLuhman16AB} the order-stitched spectra over the entire wavelength coverage of IGRINS. 
For the order stitching, we calibrated the distortions in continuum levels, mainly induced by discrepancies between different orders, so that our pseudo-continuum is in line with that of the Magellan/FIRE medium-dispersion spectra of Luhman 16AB in \citet{Kellogg2017}. 
We used each order separately for the subsequent analysis. Each order of the spectral atlas is shown in Appendix \ref{sec:atlas}.

\subsubsection{Validation of the Telluric Standard Correction}

Two of the six telluric standard stars used in the reduction, HIP~52407 and HIP~53836, displayed non-negligible residuals at the Br$\gamma$ line (2.1661 µm), leaving spurious emission features in the target spectra in order $m=$~82.
To correct this, we substituted the 2.16544--2.16707 µm wavelength range in the combined spectra with the weighted average of only the spectra unaffected by the Br$\gamma$ residuals. 
\redsep{We also found a two-dimensional ghost on every image causing a false emission line at 2.16067 µm, near the Br$\gamma$ line, in all Luhman 16B spectra. We masked 
a 2.5~{\AA}-wide region around this peak.}
The final spectrum in Figure~\ref{fig:stitchedLuhman16AB} has been thus corrected.

\begin{figure*}
	\includegraphics[width=0.95\textwidth]{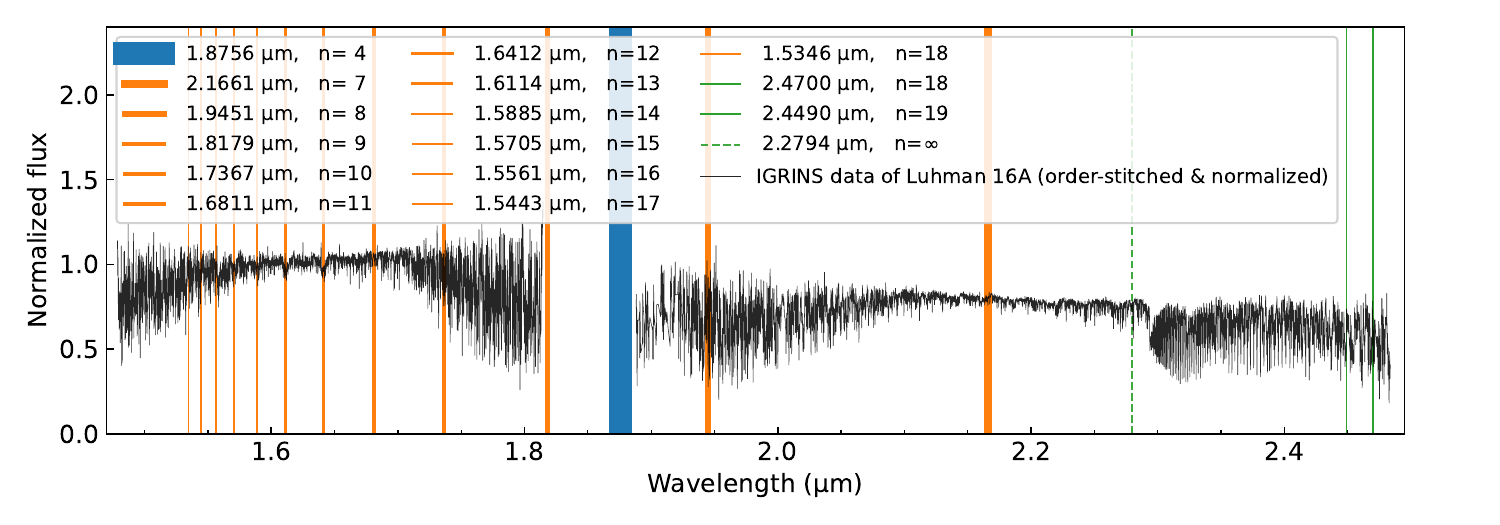} 
    \caption{The positions of hydrogen atomic lines are indicated by the vertical lines overplotted on the spectrum of Luhman 16A (\redsep{black spectrum).}
    The blue line is Paschen-$\alpha$, the orange lines are the Bracket series, and the green lines are the Pfund series.
    The $n$ in the legend is the quantum number of the upper energy level. After removing residuals centred on the 2.1661~µm Br$\gamma$ line associated with two of the telluric standards, no further correction of the telluric calibration was required. 
    } 
    \label{fig:hydrogenlines}
\end{figure*}

We further inspected all other orders in which we could anticipate similar contamination from weaker hydrogen lines in telluric standard spectra.
Several H line positions are covered in our spectral range as shown in Figure \ref{fig:hydrogenlines}. 
The Paschen series (principal quantum number $n^\prime=3$ for the lower energy level) does not affect our spectra. Paschen-$\alpha$ falls into the gap between the two bands, and the other lines are shorter than the wavelength range of IGRINS. The Bracket series ($n^\prime=4$) pervades the $H$ and $K$ bands, with transitions from upper energy levels with $n\geq7$. 
Only the Br$\gamma$ ($n=7$, 2.1661~µm) showed significant residuals in some of our spectra: corrected as outlined above. The Br~$\delta$ transition ($n=8$, 1.9451~µm) may also show a small upward residual (see panels for order $m=$~92 in Figures \ref{fig:atlasA} and \ref{fig:atlasB}), but that is less evident in the context of the surrounding noise. 
The spectrum in the corresponding order ($m=$~92) is fragmented due to a large number of telluric-line masking and several points show large residuals; the upward residuals at 1.9451~µm appear to be the largest among them (Table \ref{tab:discrepancies_both}).  
The Pfund series ($n^\prime=5$) lines are mostly redder than our wavelength range, and no residuals were detected even at the expected 2.2794~µm Pfund break where bound-free absorption begins. Hence, we found no further need to correct the telluric calibration of the Luhman 16AB spectra.

\redsep{
\subsubsection{Degraded spectral resolution}
\label{sec:degradedresolution}
}

\begin{figure}
    \centering
        \includegraphics[width=0.49\textwidth]{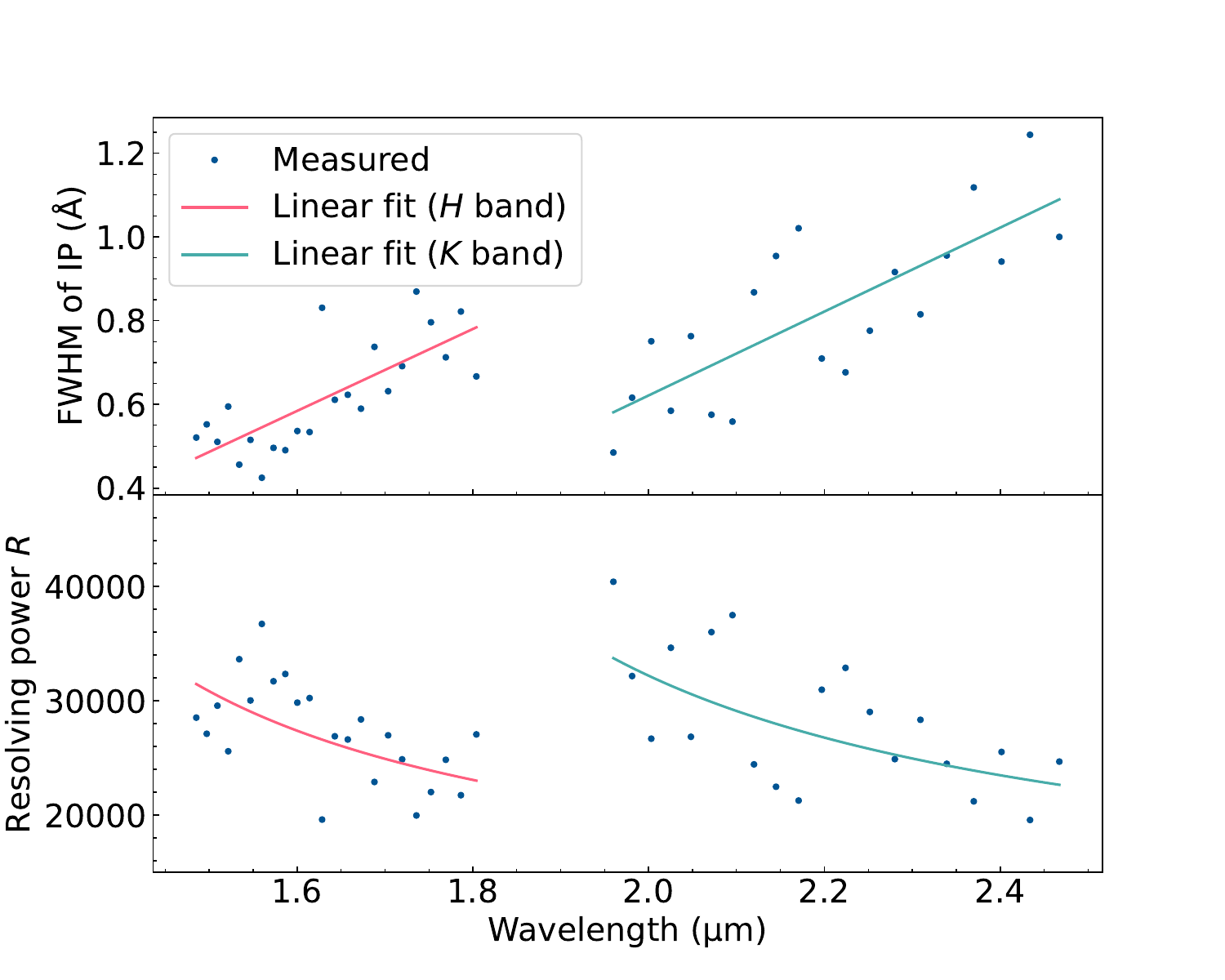} 
    \caption{\redsep{The FWHM of the instrumental broadening profile IP of our spectra (top) and its corresponding resolving power $R$ (bottom) as a function of wavelength.
    The data points indicate our measurements for the individual used clean telluric lines in the A0V spectra.
    The red and green solid lines are based on the linear fits of the FWHMs (Equations~\ref{eq:sigma_H} and \ref{eq:sigma_K}).}
    }\label{fig:degraded_resolution}
\end{figure}

\redsep{
The IGRINS team reports\footnote{\url{https://sites.google.com/site/igrinsatgemini/2018-k-band-resolution}} that the spectral resolution of IGRINS was degraded between April 2018 and May 2019 due to a defocus of the echellogram.
We found that this affected our data not only in the K band mentioned in their report but also in the H band with a resolution lower than the configured value of $\sim$40\,000.
The spectral resolution is characterised by the instrument broadening kernel, which depends on the position on the detector, i.e., wavelength.
We assume this instrumental profile (IP) to be Gaussian. We estimated its full width of half maximum (FWHM) at each wavelength using clean telluric lines recorded in the spectra of the A0V telluric standard stars. We fit two relations for the FWHM of the Gaussian (in µm), in the $H$ and $K$ bands respectively:
\begin{align}
  \mathrm{FWHM}_H(\lambda) &= 9.80 \times 10^{-5} \lambda - 9.82 \times 10^{-5} \label{eq:sigma_H} \\ 
  \mathrm{FWHM}_K(\lambda) &= 10.01 \times 10^{-5} \lambda - 13.82 \times 10^{-5} \label{eq:sigma_K}
\end{align} 
where $\lambda$ is a wavelength in µm. 
Figure~\ref{fig:degraded_resolution} shows the resolving power $R$ ($\lambda / \Delta \lambda$) as a function of wavelength, adopting these FWHMs as $\Delta \lambda$. 
The average $R$ ranges from about 20\,000 to 35\,000, with an average of $\sim$28\,000.}

\vspace{1.5\baselineskip}

\section{Photospheric analysis} \label{sec:photosphere}

\subsection{Fitting of Photospheric Models}
\label{sec:modelfits}

We compared the combined spectra of each component to theoretical spectral models, following the order-by-order fitting methodology of \citet{Tannock2021, Tannock2022} (hereafter Tann22) using the MASTIFF model fitting tool\footnote{\url{https://github.com/megantannock/Mastiff_ModelFittingTools}} to 
simultaneously determine the photospheric parameters, the projected rotational velocity $v\sin i$ and the RV.

We used an iteration of the Bobcat Alternative A model adopted in Tann22, which is based on the Sonora Bobcat model \citep{2021ApJ...920...85M} with more up-to-date molecular line lists. The molecules for which Tann22 updated line lists are $\mathrm{H_{2}O}$~\citep{2018MNRAS.480.2597P}, CH$_4$~\citep{2020ApJS..247...55H}, and NH$_3$~\citep{2019MNRAS.490.4638C}. Sonora Bobcat calculates emergent spectra from 1D cloudless radiative-convective equilibrium model atmospheres. While a cloudless model was appropriate for the T6 dwarf analysed in Tann22, the two L/T-transition components of Luhman 16 require both condensate clouds and disequilibrium chemistry.\footnote{The Sonora model series is undergoing continuous development. Disequilibrium chemistry has been introduced by \citet[][Sonora Cholla]{2021ApJ...923..269K} and \citet[][Sonora Elf Owl]{2024ApJ...963...73M} and clouds by \citet[][Sonora Diamondback]{2024ApJ...975...59M}. However, a comprehensive model combining both effects has yet to be released.} 
We added both effects self-consistently using the PICASO 1D radiative-convective equilibrium climate model~\citep{2023ApJ...942...71M, 2019ApJ...878...70B}.
The condensate clouds were computed using the VIRGA cloud model~\citep{2022ApJ...925...33R} with a uniform sedimentation efficiency parameter $f_\mathrm{sed}$ within our self-consistent modelling framework. 
The vertical eddy diffusion was parameterised with the $K_\mathrm{zz}$ coefficient in these models. 
The $K_\mathrm{zz}$ parameter relates the local scale height, the turbulent mixing length, and the convective heat flux of the atmosphere. In convective layers of the atmosphere, it is parameterised via Equation (2) in \citet{2022ApJ...938..107M}. In radiative atmospheres, the turbulent mixing length was decreased following Equation (6) in \citet{2001ApJ...556..872A}. 
To account for the inherent uncertainty of $K_\mathrm{zz}$, we employed two other cases where the nominal $K_\mathrm{zz}$ profiles were uniformly multiplied by factors of 0.1 and 100 across all atmospheric layers. We refer to this multiplication factor as $C_{K_\mathrm{zz}}$ hereafter. 
After the initial scaling, the $K_\mathrm{zz}$ profiles were iteratively re-calculated to preserve the effective temperature of the photosphere, following \citet{2023ApJ...942...71M}.
Disequilibrium chemistry was treated using the quench time approximation, also as in \citet{2023ApJ...942...71M}. 
We hereafter call the new self-consistent grid of models ``Bobcat Alternative B.'' 

The spectra from the Bobcat Alternative B model grid were generated with the DISORT radiative transfer tool~\citep{2017ascl.soft08006S}. The grid of parameters included effective temperature ($1200 \leq T_{\mathrm{eff}}/\mathrm{K} \leq1600$, step size of 100 K), surface gravity ($4.5\leq\log{g}\leq 5.5$, step size of 0.5), sedimentation efficiency ($3\leq f_\mathrm{sed}\leq5$, step size of 1), and eddy diffusion coefficient multiplication factor $C_{K_\mathrm{zz}}$. The $C_{K_\mathrm{zz}}$ multiplication factor is a scaling (0.1, 1.0, or 100) of the nominal eddy diffusion coefficient $K_\mathrm{zz}$ to account for the uncertainty of $K_\mathrm{zz}$. We further extended the dimensionality of the photosphere grid by including rotational broadening \citep[following][]{1992oasp.book.....G}
and RV offsets for each component of the binary. For Luhman 16A, these were 
$12\leq v\sin{i}/({\rm km~s^{-1}}) \leq 17$, step size of 0.2 km s$^{-1}$, and 
$16\leq {\rm RV}/({\rm km~s^{-1}}) \leq 18$, step size of 0.05 km s$^{-1}$. For Luhman 16B, these were 
$22\leq v\sin{i}/({\rm km~s^{-1}}) \leq 28$, step size of 0.2 km s$^{-1}$, and 
$19\leq {\rm RV}/({\rm km~s^{-1}}) \leq 22$, step size of 0.05 km s$^{-1}$.

The model photospheres have a higher spectral resolution ($R > 100\,000$) and 3--7 times finer wavelength sampling than the data. 
\textcolor{black}{We first convolved the models to account for rotational and IP broadening. For the IP, we used a Gaussian with a variable FWHM as a function of wavelength, as determined from Equations~\ref{eq:sigma_H} and \ref{eq:sigma_K}.}
We then resampled the model photospheres to match the wavelength sampling for the data.

We computed the reduced chi-square goodness-of-fit statistic ($\chi^2_{\mathrm{R}}$) 
for each point on the model photosphere grid. We incorporated a quadratic fit to the continuum in each order as a free parameter in the fitting, to correct for instrumental effects that curve the spectra towards the ends. As noted in Tann22, the quadratic function helps account for any systematic trends with wavelength from the blaze correction or from discrepancies between the spectra and the model photospheres.
The parameters at the grid point with the smallest $\chi^2_{\mathrm{R}}$ were considered the best-fit values for each order.

\begin{figure*}
	\includegraphics[width=0.905\textwidth]{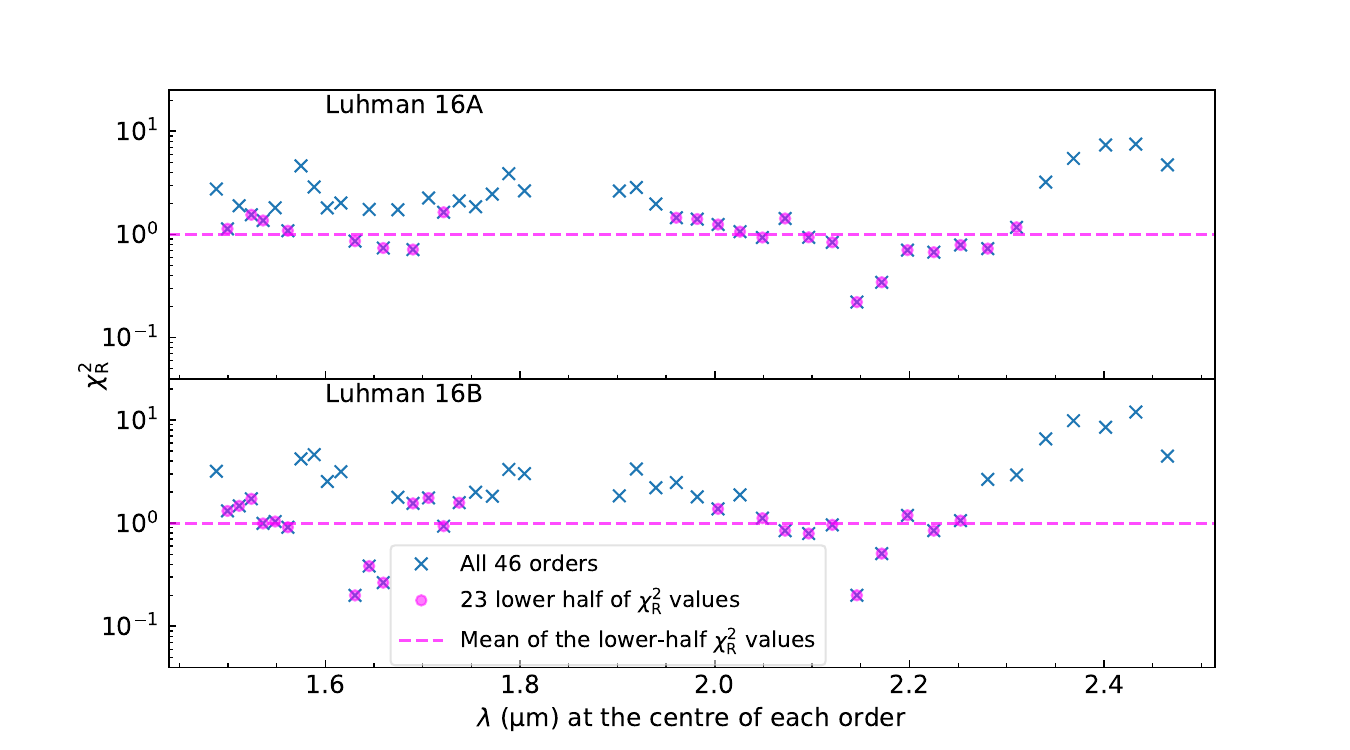} 
    \caption{Fitting of photospheric models. The two panels show best-fit $\chi^2_{\mathrm{R}}$ values in each order as a function of the central wavelength of the order, for each binary component. 
    The 23 pink circles highlight the lower half of the $\chi^2_{\mathrm{R}}$ values.
    The $\chi^2_{\mathrm{R}}$ values were re-evaluated by adding a \redsep{1.6\%}
    (for Luhman 16A) or \redsep{1.4\%}
    (for Luhman 16B) systematic uncertainty $\sigma_M$ in the model, so that the mean of these 23 values is unity (Section~\ref{sec:modelfits}).
    }  
    \label{fig:lower_half_mean}
\end{figure*}

In Tann22, the $\chi^2_{\mathrm{R}}$ values were scaled so that the $\chi^2_{\mathrm{R}}$ in the best-fit order was unity.
However, we modified this approach to avoid overweighting relatively featureless orders when calculating the global best-fit parameters (Section \ref{sec:bestfitparams}). 
For each object, the $\chi^2_{\mathrm{R}}$ values for all orders were adjusted so that the mean of the lower half of the $\chi^2_{\mathrm{R}}$ values is unity (Figure \ref{fig:lower_half_mean}).  
\redsep{
This adjustment is not based on a strict statistical justification, but rather on practical considerations.
Our primary goal is to prevent overweighting of orders with few absorption features 
\textcolor{black}{and poor diagnostic power that may}
otherwise dominate the final estimates of atmospheric parameters 
\textcolor{black}{because of their very} 
low $\chi^2_{\mathrm{R}}$ values. 
This approach allows us to balance the contribution of all orders, ensuring that spectral regions with significant absorption features 
\textcolor{black}{and good diagnostic power}
are appropriately weighted.
We calculated the adjusted $\chi^2_{\mathrm{R}}$}
by adding an extra systematic uncertainty, $\sigma_\mathrm{M}$, in quadrature to the observational uncertainty on the data. 
This $\sigma_\mathrm{M}$  
\textcolor{black}{is attributed to a combination of the scatter of the actual IP widths around the adopted (fitted) values as a function of wavelength (Figure~\ref{fig:degraded_resolution})}
and a systematic uncertainty of the model. 
The obtained values for $\sigma_\mathrm{M}$ are \redsep{1.6\% and 1.4\%}
of the normalized flux for Luhman 16A and B, respectively.
We hereafter adopt the notation $\chi^2_{\mathrm{R}}$ to denote the updated computation of the reduced chi-square.
The best-fit model for each order overplotted on the observed spectrum is shown in the spectral atlas in Appendix \ref{sec:atlas} 
across all orders for both objects. 

\subsection{Fitting Component A Using Component B as a Template} 
\label{sec:avsb} 

The two components of the binary Luhman 16A and B likely share the same formation and evolution history.
However, the slight differences in mass and temperature give rise to 
the slight differences in their atmospheric chemistry. The differences are particularly interesting in the case of this binary, as its components narrowly span the L/T transition and the associated change in atmospheric composition.

To directly compare the spectral features of the two components, we broadened and shifted the spectra of the more slowly rotating Luhman 16A to match those of B to compensate for the differences in rotational and radial velocities. 
The values for the broadening and shifting are selected to minimize the $\chi^2_{\mathrm{R}}$ of the fit for each order \redsep{after applying the quadratic correction to the continua to remove instrumental effects,}
as 
done for the model fitting in Section \ref{sec:modelfits}. 

\begin{figure*}
    \includegraphics[width=0.7590\textwidth]{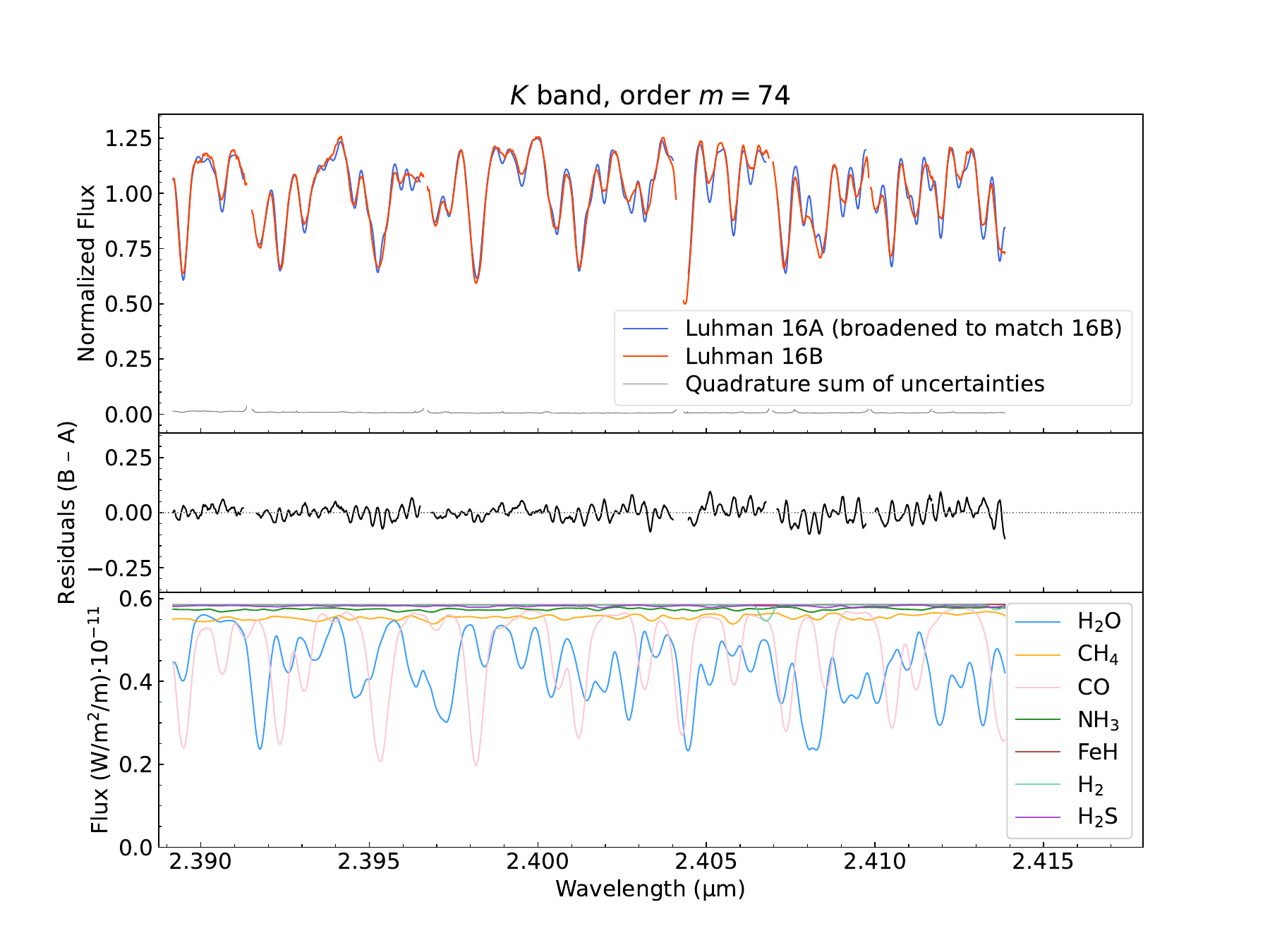} 
    \includegraphics[width=0.7590\textwidth]{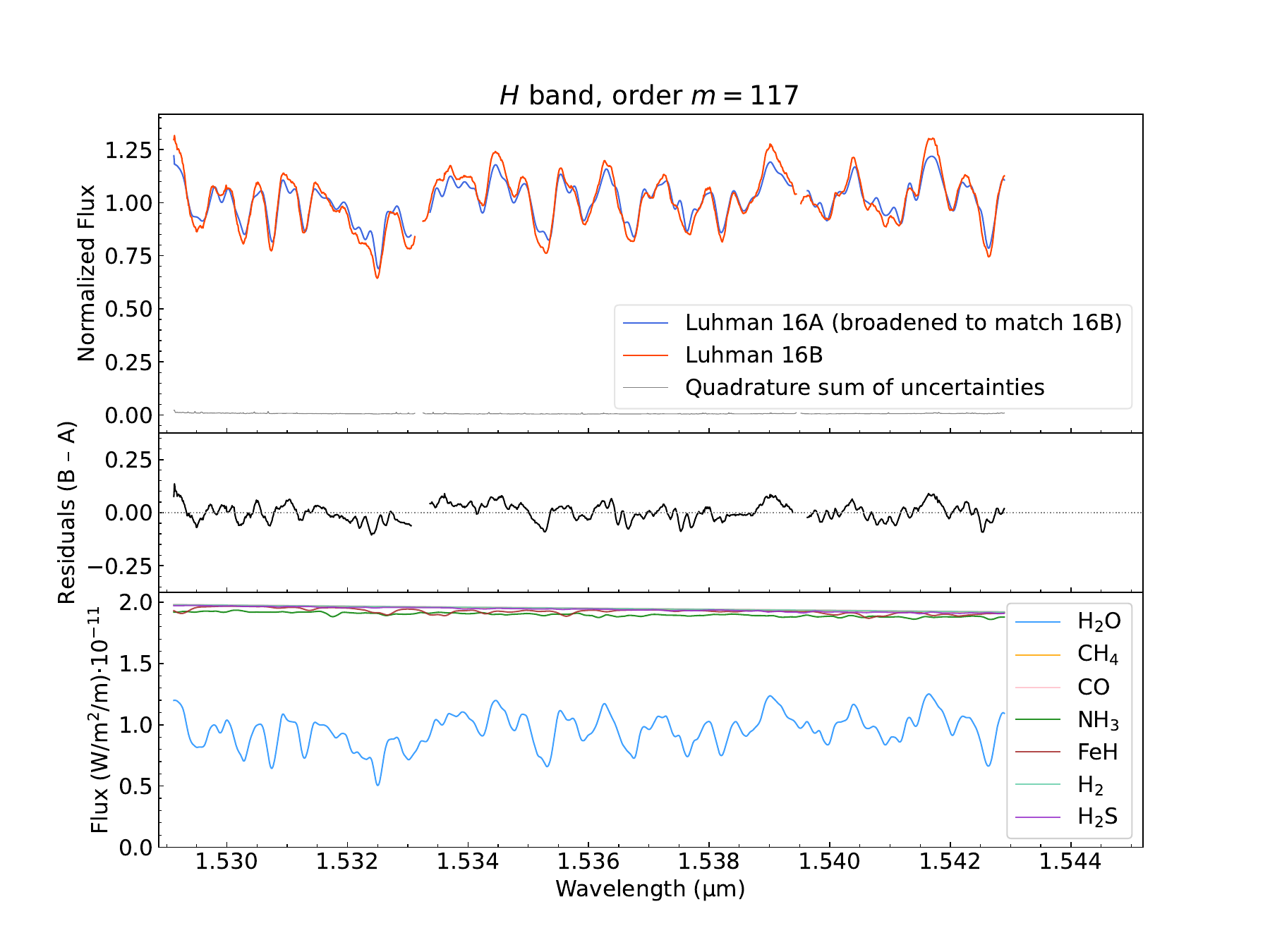} 
    \caption{Comparison of the velocity-shifted and broadened spectra of Luhman 16A in orders $m=74$ (upper set of panels, showing strong CO and H$_2$O absorption) and 117 (lower set of panels, dominated by H$_2$O) to those of Luhman 16B.
    Velocity broadening and shifting were applied to minimize the difference between the two spectra in each order. 
    The top panel in each set shows both spectra and the second panel depicts the difference between them.
    The bottom panel displays single-molecule absorption spectra that include only line opacities from the respective molecule and continuum opacity from collision-induced H$_2$ and He absorption.
    The single-molecule spectra are calculated for the global best-fit parameters for Luhman 16B (Table \ref{tab:bestfitparams}). 
        }
    \label{fig:avsb}
\end{figure*}

Example matched spectra in orders $m=74$ and 117 are plotted in Figure \ref{fig:avsb} along with the predicted absorption from individual molecules in a separate panel. 
\redsep{Differences that are noticeable in the peaks and troughs of the spectra indicate intrinsic differences between the spectra of the two objects.}
The insights for individual molecules from the comparison are discussed in Section \ref{sec:moleculedetection}.

\section{Results and Discussion} \label{sec:discussion} 

\subsection{Determination of physical parameters} \label{sec:bestfitparams} 

\begin{figure*}
        \includegraphics[trim={3.5cm 6.5cm 3.5cm 7cm},clip,width=0.870\textwidth]{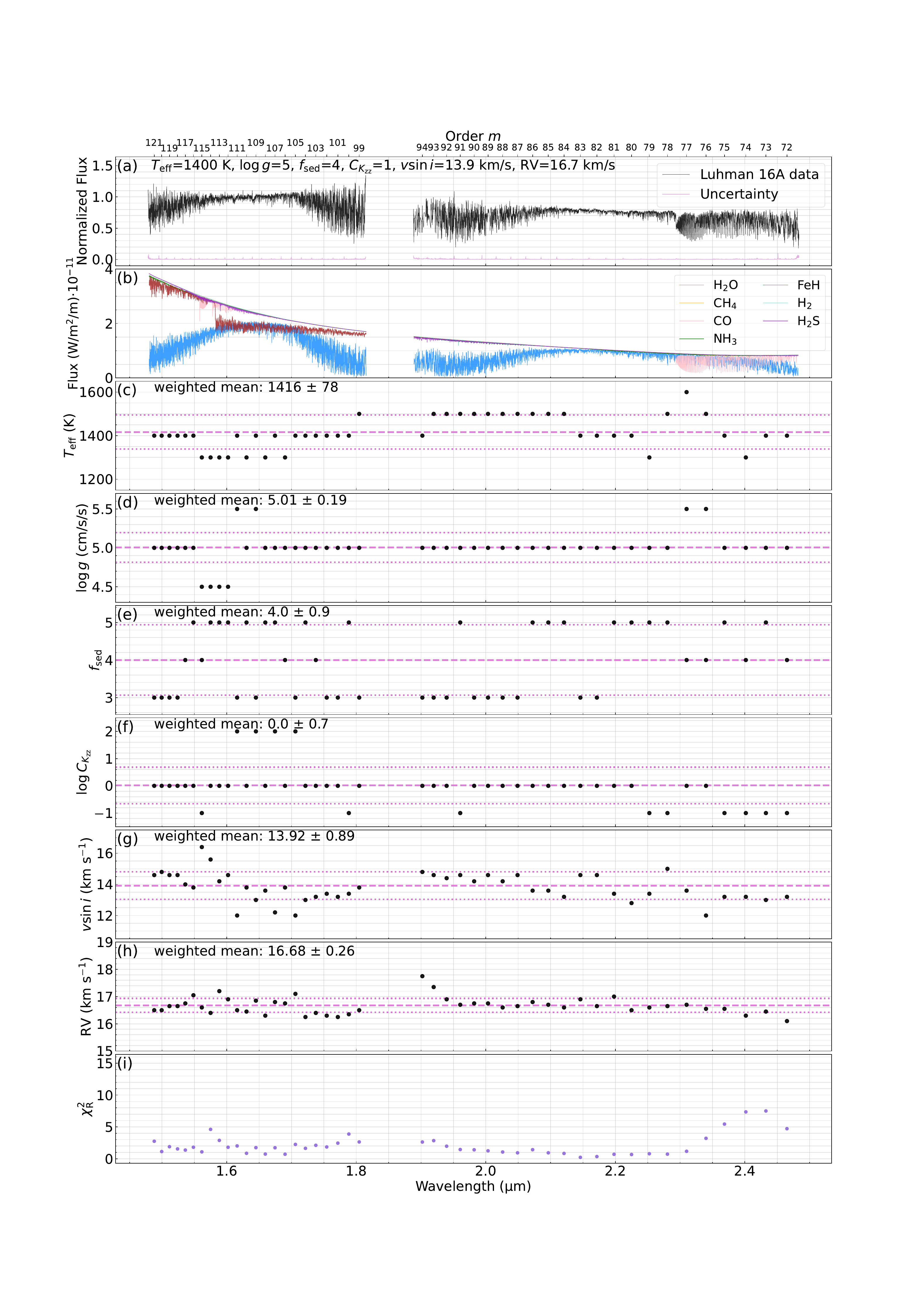} 
    \caption{Results of the parameter estimation from the model fitting for Luhman 16A. 
        Panel (a): IGRINS spectrum normalized to unity over the 1.6--1.7~µm wavelength range 
        after \redsep{masking telluric lines and}
        stitching the orders across each band.
        The IGRINS echelle order numbers ($m$) are shown along the top axis.
        Panel (b): Individual molecular absorption spectra with CIA continuum opacity due to H$_2$ and He included, resampled to match the spectral resolution of IGRINS. 
        The $y$ axis indicates the model flux expected at the surface of the object. Because the individual absorber spectra exclude significant opacity from other molecules, they are used only as an indication where absorption may be expected, rather than as a representation of the actual absorption strength.
        Six panels from (c) to (h): the stellar parameters of the best-fit model for each order. 
        From the top to bottom, the parameters are as follows: $T_{\mathrm{eff}}$, $\log{g}$, $f_\mathrm{sed}$, $\log{C_{K_\mathrm{zz}}}$, $v\sin{i}$, and RV. 
        The weighted average of each parameter is given in the upper left corner.     
        The dashed and dotted horizontal lines indicate the weighted average and 1$\sigma$ range.
        Panel (i): The $\chi^2_{\mathrm{R}}$ value for each order as an indicator of goodness of fit, adjusted as described in Section~\ref{sec:modelfits}. 
    }
    \label{fig:wavvsgbp_A}
\end{figure*}

\begin{figure*}
	\includegraphics[trim={3.5cm 6.5cm 3.5cm 7cm},clip,width=0.870\textwidth]{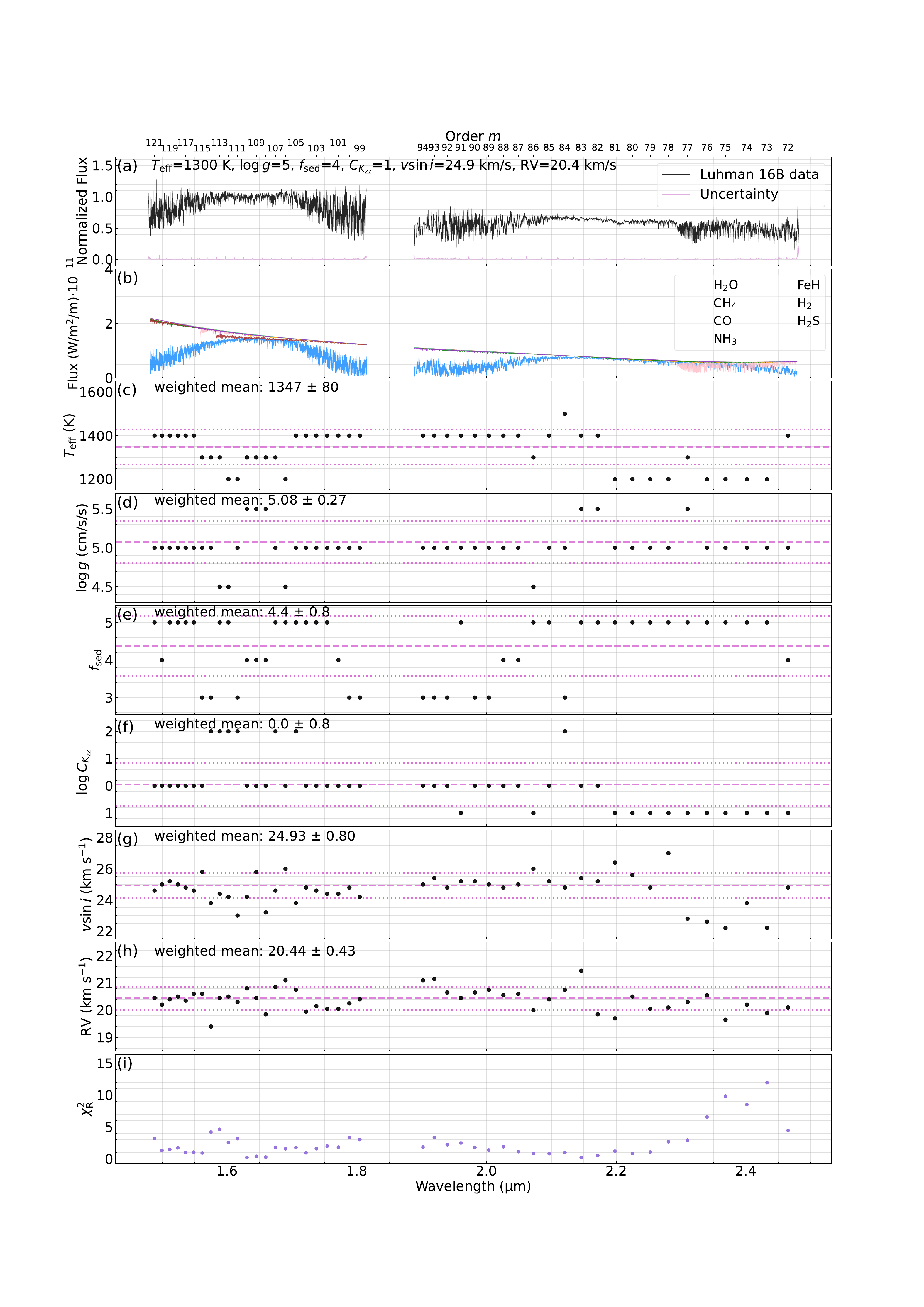} 
    \caption{Same as Figure~\ref{fig:wavvsgbp_A} but for the cooler component, Luhman 16B.
    While most parameters are estimated with similar precision to those of Luhman 16A, the uncertainty of the RV is larger here due to the broader line width caused by the faster rotation.
    The wavelength dependence of each estimated parameter or $\chi^2_{\mathrm{R}}$ shows a similar distribution to that of the hotter counterpart. 
    }
    \label{fig:wavvsgbp_B}
\end{figure*}

We show the best-fit parameters and $\chi^2_{\mathrm{R}}$ values across all orders for Luhman 16A and B, in panels (c) to (i) in Figures \ref{fig:wavvsgbp_A} and \ref{fig:wavvsgbp_B}, respectively. 
The weighted average and 1$\sigma$ standard deviation range for each parameter are indicated by the dashed and dotted horizontal lines in each panel. 
The noticeable trends for each parameter and $\chi^2_{\mathrm{R}}$ are discussed in the following paragraphs. 
For reference, panel (a) shows the order-stitched IGRINS spectrum, and panel (b) shows the model spectra calculated at the global best-fit parameters with individual molecular species only, in addition to continuum opacity from collision-induced absorption (CIA) by molecular hydrogen and helium. 
This helps to identify the dominant line absorbers in each wavelength region as reported in Table \ref{tab:orderswavelengths}. However, because the individual molecule spectra exclude significant opacity from other molecular absorbers, the predicted \redsep{absolute}
strengths are not reliable.

Small $\chi^2_{\mathrm{R}}$ values are observed particularly in wavelength ranges where molecular absorption is weak. 
As we approach the edges of both $H$ and $K$ bands, 
H$_2$O absorption becomes stronger, and $\chi^2_{\mathrm{R}}$ increases. 
Additionally, around 1.57 µm ($m=114$ for both objects and also $m=113$ for Luhman 16B), relatively large $\chi^2_{\mathrm{R}}$ values are shown, which coincide with a bandhead position of FeH. 

The best $\chi^2_{\mathrm{R}}$ values for the individual orders were used in the same manner as Tann22 to estimate the global best-fit parameters for $T_{\mathrm{eff}}$, $\log{g}$, $f_\mathrm{sed}$, $\log{C_{K_\mathrm{zz}}}$, $v\sin{i}$, and RV, of each object.
We took the weighted average of 46 best-fit values across all 23 $H$- and 23 $K$-band orders with the weights calculated as $\exp{( -\frac{1}{2} \chi^2_{\mathrm{R}} )}$. 
The global best-fit parameters are shown in Table \ref{tab:bestfitparams}.
The uncertainty for each parameter is obtained as the unbiased weighted standard deviation of the best-fit values for individual orders. 

\begin{figure*}
    \centering
        \includegraphics[width=0.95\textwidth]{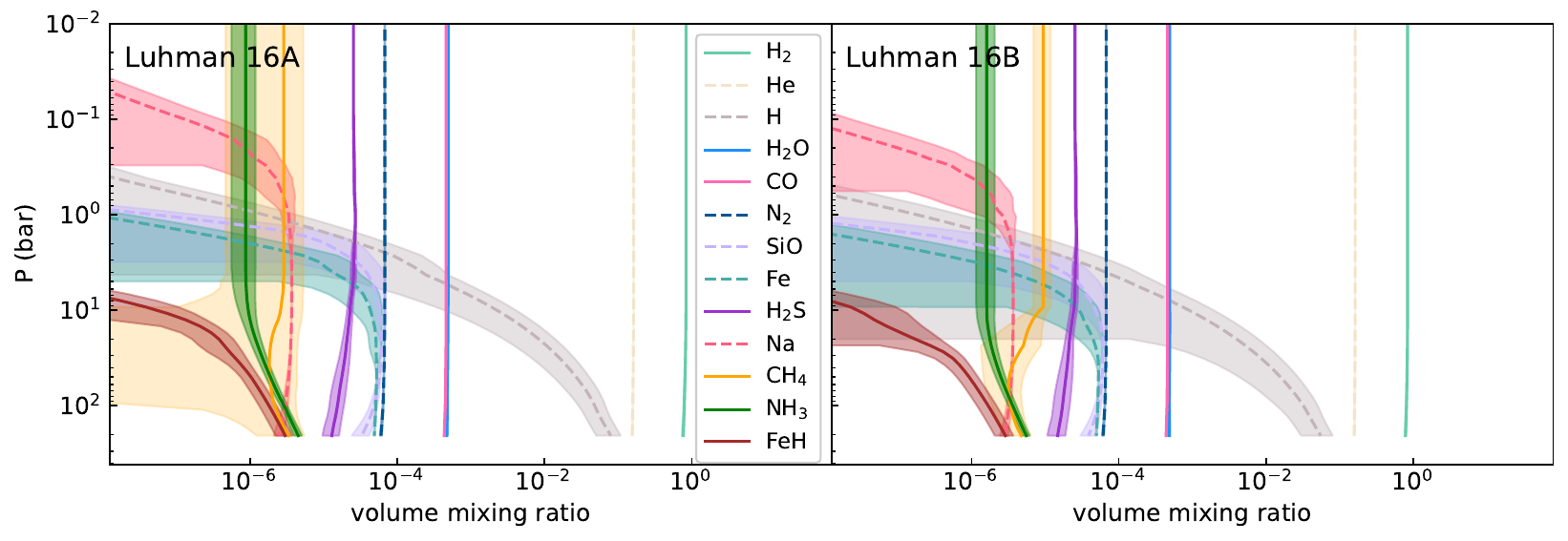} 
    \caption{\redsep{The vertical profiles of the chemical abundances expected for the Luhman 16A (left) and B (right) atmospheres based on the model fitting. 
    The solid line shows the average profile of the best-fit models of 46 individual orders, and the shaded area shows the range of 1$\sigma$ above and below it.
    The plot shows the VMR of the top 13 most abundant chemical species in the $P$=10-bar layer. 
    This illustrates their relative abundances across different pressure levels.}
    }\label{fig:chemicalprofile}
\end{figure*}

The $T_{\mathrm{eff}}$ estimates for Luhman 16A and B have been known to be comparable within the uncertainties (the difference is about 15--30 K) based on the bolometric luminosities~\citep{2014ApJ...790...90F, 2015A&A...581A..73L}. 
Our global best-fit parameters show a difference of \redsep{70}
K with uncertainties of \redsep{80}
K. 
However, 
the spectral fitting in \redsep{nine}
orders, \redsep{$m=73$, 75--78, 80, 81, 86,}
and 111, results in $T_{\mathrm{eff}}$ differences of \redsep{200 K or more}
(compare Figures \ref{fig:wavvsgbp_A} and \ref{fig:wavvsgbp_B}). 
These orders coincide with wavelength regions of CH$_4$ absorption. 
\textcolor{black}{This is to be expected, since the volume mixing ratio (VMR) of CH$_4$ is most sensitive to temperature at the L-to-T transition, compared to other species (Figure~\ref{fig:chemicalprofile}).}
To be sure, CH$_4$ lines are weak in both objects and do not significantly affect the overall shape of the spectra (Section~\ref{sec:ch4}). Nonetheless, our high-SNR data with high spectral resolution capture their $T_{\mathrm{eff}}$ dependence. 
Elsewhere in the $H$ band, the difference in $T_{\mathrm{eff}}$ estimates between the two objects is generally small, suggesting that  $\mathrm{H_{2}O}$ absorption lines in the $H$ band are relatively insensitive to $T_{\mathrm{eff}}$. 
The $K$-band spectra, and in particular their CH$_4$ line content, seem to be more sensitive to $T_{\mathrm{eff}}$ for brown dwarfs at the L/T transition. 

Results for $\log{g}$ are generally consistent across orders, indicating $\log{g} = 5.0$ for both objects. Some orders are fit by a value of 4.5 or 5.5 dex for both objects, which points to systematics in the models. 
The $f_\mathrm{sed}$ parameter was not strongly constrained by our modelling analysis. 

More than half of the orders support nominal values for $K_\mathrm{zz}$. However, the middle of the $H$ band prefers larger $K_\mathrm{zz}$ and the redder part of the $K$ band prefers smaller $K_\mathrm{zz}$. 
The former region is sensitive to FeH features, and the latter to CO features, from which we infer that the volume mixing ratios (VMR) of these molecules are affected by disequilibrium processes. 

The results for $v\sin{i}$ and RV are consistent across all wavelengths.
The reason for the larger dispersion in RV for Luhman 16B is the greater line broadening by faster rotation. 
Our global best-fit value of $v\sin{i}$ for Luhman 16A is slightly smaller than previous estimates 
\redsep{\citep{Crossfield2014, 2022ApJS..258...31K}. 
This is likely due to 
our lower spectral resolution, which is insufficient to measure the $v\sin{i}$ of this object accurately, and so our analysis may underestimate the value.}

\redsep{Our estimates of the stellar parameters are generally consistent with the recent determinations by \citet{2024MNRAS.533.3114C} and more precise. 
\citet{2024MNRAS.533.3114C} used the Sonora Diamondback models, 
which include clouds but do not treat chemical disequilibrium,
so the difference of best-fit values in $f_\mathrm{sed}$ could be attributed to the disequilibrium effects that are incorporated in our custom Sonora Bobcat Alternative B models.}

Combining our $v\sin{i}$ for Luhman 16B of 24.9 $\pm$ 0.8
km\,s$^{-1}$ with its radius of 1.02 $\pm$ 0.07 Jupiter radii ($R_{\mathrm{J}}$) estimated by \citet{2015ApJ...810..158F} 
and its rotation period of 5.28 hr derived photometrically by \citet[][5.25 hr in the follow-on study by \citealt{2024ApJ...965..182F}]{2021ApJ...906...64A},  
we evaluated its spin axis inclination to be $>73$\degree, 
supporting the \citet{2021ApJ...906...64A} estimate of an almost equator-on view. 
This also agrees with the orbital inclination of 
\redmar{$80.15^{+0.22}_{-0.36}$\degree }
\citep{2024AN....34530158B}, 
indicating that the rotation axis of Luhman 16B is aligned with the orbital axis of the binary system.
The 6.94-hr rotation period of Luhman 16A tentatively suggested by \citet{2021ApJ...906...64A} combined with our $v\sin{i}$ (13.9 $\pm$ 0.9 km\,s$^{-1}$) and its radius (1.01 $\pm$ 0.07 $R_{\mathrm{J}}$) 
from \citet{2015ApJ...810..158F} yields a spin axis inclination of $50_{-6}^{+7}$\degree, in modest misalignment with the binary orbit, \redsep{while it should be noted that the suggested underestimation of the $v\sin{i}$ of Luhman 16A may change the conclusion.}

\begin{table*}
	\centering
	\caption{Best-fit photospheric model parameters determined for the IGRINS spectra of Luhman 16A and B. 
	}
	\label{tab:bestfitparams}
	\begin{tabular}{lcc} 
		\hline
		Parameter & Luhman 16A & Luhman 16B \\
		\hline
        Effective temperature, $T_{\mathrm{eff}}/ {\rm K}$& $1416 \pm 78$ & $1347 \pm 80$ \\
        Surface gravity, $\log (g/{\rm cm\,s^{-2}})$ & $5.0 \pm 0.2$ & $5.1 \pm 0.3$ \\
        Sedimentation efficiency parameter, $f_\mathrm{sed}$ & $4.0 \pm 0.9$ & $4.4 \pm 0.8$ \\
        Vertical eddy diffusion coefficient,\tablenotemark{a} ~$\log{C_{K_\mathrm{zz}}}$ & $0.0 \pm 0.7$ & $0.0 \pm 0.8$ \\ 
        Projected rotational velocity, $v\sin{i} / ({\rm km\,s^{-1}})$ & $13.9 \pm 0.9$ & $24.9 \pm 0.8$  \\
        Radial velocity, RV / (km\,s$^{-1}$) & $16.7 \pm 0.3$ & $20.4 \pm 0.4$ \\
		\hline
	\end{tabular}
\tablenotetext{a}{$K_\mathrm{zz} =$ nominal value $\times C_{K_\mathrm{zz}}$, where the units of $K_\mathrm{zz}$ are cm$^{2}$\,s$^{-1}$. See Section~\ref{sec:modelfits} for more details.}
\end{table*}

\subsection{Features of individual molecular species} \label{sec:moleculedetection}

Table \ref{tab:orderswavelengths} lists the molecular line absorbers in each IGRINS order along with the correspondence of the orders and wavelengths.
Most parts of our spectra are dominated by H$_2$O and in some regions CO also exhibits strong features. 
Absorption by FeH, CH$_4$, and NH$_3$ produces more subdued effects in some wavelength regions. 
We also detect absorption by individual lines of H$_2$ at 2.1218~µm, 2.2233~µm, and 2.4066~µm, and H$_2$S at 1.5900~µm. 
Tann22 were the first to detect H$_2$ or H$_2$S in a star-like object outside the solar system: a T6 dwarf. \redsep{\citet{2023ApJ...953..170H} and \citet{lew_etal24} subsequently reported H$_2$S in a T9 dwarf and a Y dwarf, respectively.}
The detections here show that H$_2$ and H$_2$S are seen at high spectral resolution in brown dwarfs as early as L7.5.

\redsep{The vertical profiles of the VMR of the major chemical species expected in the target atmospheres are shown in Figure~\ref{fig:chemicalprofile} based on the model atmospheres with the best-fit parameters for individual orders. 
Species whose profiles are plotted as solid lines here have their spectral features discussed in} 
the following sections.

\subsubsection{Water} \label{sec:h2o}

The most prominent molecular absorber is water (H$_2$O), which contributes most of the line absorption in the $H$ and $K$ bands and forms a pseudo-continuum in many parts. The direct comparison of Luhman 16A and 16B shows the water lines to be nearly identical in strength in the two objects (e.g., Figure~\ref{fig:avsb}), even if at some wavelengths the absorption is slightly stronger in component A (e.g., the 2.408--2.414~µm region in Figure~\ref{fig:avsb}, top) while at others is it slightly stronger in B (e.g., the 1.529--1.543~µm region in Figure~\ref{fig:avsb}, bottom). 

Tann22 found the ExoMol/POKAZATEL water line list from \citet{2018MNRAS.480.2597P} to be the most reliable in fitting the IGRINS spectrum of a T6 dwarf with their Bobcat Alternative A model. This is also the water line list used in the present Bobcat Alternative B model, and we confirm that the fits to the spectra are satisfactory. Nevertheless, we do find that in some spectral regions dominated by water, the residuals between the observed spectra and the best-fit model exhibit P Cygni profile-like features, indicative of slight ($<$1 nm) inaccuracies of some water line wavelengths. A comparison to the T6 dwarf spectrum and modelling in Tann22 confirms that they are present for that object, too. A discussion of these residuals in the context of other potential new features or contributions from isotopologues is presented in Section \ref{sec:waterwiggle}.

\subsubsection{Carbon monoxide} \label{sec:co}

\begin{figure*}
	\includegraphics[width=0.7590\textwidth]{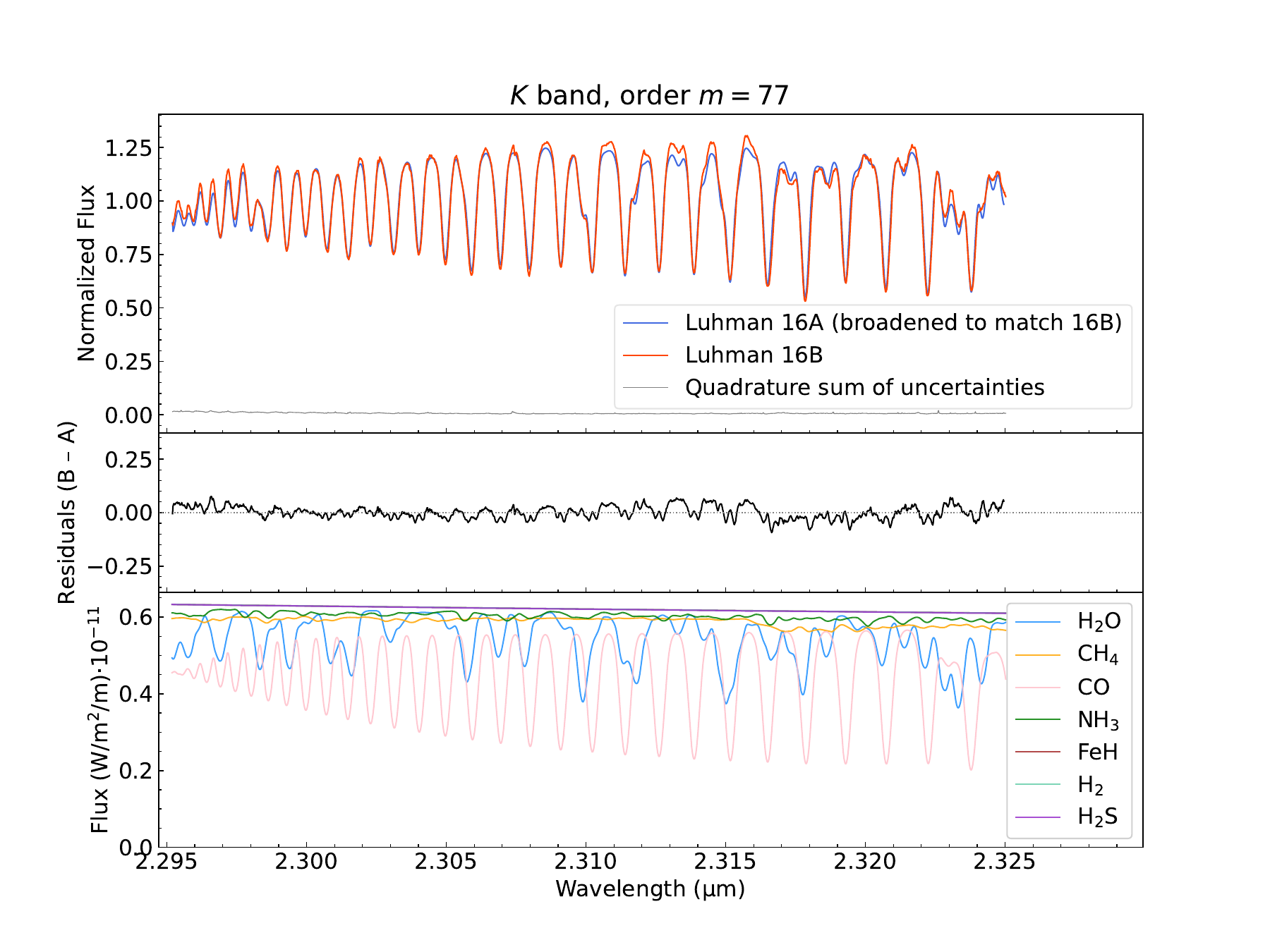} 
	\includegraphics[width=0.7590\textwidth]{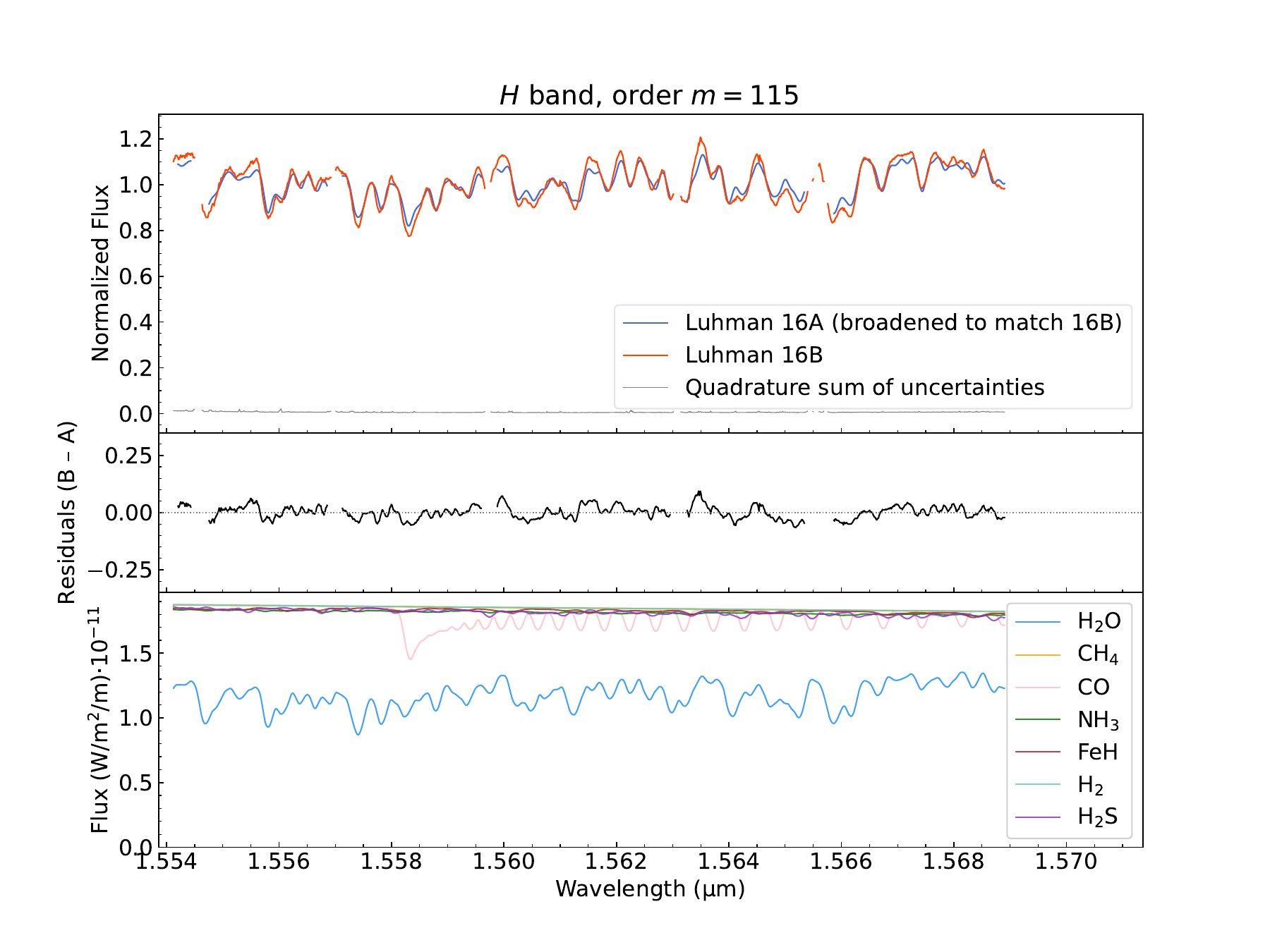} 
    \caption{Comparison of Luhman 16A and B as in Figure~\ref{fig:avsb} but for orders $m=77$ ($K$ band) and $m=115$ ($H$ band)
    where the CO absorption lines are prominent.
    }
    \label{fig:co}
\end{figure*}

The second most prominent absorber in the spectra is carbon monoxide (CO), with three bandheads at 1.558~µm, 1.578~µm, and 2.294~µm. 
The absorption lines in the $K$ band are particularly strong, comparable to those of $\mathrm{H_{2}O}$, and depress the continuum in orders $m=$~72--78.
The comparison between Luhman 16A and B in Figure \ref{fig:avsb} (top panel) and in Figure \ref{fig:co} shows that the difference in the normalized depths of CO lines between the two components is almost negligible. 
CO absorption strength in Luhman 16B is only $\lesssim$5\% stronger at the 2.3~µm bandhead. 
The $H$-band CO lines show no significant difference between the two objects.

This follows expectations from the Sonora Bobcat models. The CO mixing ratio decreases with decreasing temperature from the L dwarfs into the T dwarfs as CH$_4$ consumes more carbon. However, this is offset by a predicted increase in the CO absorption line strength with decreasing $T_{\mathrm{eff}}$ at the L/T transition. The trade-off between these two effects results in the normalized depth of the CO lines being insensitive to the $T_{\mathrm{eff}}$ across the L/T transition.

\subsubsection{Iron hydride} \label{sec:feh}

\begin{figure*}
	\includegraphics[width=0.905\textwidth]{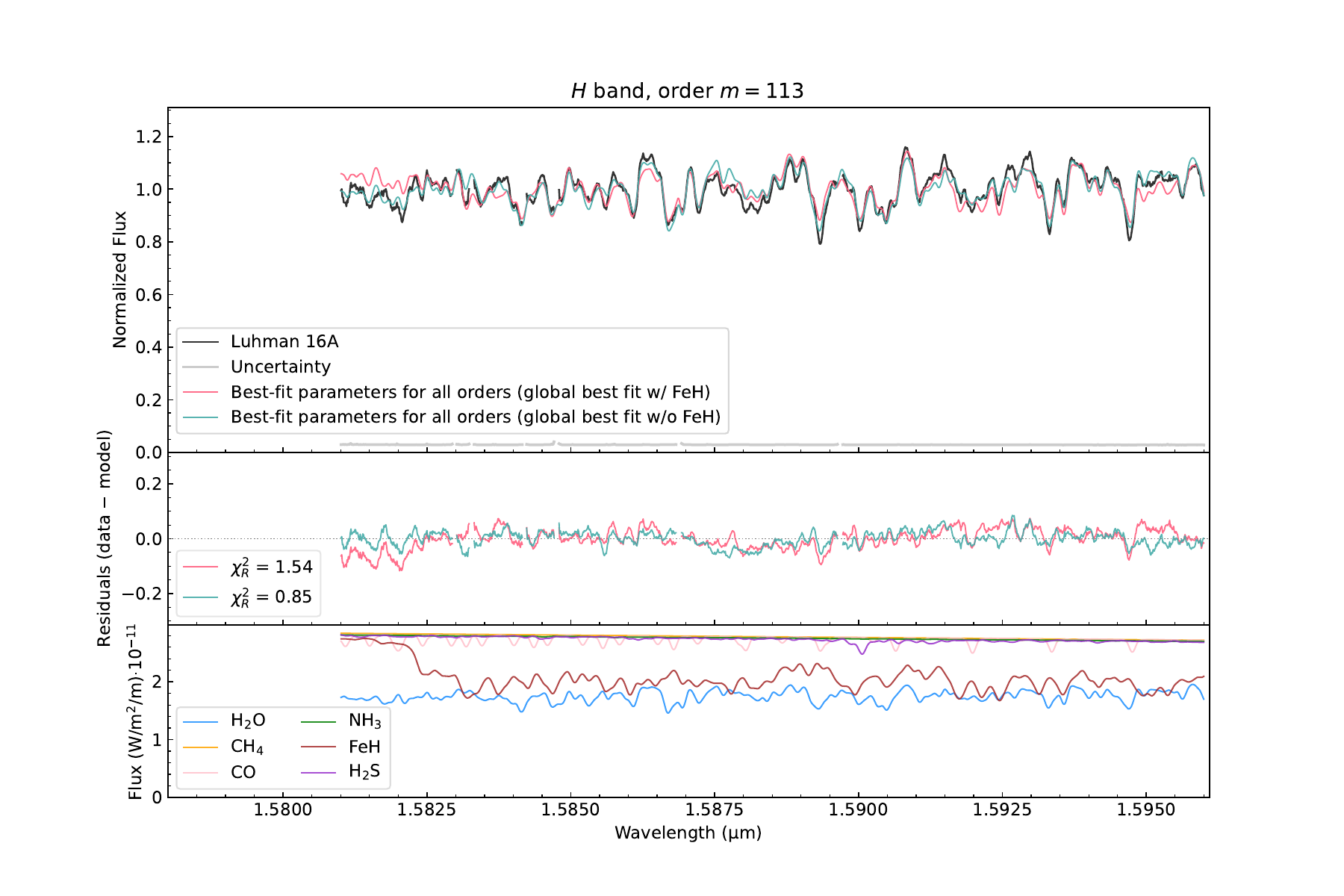} 
    \caption{\textcolor{black}{A test of the required FeH opacity contribution in photospheric models of Luhman 16A. Order $m=113$ contains the FeH bandhead at 1.582~µm, which is expected to nearly double the atmospheric line opacity dominated by water.}
    \redsep{The data (in black) and its uncertainty (in gray) are plotted in the top panel, along with} 
    \textcolor{black}{two global best-fit model photospheres ($T_{\mathrm{eff}} = 1400$ K, $\log{g} = 5.0$, $f_\mathrm{sed} = 4$, $C_{K_\mathrm{zz}} = 1$): one including FeH opacity (red spectrum) and one without FeH opacity (green spectrum).} 
    The second panel shows the residuals between the data and the respective models, where the legend indicates the $\chi^2_{\mathrm{R}}$ values.
    The bottom panel shows the models for individual molecular species and CIA by H$_2$ and He. 
    They are calculated with the global best-fit parameters for Luhman 16A. 
    }
    \label{fig:feh}
\end{figure*}

Iron hydride (FeH) has absorption lines in the wavelength range of 1.58--1.72 µm (orders $m=$~104--114) in the $H$ band. 
Our model, which assumes solar chemical composition, predicts a strong contribution from FeH 
in the spectrum of Luhman 16A (Figure~\ref{fig:wavvsgbp_A}b) and a more subdued one in Luhman 16B (Figure~\ref{fig:wavvsgbp_B}b). 
\textcolor{black}{The strongest effect of FeH absorption should be detectable at the 1.582~µm bandhead, where its contribution should jump from zero to nearly half of the atmospheric line opacity in Luhman 16A (Figure~\ref{fig:wavvsgbp_A}b and bottom panel of Figure~\ref{fig:feh}). However, no FeH bandhead is visible in our spectra of either Luhman 16 component (Figures~\ref{fig:wavvsgbp_A}a and \ref{fig:wavvsgbp_B}b). A comparison of the best-fit models to Luhman 16A with and without FeH shows that the addition of FeH is not favoured with the global best-fit parameters (Figure~\ref{fig:feh}, middle panel).}

\begin{figure*}
	\includegraphics[width=0.905\textwidth]{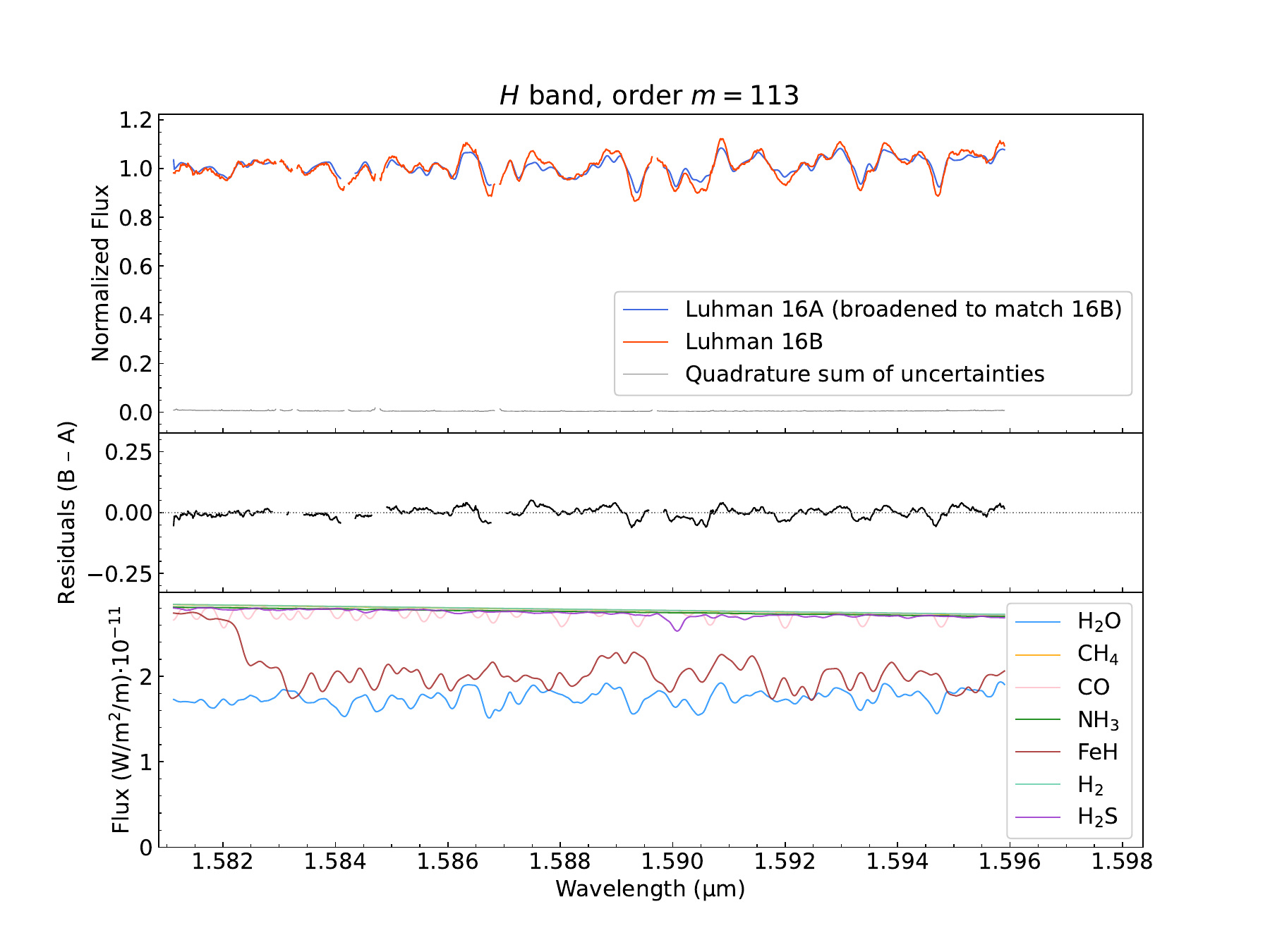} 
    \caption{Comparison of Luhgman 16A and B as in Figure~\ref{fig:avsb}, but for order $m=113$, where \textcolor{black}{the FeH absorption should be most prominent.} 
    \redsep{The single-molecule spectra in the bottom panel are based on Luhman 16A's global best-fit parameters.}
    There is no difference between the spectra of the two objects \redsep{even at the 1.582~µm onset of the FeH bandhead,}
    contrary to expectations from the global best-fit photospheres (Section~\ref{sec:feh}). 
    }
    \label{fig:feh_avsb}
\end{figure*}

There is also no appreciable difference between Luhman 16A and 16B \textcolor{black}{at the FeH bandhead}
(Figure~\ref{fig:feh_avsb}), \textcolor{black}{despite expectations based on their slightly different spectral types and effective temperatures.} 
\redsep{The FeH-only model calculated with the global best-fit parameters predicts
stronger FeH absorption in the spectrum of Luhman 16A compared to B (panel (b) in Figures \ref{fig:wavvsgbp_A} and \ref{fig:wavvsgbp_B}), while such is not seen in the difference of the two spectra.}

\redsep{The local best-fit photospheric models favour lower $T_{\mathrm{eff}}$ and higher $K_\mathrm{zz}$ than those in the global best-fit models in the FeH-affected orders ($m=109$--114) for both Luhman 16A and B (panels c and f of Figures~\ref{fig:wavvsgbp_A} and \ref{fig:wavvsgbp_B}).}
Both lower $T_{\mathrm{eff}}$ and stronger vertical mixing result in a cooler temperature-pressure profile for the entire atmosphere, which would agree with  \redsep{weaker}
FeH absorption. We thus suspect that the model may not fully capture the efficiency of the iron rainout~\citep[e.g.,][]{2002ApJ...571L.151B, 2003ApJ...582.1066C} that removes iron from the atmosphere at temperatures below $\sim$1500--2000 K \citep{2010ApJ...716.1060V}. 
\textcolor{black}{The Sonora Bobcat model already has the FeH line strengths \citep[from][]{2010AJ....140..919H} scaled down by a factor of 1/3 to account for discrepancies with early-L dwarf spectra \citep{2021ApJ...920...85M}. We find that an even further decrease of the overall FeH opacity would be needed---amounting to a
near-complete condensation of iron at the L/T transition---to} explain the observation that neither Luhman 16A nor B shows 
signs of FeH absorption.

\subsubsection{Methane} \label{sec:ch4}

\begin{figure*}
	\includegraphics[width=0.7770\textwidth]{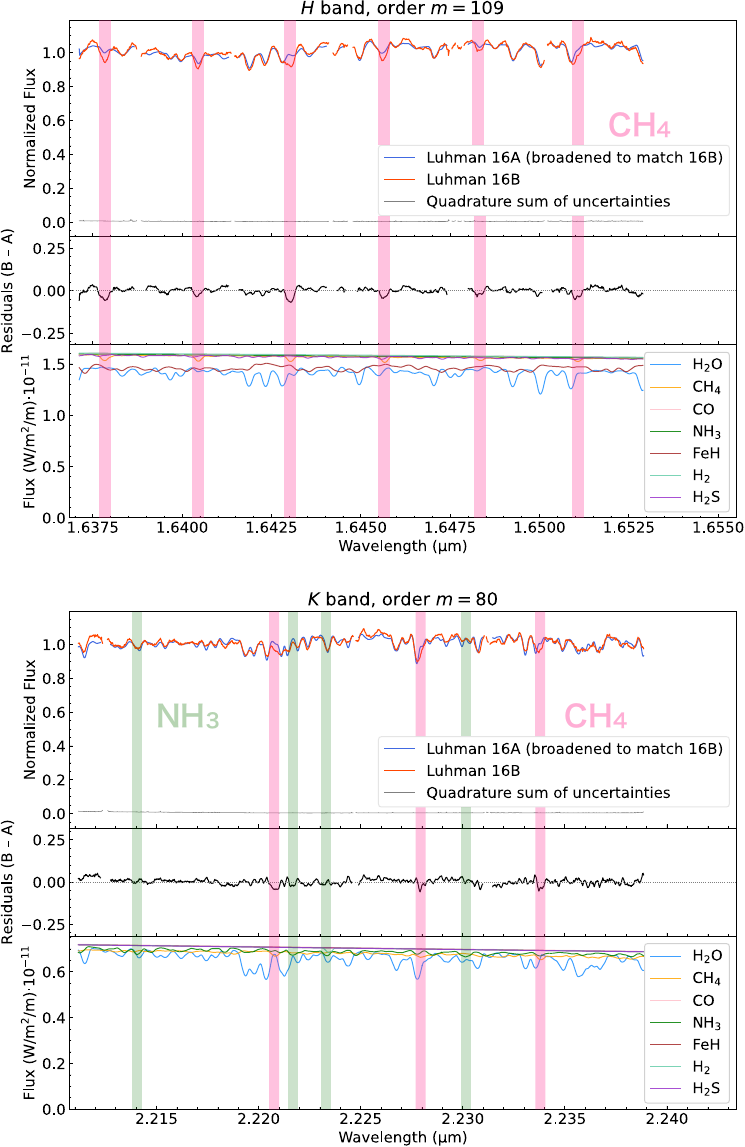} 
    \caption{
    Methane absorption as a diagnostic of the L/T transition. The figure follows
    the same format as Figure~\ref{fig:avsb} but for orders $m=109$ and 80 in the upper and lower panels, respectively.
    The pink-shaded regions highlight CH$_4$ absorption and the green-shaded regions highlight NH$_3$. 
    CH$_4$ shows deeper absorption in the spectra of Luhman 16B than those of Luhman 16A, while \redsep{differences in the depths of}
    NH$_3$ lines (or any of the other molecular absorbers) are 
    less pronounced.
    The model spectra of individual molecules in the bottom panels are calculated based on the global best-fit parameters for Luhman 16B. 
    }
    \label{fig:ch4lines}
\end{figure*}

Methane (CH$_4$) lines are not strongly expressed in the spectra of our L7.5 and T0.5 targets. 
Nonetheless, we do detect significant differences between the spectra of Luhman 16A and B in the wavelength regions where CH$_4$ absorption is expected. 
Around 1.64 µm (orders $m=109,$ 110) and 2.20 µm (orders $m=78$--81), comb-like residuals appear in the difference between the spectra of Luhman 16B and 16A. 
Figure \ref{fig:ch4lines} shows that these features are aligned with the wavelengths of CH$_4$ lines, even though most parts of the nearby pseudo-continuum are determined by $\mathrm{H_{2}O}$.
Methane absorption is thus weak but diagnostic at the L/T transition: it is not the strongest absorber at the 1.6~µm and 2.2~µm methane bandheads even in the T0.5 dwarf, but shows the strongest sensitivity to temperature (Section~\ref{sec:bestfitparams}).

\subsubsection{Ammonia} \label{sec:nh3}

Although ammonia (NH$_3$) is characteristic of the much colder Y spectral type and is known to exhibit prominent features at lower temperatures \citep[e.g.,][]{delorme_etal08, cushing_etal11}, it has also been detected in a high-SNR IGRINS spectrum of a T6 dwarf by Tann22.

\begin{figure*}
	\includegraphics[width=0.95\textwidth]{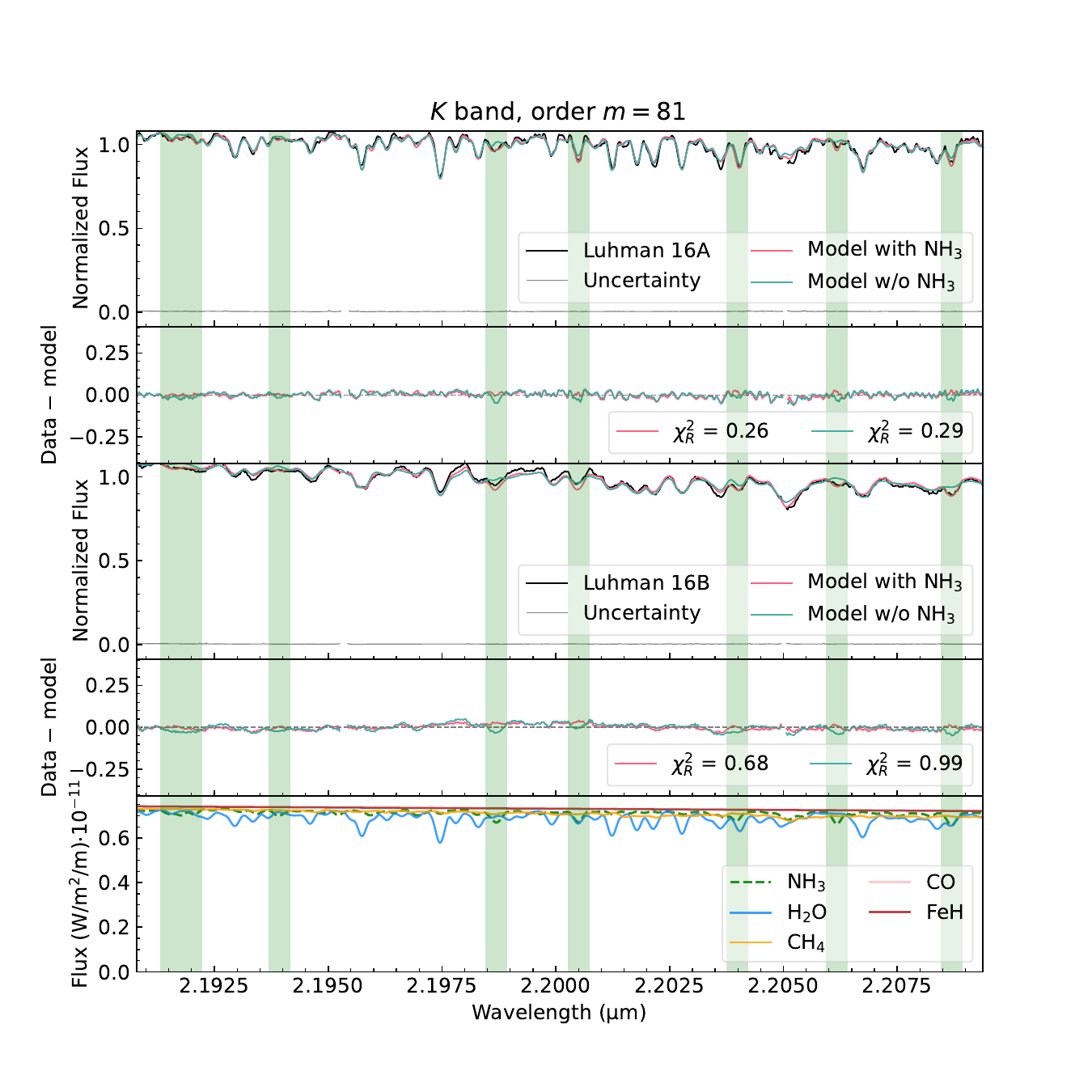} 
    \caption{
    The $m=81$ IGRINS order of the Luhman 16A and B spectra, \redsep{demonstrating the NH$_3$ line detection.}
    The \redsep{data and uncertainties}
    for Luhman 16A and B \redsep{spectra, their}
    local best-fit models \redsep{with or without NH$_3$ opacity,}
     and the residuals between the data and models \redsep{(the legend indicates the $\chi^2_{\mathrm{R}}$ of each model fit)}
     are shown in the top four panels. 
    The bottom panel shows the spectra for each molecule alone with the CIA continuum, using the global (rather than local) best-fit parameters for Luhman 16B. 
    \redsep{The wavelength regions where the contribution of NH$_3$ is expected to be strongest are shaded in green.}
    }
    \label{fig:nh3lines}
\end{figure*}

\redsep{In our spectra,}
there is only a very limited region in orders $m=78$--82 where the NH$_3$ opacity exceeds that of other molecules. 
The NH$_3$ opacity is small and \redsep{most of the}
lines are \redsep{veiled}
by H$_2$O lines. 
\redsep{Figure \ref{fig:nh3lines} shows the 
the IGRINS order in which the NH$_3$ lines can be most easily discerned.
Models with NH$_3$ opacity reproduce the spectra for both objects slightly better, especially for Luhman 16B, than models without NH$_3$ opacity. Notably, the strongest residuals from models lacking NH$_3$ opacity are at the wavelengths where NH$_3$ absorption is expected. The effect is more strongly noticeable for Luhman 16B, but the same wavelengths also show NH$_3$-related residuals in Luhman 16A.
This makes Luhman 16A one of the warmest ($1416\pm78$K; Table~\ref{tab:bestfitparams}) atmospheres with robust NH$_3$ detection along with DENIS J025503.3$-$470049 \citep[L9;][]{2024A&A...688A.116D}.}

\subsubsection{Molecular hydrogen} \label{sec:h2}

Both Luhman 16A (L7.5) and 16B (T0.5) spectra show unambiguous detections of molecular hydrogen (H$_2$) lines. This is only the second report of H$_2$ detection in substellar objects outside the solar system, following the discovery of the 1--0 S(1) quadrupole transition 2.1218~µm H$_2$ line in the IGRINS spectrum of a T6 dwarf by Tann22.

\begin{figure*}
	\includegraphics[width=0.905\textwidth]{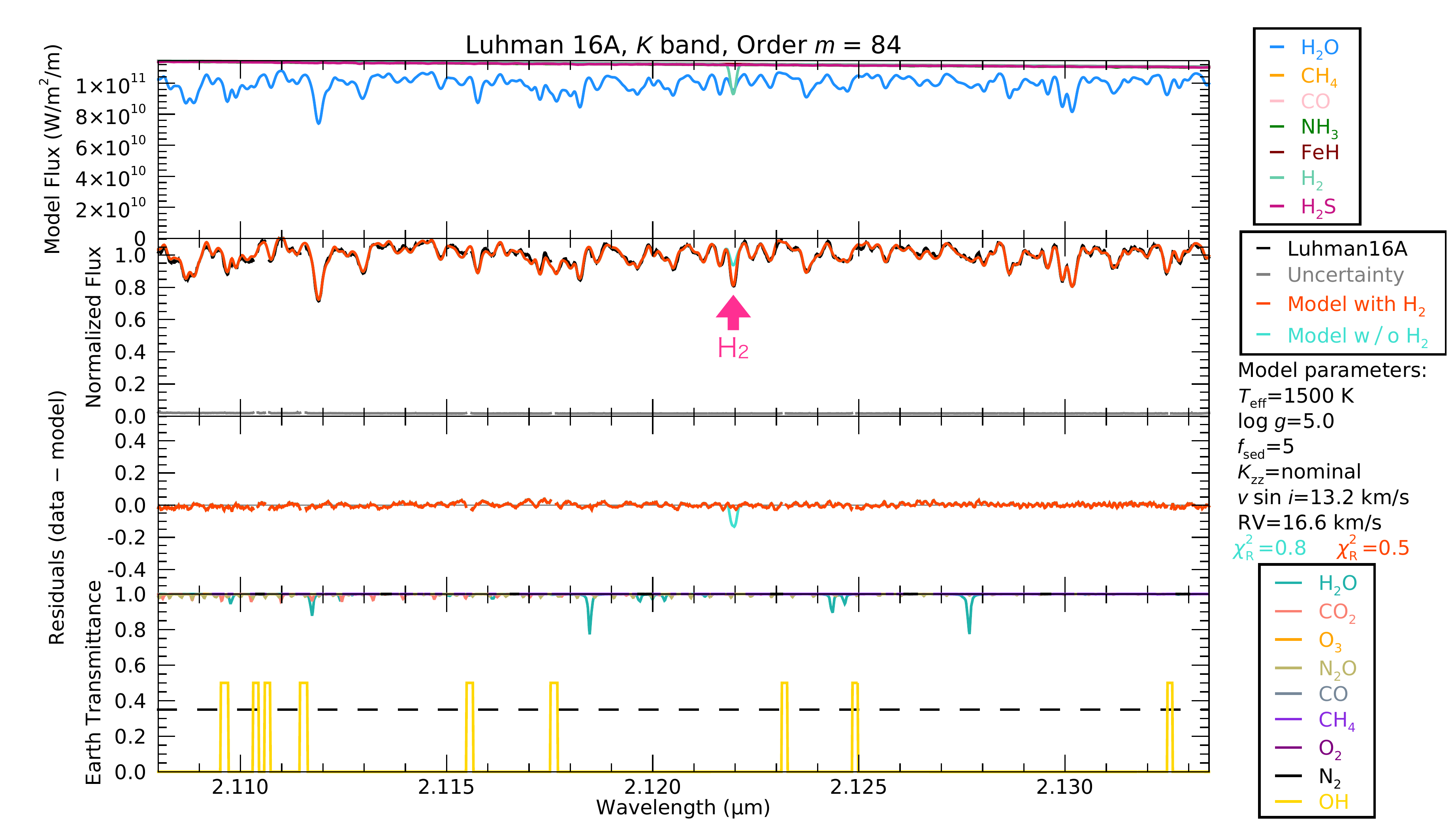} 
	\includegraphics[width=0.905\textwidth]{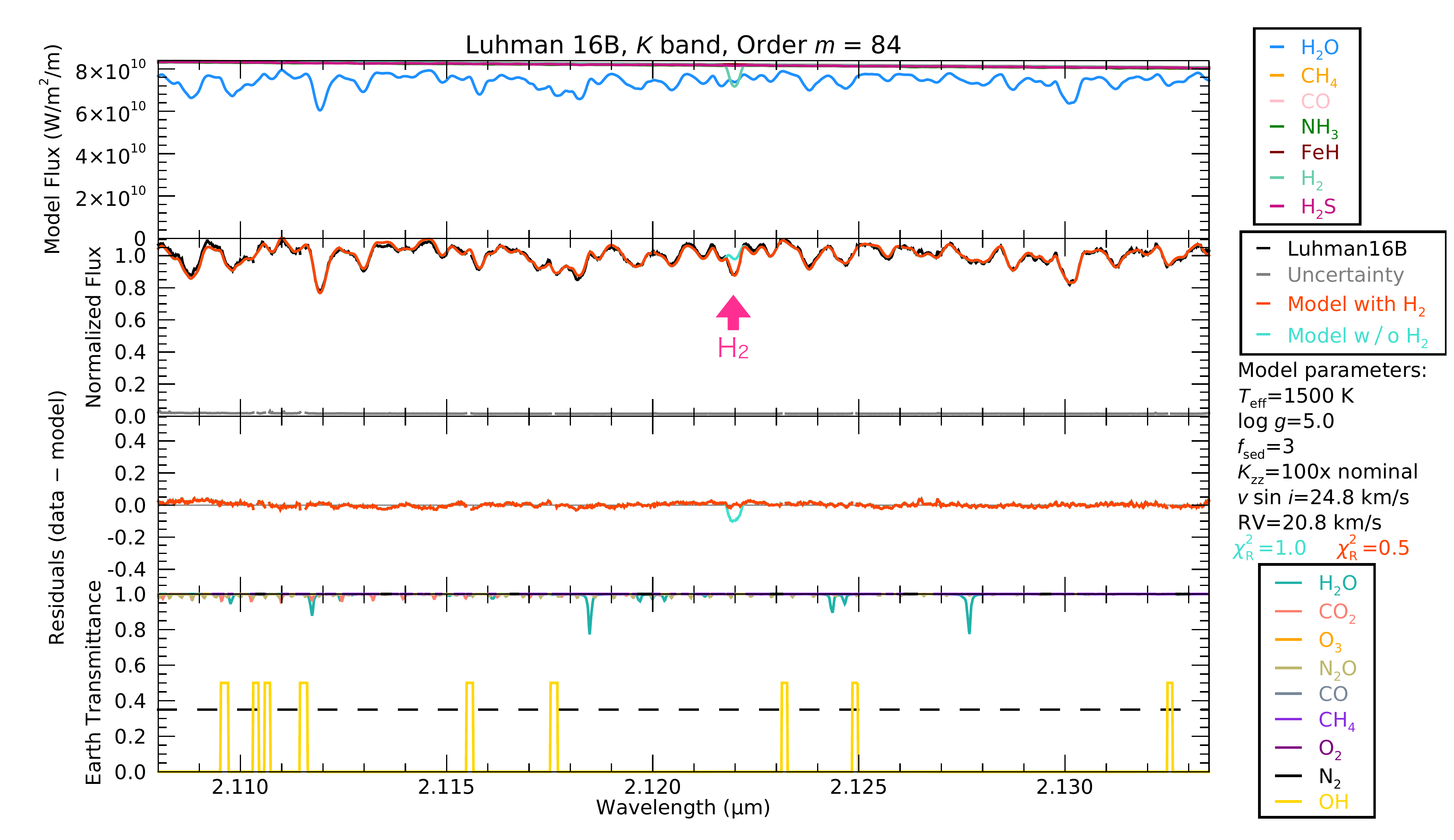} 
    \caption{
    Detection of the H$_2$ 2.1218~µm quadrupole line in the spectra of Luhman 16A (top set of four panels) and 16B (bottom set of four panels).
    The top panel of each set of four shows the contribution of each molecular species to the model, which indicates that this spectral region is shaped mostly by $\mathrm{H_{2}O}$ absorption features. 
    The IGRINS spectrum and its best-fit models with and without H$_2$ line opacity are plotted in the second panel from the top.
    Their differences (data $-$ model) are plotted in the third panel from the top. The differences show a clearly visible residual at the H$_2$ wavelength.
    The bottom panel shows the PSG Earth’s transmittance to help assess the telluric lines in our spectra. 
    The positions of OH emission lines are indicated with the yellow boxes, although box height is not proportional to line strength.
    The horizontal dashed line indicates the 35 percent transmittance threshold used to mask out the telluric-affected region (see Section \ref{sec:reductions}), although no such contamination is found in this echelle order.
    The physical parameters of the best-fit model for each order are written in the margins between the legend boxes on the right side of the panels. 
    }
    \label{fig:h2detection}
\end{figure*}

Figure \ref{fig:h2detection} shows the spectral fit for order $m=84$ where the 2.1218~µm H$_2$ line appears in both objects.
It illustrates that the addition of H$_2$ in the model makes the residuals flatter and improves the $\chi^2_{\mathrm{R}}$ value of the fit.
While this is the most significant detection, we also found indications of other absorption lines by H$_2$. 
The 1--0 S(0) line at 2.2233 µm ($m=80$, Figure~\ref{fig:atlasA}) is relatively shallow, but thanks to the excellent reproduction of the surrounding H$_2$O lines in the model, we found that the addition of H$_2$ to the model improves the $\chi^2_{\mathrm{R}}$ value. 
The 1--0 Q(1) line at 2.4066 µm ($m=74$, Figure~\ref{fig:atlasB} is in a wavelength range crowded with telluric lines, but both visually and in terms of the $\chi^2_{\mathrm{R}}$, the fit also improves with the addition of H$_2$.
Some other promising H$_2$ lines either fall into the gap between the $K$ and $H$ bands (e.g., 1.8358, 1.8919 µm) or are obscured by stronger water absorption (e.g., 1.7480 µm in order $m=102$, 1.9576 µm in order $m=91$, 2.0337 µm in order $m=88$) and are not detected.

Despite being the most abundant gas in substellar atmospheres, H$_2$ is difficult to detect because it has no dipole transitions because of its symmetric structure. 
We detect quadrupole transitions, which are weak, and prior to Tann22 have only been seen in solar system giant planets. 
The observation of the quadrupole lines of H$_2$ was predicted by \citet{1938ApJ....87..428H} and identified in the spectra of Uranus and Neptune by \citet{1952ApJ...115..337H}. 
These lines have been used to infer hydrogen abundances, ortho-para ratios, atmospheric vertical structures and cloud distributions
\citep[e.g.,][]{2017Icar..286..223F, 1992ApJ...393..357C, 1966PASJ...18..339T}. 
More recently, \citet{2021KPCB...37..230A} and \citet{2021AzAJ...16a..83A} have detected H$_2$ quadrupole lines at optical wavelengths in the atmospheres of all four solar-system giant planets 
and use these to estimate the physical parameters of the absorption layer and the content of molecular hydrogen.

\begin{figure*}
    \includegraphics[width=0.7590\textwidth]{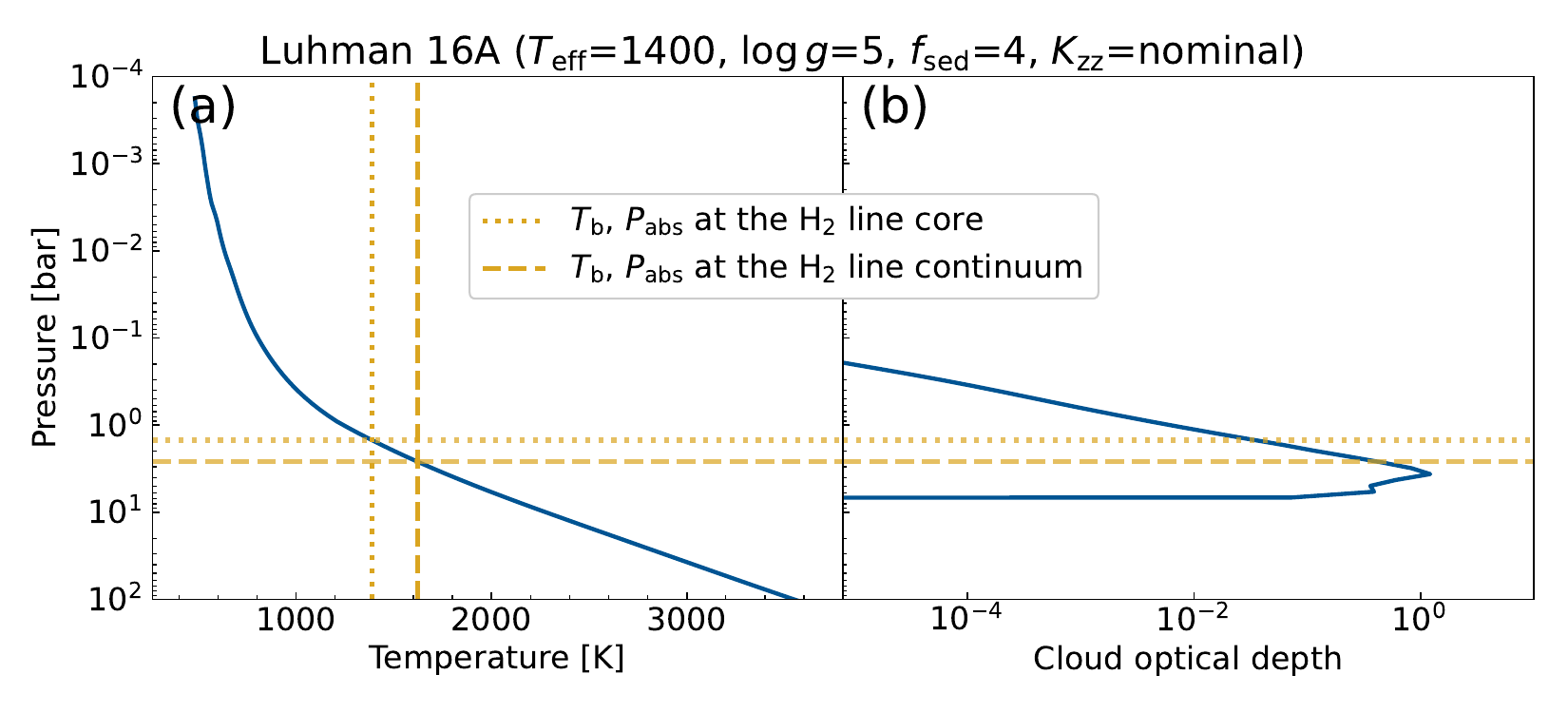} 
    \includegraphics[width=0.7590\textwidth]{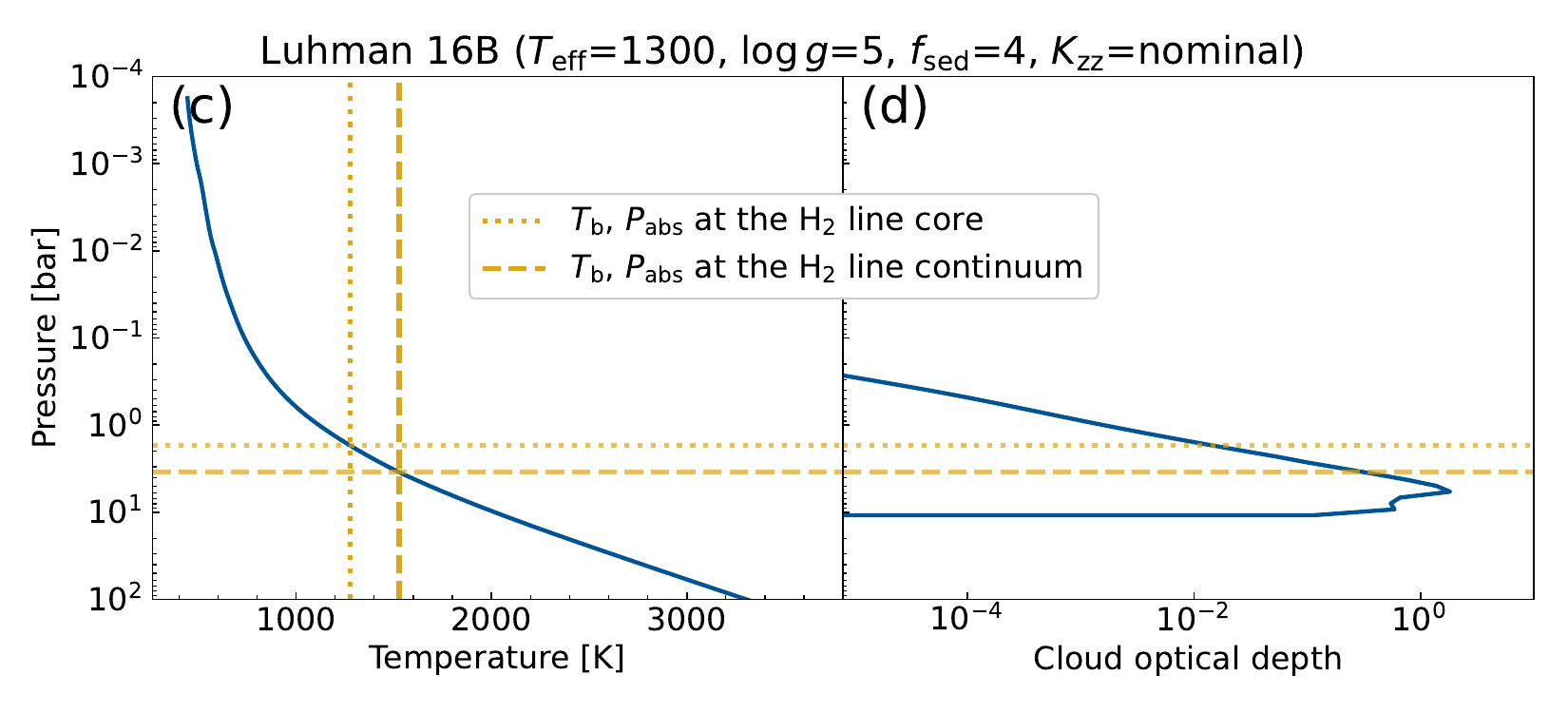} 
    \caption{
    Temperature-pressure ($T$-$P$) profiles (panels a and c) and cloud optical depth (panels b and d) \redsep{for Luhman 16A (top) and B (bottom) in the global best-fit atmospheric models.}
    The yellow vertical dotted and dashed lines in panels (a) and (c) indicate the brightness temperatures ($T_{\mathrm{b}}$) corresponding to the 2.1218~µm centre of the H$_2$ line and the nearby pseudo-continuum, respectively.
    The horizontal dotted and dashed yellow lines indicate the pressure at which the $T$-$P$ profile and the corresponding vertical lines intersect.  \textcolor{black}{The cloud optical depth peaks at a comparable or higher pressure than the pressure of the absorbing layer corresponding to the centre of the H$_2$ line, consistent with our inference of a cloud-free upper atmosphere, from which we detect H$_2$ absorption.} 
    }
    \label{fig:h2_cloudheights} 
\end{figure*}

As done in Tann22, we estimate the H$_2$ column density ($N_{\mathrm{H_2}}$\redsep{)}
and dust extinction in the H$_2$ absorbing layer by comparing to ISM-like conditions. We assume that the top of the absorbing layer is set by the brightness temperature ($T_{\mathrm{b}}$) of the continuum around the H$_2$ line in the best-fit model spectra. The temperature-pressure ($T-P$) profile of the atmospheric model can then be used to infer the pressure of the absorbing layer (Figure \ref{fig:h2_cloudheights}): $P_{\mathrm{abs}}=2.6$ bar for Luhman 16A and $P_{\mathrm{abs}}=3.5$ bar for Luhman 16B. 
We then computed the local number density of H$_2$ molecules for each atmospheric layer with the ideal gas law and a model-informed VMR of H$_2$, which is approximately constant of 0.84 in all atmospheric layers of interest in both objects.
Since the physical distance between each layer can be calculated from the density and pressure difference, we integrated the H$_2$ number density from $P_{\mathrm{abs}}$ to the top of the atmosphere to obtain column densities of $N_{\mathrm{H_2}} \sim 4.7 \times 10^{25}$~cm$^{-2}$ for Luhman 16A and $N_{\mathrm{H_2}} \sim 6.5 \times 10^{25}$~cm$^{-2}$ for Luhman 16B. 
As done in Tann22, we adopt the $K$-band extinction law of \citet{1989ApJ...345..245C} and assume that all the hydrogen is comprised of H$_2$ molecules, which yields H$_2$ column density per magnitude of ISM-like $K$-band extinction of $N_{\mathrm{H_2}}/A_{K} = 0.97 \times 10^{22}$~cm$^{-2}$~mag$^{-1}$.
If the gas-to-dust ratios in our targets were ISM-like, these $N_{\mathrm{H_2}}$ values would correspond to  $K$-band extinctions of $A_K \sim$~4900~mag and $\sim$~6700~mag for Luhman 16A and B, respectively.
The fact that the H$_2$ lines are clearly detected indicates that the actual dust concentration is several thousand times smaller than in the ISM. 
That is, the dust has separated from the gas to form clouds that reside mostly below the H$_2$ gas absorption layer.

A comparison to the pressure level of the cloud base in our adopted model confirms that it is at a similar or higher pressure than $P_{\mathrm{abs}}$
(Figure \ref{fig:h2_cloudheights}).
This is consistent with the expected sedimentation of dust clouds below the photosphere at $T_{\mathrm{eff}}<1400$ K \citep[e.g.,][]{2021ApJ...914..124B, 2008ApJ...689.1327S, 2010ApJ...723L.117M}.

\subsubsection{Hydrogen sulfide} \label{sec:h2s}

\begin{figure*}
	\includegraphics[width=0.905\textwidth]{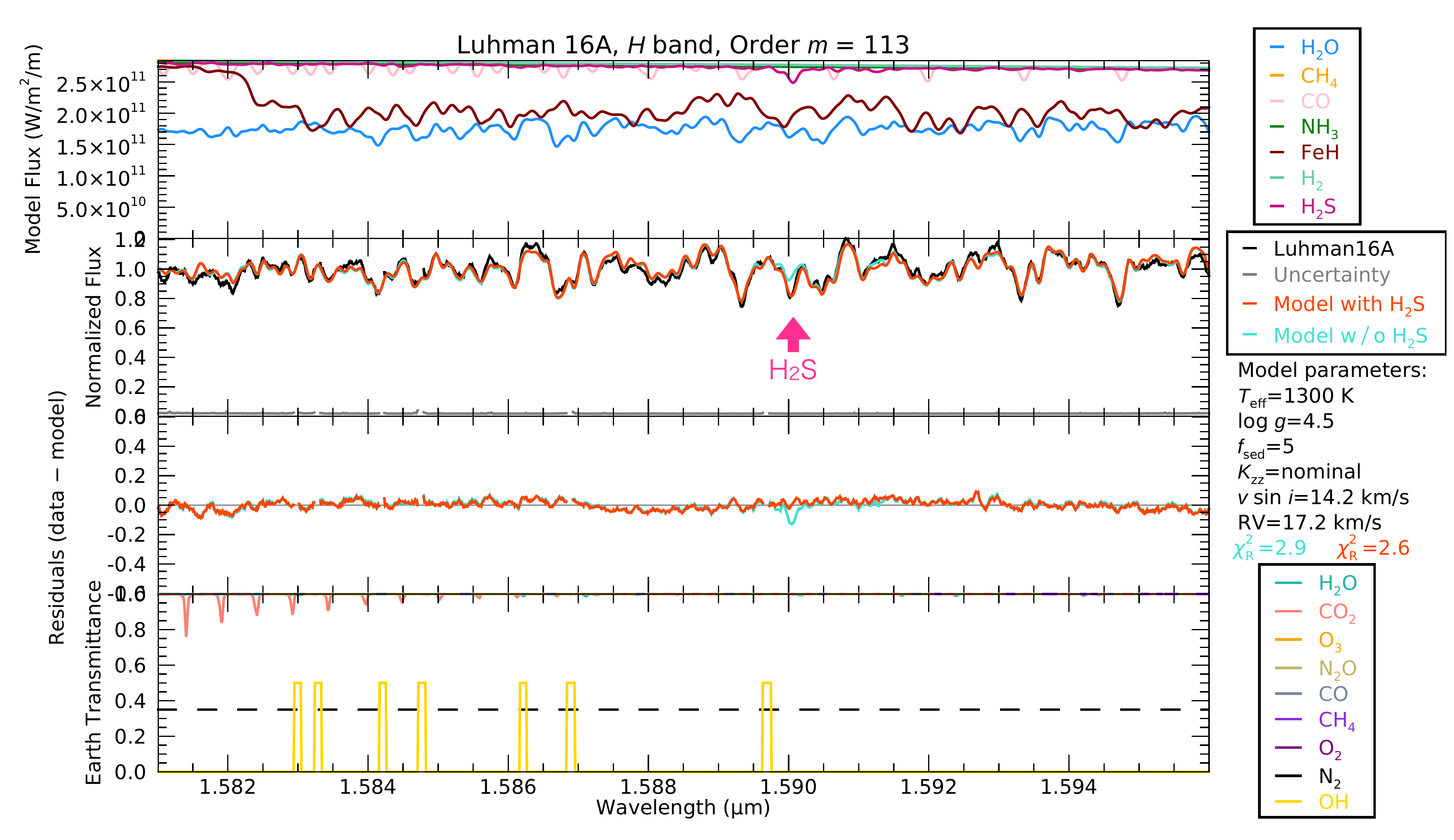} 
	\includegraphics[width=0.905\textwidth]{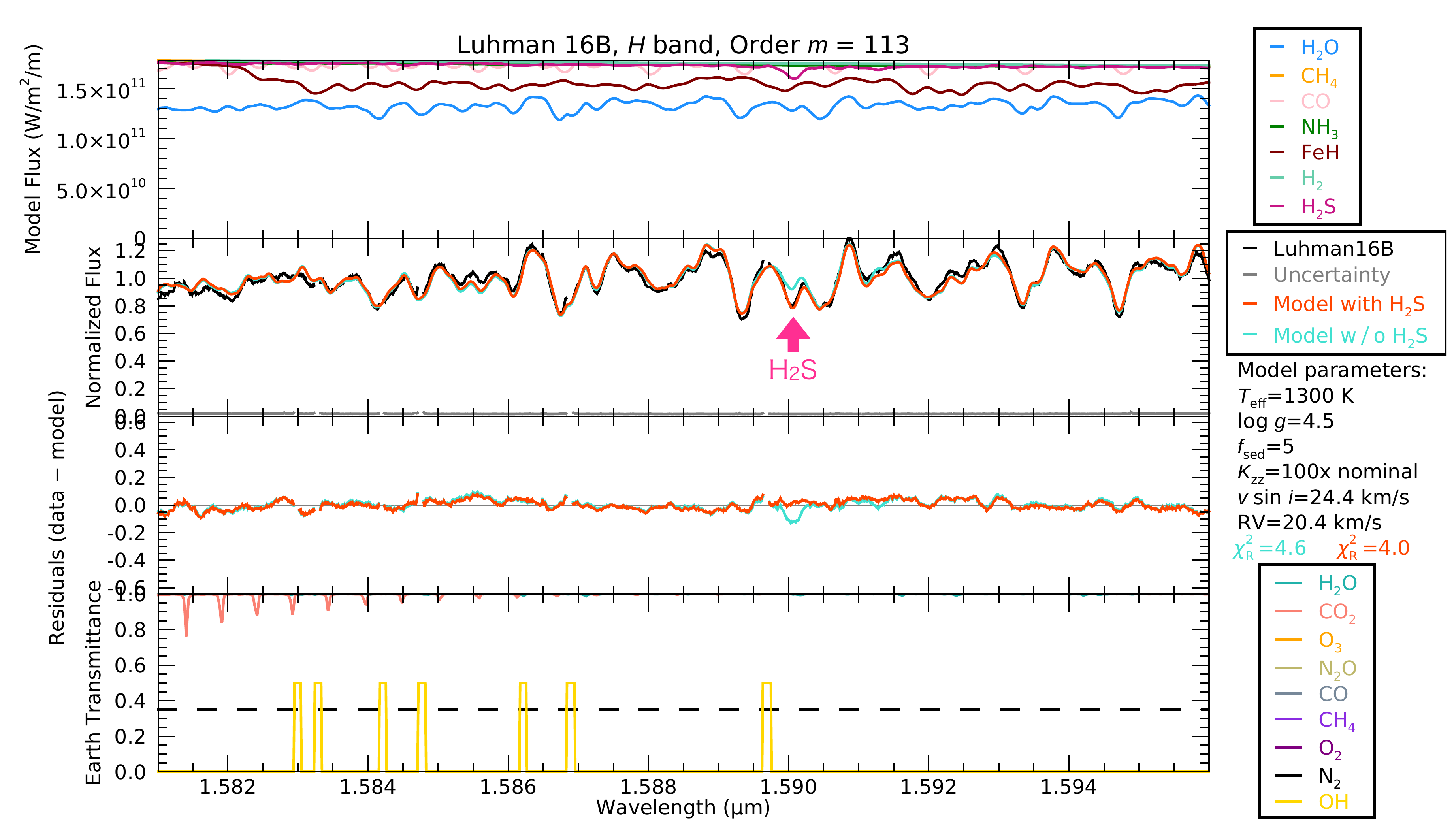} 
    \caption{
     Detection of the H$_2$S 1.5900~µm line in IGRINS order $m=113$ the spectra of Luhman 16A (top set of four panels) and 16B (bottom set of four panels).
     The layout is the same as in Figure \ref{fig:h2detection}.
    }
    \label{fig:h2sdetection}
\end{figure*}

As with the detection of H$_2$, both Luhman 16A and 16B spectra show unambiguous detections of hydrogen sulfide (H$_2$S). The H$_2$S 1.5900~µm line in order $m=113$ is shown in Figure \ref{fig:h2sdetection}.
We also tentatively found H$_2$S contributions at wavelengths of 1.5687, 1.5700, 1.6163, 1.6165, 1.6265, 1.6435, 1.9094, 1.9815 and 2.4634~µm. Since the best-fit models in the spectral atlas in Appendix \ref{sec:atlas} do not include H$_2$S opacity unless otherwise stated, the small dips at those wavelengths seen in the (data $-$ model) residuals can be attributed to H$_2$S. 
The 2.1084 µm feature predicted by \citet{2000ApJ...541..374S} was not detected in our target or in the T6 spectrum of Tann22, nor was it reproduced in our model.

The only previous bona fide near-infrared detection of H$_2$S is by Tann22 in the IGRINS spectrum of a T6 dwarf. A follow-on near-infrared detection is reported tentatively in a T9 dwarf by \citet{2023ApJ...953..170H} \redsep{and in a Y dwarf by \citet{lew_etal24}.}
H$_2$S is fully expected in our atmospheric model at temperatures $\lesssim$2000 K. However, its detection in low-dispersion near-infrared spectra is challenging because of surrounding H$_2$O absorption. At $>$3 µm thermal infrared wavelengths, H$_2$S has a more prominent absorption signature between 3.5--4.5 µm, which has been detected in JWST NIRSpec spectra: in the Y dwarf WISEPA J182831.08+265037.8 \citep{lew_etal24}\redsep{.}

H$_2$S is the dominant sulfur-bearing species in most substellar atmospheres~\citep[e.g.,][]{2006ApJ...648.1181V}. 
It is relatively under-studied, as it is challenging to detect outside the solar system. Its importance has been discussed in the literature in two cases. On the one hand, sulfur abundance can be a differentiator among planet-formation scenarios, distinguishing between pebble- or planetesimal accretion formation modes \citep{2023ApJ...952L..18C}. Since sulfur compounds are more volatile than carbon- or oxygen-bearing compounds, the C/S and O/S ratios in the atmosphere of a planet formed via pebble accretion are predicted to be higher than those of a planet formed via planetesimal accretion, for which the C/S and O/S ratios should be similar to the host star. On the other hand, H$_2$S is also important as a precursor of SO$_2$, which is an indicator molecule for photochemical reactions \citep[e.g.,][]{Tsai_etal23, Sing_etal24}. Photochemistry is a contributing factor in setting the temperature structure and cloud composition in the upper atmospheres of irradiated planets.

\begin{figure*}
    \includegraphics[width=0.755\textwidth]{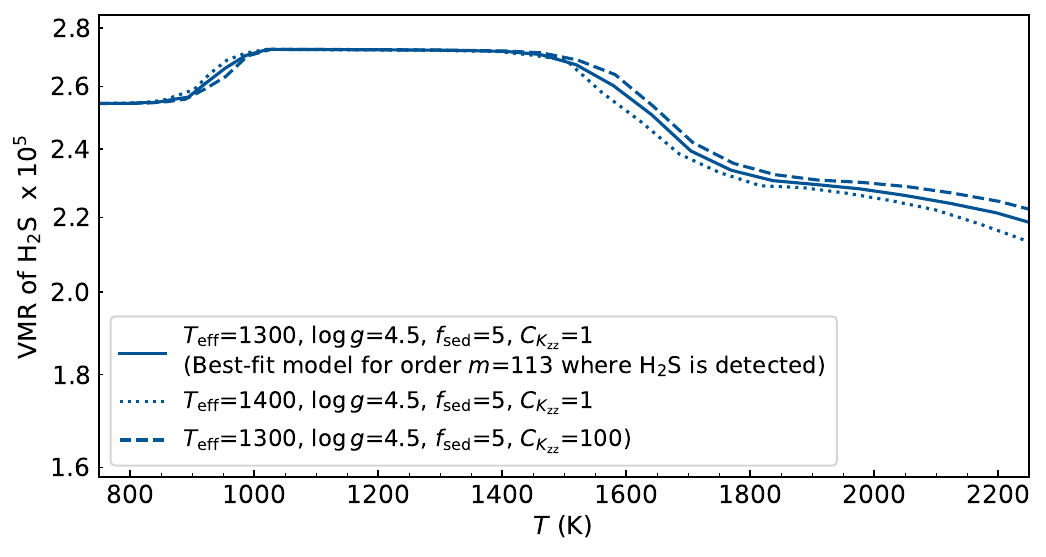} 
    \caption{
    Dependence of the H$_2$S VMR (solid line) on temperatures in Luhman 16A based on the best-fit model atmosphere for order $m=113$, where the H$_2$S line is detected.
    The y-axis is on a logarithmic scale.
    The VMR changes only by a factor of 1.2 between 1000 and 2000 K, throughout the photosphere. The dotted and dashed lines show that this is also the case for models with different $T_{\mathrm{eff}}$ and $K_\mathrm{zz}$, respectively. This illustrates that the strength of  \redsep{the H$_2$S line}
    is sensitive mostly to the S abundance, and not to \redsep{temperature}
    or disequilibrium effects. 
    } 
    \label{fig:vmr_h2s}
\end{figure*}

The high-SNR H$_2$S detections in L7.5 and T0.5 dwarfs reported here, in a T6 in Tann22, and in a Y dwarf in \citet{lew_etal24} can be used as a baseline for future analyses of substellar atmospheres. Importantly, the atmospheric concentration of H$_2$S is relatively insensitive to temperature or convection.
Our model shows that the VMR of H$_2$S is constant to within 20\% at $2.5 \times 10^{-5}$ for both objects in atmospheric layers with temperatures in 1000--2000 K range (Figure \ref{fig:vmr_h2s}). 
We calculate the H$_2$S column densities $N_{\mathrm{H_{2}S}}$ from the flux around the 1.5900~µm H$_2$S line, as we did for H$_2$ (Section~\ref{sec:h2}).
The computed $N_{\mathrm{H_{2}S}}$ values are $\sim 1.6 \times 10^{21}$~cm$^{-2}$ and $\sim 2.8 \times 10^{21}$~cm$^{-2}$ for Luhman 16A and B, respectively.
H$_2$S is known to be one of the main components of the clouds of Uranus \citep{2018NatAs...2..420I} and Neptune \citep{2019Icar..321..550I}.
Without any anticipated H$_2$S condensation at the $\gtrsim$1300~K temperatures of Luhman 16A or B, our $N_{\mathrm{H_{2}S}}$ estimates are respectively 6--270 times and 10--470 times higher than those for the solar system ice giants. 

\subsubsection{Hydrogen fluoride} \label{sec:hf}
\begin{figure*}
	\includegraphics[width=0.905\textwidth]{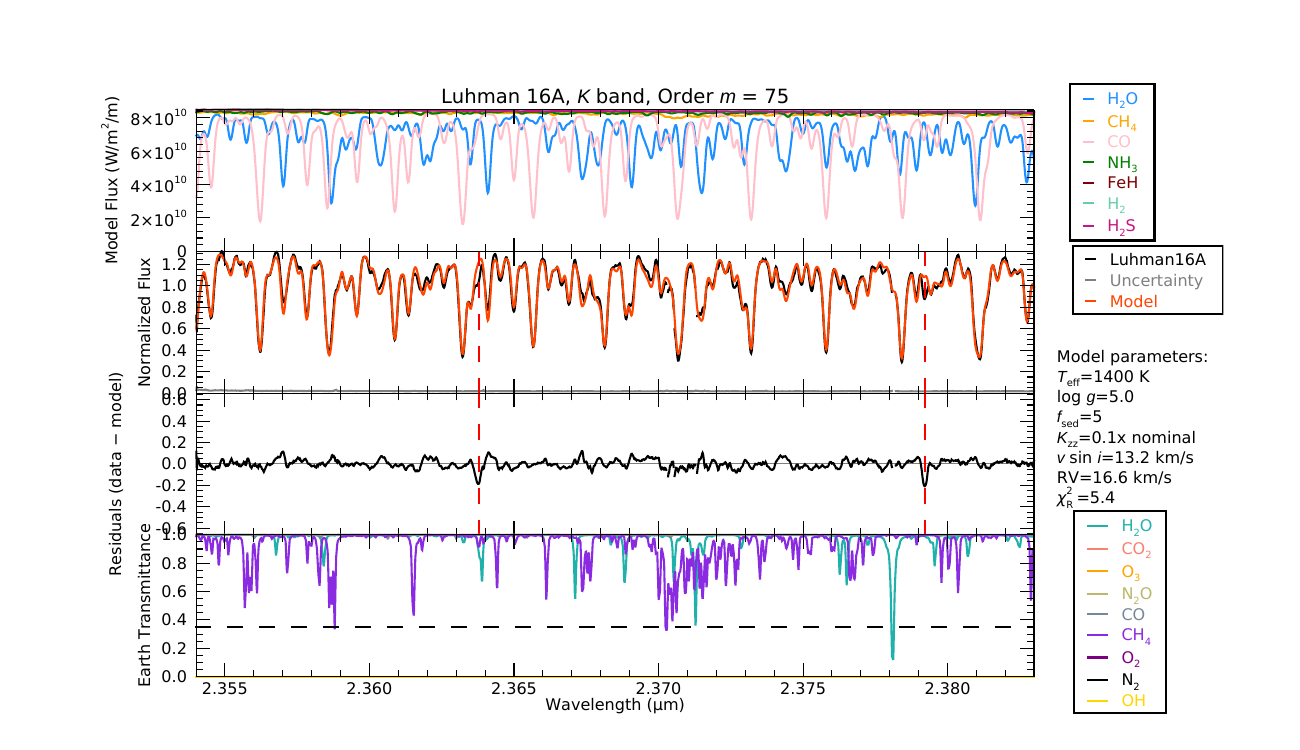} 
	\includegraphics[width=0.905\textwidth]{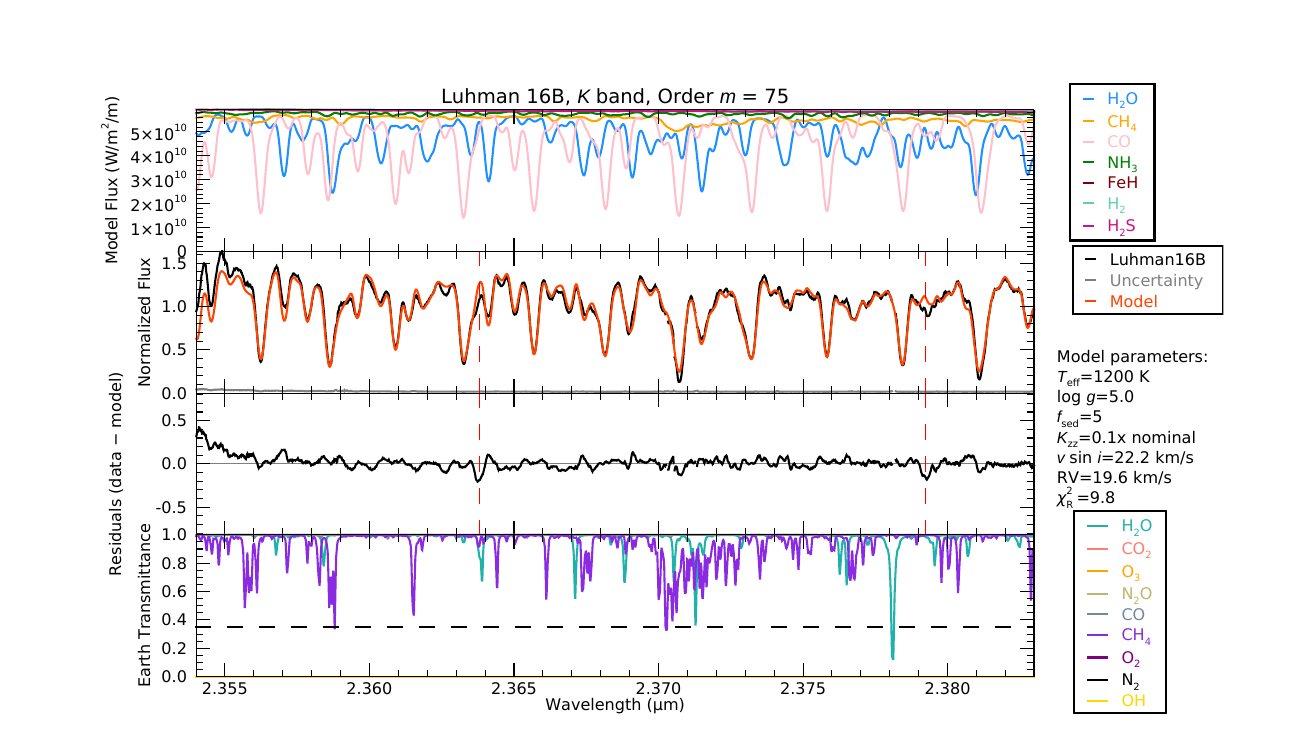}
    \caption{
    Example detections of HF lines at 2.3636~µm and 2.37905~µm in order $m=75$ of both Luhman 16A (top four panels) and B (bottom four panels) spectra.
    The layout is the same as Figure \ref{fig:h2detection}.
    These are the most prominent examples, but we detect multiple HF lines in orders 72--76 (Table~\ref{tab:hflines}).
    Our models do not include HF opacity. The red vertical dashed lines indicate where the HF lines are predicted \citep{2015JQSRT.151..133C}, coinciding with the positions of the noticeable dips in the residual plot (data $-$ model).
    }
    \label{fig:hfdetection}
\end{figure*}

We extend the line search to hydrogen fluoride (HF) following 
the reports of HF detections in young late-M dwarfs \citep{2024A&A...689A.212G}, a directly imaged young super-Jupiter \citep{2024AJ....168..246Z}, and isolated, likely 0.02--3 Gyr, L4--L5 dwarfs \citep{2025A&A...694A.164M}.
We compared the locations of the expected HF lines \citep{2015JQSRT.151..133C} to our residuals (data $-$ model) and confirmed that we see them in both components of Luhman 16. HF is thus detectable over a range of substellar temperatures and into at least early T dwarfs.
Table~\ref{tab:hflines} lists the detected HF lines and Figure~\ref{fig:hfdetection} shows the clearest examples.
These HF lines are not detected in the T6 spectrum in Tann22 because the CH$_4$ opacity is much larger and obscures contributions from most other species in this wavelength region.
\begin{table}
    \centering
    \caption{
        Detected HF lines}
    \label{tab:hflines}
	\begin{tabular}{ccl} 
        \hline
        Order & Wavelength & Notes \\
        $m$ & (µm) &  \\
        \hline
             72 &  2.4759 &  \\ 
             72 &  2.4556 &  \\ 
             72 &  2.4538 &  \\ 
             73 &  2.4331 & overlapped by a telluric mask  \\ 
             74 &  2.4138 & unclear as this is on the order edge \\ 
             74 &  2.3958 &  \\
             75 &  2.3791 &  \\
             75 &  2.3636 &  \\
             76 &  2.3494 & detected only in Luhman 16A \\
             76 &  2.3365 &  \\
        \hline
        \end{tabular}
\end{table}

\subsection{Unidentified features} \label{sec:mysterylines}

While the spectra of both binary components were well reproduced by the model, the high SNR and high wavelength resolution of our data revealed several discrepancies with the best-fit model photospheres. 
Compiling them could help recognize new absorbing species not included in the current model or identify incompleteness in the molecular line lists (e.g., as in Tann22).

Following Tann22, we adopted two criteria to identify discrepancies between the model and data.  
For the first criterion, we calculated the standard deviation ($\sigma$) of the residuals (data $-$ model) for each order and selected regions where the residuals exceed 3$\sigma$ for more than five consecutive data points. 
For the second criterion, we used matched filtering with a line profile kernel. The kernel consists of the instrumental broadening and rotational broadening profiles. The wavelength-dependent \redsep{IPs}
at $H$ and $K$ were as determined in Section~\ref{sec:modelfits}: we used Gaussian with the \redsep{FWHM}
given by Equations~\ref{eq:sigma_H} and \ref{eq:sigma_K}. For the $v\sin{i}$ of rotation broadening, we adopted the global best-fit values for each object in Table~\ref{tab:bestfitparams}. 
We convolved the (data $-$ model) residual spectra with the line profile kernel to obtain filtered residuals.
We selected spectral regions where the square of the filtered residuals exceeded 3$\sigma$.
The regions selected by both of these criteria were further examined for potential unidentified features. 

The quadratic fit of the data to the continuum in the reduction process (Section~\ref{sec:reductions}) can produce unintended deviations in about 1 nm ranges at each edge of an IGRINS order.
Candidate unidentified features detected in these edge regions were checked against the adjacent order to eliminate false detections.  

With the availability of two simultaneously observed spectra of similar objects, and the previously published T6 spectrum in Tann22, we adopted a third criterion for confirming candidate unidentified features. Features that appear in the residuals in at least two, or in all three of the objects, are highly likely to be of astrophysical origin. In all cases, we find that their widths match the rotational widths of other spectroscopic features in the respective objects. 

\begin{figure*}
	\includegraphics[width=0.755\textwidth]{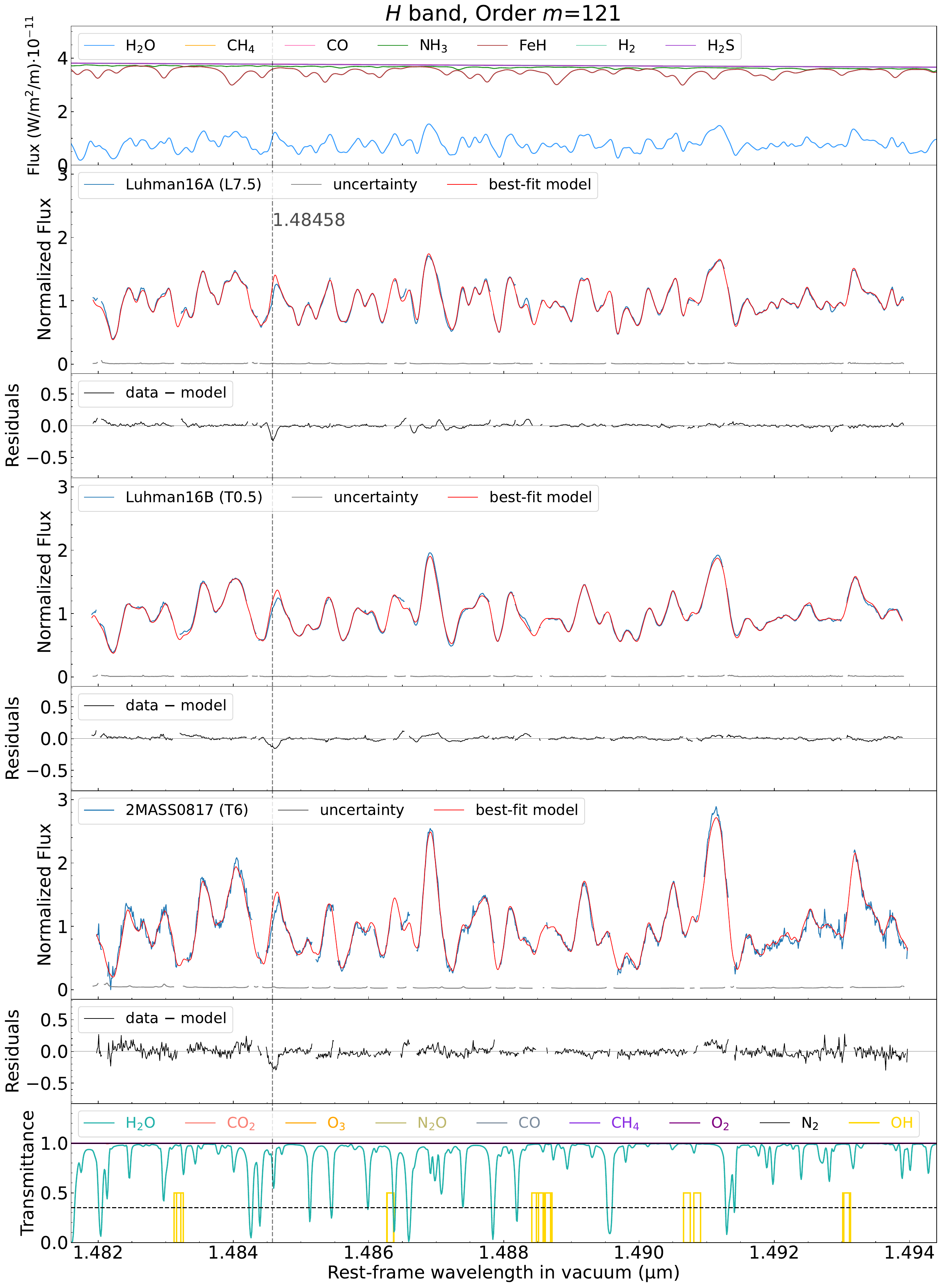} 
    \caption{
    Example of a previously unidentified absorption feature in the spectra of Luhman 16A (L7.5), Lumhan 16B (T0.5), and 2MASS J08173001$-$6155158 (T6; Tann22).
    The top panel shows the absorption contributions for individual molecular species and CIA for the global best-fit model to Luhman 16A.
    For each object, the upper panel shows the IGRINS spectrum and the best-fit photospheric model for this order, and the lower panel shows the residuals between them. 
    The vertical dashed line indicates the position of the newly detected unidentified feature. 
    The bottom panel shows the Earth's transmittance to illustrate the masking of telluric-affected regions. 
    The format is the same as the bottom panel of Figure \ref{fig:h2detection}.
    } 
    \label{fig:mysteryline_1.4846}
\end{figure*}

We list and explore the unidentified regions in Sections~\ref{sec:absorption}--\ref{sec:waterwiggle}.
Features detected in both \redsep{or either of}
Luhman 16A and B are summarized in Table \ref{tab:discrepancies_both} and their presence in the T6 dwarf spectra of Tann22 is added in the \redsep{fifth}
column.
Even if a feature met the first two criteria in only Luhman 16A, it was deemed to be detected in both objects when it met the criteria for Luhman 16B using 1$\sigma$ cutoff instead of 3$\sigma$ on account of the faster rotation of Luhman 16B that further smeared features. 

In the T6 spectra in Tann22, the pseudo-continuum over 1.60--1.73 µm and 2.10--2.48 µm is dominated by CH$_4$ absorption, unlike in our targets.
No common unidentified features are detected at all in these wavelength ranges, as any Luhman 16AB features are likely obscured by CH$_4$ in the T6 dwarf.
On the other hand, most of the unidentified features detected at these wavelengths by Tann22 are not detected in our targets, suggesting that they may be attributed to inaccuracies in the CH$_4$ line list.

\subsubsection{Absorption-like features}
\label{sec:absorption}

The most numerous of the newly detected features are those that appear as absorption (i.e., troughs) in the (data $-$ model) residuals, as noted in Table~\ref{tab:discrepancies_both}. 
We show an example of such regions in Figure \ref{fig:mysteryline_1.4846} with the vertical dashed line indicating the region of interest.
This feature is found in the same wavelength position in all three objects including the T6 dwarf analysed in Tann22.

We searched for spectral lines that matched the wavelengths of these features using the line lists from the compilation of the ExoMol database \citep{2016JMoSp.327...73T} or the Vienna Atomic Line Database (VALD; \citealt{1999A&AS..138..119K}, \citealt{2015PhyS...90e4005R}). Specifically, we checked for absorption by H$_2$O, H$_2$, CO, or CH$_4$ isotopologues, OH, CH, NH, C$_2$, MgH, NaH, CaH, PH$_3$, C$_2$H$_2$, H$_2^+$, HeH$^+$, H$_3^+$, OH$^+$, and CH$^+$.
We did not find any matches among the known lines. All of these unidentified features fall in wavelength regions dominated by water absorption, although in some cases there is also strong absorption by CO (Table \ref{tab:discrepancies_both}). 
We therefore suspect that the unidentified absorption features are either new molecular lines or the result of slight inaccuracies in the wavelengths or strengths of known transitions of the water or possibly the CO molecule. 

\begin{table*} 
    \centering
    \caption{
        Unidentified features: discrepancies between data and best-fit models. 
	}
    \label{tab:discrepancies_both}
	\begin{tabular}{cclllp{0.35\linewidth}} 
        \hline
        Order & Wavelength & Prominent & Type of  & Detected in & Notes \\ 
        $m$ & (µm) & Absorbers & Feature & L7.5/T0.5/T6\tablenotemark{\dagger}\;? &  \\
        \hline

             73 &  2.44152 & H$_2$O, CO & emission-like residual & Y/Y/N &  \\
             76 &  2.34465 & H$_2$O, CO & absorption  & Y/Y/N &  \\
             76 &  2.34201 & H$_2$O, CO & absorption  & Y/Y/N &  \\
             77 &  2.32392 & H$_2$O, CO & emission-like residual  & Y/Y/N &  \\
             77 &  2.32248 & H$_2$O, CO & emission-like residual  & Y/Y/N &  \\
             78 &  2.29393 & H$_2$O, CO & absorption  & Y/Y/N & CO bandhead \\
             78 &  2.29338 & H$_2$O, CO & emission-like residual  & Y/Y/N & CO bandhead \\
             79 &  2.25475 & H$_2$O & emission-like residual  & Y/N/N &  \\ 
             79 &  2.24378 & H$_2$O & emission-like residual  & Y/Y/N & might be a part of P Cygni-like residual\tablenotemark{\ddagger}\ \\
             81 &  2.18466 & H$_2$O & emission-like residual & Y/Y/N & confirmed in the adjacent order $m=$~82.  \\ 
             82 &  2.16681 & H$_2$O & emission-like residual & Y/N/N & residual effect of telluric Br$\gamma$ \\
             83 &  2.14736 & H$_2$O & absorption  & Y/N/N &  \\ 
             85 &  2.09493 & H$_2$O & P Cygni-like residual  & Y/Y/Y &  \\
             85 &  2.08698 & H$_2$O & P Cygni-like residual  & Y/Y/Y &  \\  
             85 &  2.08420 & H$_2$O & absorption\  & Y/Y/N & confirmed in the adjacent order $m=$~86. \\
             86 &  2.07405 & H$_2$O & absorption  & Y/N/N &  \\
             88 &  2.02351 & H$_2$O & emission-like residual  & Y/N/N & might be affected by a CO$_2$ line in telluric standard stars \\
             90 &  1.98139 & H$_2$O & absorption  & Y/Y/Y &  \\ 
             92 &  1.94519 & H$_2$O & emission-like residual  & Y/Y/N & might be affected by a hydrogen line in telluric standard stars  \\ 
            100 &  1.79740 & H$_2$O & P Cygni-like residual  & Y/Y/Y & confirmed in the adjacent order $m=$~99  \\
            100 &  1.78683 & H$_2$O & P Cygni-like residual  & Y/Y/Y &  \\
            100 &  1.78667 & H$_2$O & P Cygni-like residual  & Y/Y/Y &  \\
            101 &  1.78081 & H$_2$O & absorption  & Y/Y/N &  \\
            102 &  1.75665 & H$_2$O & absorption  & Y/Y/N &  \\
            102 &  1.75064 & H$_2$O & emission-like residual  & Y/N/N &  \\
            103 &  1.74238 & H$_2$O & P Cygni-like residual  & Y/Y/Y &  \\
            103 &  1.74223 & H$_2$O & P Cygni-like residual  & Y/Y/Y &  \\
            103 &  1.73535 & H$_2$O & absorption  & Y/N/N &  \\ 
            104 &  1.72719 & H$_2$O & P Cygni-like residual  & Y/Y/N &  \\
            106 &  1.69776 & H$_2$O & absorption  & Y/Y/N &  \\
            110 &  1.62739 & H$_2$O & absorption  & Y/N/N &  \\
            114 &  1.56890 & H$_2$O & P Cygni-like residual  & Y/Y/Y &  \\
            115 &  1.56397 & H$_2$O, CO & P Cygni-like residual  & Y/Y/Y &  \\ 
            115 &  1.56383 & H$_2$O, CO & P Cygni-like residual  & Y/Y/Y &  \\
            116 &  1.55114 & H$_2$O & P Cygni-like residual  & Y/Y/Y &  \\
            116 &  1.54166 & H$_2$O & P Cygni-like residual  & Y/Y/Y &  \\
            117 &  1.53842 & H$_2$O & P Cygni-like residual  & Y/Y/Y &  \\
            118 &  1.52595 & H$_2$O & P Cygni-like residual  & Y/Y/Y &  \\
            118 &  1.52085 & H$_2$O & emission-like residual  & Y/Y/Y &  \\
            119 &  1.51877 & H$_2$O & absorption & Y/Y/N & barely verified in the adjacent order $m=$~118.  \\ 
            118 &  1.51700 & H$_2$O & P Cygni-like residual  & Y/Y/Y & confirmed in the adjacent order $m=$~119. \\
            119 &  1.51689 & H$_2$O & P Cygni-like residual  & Y/Y/Y &  \\
            119 &  1.51700 & H$_2$O & P Cygni-like residual  & Y/Y/Y &  \\
            120 &  1.50227 & H$_2$O & absorption  & Y/Y/Y &  \\
            121 &  1.48459 & H$_2$O & absorption  & Y/Y/Y &  \\

        \hline
        \end{tabular}
    \tablenotetext{\dagger}{T6 dwarf 2MASS J08173001--6155158 in Tann22. }
    \tablenotetext{\ddagger}{See Section \ref{sec:waterwiggle} for discussion of all P Cygni profile-like residuals. }
\end{table*}

\subsubsection{Emission-like features} \label{sec:emission}

\begin{figure*}
	\includegraphics[width=0.755\textwidth]{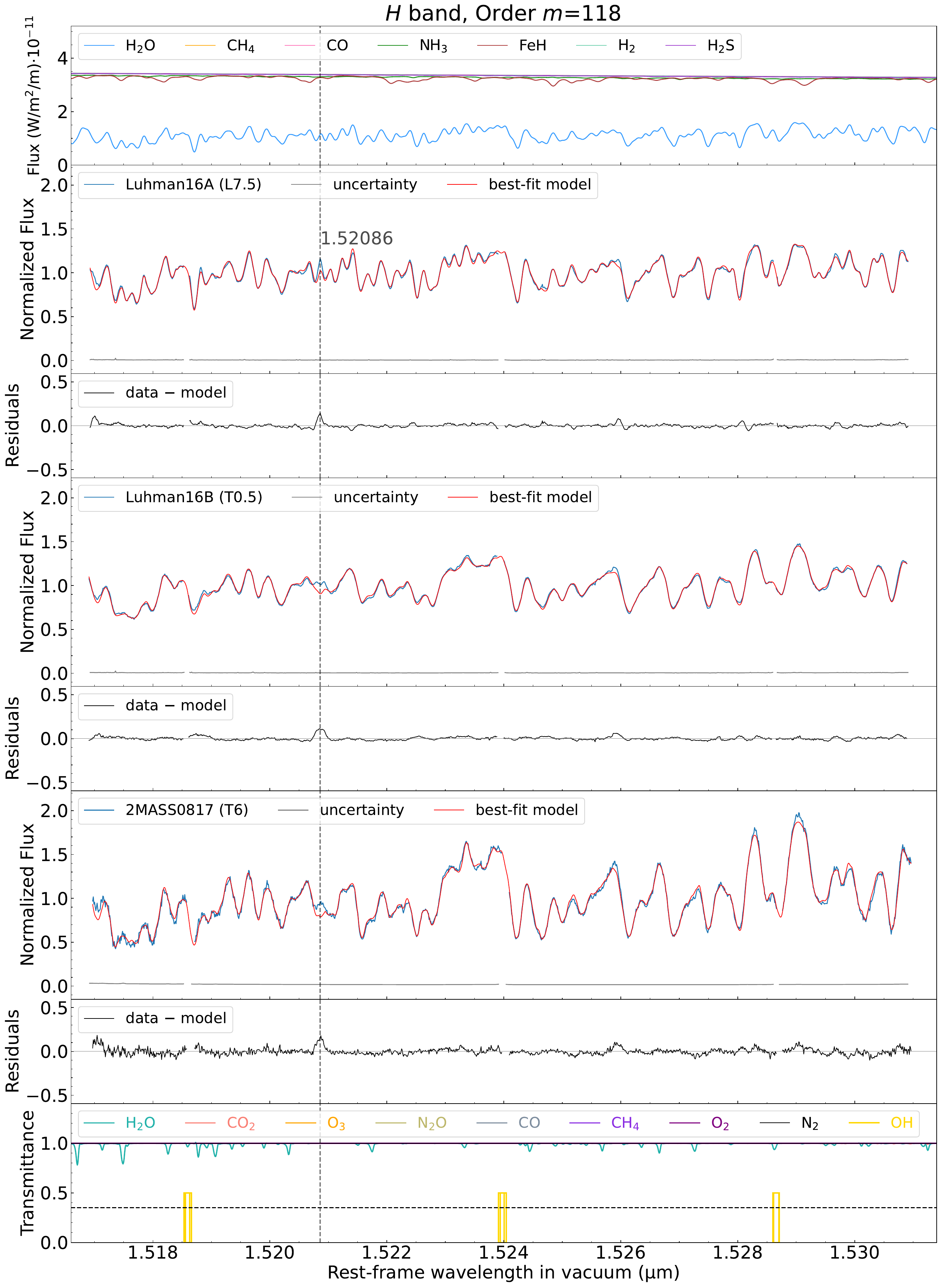} 
    \caption{
    The clearest 
    example of a previously unidentified emission-like feature---at 1.52085~µm---in the (data $-$ model) residual spectra of Luhman 16A (L7.5), Lumhan 16B (T0.5), and 2MASS J08173001$-$6155158 (T6; Tann22). The figure format is the same as of Figure \ref{fig:mysteryline_1.4846}.
    As discussed in Section~\ref{sec:emission}, this feature is possibly the result of over-estimated water line absorption in the atmospheric model.  
    }
    \label{fig:mysteryline_1.5209}
\end{figure*}

An upward bump in the (data $-$ model) residual at 1.52085~µm (Figure \ref{fig:mysteryline_1.5209}) is one of the most intriguing unidentified features. 
It is the clearest emission-like feature without any other contaminating factors in the "Notes" column of Table \ref{tab:discrepancies_both}, such as a potentially miss-estimated CO line absorption (e.g., 2.29338~µm), part of a P Cygni profile-like residual (e.g., 2.24378~µm; Section \ref{sec:waterwiggle}), or remnants from telluric line contamination (e.g., 1.94519~µm).
It is seen in the IGRINS spectra of all three objects, presented here and in Tann22. The fact that it appears broader in Luhman 16B than in 16A, like other stellar absorption lines, confirms that it is intrinsic in origin. 
An upward profile suggests that either an emission line was observed or that the model is overestimating absorption. 

We found a FeH line from a $F^4\Delta-X^4\Delta$ electronic transition in the MoLLIST \citep{2020JQSRT.24006687B} recorded in the ExoMol database that matches this wavelength.
However, there is no reason to think that FeH would exhibit an emission line as the rest of the nearby FeH lines 
\redsep{are in absorption, if at all present
(Section~\ref{sec:feh}).}
Moreover, Tann22 reported a similar discrepancy at the same wavelength in the analysis of a T6 object, where the contribution of FeH should be even smaller. We therefore conclude that FeH is not the cause. 

In this wavelength range the shape of the model spectrum is determined almost entirely by H$_2$O absorption (see panels of order $m=118$ in Figures \ref{fig:atlasA} and \ref{fig:atlasB}), so if the model's overestimation of water absorption is the cause, the feature would be attributed to an inaccuracy in the water line list.

\subsubsection{P Cygni profile-like features in the residuals} \label{sec:waterwiggle} 

\begin{figure*}
	\includegraphics[width=0.8249\textwidth]{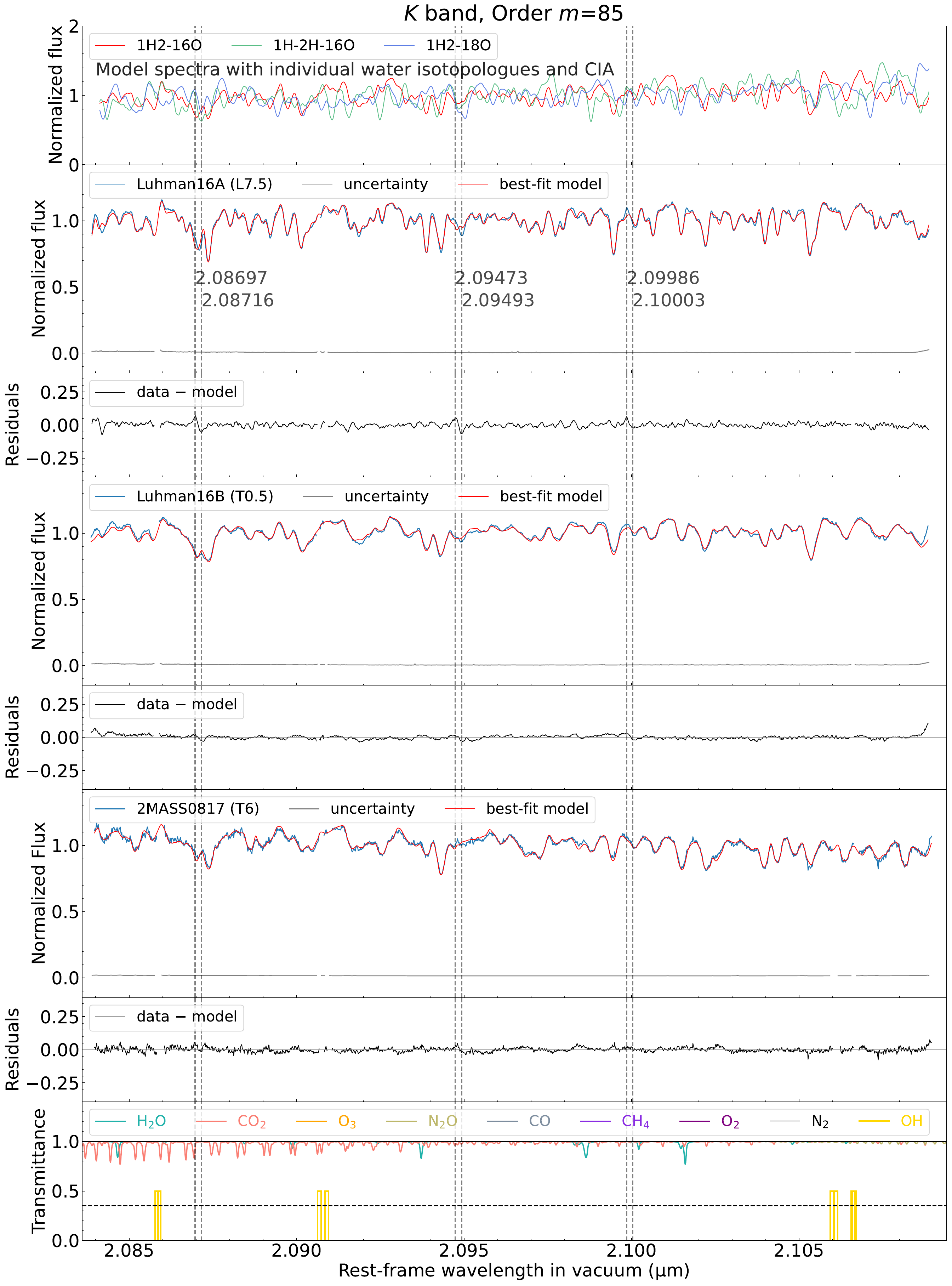} 
    \caption{P Cygni-like features observed in the (data $-$ model) residuals in the wavelength region dominated by H$_2$O absorption are indicated by vertical dashed lines.
    This is an example in order $m=85$ in the $K$ band.
    The top panel shows normalized model spectra of three water isotopologues ($\mathrm{^{1}H_{2}^{16}O}$, $\mathrm{^{1}H^{2}H^{16}O}$, $\mathrm{^{1}H_{2}^{18}O}$), that also include CIA.
    The following six panels, in three pairs of two panels each, show the IGRINS spectra and their best-fit models for Luhman 16A, Luhman 16B, and 2MASS J08173001--6155158 (T6), respectively, and the (data $-$ model) residuals.
    The residuals of the three objects show common P Cygni-like patterns at the same wavelengths.
    The bottom panel shows the Earth’s transmittance to see the effects of the masking of telluric-affected regions, as in Figure \ref{fig:mysteryline_1.4846}. 
    } 
    \label{fig:waterwiggle_K}
\end{figure*}

\begin{figure*}
	\includegraphics[width=0.8349\textwidth]{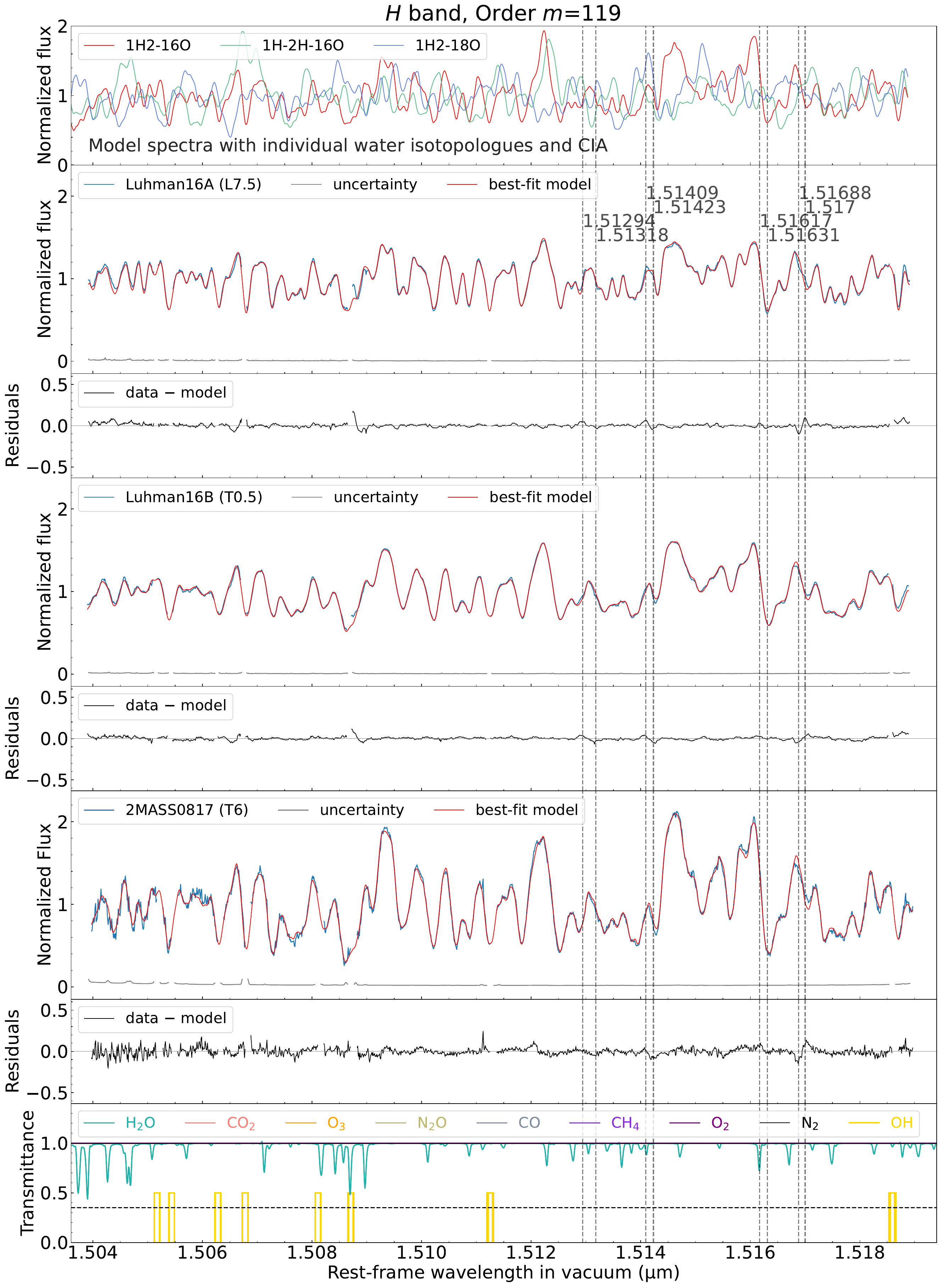} 
    \caption{Same as Figure~\ref{fig:waterwiggle_K} but for order $m=119$ in the $H$ band.
    Most P Cygni-like residuals have bluer peaks and redder troughs, although an exception is seen around 1.517 µm here. 
    } 
    \label{fig:waterwiggle_H}
\end{figure*}

A notable class of feature in several regions dominated by water is a P Cygni-like profile in the residuals, where a 
peak and a 
trough are found adjacent to each other. 
Figure \ref{fig:waterwiggle_K} shows examples in a part of the $K$ band and Figure \ref{fig:waterwiggle_H} shows some in the $H$ band.
Most of them show bluer peaks and redder troughs in the residuals, although some exhibit the opposite pattern (e.g., 1.517 µm; see Figure \ref{fig:waterwiggle_H}). 
They are always found in both components of Luhman 16 and also in the T6 dwarf of Tann22 in most cases. 

Figures~\ref{fig:waterwiggle_K} and \ref{fig:waterwiggle_H} higlight the strongest P Cygni-like residuals from the model photosphere subtraction. All such features that pass our 3$\sigma$ threshold (Section~\ref{sec:mysterylines}) are listed in Table \ref{tab:discrepancies_both}. However, many more such features exist below the 3$\sigma$ threshold.
These features are ubiquitous in broad regions where $\mathrm{H_{2}O}$ sets the pseudo-continuum.
It is often difficult to attribute them to any particular absorption line because they do not necessarily appear around prominent $\mathrm{H_{2}O}$ lines.  

We tested the possibility that the P Cygni-like residuals are caused by isotopologues of water by plotting the spectra of three isotopologues in the top panels of Figures \ref{fig:waterwiggle_K} and \ref{fig:waterwiggle_H}: $\mathrm{^{1}H_{2}^{16}O}$~\citep{2018MNRAS.480.2597P}, $\mathrm{^{1}H^{2}H^{16}O}$~\citep{2010MNRAS.402..492V}, and $\mathrm{^{1}H_{2}^{18}O}$~\citep{2017MNRAS.466.1363P}.
The plotted spectra were normalized to their respective averages because the actual isotopic ratios are unknown and our purpose is to check for correspondence among the isotopologue lines and the P Cygni-like residuals. 
Most residual features do not line up with the wavelengths where an isotopologue has a strong line.
While in a few cases alignment exists, realistic isotopic ratios (e.g., on Earth, $^{18}$O is about 0.2\% of $^{16}$O and $^{2}$H is less than 0.02\% of $^{1}$H) dictate that the contributions from the minor isotopologues should be significantly lower than the main isotopologue $\mathrm{^{1}H_{2}^{16}O}$, and so should not produce detectable residuals. 
We conclude that the contribution of the two minor water isotopologues is negligible and they do not cause the P Cygni-like residuals.

Instead, as a likely explanation, we suspect inaccuracies in the H$_2$O line list. 
It is possible that absorption lines with specific quantum parameters share similar wavelength shifts, which could explain the consistent blue peak / red trough for the majority of the P Cygni-like residuals, but no further conclusions can be drawn in this study.

\subsection{Disequilibrium chemistry} \label{sec:disequilibrium}

Our models incorporate disequilibrium, and as already seen for the best-fit models for Luhman 16A (Figure~\ref{fig:wavvsgbp_A}) and B (Figure~\ref{fig:wavvsgbp_B}), the $C_{K_{zz}}$ eddy diffusion coefficient parameter is around unity, although with an up to an order of magnitude scatter across the spectral orders. A clear example of the need for chemical disequilibrium is seen in the fitting of the 1.58--1.75~µm $H$-band orders, 
where FeH is expected to be an important absorber. The average of the best-fit photospheric models across all orders requires a moderate amount of eddy diffusion ($\log C_{K_{zz}}=0.1$). However, some of the $H$-band FeH orders are fit with a value two orders of magnitude higher, $\log C_{K_{zz}}=2$, for both Luhman 16A and B. The effect of the disequilibrium \redsep{here}
is to decrease the contribution of FeH through changes in the atmospheric temperature-pressure profiles,
so that iron has rained out \redsep{more at lower altitude}
and FeH 
is mostly undetectable 
(Figure~\ref{fig:feh}, Section~\ref{sec:feh}).

\begin{figure*}
	\includegraphics[width=0.95\textwidth]{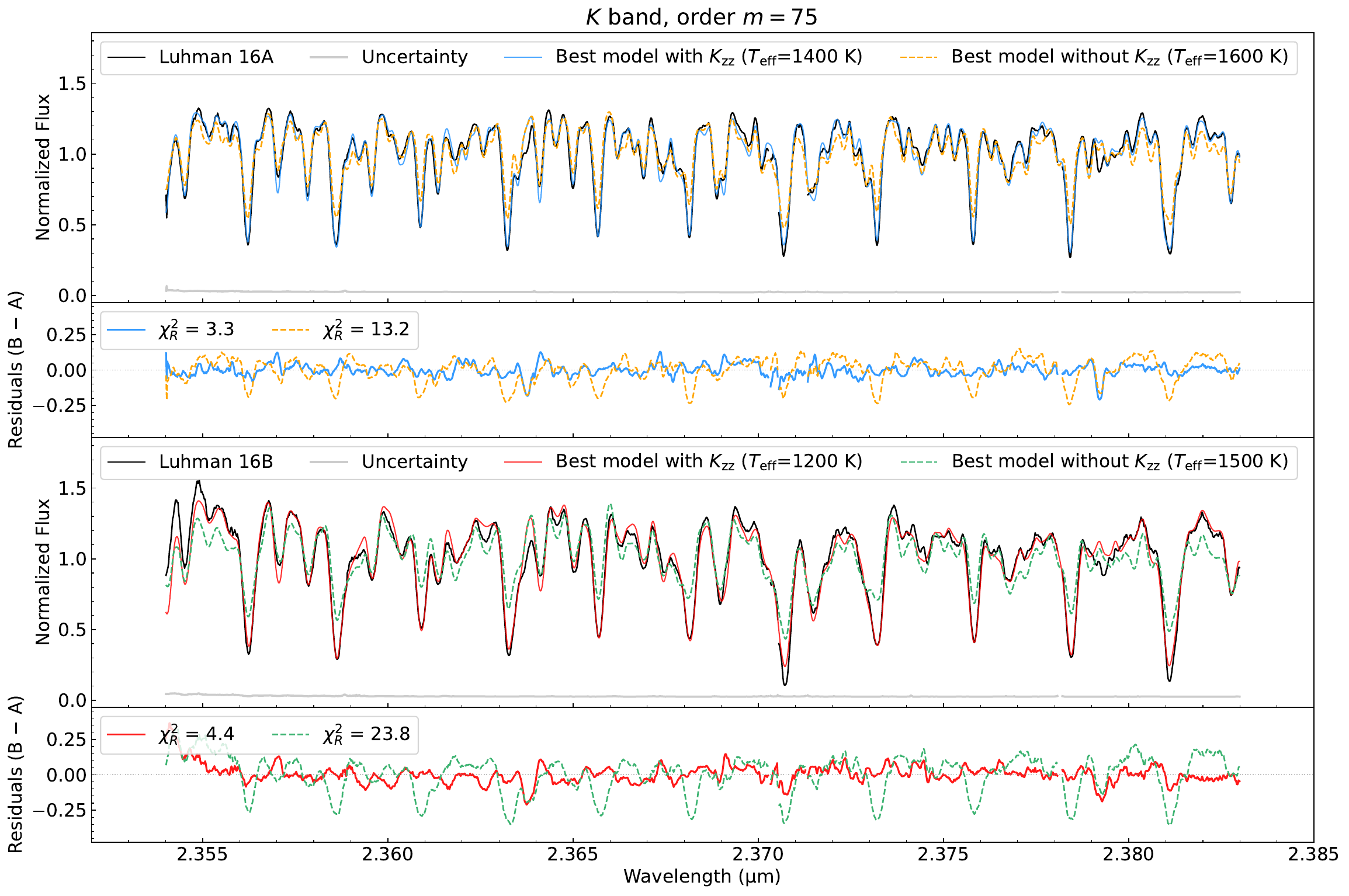} 
    \caption{Comparison of the inclusion of vertical mixing as a cause for disequilibrium chemistry in a wavelength region where CO is the primary absorber (order $m=75$). 
    The top two panels show the spectral fitting and residuals for Luhman 16A and the bottom two for Luhman 16B.
    Models that incorporate vertical mixing ($K_\mathrm{zz}$ parameter) reproduce the data better. 
    }
    \label{fig:equilibrium_compared}
\end{figure*}
A more typical manifestation of disequilibrium chemistry at the L/T transition is stronger-than-expected CO absorption. Figure \ref{fig:equilibrium_compared} demonstrates this by comparing best-fit models without and with disequilibrium chemistry for both L7.5 and T0.5 components of Luhman 16. The best-fit equilibrium models require an effective temperature that is too high ($T_{\mathrm{eff}}=1600$ K) and still underestimate the normalized depths of the CO lines. Their $\chi^2_{\mathrm{R}}$ values are also significantly higher than for the disequilibrium model fits. Even if the $C_{K_\mathrm{zz}}$ values of the best-fit models are in this case modest (0.1), the need for disequilibrium treatment is evident.

\section{Planet Search} 
\label{sec:planetsearch} 
Our series of high-dispersion spectroscopic observations spans 70 days with a cadence as short as 0.15 days
but most commonly 1 day. We use these to set upper limits on the RV variations due to unseen planets around either component.

\subsection{Radial velocity measurements}
\label{sec:rvmethods}

Precise RV measurements involve using a spectroscopic template as a reference spectrum to be shifted to the spectra at each epoch.
We considered two different approaches to creating spectroscopic templates.
The first approach uses the high-SNR barycentric motion-corrected, combined spectrum of Luhman~16A or Luhman~16B as a self-reference template for itself at every epoch. 
The second approach uses the spectrum of Luhman~16A as a template for the spectrum of Luhman~16B. 
Since the spectra of the two components are observed at the same time and under the same conditions and instrumental setup, the relative RVs are expected to be free from observational systematic effects. Because the spectral types of the components are similar, L7.5 and T0.5, either component is a good template for the other. Luhman~16A is the slower rotator and exhibits sharper absorption lines than B. Hence, we convolved the spectra of Luhman~16A with a rotational broadening kernel to match those of Luhman~16B. 
The value of the relative $v\sin{i}$ for the broadening kernel is fixed to 14.0  km\,s$^{-1}$, which was the average determined from the order-by-order matching of the combined spectra of the two components (Section \ref{sec:avsb}).

We thus analyse three different RV series:
\begin{enumerate}
    \item individual epochs of Luhman 16A velocity-shifted to match
    the high-SNR combined spectrum of Luhman 16A;
    \item individual epochs of Luhman 16B 
     velocity-shifted to match the high-SNR combined spectrum of Luhman 16B;
    \item individual epochs of Luhman 16B 
    velocity-shifted to the corresponding simultaneous Luhman 16A spectra.
\end{enumerate}

\begin{figure*}
	\includegraphics[width=0.95\textwidth]{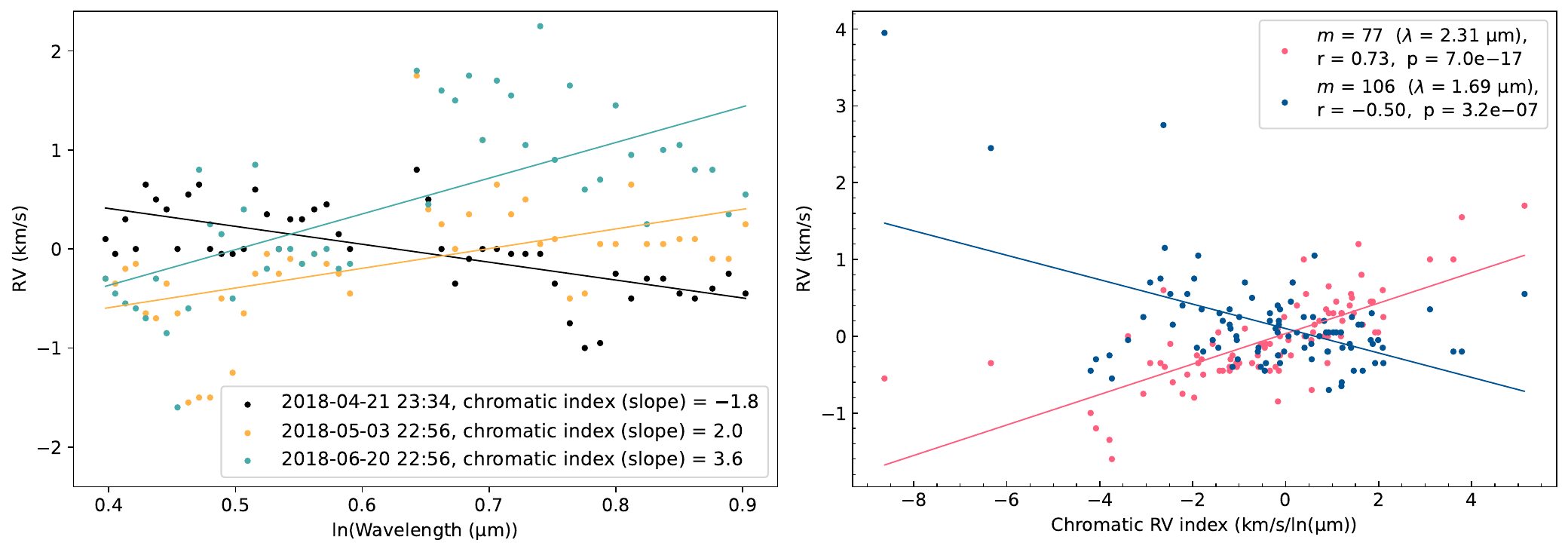} 
    \caption{Examples of the derivation of the chromatic RV indices (left panel) and the correlation between these indices and RVs (right panel). The chromatic index is derived, for the entire $H$- and $K$-band spectrum at each epoch, as the slope of a straight line fitted to the RV as a function of the natural logarithm of wavelengths, following the definition in \citet{2018A&A...609A..12Z}.
    The left panel shows examples of three different epochs for Luhman 16A. 
    The solid lines represent linear fits to data points with the corresponding colours, and their slopes are the chromatic indices. 
    The right panel exemplifies the correlation between the RVs of Luhman 16A and the chromatic indices in two different echelle orders, $m=77$ and $m=106$.
    The legend shows the Pearson correlation coefficient $r$ with the associated p-value, which confirms the correlation between RV and chromatic index.
    The correlation indicates that the RV variability is due to surface inhomogeneities rather than a planetary companion. 
    }
    \label{fig:crx}
\end{figure*}

For each of the three RV series, we detrended for possible activity-induced RV variations using the following procedure. 
We calculated a chromatic RV index \citep{2018A&A...609A..12Z}
as the slope of a linear fit to the RV values from each echelle order as a function of the natural logarithm of the central wavelength of each order (left panel in Figure~\ref{fig:crx}).
The RV series of some orders shows a correlation between RVs and chromatic indices at individual epochs (right panel in Figure~\ref{fig:crx}).
This indicates RV variation caused by weather-induced phenomena such as surface spots rather than planetary companions.
We removed this correlation 
by subtracting a linear fit to the RV values in each order and adding back the average.
After detrending, we took a weighted average of RVs for all individual orders at each epoch to get an overall RV time series for later analysis.

\begin{figure*}
	\includegraphics[width=0.95\textwidth]{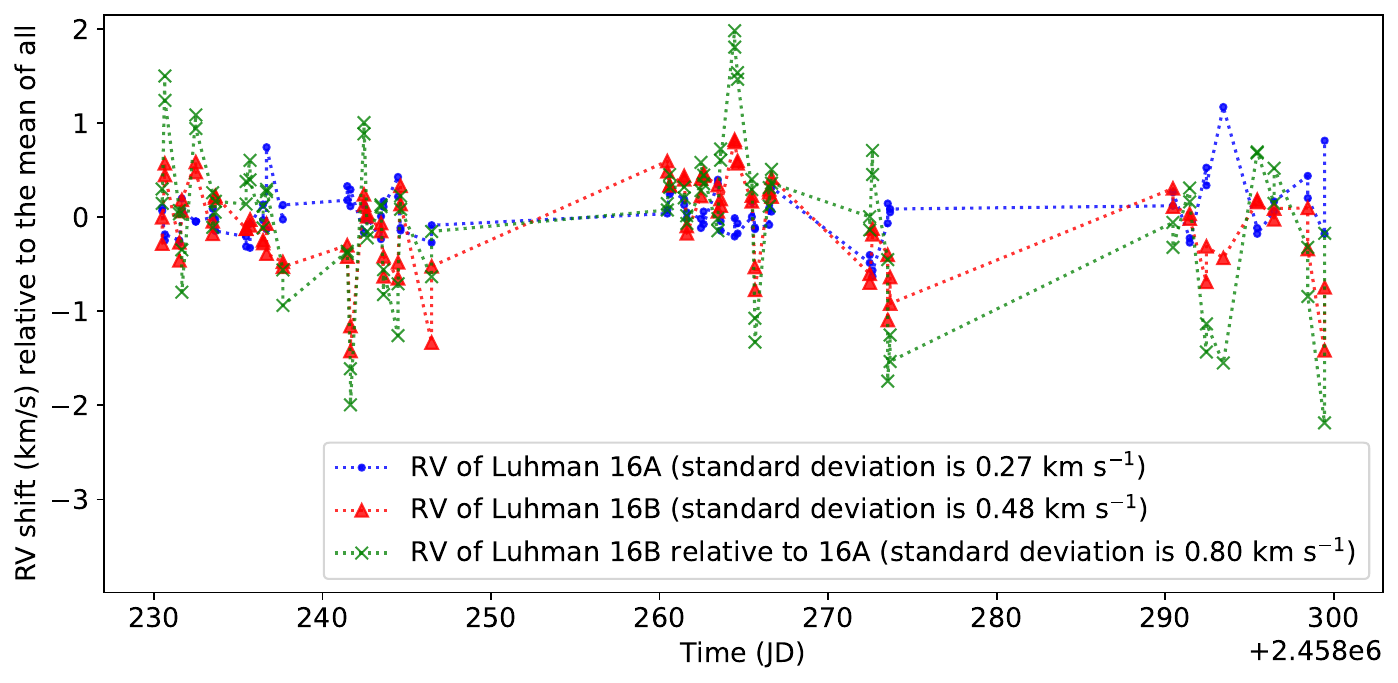} 
    \caption{Time series of RV shifts relative to the mean of all data points for Luhman 16A and 16B as a function of time in Julian days.
    The blue circles and the red triangles indicate the RV series for each binary component with its own multi-epoch combined spectrum as a reference.
    The green crosses show the relative RV of Luhman 16B with respect to 16A, 
    measured by shifting 16A's broadened spectrum to match that of 16B
    at each epoch. 
    }
    \label{fig:rvs}
\end{figure*}

The three order-averaged RV series are shown in Figure~\ref{fig:rvs}.
The scatter of relative RVs as a function of time were consistent among the three approaches. However, the 0.80~km\,s$^{-1}$ standard deviation of the case (iii) RV series is larger than the 0.55~km\,s$^{-1}$ quadrature sum of those for cases (i) 0.27~km\,s$^{-1}$ and (ii) 0.48~km\,s$^{-1}$ .
That is, systematic uncertainties arising from the intrinsic spectroscopic differences of these two similar but not identical objects have 
a greater effect on the RV 
than systematics arising from the observing or instrumental environment.
Hence, the approach that our custom observations were intended to benefit from, namely measuring RVs using binary components observed simultaneously as mutual references to each other, is not necessarily advantageous.

\subsection{Periodogram Analysis}
\label{sec:periodogram}

\begin{figure*}
	\includegraphics[width=0.95\textwidth]{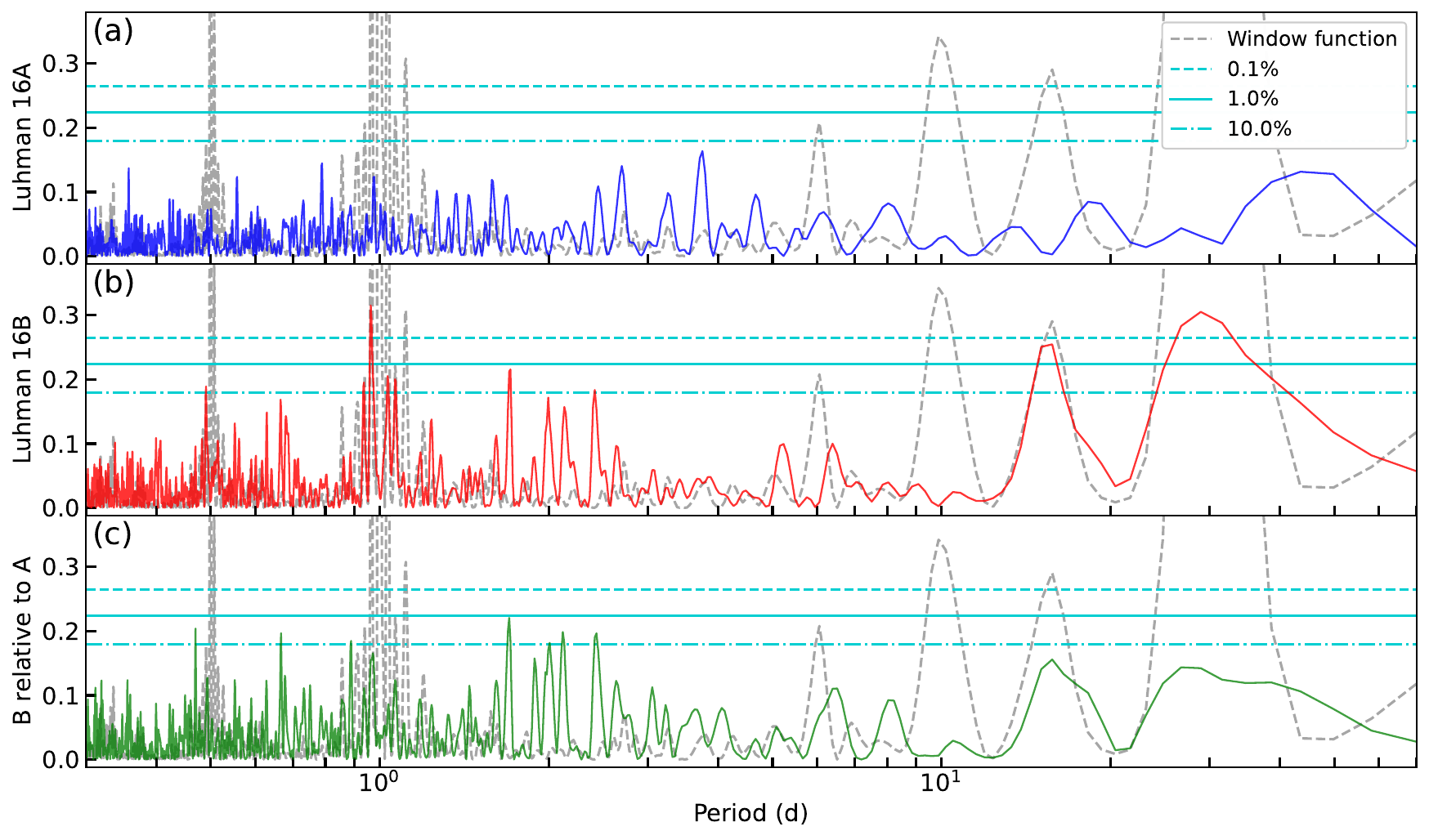} 
    \caption{Generalized LS periodograms for the RV data points in Figure \ref{fig:rvs}. 
    From top to bottom, the panels depict the power spectrum of the RVs of Luhman~16A, that for Luhman~16B, and that for the relative RVs of Luhman~16B with respect to 16A. 
    The dashed grey lines in all panels show the window function.
    The three horizontal lines indicate FAP levels from 0.1 to 10\%. 
    }
    \label{fig:periodogram}
\end{figure*}

We searched for periodic RV signals by computing the generalised Lomb-Scargle (GLS) periodograms \citep{2009A&A...496..577Z} for each binary component, and for the relative RVs of B with respect to A, for the period range 0.3--70 days. 
We also calculated the window function for our data, to avoid spurious signals arising from the sampling. 
The resulting power spectra are shown in Figure~\ref{fig:periodogram} with the horizontal lines marking False Alarm Probability (FAP) levels between 0.1--10\%. The window function shows prominent peaks at 1 day and 0.5 days, corresponding to the most common spacing between the observations and its first overtone frequency.
We do not identify any notable signals with FAP $<$ 0.1\% away from prominent peaks of the window function.

\begin{figure*}
	\includegraphics[width=0.95\textwidth]{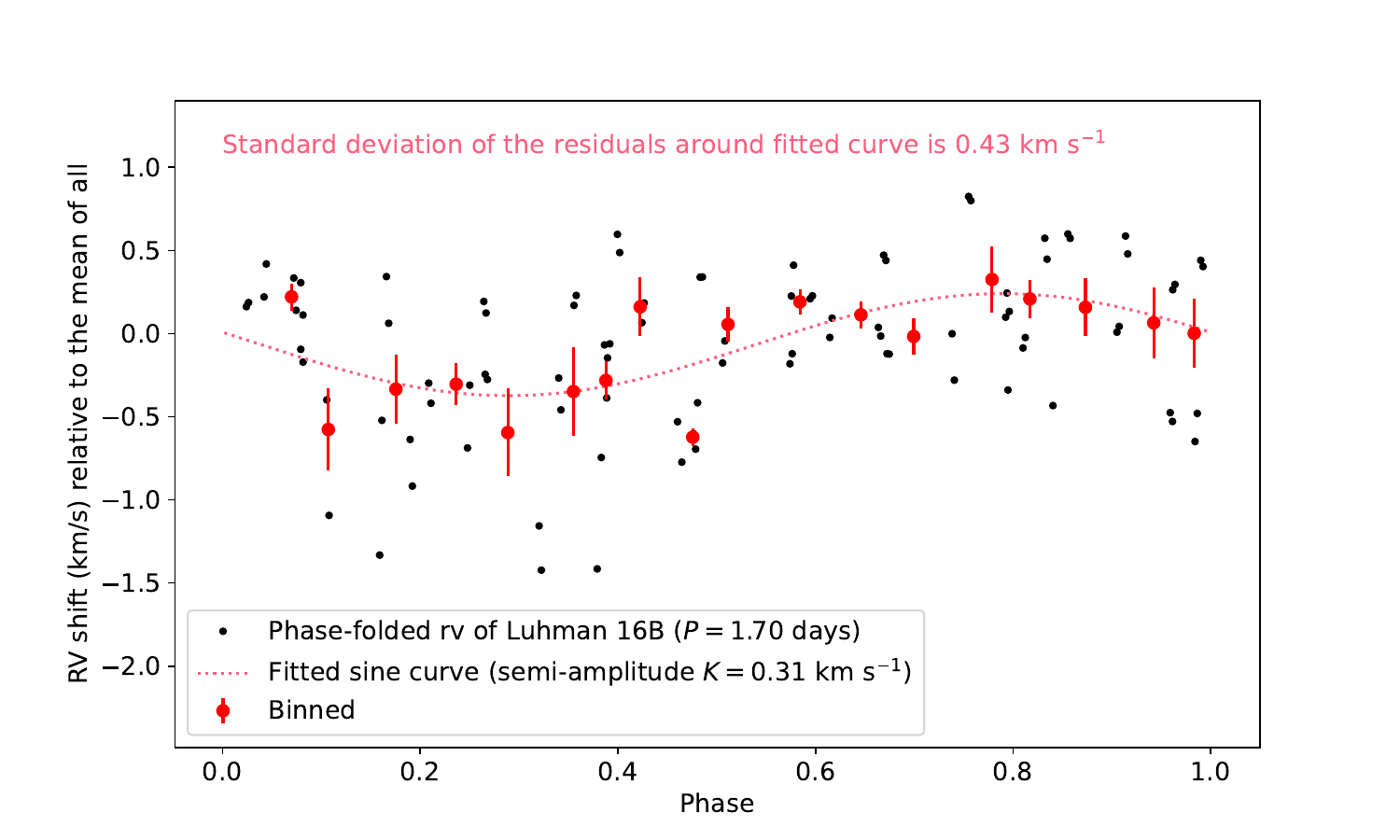} 
    \caption{
    Time series of RVs of Luhman 16B folded on the 1.70-day period that corresponds to the highest peak (FAP = 1.7\%) in the Luhman 16B periodogram (Figure \ref{fig:periodogram}, panel (b)). 
    The black points represent the data from all 93 exposures, and the red points are the binning of every 5 data points.
    The dotted line is the best-fit sine curve.
   Its semi-amplitude is 0.31 km\,s$^{-1}$. 
    }
    \label{fig:phasefolded_170}
\end{figure*}

The highest periodogram peak that is distinct from a peak in the window function is found in the RV series of Luhman 16B: 
at 1.70 days (panels b and c in Figure~\ref{fig:periodogram}). 
The phase-folded RV data for this period are shown in Figure \ref{fig:phasefolded_170}.
If this were caused by a real planet around Luhman 16B, the minimum mass is estimated to be 0.16 $M_{\mathrm{J}}$ with a semi-major axis of 0.008 au. 
Here we have adopted the mass estimates for the individual binary components from \citet{2018A&A...618A.111L}: 33.5 $M_{\mathrm{J}}$ for Luhman 16A and 28.6 $M_{\mathrm{J}}$ for 16B.

Our RV analysis does not yield a significant planet detection. We discuss
the detectability and upper limits on $M\sin i$ for potential planets in Section \ref{sec:planetlimitation}.  

\subsection{Constraints on planetary companions} \label{sec:planetlimitation}

\begin{figure*}
    \includegraphics[width=0.48\textwidth]{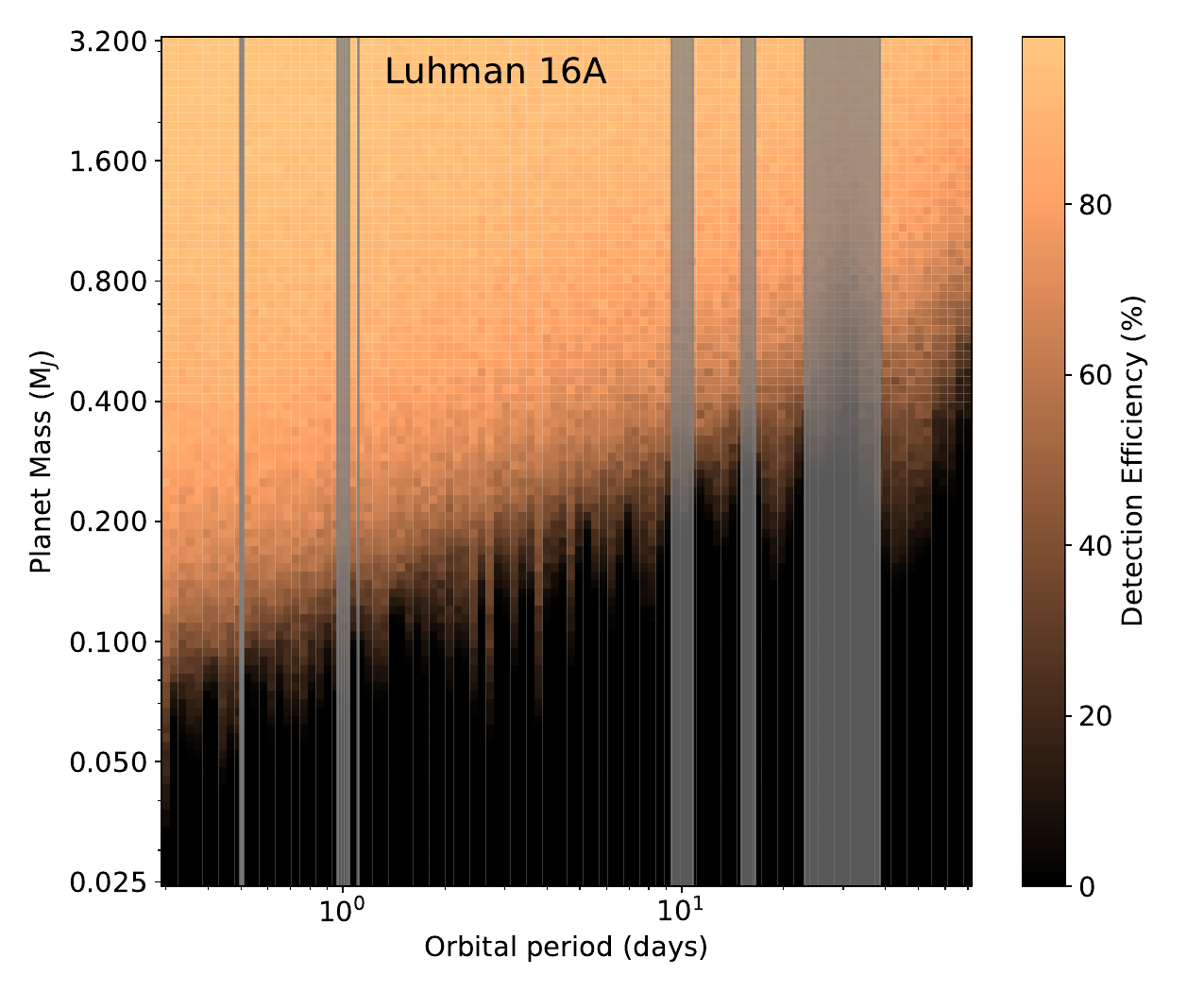} 
    \includegraphics[width=0.48\textwidth]{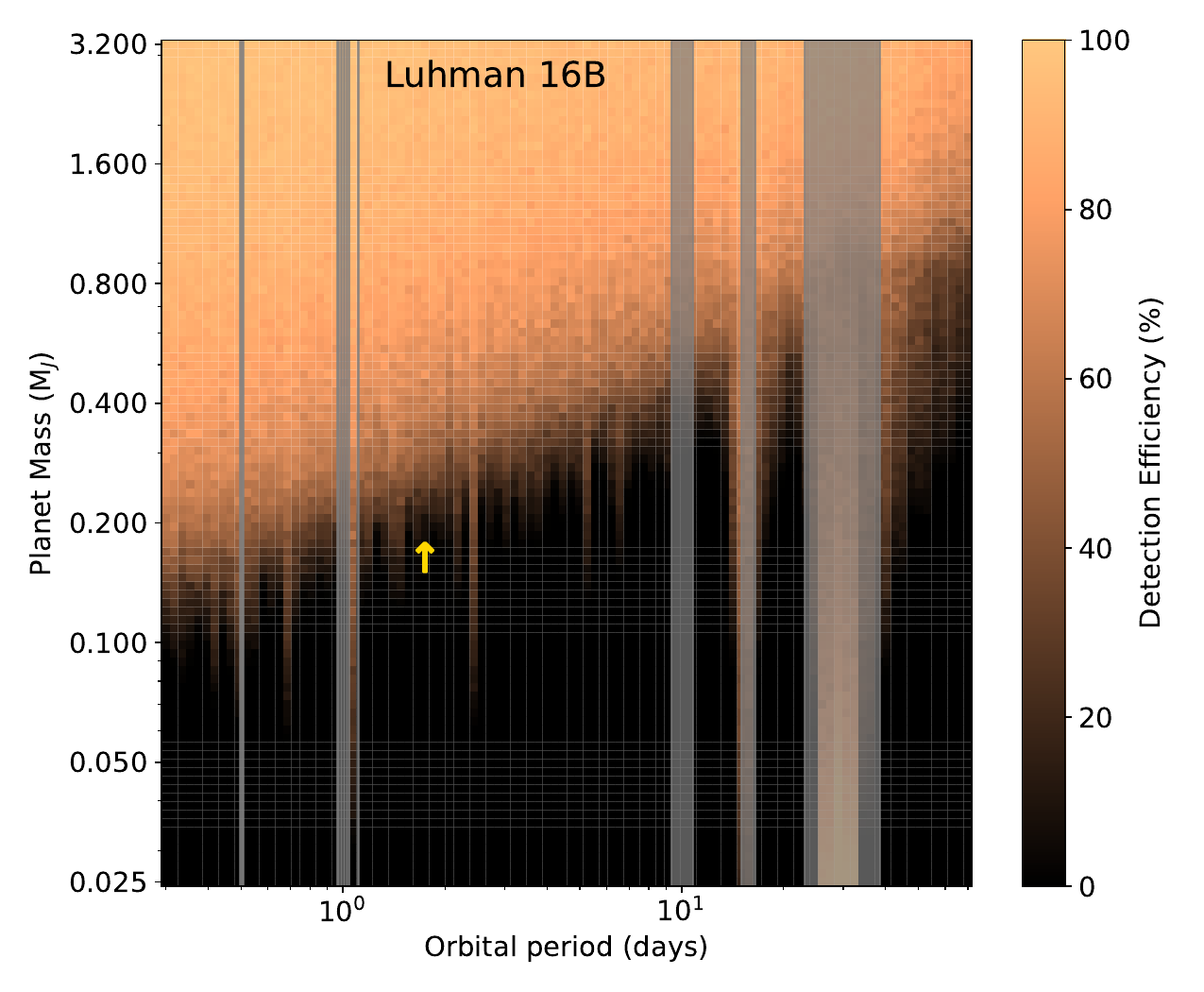} 
    \caption{Detection efficiency as a function of orbital period and planet mass of potential planets around Luhman 16A (left panel) and B (right panel).  
    The colour map represents the fraction of injected signals that were successfully recovered in the injection-recovery tests. 
    The $x$ axis shows the orbital period in days, while the $y$ axis indicates the planet mass in Jupiter masses. 
    The shades in grey indicate period ranges where the power of the window function is greater than the horizontal line indicating FAP=10\% in Figure \ref{fig:periodogram}.
    The possible planet signal at $M\sin i=0.16 M_{\mathrm{J}}$ and $P=1.70$ days around Luhman 16B discussed in Section \ref{sec:periodogram} is marked with a yellow upward arrow, indicating a lower limit of $M$. 
    Regions in a bright colour indicate high detection efficiency, while darker regions signify lower efficiency. 
    While shorter-period planets would be more readily evident in our data, any $<$2-day periods would be under-sampled by our $\sim$1-day observing cadence.
    }
    \label{fig:injectiontest}
\end{figure*}

To assess the limitations of our current data for detecting planets, we injected sinusoidal signals with periods 0.3--70
days and semi-amplitudes corresponding to planets with masses of 0.025--3.2 $M_{\mathrm{J}}$ around each binary component,
and performed the same periodogram analysis as in Section \ref{sec:periodogram}. 
We 
randomly sampled eccentricities from a Rayleigh distribution with $\sigma = 0.19$ \citep{2023PNAS..12017398S} and inclinations from a uniform distribution between 0--2$\pi$. 
We deemed the injected signals to be recovered when the power at the injected period was above a FAP threshold of 0.1\%. 
The detection efficiency was calculated at each grid point of periods and masses, 
as the fraction of injected signals that were successfully recovered in the periodogram. 
Figure~\ref{fig:injectiontest} shows the results. 
Barring extremely unfavourable orbital inclinations, the data are sensitive to planets of masses above 
$\sim$0.2 $M_{\mathrm{J}}$
up to periods of 1 d and 
$\sim$0.4 $M_{\mathrm{J}}$
up to 10 d for Luhman 16A, and $\sim$0.3 $M_{\mathrm{J}}$
up to periods of 1 d and 
$\sim$0.7 $M_{\mathrm{J}}$
up to 10 d for Luhman 16B 
except for the period ranges affected by the window function.
We rule out the presence of the planets in these parameter spaces, although planets that are smaller or on near-face-on orbits may still be hidden. 
Nevertheless, it should be noted that signals with $<$2-day periods would have been under-sampled, given that our typical observation frequency is 1 day.

\section{Conclusions}
\label{sec:Conclusions}

We have presented a comprehensive 1.48--2.48 µm $R=\redsep{28\,000}$ spectroscopic atlas of the L7.5 and T0.5 components of the Luhman 16AB binary system: the closest pair of brown dwarfs to Earth. 
Our high-SNR combined spectra allow us to test the accuracy of photospheric models and their adopted line lists, to identify previously undetected molecules in L/T-transition dwarfs. 
Our \redsep{70-day-long}
spectroscopic campaign also enables us to place upper limits on the presence of planets around either component of the binary. 

Our specific findings are:
\begin{itemize}
    \item The Sonora Bobcat atmospheric models \citep{2021ApJ...920...85M} can fit the high-dispersion 1.48--2.48 µm spectra of the L7.5 and T0.5 components sufficiently well, provided that they are augmented to include disequilibrium chemistry and clouds. We named this model family Sonora Bobcat Alternative B, iterating on the cloudless Sonora Bobcat Alternative A models discussed in \citet{Tannock2022}. 
    
    \item The well-established CO~\citep[HITEMP2019: ][]{2010JQSRT.111.2139R, 2015ApJS..216...15L} and the updated CH$_4$~\citep{2020ApJS..247...55H} line lists match the high-dispersion spectra well. The $\mathrm{H_{2}O}$ line list from \citet{2018MNRAS.480.2597P} is noted as an improvement over previous $\mathrm{H_{2}O}$ line lists by Tann22. However, an ensemble comparison of the present data together with the spectrum of a T6 dwarf in Tann22 shows that wavelength discrepancies in the water lines remain, manifested as P Cygni-like profiles in the (data $-$ model) residuals. 
    
    \item FeH lines, while expected to be strong in the spectra of the two components under chemical equilibrium conditions, especially in the L7.5 component A, are not 
    detected. 
    Wavelength regions \redsep{that should be}
    affected by FeH favour lower-$T_{\mathrm{eff}}$ or higher-$C_{K_\mathrm{zz}}$ models 
    and more efficient iron rainout than models with the global best-fit parameters. 

    \item We \redsep{detected NH$_3$ lines in both Luhman 16A and B spectra, which makes them the warmest objects, along with the L9 dwarf DENIS J025503.3$-$470049, for which NH$_3$ is detected.}
    
    \item We detected H$_2$ and H$_2$S absorption lines in L/T transition objects for the first time, following similar detections of both molecules in a T6 dwarf by Tann22 and reports of tentative or significant H$_2$S detections in a T9 dwarf \citep{2023ApJ...953..170H} 
    \redsep{or}
    in a Y dwarf \citep{lew_etal24}\redsep{.}
    Such detections have been challenging in low-dispersion data because of contamination from other strong molecular bands---mainly $\mathrm{H_{2}O}$ or CH$_4$---despite the significant abundance of H$_2$ and H$_2$S. 

    \item The detection of H$_2$---to date only with IGRINS---is a testament to the exquisite sensitivity attainable in the near-infrared at high spectral resolution with IGRINS on bright brown dwarfs. We employ the H$_2$ column density measurement to set a lower limit on the atmospheric pressure of the cloud deck. We conclude that the clouds in Luhman 16A and B reside at pressure levels $P_{\mathrm{abs}}\gtrsim 3$ bar.

    \item H$_2$S is important as the primary sulfur reservoir in hydrogen/helium-dominated atmospheres and, in combination with CO and H$_2$O measurements~\citep[][]{2023A&A...670A.161P},  
    can be used to estimate the bulk sulfur abundance. This is relevant for informing models of planet formation and evolution.

    \item \textcolor{black}{We detect HF lines in L/T transition objects for the first time, extending the regime in which HF has previously been detected: a young directly imaged super-Jupiter, young late M dwarfs and field L4--L5 brown dwarfs \citep{2024A&A...689A.212G, 2024AJ....168..246Z, 2025A&A...694A.164M}.}

    \item Our 70-day RV monitoring campaign of the Luhman 16 binary did not yield any significant planet detections. We rule out planets with $M\sin{i}>0.2 M_{\mathrm{J}}$ in $<$1-day orbital periods $>$0.4$M_{\mathrm{J}}$ in $<$10-day periods around Luhman 16A, and $M\sin{i}>0.3 M_{\mathrm{J}}$ in $<$1-day orbital periods $>$0.7$M_{\mathrm{J}}$ in $<$10-day periods around Luhman 16B. 
\end{itemize}

\section*{Acknowledgements}

We acknowledge the support of the Canadian Space Agency through the Flights and Fieldwork for the Advancement of Science and Technology (FAST) program (reference No. 21FAUWOB12) and the Natural Sciences and Engineering Research Council of Canada (NSERC) Discovery Grants program, RGPIN-2019-06822.

This paper contains data based on observations obtained at the international Gemini Observatory (Program ID GS-2018A-Q-304), a program of NSF's NOIRLab, which is managed by the Association of Universities for Research in Astronomy (AURA) under a cooperative agreement with the National Science Foundation on behalf of the Gemini Observatory partnership: the National Science Foundation (United States), National Research Council (Canada), Agencia Nacional de Investigaci\'{o}n y Desarrollo (Chile), Ministerio de Ciencia, Tecnolog\'{i}a e Innovaci\'{o}n (Argentina), Minist\'{e}rio da Ci\^{e}ncia, Tecnologia, Inova\c{c}\~{o}es e Comunica\c{c}\~{o}es (Brazil), and Korea Astronomy and Space Science Institute (Republic of Korea).

This work used the Immersion Grating Infrared Spectrometer (IGRINS) that was developed under a collaboration between the University of Texas at Austin and the Korea Astronomy and Space Science Institute (KASI) with the financial support of the US National Science Foundation under grants AST-1229522 and AST-1702267, of the University of Texas at Austin, and of the Korean GMT Project of KASI. We thank the IGRINS team for their continued support throughout this project.

PAMP acknowledges the use of the grant RYC2021-031173-I funded by MCIN/AEI/10.13039/501100011033 and by the 'European Union NextGenerationEU/PRTR' as well as by the grant MDM-2017-0737.
RK acknowledges partial support from ANID’s FONDECYT Regular grant \#1240249 and ANID’s Millennium Science Initiative through grants ICN12 009 and AIM23-0001.

\redsep{The authors declare no competing interests.}

\section*{Data Availability}

The raw IGRINS data for Luhman 16AB are available on the Gemini Archive under Program ID GS-2018A-Q-114. 
The data underlying this article are available in [repository name, e.g. the Dryad Digital Repository], at https://dx.doi.org/[doi]
All reduced spectra and the best-fit Bobcat Alternative B model spectra\redsep{, along with best-fit parameters for each order in Figures \ref{fig:wavvsgbp_A} and \ref{fig:wavvsgbp_B} and RV analysis results in Figures \ref{fig:crx} and \ref{fig:rvs},}
are available as supplementary files, 
in Zenodo at \url{https://doi.org/10.5281/zenodo.15001024}, or
in Harvard Dataverse at \url{https://doi.org/10.7910/DVN/TKY3KC}.

\bibliographystyle{mnras}


\appendix

\section{Information of each exposure} \label{sec:observationtable}
Table~\ref{tab:observations} contains the details of each exposure during the \redsep{70-day}
observation.

\onecolumn 
\begin{longtable}{lcccccccc}
    \caption{ 
        Gemini South/IGRINS Spectroscopic Observations of Luhman 16AB under Gemini program ID GS-2018A-Q-114 (PI: S. Metchev). The given SNR values are for the final, combined spectra.	
	All exposure times are 300 s, except for the latter two exposures of 2018 05 26, which were 600 s each.
	}\label{tab:observations}\\  
    \hline
      & & & & \multicolumn{2}{c}{Luhman 16A} & \multicolumn{2}{c}{Luhman 16B} & \\
    Date & Slit & Target & Telluric & $H$-band & $K$-band &  $H$-band & $K$-band & Julian day of \\
    observed & pos. & airmass & A0V & SNR (at & SNR (at &  SNR (at & SNR (at & mid-exposure \\
    & & & standard & 1.589 µm) & 2.101 µm) & 1.589 µm) & 2.101 µm) &  \\
    \hline
    \endhead
        2018 04 21 & A & 1.17 & HIP 41373 & 73 & 71 & 76 & 66 & 2458230.482323 \\ 
        2018 04 21 & B & 1.16 & HIP 41373 & 76 & 74 & 79 & 70 & 2458230.486176 \\ 
        2018 04 21 & A & 1.17 & HIP 55019 & 70 & 68 & 73 & 63 & 2458230.641519 \\ 
        2018 04 21 & B & 1.18 & HIP 55019 & 64 & 61 & 67 & 56 & 2458230.645374 \\ 
        2018 04 22 & A & 1.13 & HIP 61557 & 99 & 97 & 100 & 89 & 2458231.503751 \\ 
        2018 04 22 & B & 1.12 & HIP 61557 & 105 & 101 & 105 & 93 & 2458231.507600 \\ 
        2018 04 22 & A & 1.18 & HIP 52407 & 56 & 54 & 59 & 50 & 2458231.646716 \\ 
        2018 04 22 & B & 1.19 & HIP 52407 & 74 & 73 & 79 & 72 & 2458231.650565 \\ 
        2018 04 23 & A & 1.17 & HIP 53836 & 72 & 70 & 73 & 63 & 2458232.481006 \\ 
        2018 04 23 & B & 1.16 & HIP 53836 & 77 & 76 & 77 & 68 & 2458232.481006 \\ 
        2018 04 24 & A & 1.15 & HIP 53836 & 80 & 77 & 85 & 74 & 2458233.482268 \\ 
        2018 04 24 & B & 1.15 & HIP 53836 & 82 & 78 & 86 & 75 & 2458233.486127 \\ 
        2018 04 24 & A & 1.17 & HIP 79229 & 73 & 74 & 77 & 68 & 2458233.632515 \\ 
        2018 04 24 & B & 1.17 & HIP 79229 & 77 & 77 & 81 & 71 & 2458233.636476 \\ 
        2018 04 26 & A & 1.18 & HIP 53836 & 73 & 69 & 76 & 65 & 2458235.461515 \\ 
        2018 04 26 & B & 1.18 & HIP 53836 & 74 & 71 & 77 & 67 & 2458235.465363 \\ 
        2018 04 26 & A & 1.40 & HIP 53836 & 52 & 48 & 56 & 48 & 2458235.695114 \\ 
        2018 04 26 & B & 1.42 & HIP 53836 & 60 & 55 & 66 & 55 & 2458235.698980 \\ 
        2018 04 27 & A & 1.16 & HIP 53836 & 79 & 77 & 81 & 72 & 2458236.472315 \\ 
        2018 04 27 & B & 1.16 & HIP 53836 & 82 & 79 & 83 & 74 & 2458236.472315 \\ 
        2018 04 27 & A & 1.31 & HIP 53836 & 66 & 61 & 65 & 54 & 2458236.673351 \\ 
        2018 04 27 & B & 1.33 & HIP 53836 & 72 & 66 & 60 & 55 & 2458236.677218 \\ 
        2018 04 28 & A & 1.22 & HIP 53836 & 34 & 34 & 33 & 29 & 2458237.644928 \\ 
        2018 04 28 & B & 1.23 & HIP 53836 & 23 & 22 & 22 & 17 & 2458237.648777 \\ 
        2018 05 02 & A & 1.15 & HIP 53836 & 62 & 59 & 63 & 57 & 2458241.462518 \\ 
        2018 05 02 & B & 1.14 & HIP 53836 & 66 & 65 & 70 & 63 & 2458241.466366 \\ 
        2018 05 02 & A & 1.28 & HIP 53836 & 52 & 47 & 53 & 42 & 2458241.652099 \\ 
        2018 05 02 & B & 1.30 & HIP 53836 & 49 & 44 & 50 & 39 & 2458241.655941 \\ 
        2018 05 03 & A & 1.16 & HIP 53836 & 85 & 80 & 85 & 74 & 2458242.455635 \\ 
        2018 05 03 & B & 1.15 & HIP 53836 & 84 & 79 & 85 & 74 & 2458242.459502 \\ 
        2018 05 03 & A & 1.26 & HIP 53836 & 81 & 79 & 84 & 75 & 2458242.644415 \\ 
        2018 05 03 & B & 1.28 & HIP 53836 & 79 & 78 & 82 & 73 & 2458242.648266 \\ 
        2018 05 04 & A & 1.13 & HIP 53836 & 83 & 79 & 85 & 75 & 2458243.466576 \\ 
        2018 05 04 & B & 1.13 & HIP 53836 & 84 & 80 & 87 & 76 & 2458243.470438 \\ 
        2018 05 04 & A & 1.19 & HIP 53836 & 80 & 75 & 80 & 69 & 2458243.617409 \\ 
        2018 05 04 & B & 1.20 & HIP 53836 & 88 & 82 & 91 & 78 & 2458243.621260 \\ 
        2018 05 05 & A & 1.12 & HIP 53836 & 52 & 49 & 63 & 44 & 2458244.475526 \\ 
        2018 05 05 & B & 1.11 & HIP 53836 & 60 & 57 & 61 & 52 & 2458244.479377 \\ 
        2018 05 05 & A & 1.22 & HIP 53836 & 74 & 70 & 74 & 64 & 2458244.624893 \\ 
        2018 05 05 & B & 1.23 & HIP 53836 & 72 & 68 & 72 & 62 & 2458244.628746 \\ 
        2018 05 07 & A & 1.12 & HIP 53836 & 36 & 34 & 35 & 30 & 2458246.473160 \\ 
        2018 05 07 & B & 1.11 & HIP 53836 & 40 & 38 & 40 & 33 & 2458246.473160 \\ 
        2018 05 21 & A & 1.10 & HIP 53836 & 84 & 79 & 87 & 76 & 2458260.453650 \\ 
        2018 05 21 & B & 1.09 & HIP 53836 & 86 & 82 & 88 & 78 & 2458260.457504 \\ 
        2018 05 21 & A & 1.27 & HIP 53836 & 78 & 73 & 82 & 71 & 2458260.599531 \\ 
        2018 05 21 & B & 1.28 & HIP 53836 & 73 & 67 & 74 & 64 & 2458260.599531 \\ 
        2018 05 22 & A & 1.09 & HIP 53836 & 83 & 75 & 87 & 73 & 2458261.455328 \\ 
        2018 05 22 & B & 1.09 & HIP 53836 & 80 & 74 & 84 & 71 & 2458261.459198 \\ 
        2018 05 22 & A & 1.32 & HIP 53836 & 75 & 67 & 67 & 56 & 2458261.610750 \\ 
        2018 05 22 & B & 1.33 & HIP 53836 & 75 & 67 & 67 & 56 & 2458261.610750 \\ 
        2018 05 23 & A & 1.10 & HIP 53836 & 77 & 77 & 76 & 72 & 2458262.449235 \\ 
        2018 05 23 & B & 1.09 & HIP 53836 & 68 & 71 & 67 & 65 & 2458262.453122 \\ 
        2018 05 23 & A & 1.33 & HIP 53836 & 71 & 67 & 74 & 63 & 2458262.611687 \\ 
        2018 05 23 & B & 1.35 & HIP 53836 & 78 & 72 & 82 & 69 & 2458262.611687 \\ 
        2018 05 24 & A & 1.09 & HIP 53836 & 92 & 83 & 96 & 81 & 2458263.454796 \\ 
        2018 05 24 & B & 1.09 & HIP 53836 & 89 & 80 & 93 & 78 & 2458263.454796 \\ 
        2018 05 24 & A & 1.39 & HIP 53836 & 78 & 74 & 82 & 70 & 2458263.618141 \\ 
        2018 05 24 & B & 1.41 & HIP 53836 & 77 & 72 & 82 & 68 & 2458263.622030 \\ 
        2018 05 25 & A & 1.09 & HIP 53836 & 76 & 73 & 77 & 67 & 2458264.454588 \\ 
        2018 05 25 & B & 1.09 & HIP 53836 & 78 & 74 & 79 & 68 & 2458264.454588 \\ 
        2018 05 25 & A & 1.42 & HIP 53836 & 68 & 63 & 70 & 57 & 2458264.621200 \\ 
        2018 05 25 & B & 1.44 & HIP 53836 & 79 & 73 & 82 & 68 & 2458264.625079 \\ 
        2018 05 26 & A & 1.09 & HIP 53836 & 72 & 69 & 76 & 66 & 2458265.469989 \\ 
        2018 05 26 & B & 1.09 & HIP 53836 & 67 & 65 & 72 & 62 & 2458265.473883 \\ 
        2018 05 26 & A & 1.62 & HIP 53836 & 74 & 63 & 77 & 57 & 2458265.655182 \\ 
        2018 05 26 & B & 1.68 & HIP 53836 & 72 & 61 & 75 & 55 & 2458265.655182 \\ 
        2018 05 27 & A & 1.10 & HIP 53836 & 59 & 55 & 59 & 50 & 2458266.498218 \\ 
        2018 05 27 & B & 1.11 & HIP 53836 & 59 & 55 & 61 & 51 & 2458266.502085 \\ 
        2018 05 27 & A & 1.54 & HIP 53836 & 54 & 48 & 52 & 41 & 2458266.634776 \\ 
        2018 05 27 & B & 1.56 & HIP 53836 & 54 & 47 & 53 & 40 & 2458266.638647 \\ 
        2018 06 02 & A & 1.09 & HIP 53836 & 39 & 39 & 42 & 36 & 2458272.463061 \\ 
        2018 06 02 & B & 1.09 & HIP 53836 & 50 & 49 & 52 & 45 & 2458272.466923 \\ 
        2018 06 02 & A & 1.61 & HIP 53836 & 47 & 44 & 49 & 39 & 2458272.628868 \\ 
        2018 06 02 & B & 1.64 & HIP 53836 & 48 & 44 & 50 & 39 & 2458272.632761 \\ 
        2018 06 03 & A & 1.18 & HIP 53836 & 56 & 51 & 57 & 47 & 2458273.530479 \\ 
        2018 06 03 & B & 1.19 & HIP 53836 & 44 & 41 & 47 & 36 & 2458273.534409 \\ 
        2018 06 03 & A & 2.11 & HIP 53836 & 45 & 41 & 50 & 39 & 2458273.673587 \\ 
        2018 06 03 & B & 2.16 & HIP 53836 & 39 & 36 & 45 & 34 & 2458273.677470 \\ 
        2018 06 20 & A & 1.13 & HIP 53836 & 41 & 36 & 41 & 32 & 2458290.455811 \\ 
        2018 06 20 & B & 1.13 & HIP 53836 & 40 & 33 & 39 & 29 & 2458290.459661 \\ 
        2018 06 21 & A & 1.12 & HIP 53836 & 62 & 59 & 67 & 57 & 2458291.447734 \\ 
        2018 06 21 & B & 1.13 & HIP 53836 & 71 & 68 & 77 & 68 & 2458291.451685 \\ 
        2018 06 22 & A & 1.11 & HIP 53836 & 43 & 39 & 45 & 39 & 2458292.443055 \\ 
        2018 06 22 & B & 1.12 & HIP 53836 & 43 & 40 & 47 & 40 & 2458292.443055 \\ 
        2018 06 23 & B & 1.12 & HIP 53836 & 29 & 27 & 33 & 27 & 2458293.444549 \\ 
        2018 06 25 & A & 1.15 & HIP 53836 & 65 & 60 & 69 & 54 & 2458295.453290 \\ 
        2018 06 25 & B & 1.15 & HIP 53836 & 72 & 66 & 78 & 63 & 2458295.457146 \\ 
        2018 06 26 & A & 1.15 & HIP 53836 & 83 & 78 & 89 & 77 & 2458296.455378 \\ 
        2018 06 26 & B & 1.16 & HIP 53836 & 84 & 78 & 90 & 76 & 2458296.459242 \\ 
        2018 06 28 & A & 1.16 & HIP 53836 & 38 & 33 & 42 & 32 & 2458298.454360 \\ 
        2018 06 28 & B & 1.17 & HIP 53836 & 37 & 34 & 42 & 33 & 2458298.458221 \\ 
        2018 06 29 & A & 1.16 & HIP 53836 & 31 & 27 & 33 & 26 & 2458299.453755 \\ 
        2018 06 29 & B & 1.17 & HIP 53836 & 46 & 35 & 35 & 21 & 2458299.456855 \\ 
    \hline
\end{longtable}
\twocolumn

\section{The 1.48--2.48 micron \texorpdfstring{$R$=28\,000}{R=28000} spectral atlas and best-fit models of Luhman 16A and B} \label{sec:atlas}
The high-SNR weighted-average-combined IGRINS spectrum and the best-fit model for all orders are presented in Figures ~\ref{fig:atlasA} for Luhman 16A and in Figures ~\ref{fig:atlasB} for Luhman 16B.

\begin{figure*}
    \centering
    \includegraphics[height=0.43\textheight]{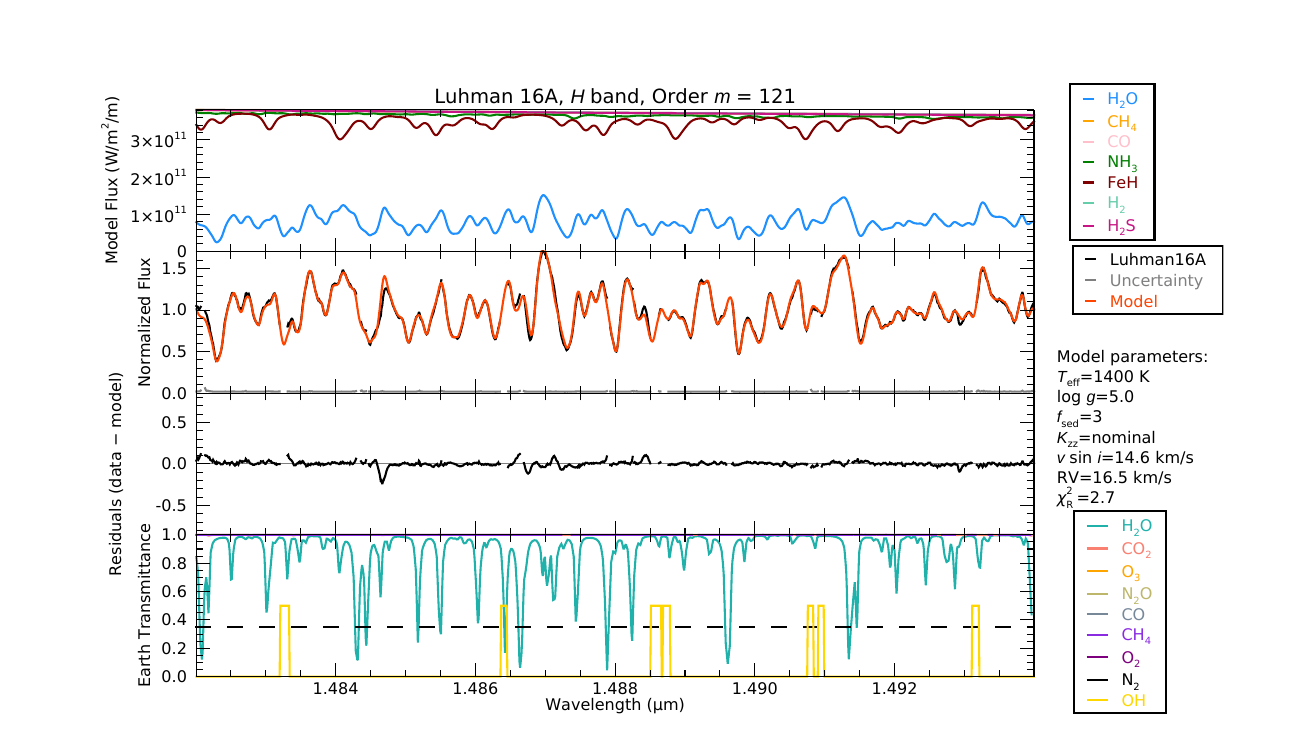} 
    \includegraphics[height=0.43\textheight]{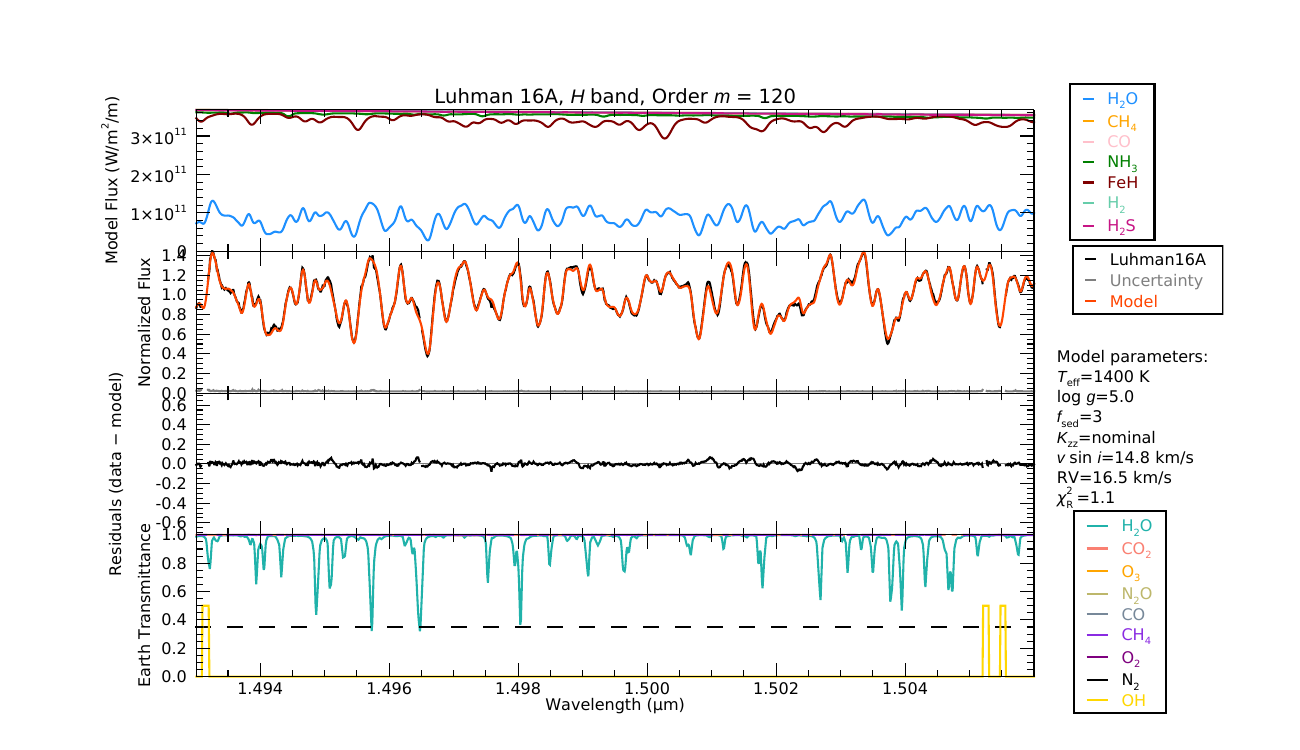}
    \caption{The spectral atlas of Luhman 16A over the full wavelength range of IGRINS from 1.48 to 2.48 µm.
        Each figure shows the spectrum in each echelle order. 
        The physical parameters of the best-fit model for each order are written in the margins between the legend boxes on the right side of the panels.
        The top panel indicates the contribution of individual molecular species to the model spectrum  (CIA is also included). 
        They are calculated with the global best-fit parameters for the corresponding object, not the local best-fit parameters for that echelle order.
        The second panel from the top shows the IGRINS data with the best-fit Bobcat Alternative B model for the order. 
        The third panel from the top shows the residuals (data $-$ model) on the same $y$-axis scale as the panel above it.
        The bottom panel shows the PSG Earth’s transmittance to help assess the telluric lines in our spectra. 
        The positions of OH emission lines are shown as yellow boxes with wide boxes indicating blended OH lines. The box height is uniform, not indicating line strength. 
        The horizontal dashed line indicates the 35 percent trasmittance threshold used to select the telluric-affected region to mask (see Section \ref{sec:modelfits}).
        In the wavelength regions where detected or potential H$_2$ and H$_2$S absorption lines are present (e.g., orders $m = $84 and 113), the two central panels include both cases where the models are with and without the line opacity of the molecule. 
        }\label{fig:atlasA}
\end{figure*} 

\begin{figure*}
    \ContinuedFloat
    \centering
    \includegraphics[height=0.43\textheight]{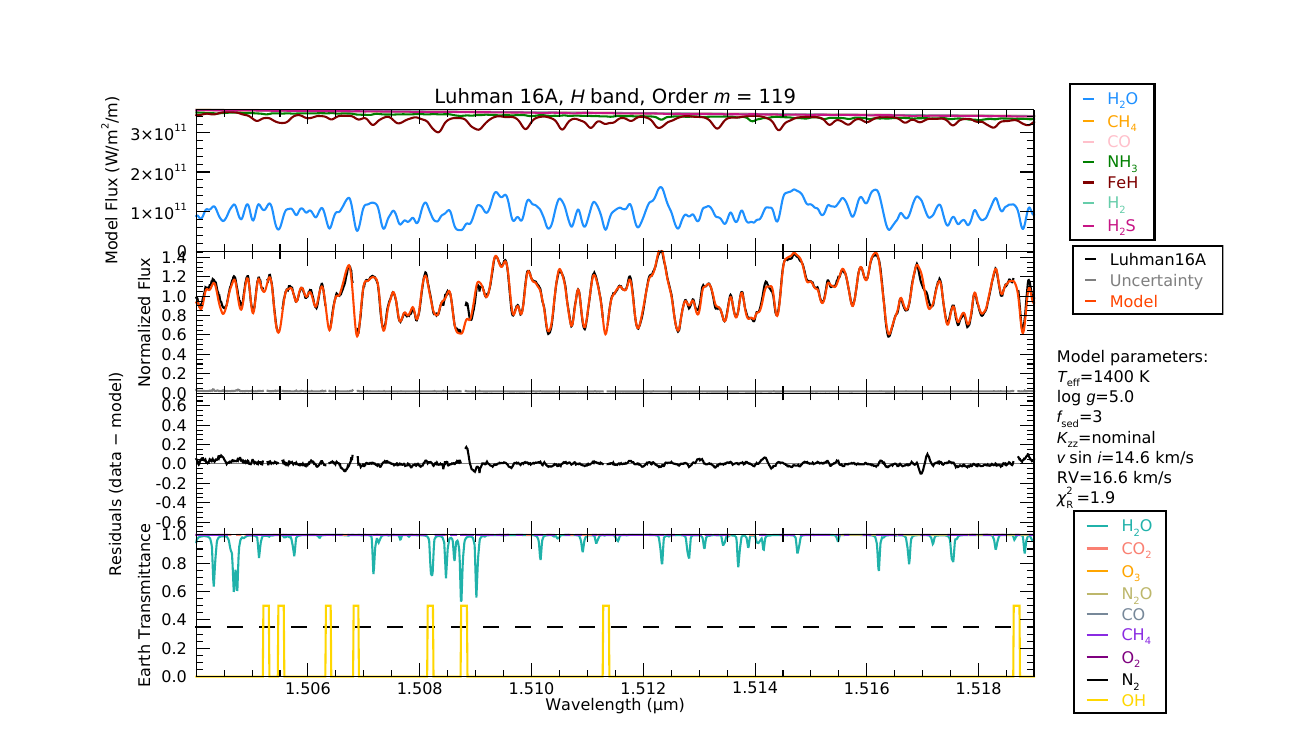}
    \includegraphics[height=0.43\textheight]{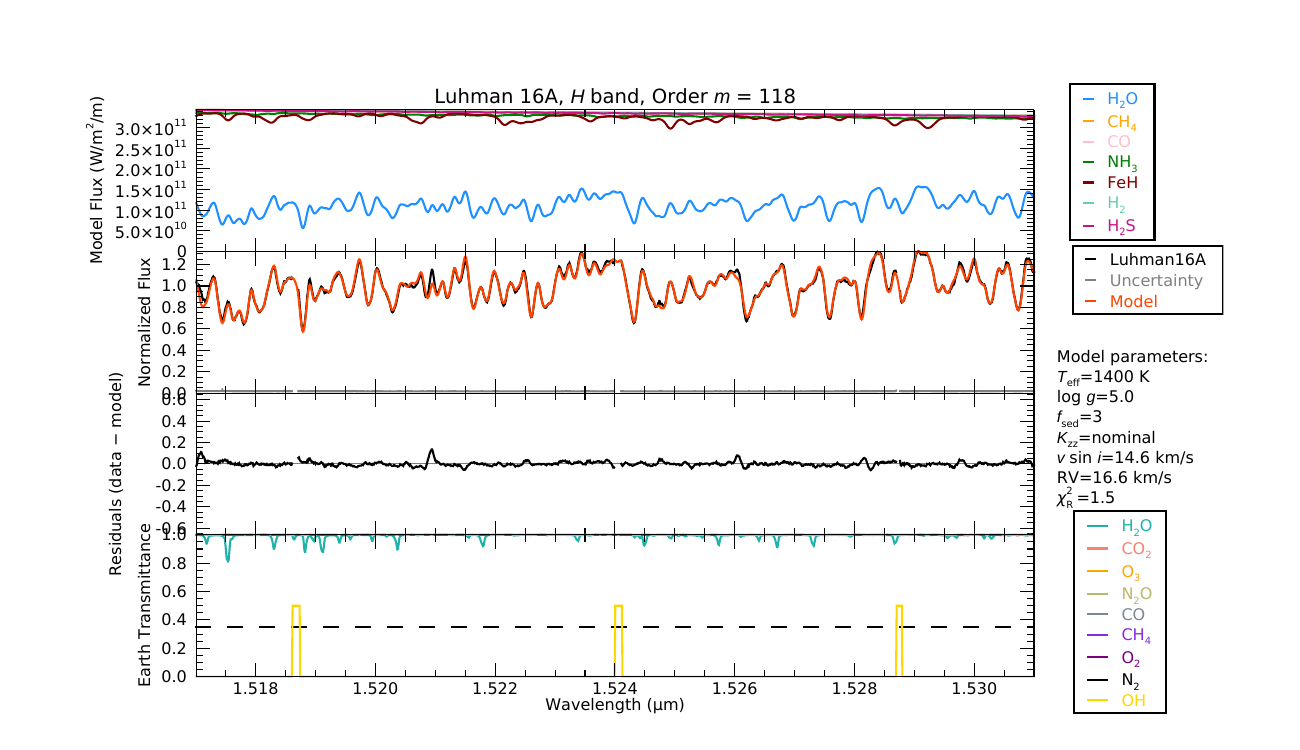}
    \caption{Continued.}
\end{figure*}

\begin{figure*}
    \ContinuedFloat
    \centering
    \includegraphics[height=0.43\textheight]{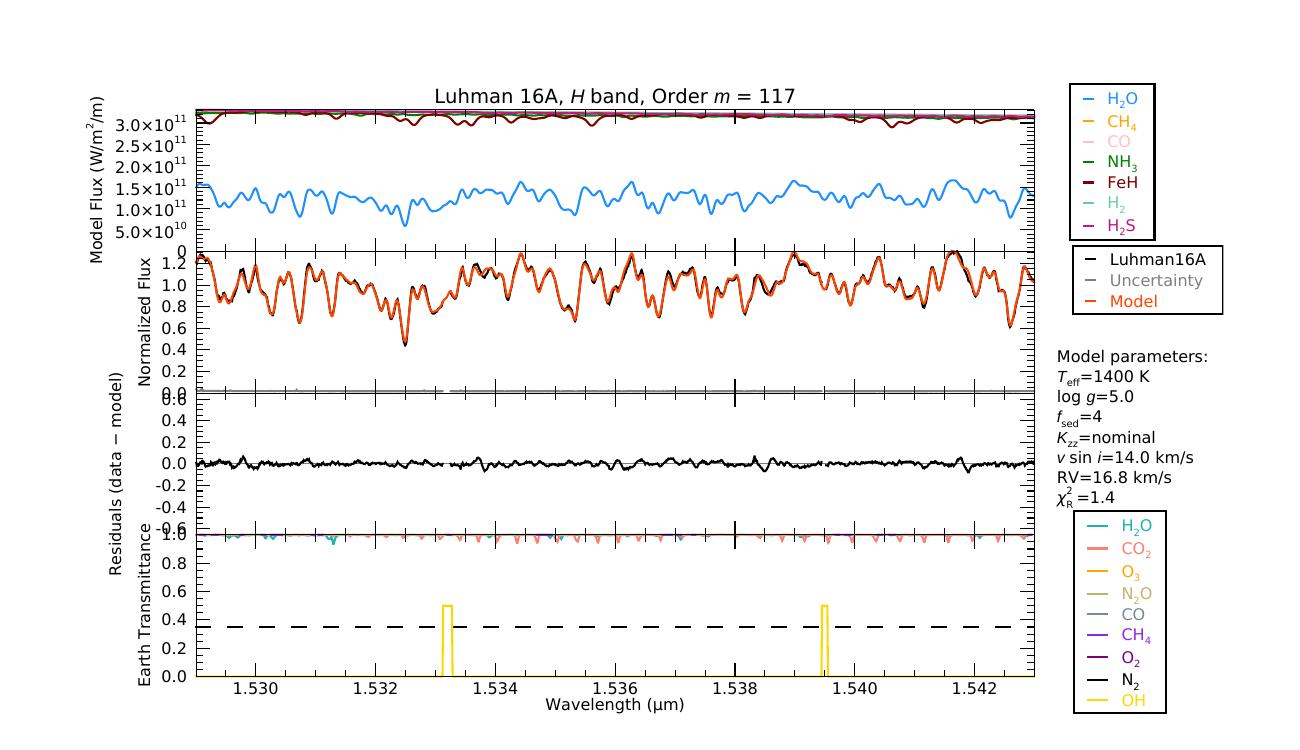}
    \includegraphics[height=0.43\textheight]{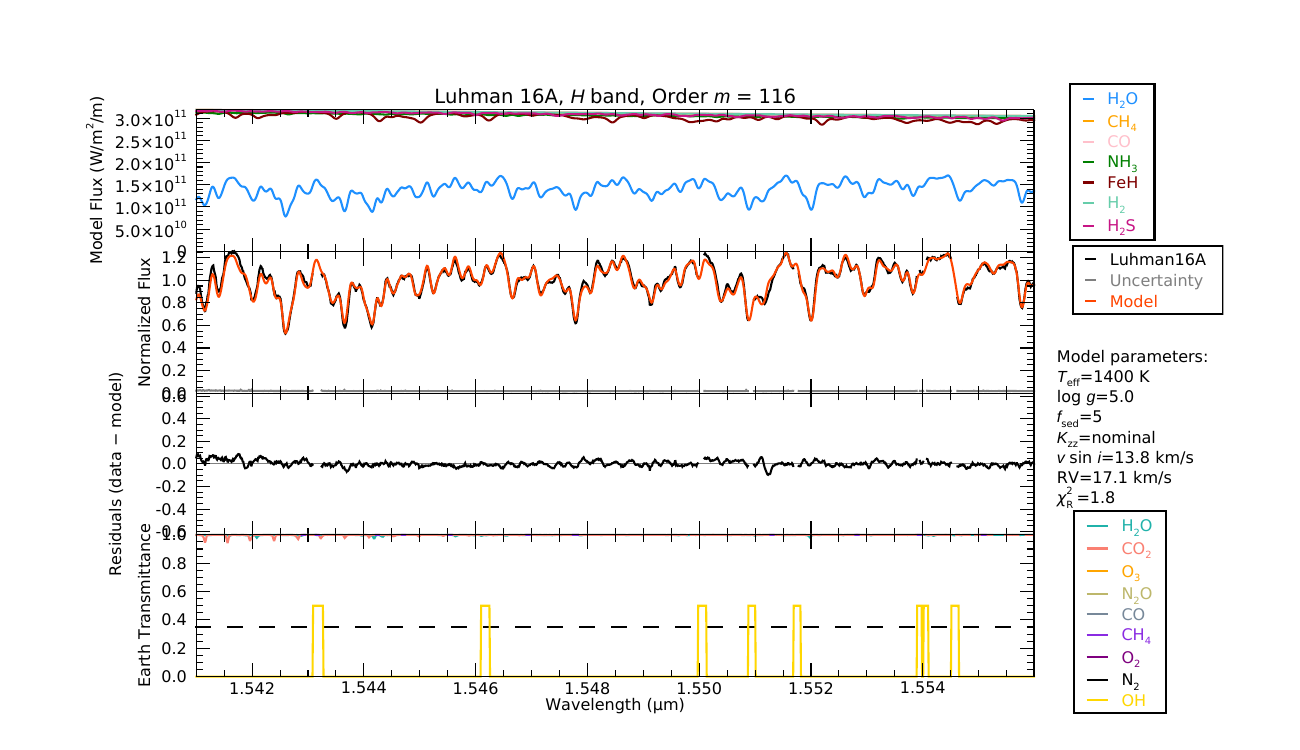}
    \caption{Continued.}
\end{figure*}

\begin{figure*}
    \ContinuedFloat
    \centering
    \includegraphics[height=0.43\textheight]{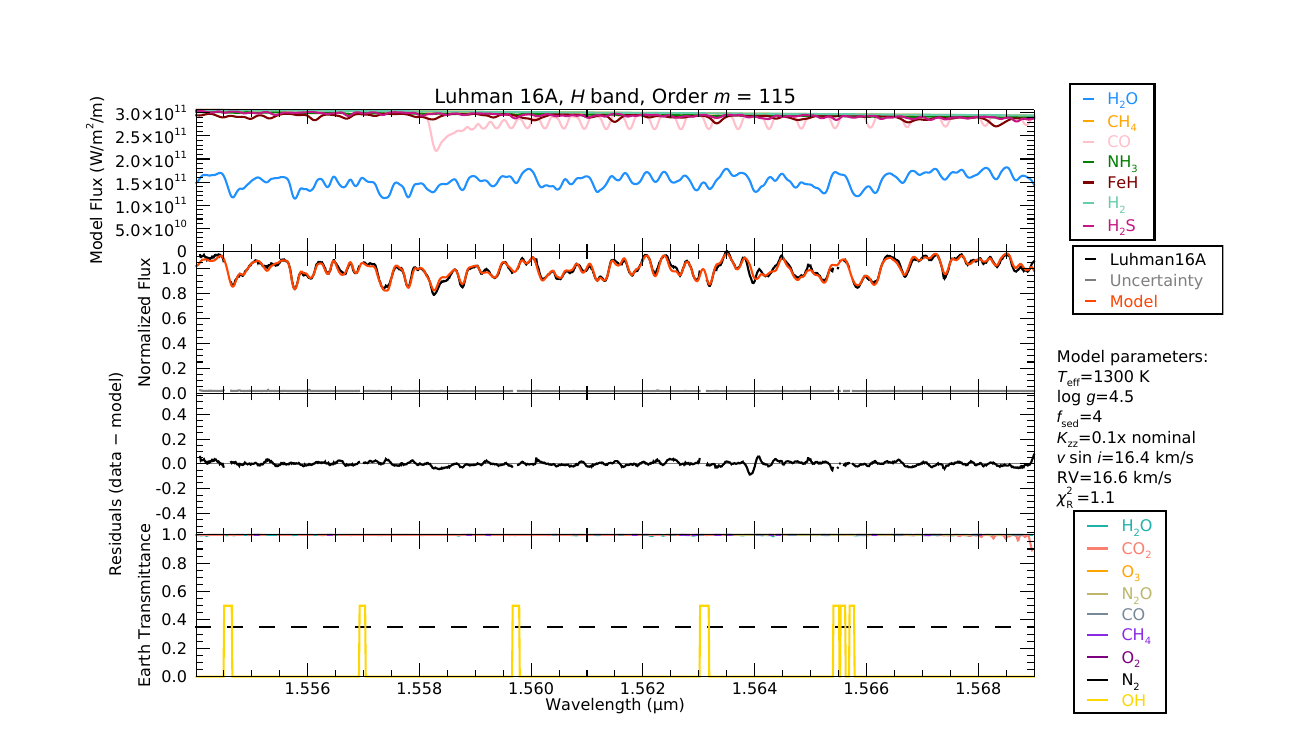}
    \includegraphics[height=0.43\textheight]{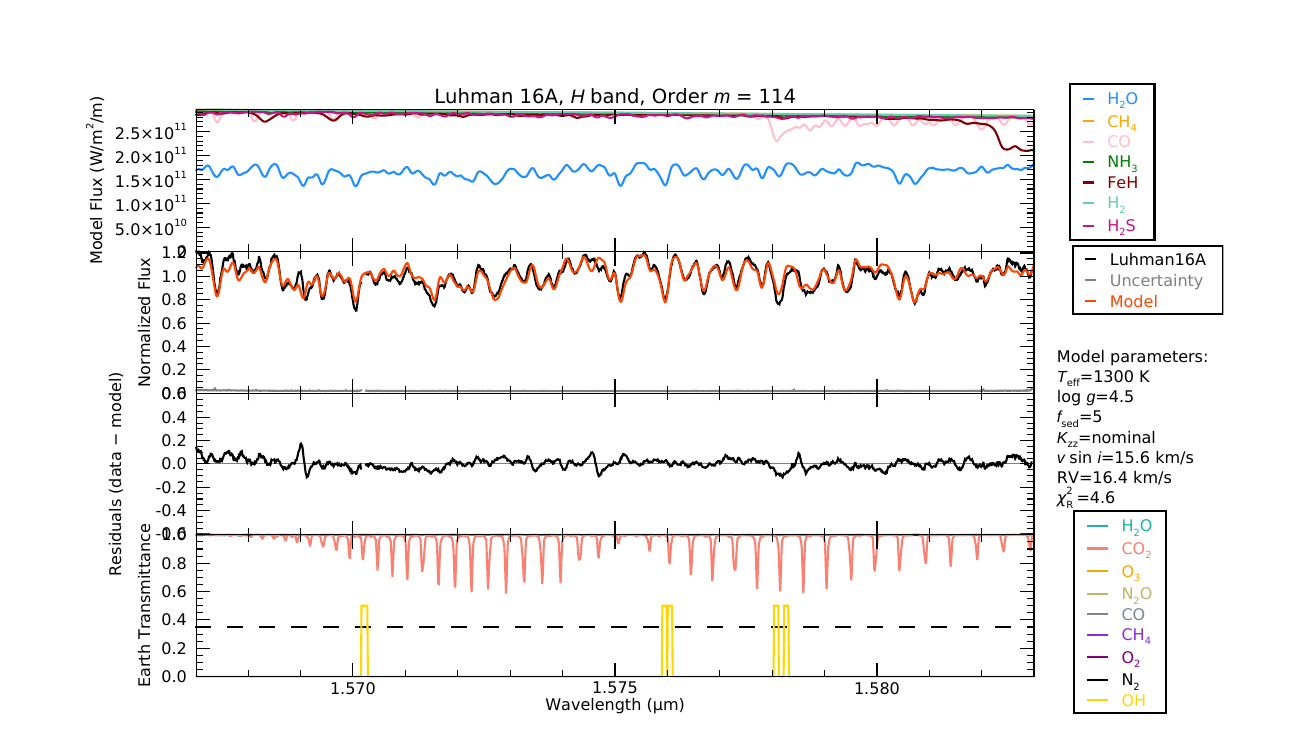}
    \caption{Continued.}
\end{figure*}

\begin{figure*}
    \ContinuedFloat
    \centering
    \includegraphics[height=0.43\textheight]{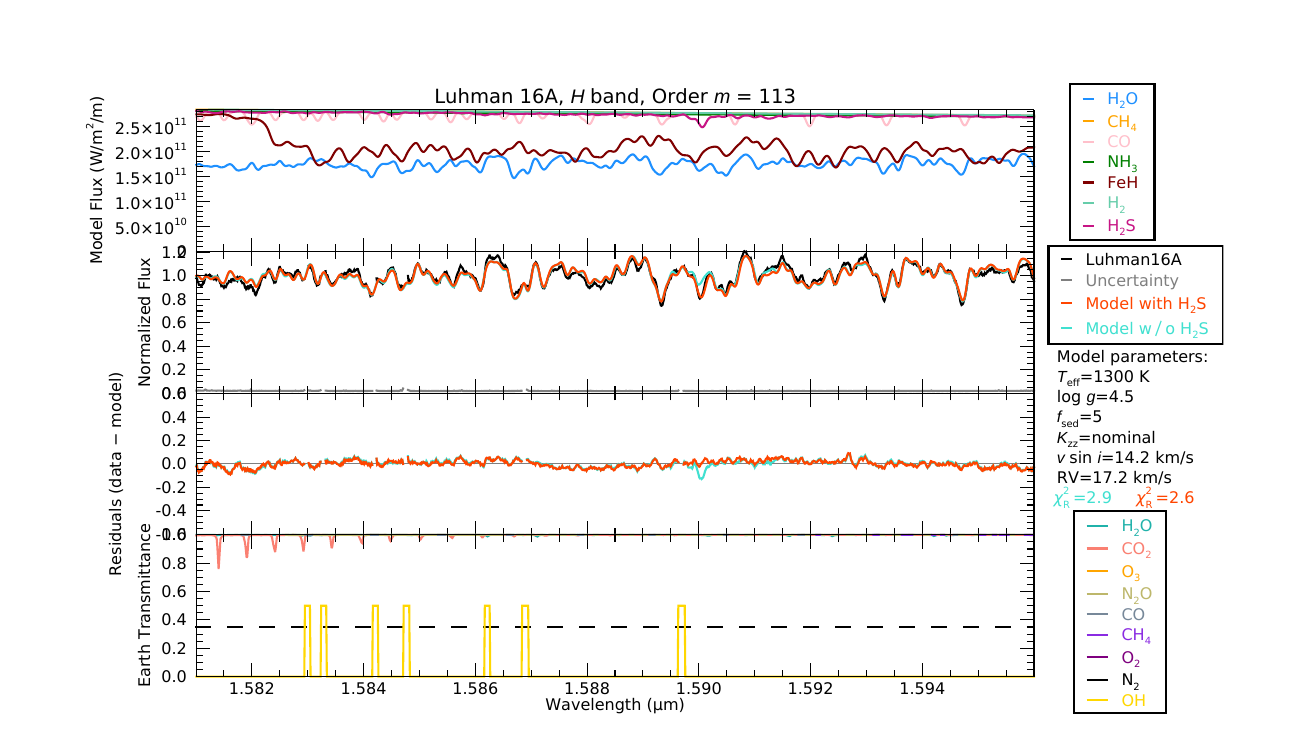}
    \includegraphics[height=0.43\textheight]{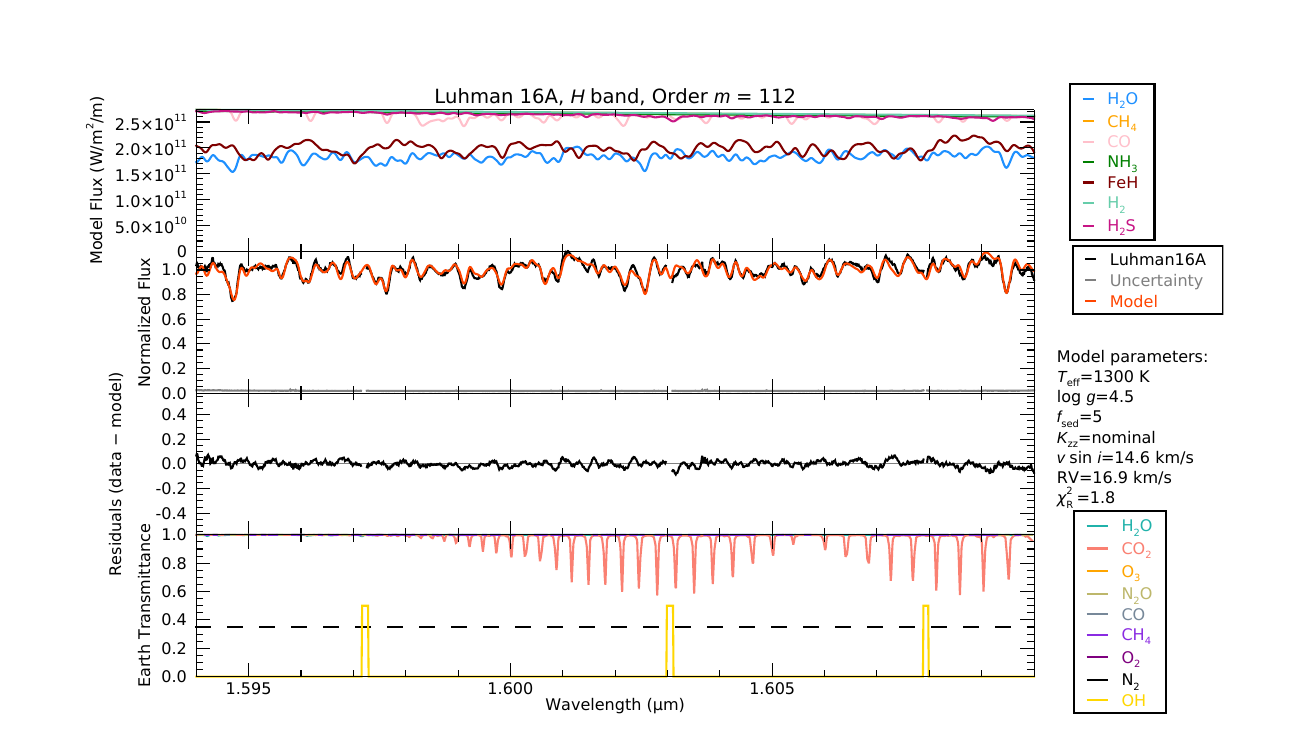}
    \caption{Continued.}
\end{figure*}

\begin{figure*}
    \ContinuedFloat
    \centering
    \includegraphics[height=0.43\textheight]{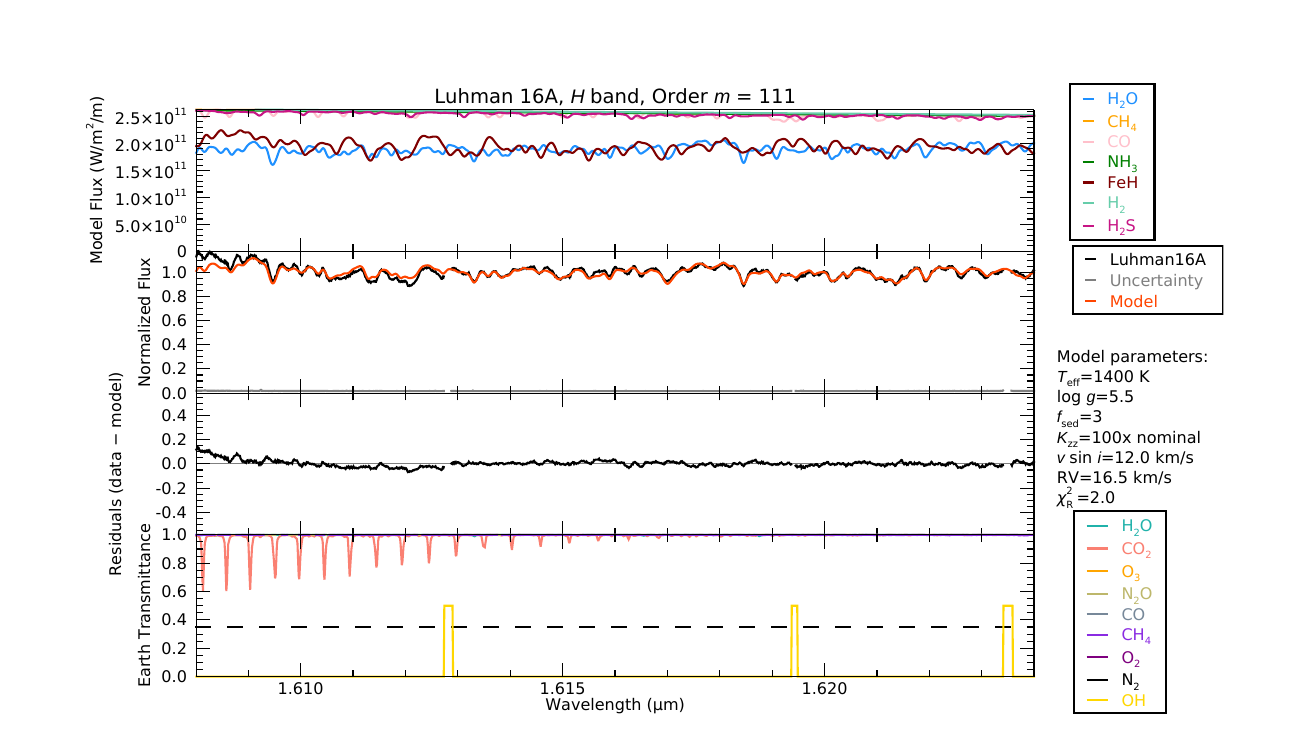}
    \includegraphics[height=0.43\textheight]{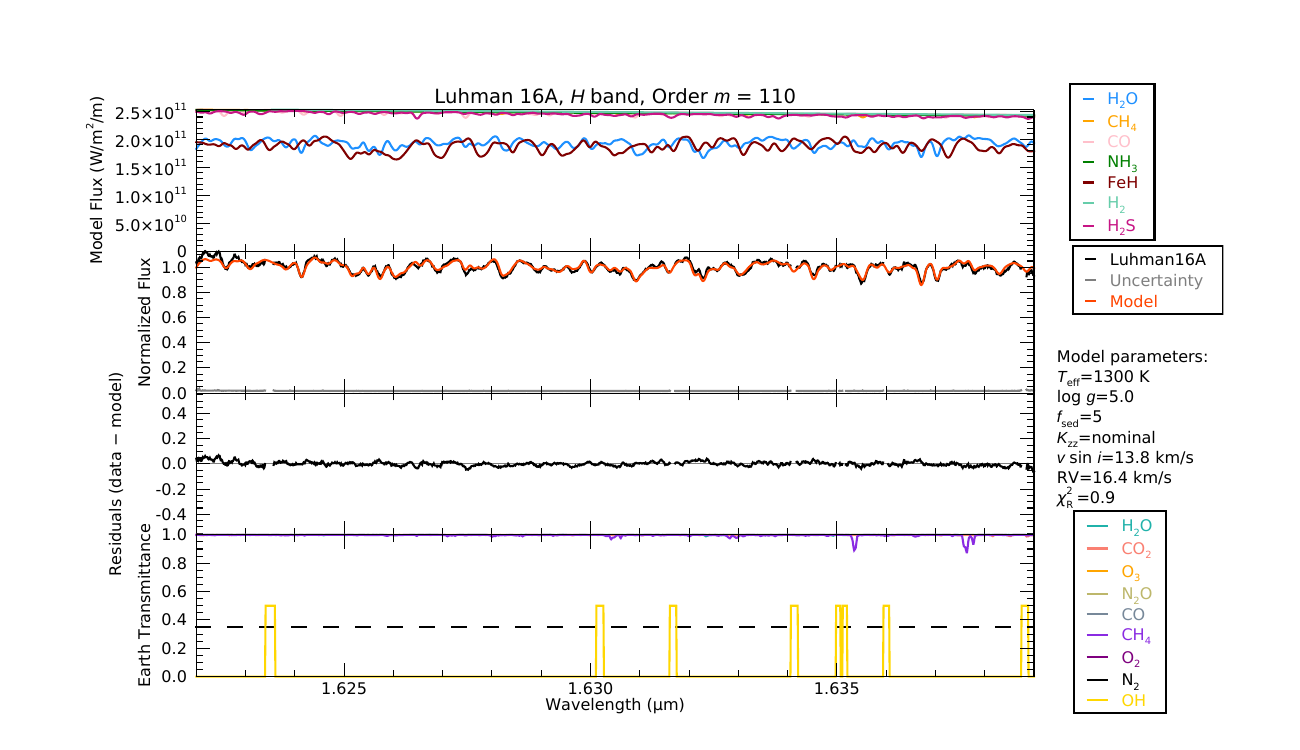}
    \caption{Continued.}
\end{figure*}

\begin{figure*}
    \ContinuedFloat
    \centering
    \includegraphics[height=0.43\textheight]{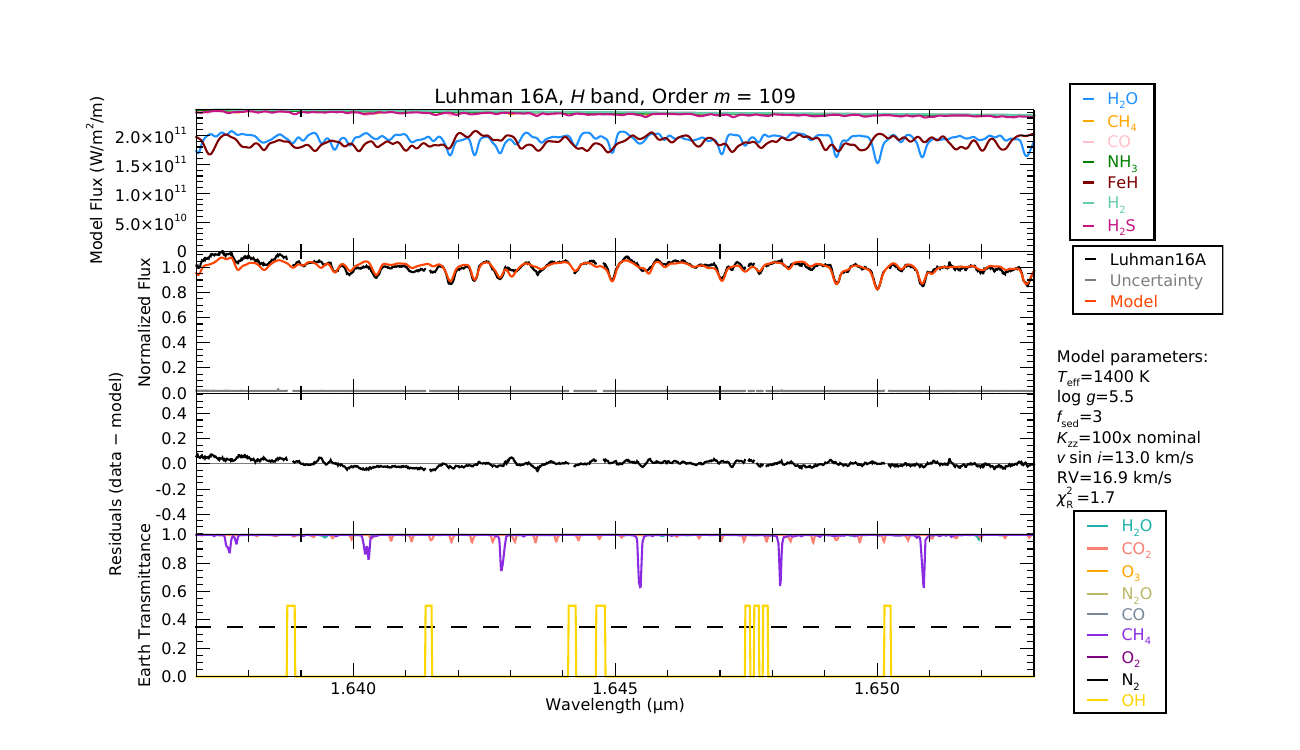}
    \includegraphics[height=0.43\textheight]{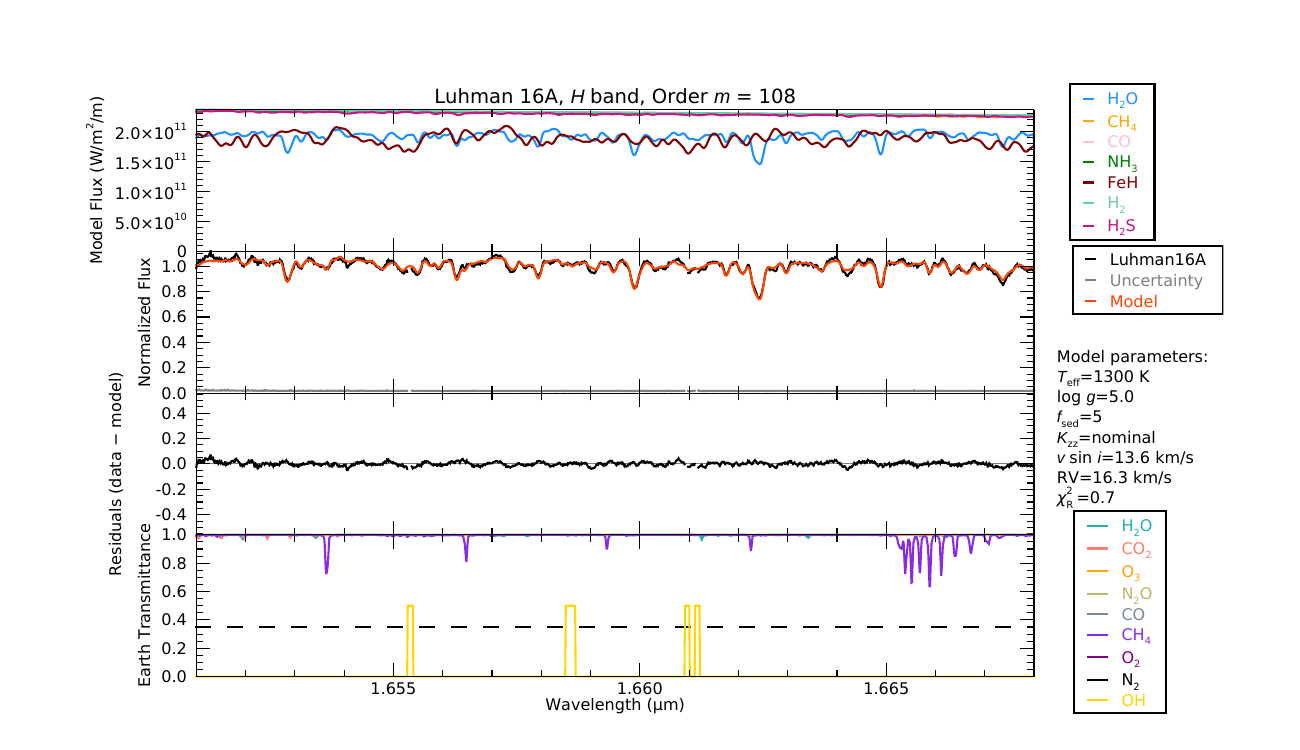}
    \caption{Continued.}
\end{figure*}

\begin{figure*}
    \ContinuedFloat
    \centering
    \includegraphics[height=0.43\textheight]{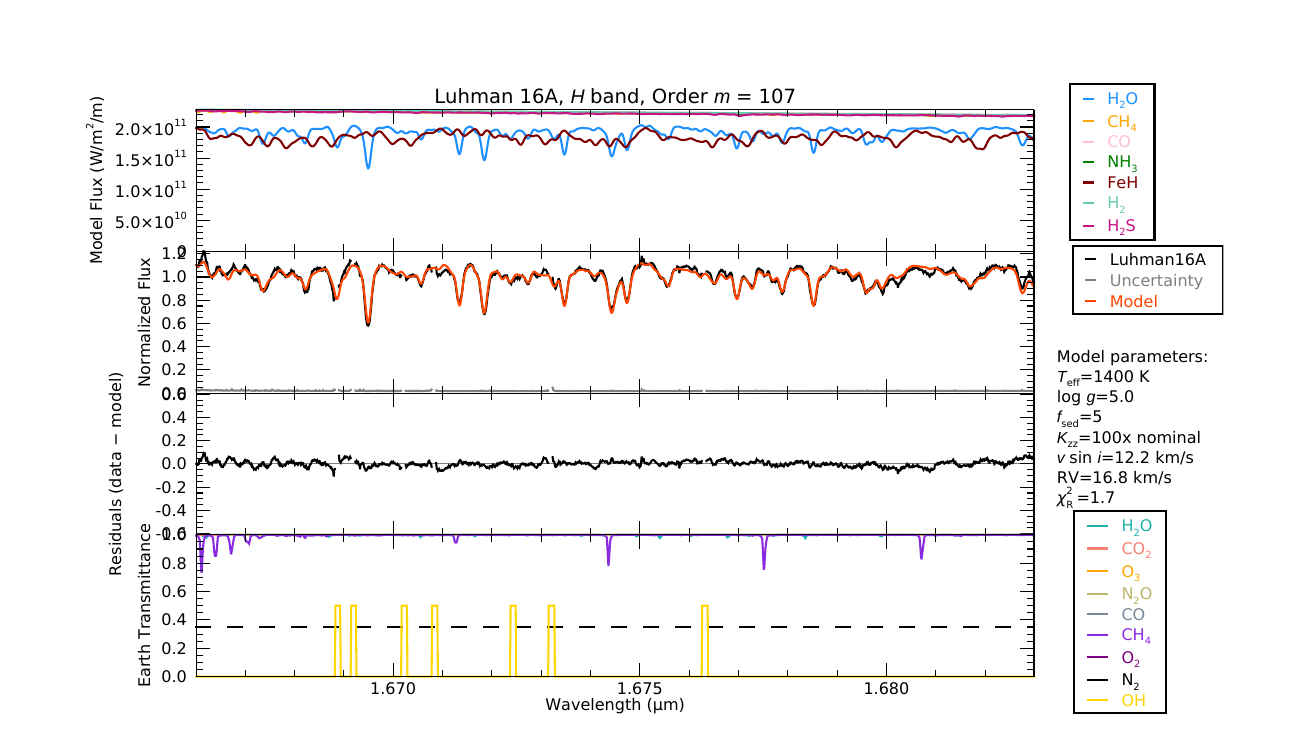}
    \includegraphics[height=0.43\textheight]{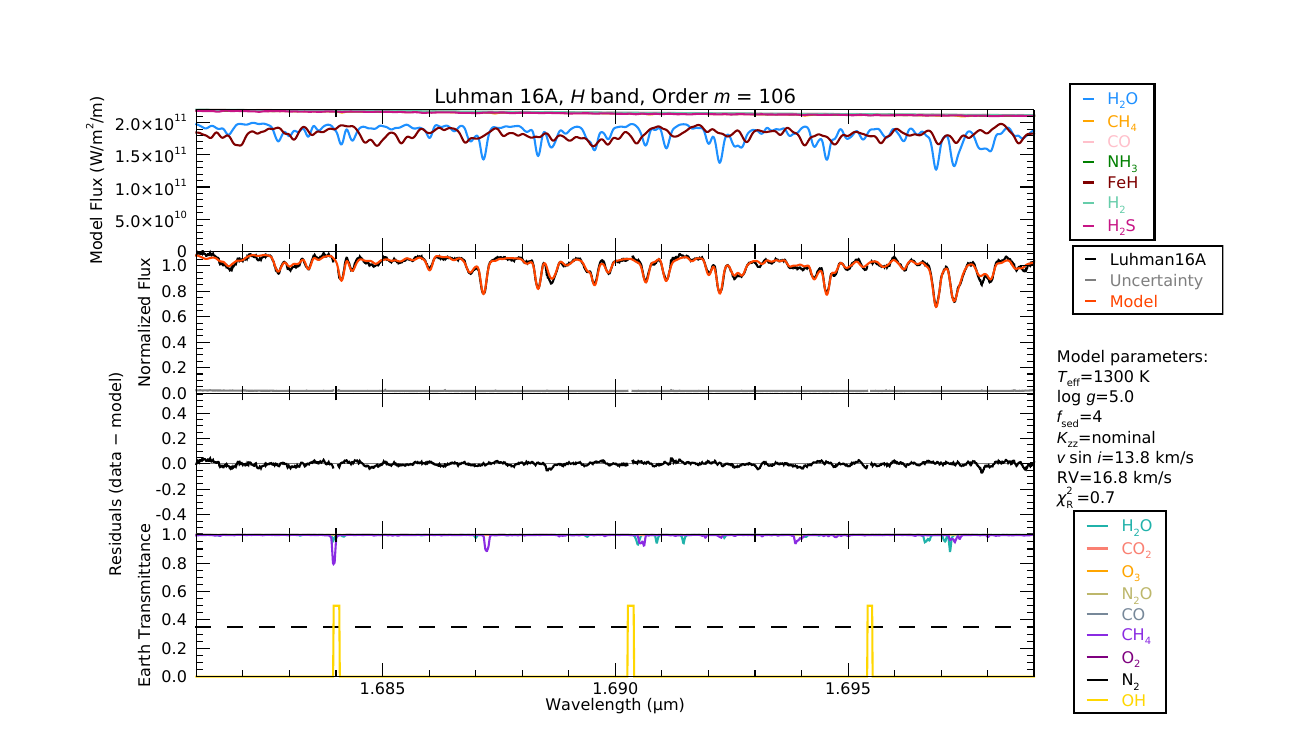}
    \caption{Continued.}
\end{figure*}

\begin{figure*}
    \ContinuedFloat
    \centering
    \includegraphics[height=0.43\textheight]{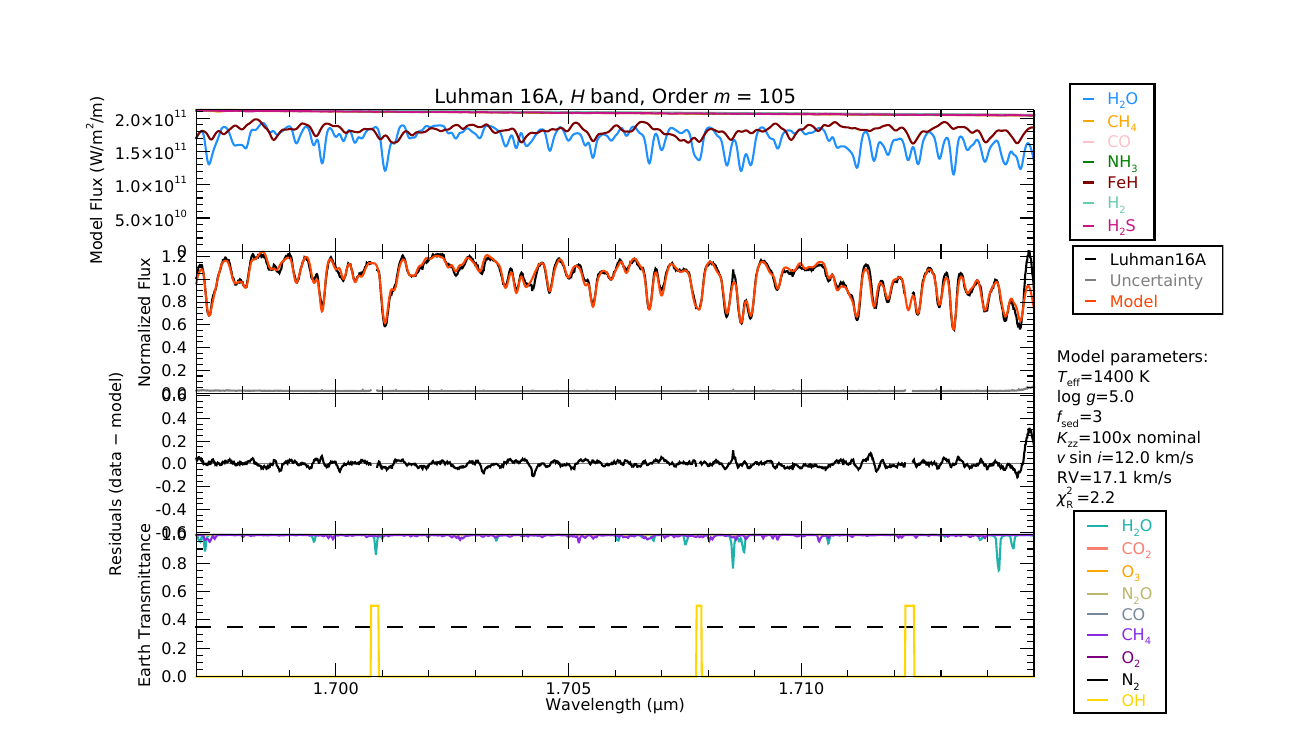}
    \includegraphics[height=0.43\textheight]{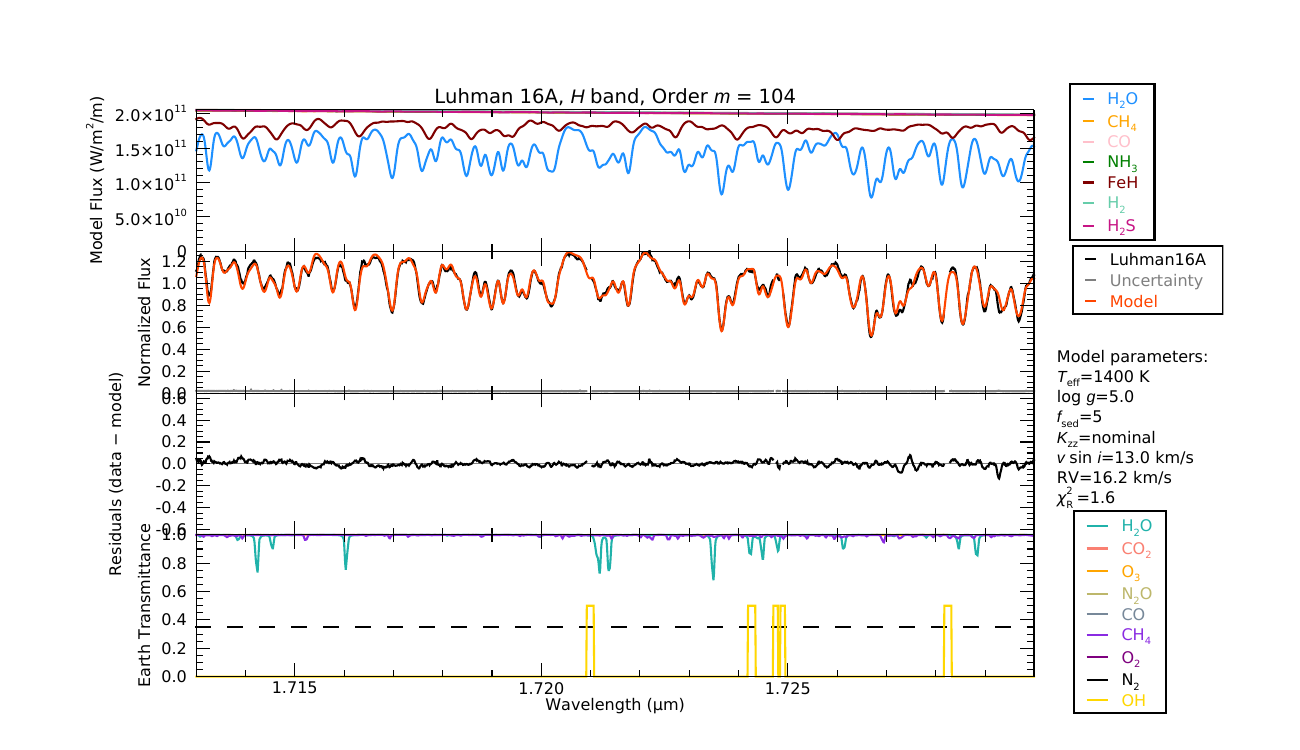}
    \caption{Continued.}
\end{figure*}

\begin{figure*}
    \ContinuedFloat
    \centering
    \includegraphics[height=0.43\textheight]{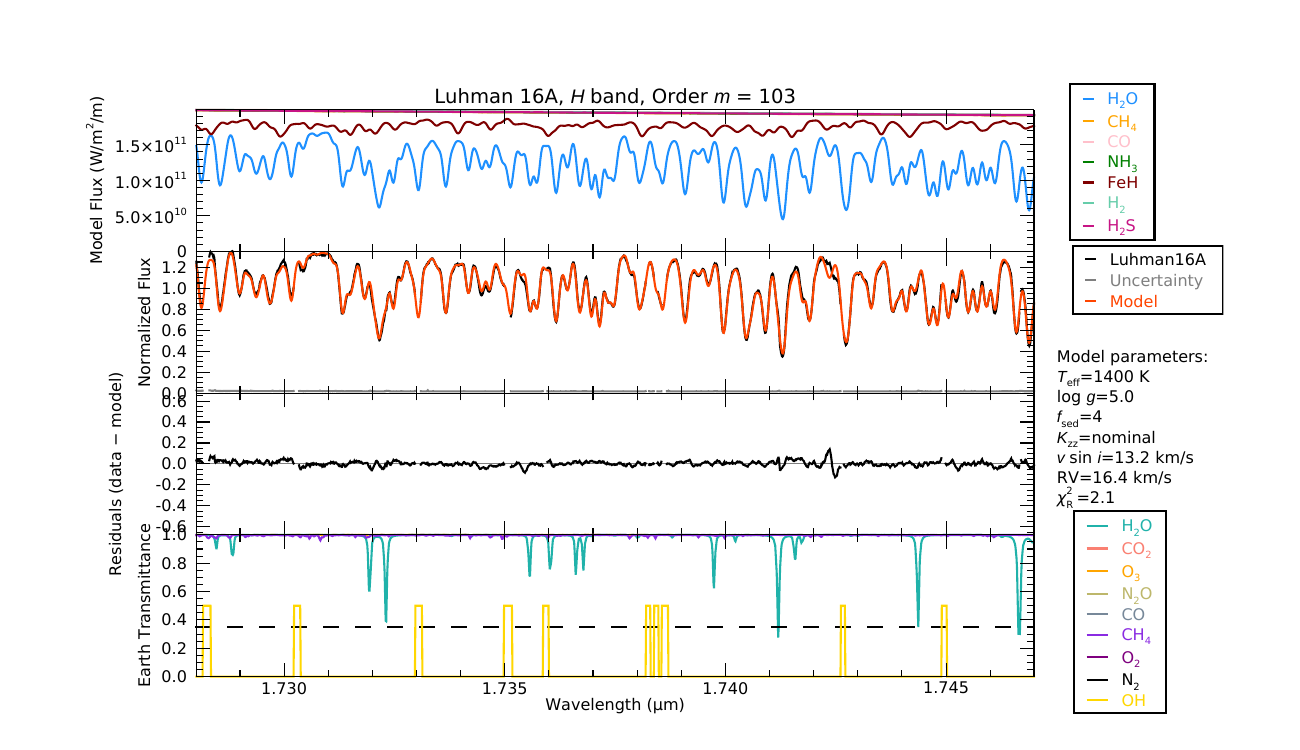}
    \includegraphics[height=0.43\textheight]{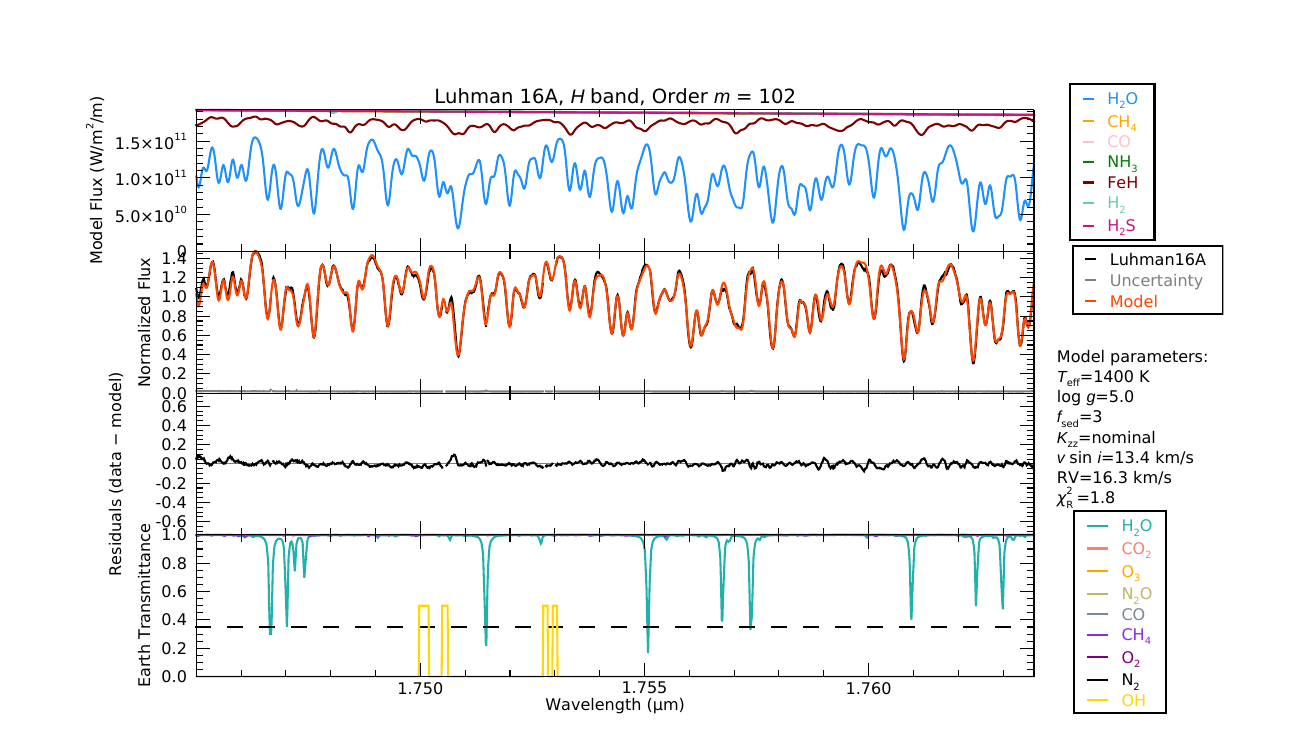}
    \caption{Continued.}
\end{figure*}

\begin{figure*}
    \ContinuedFloat
    \centering
    \includegraphics[height=0.43\textheight]{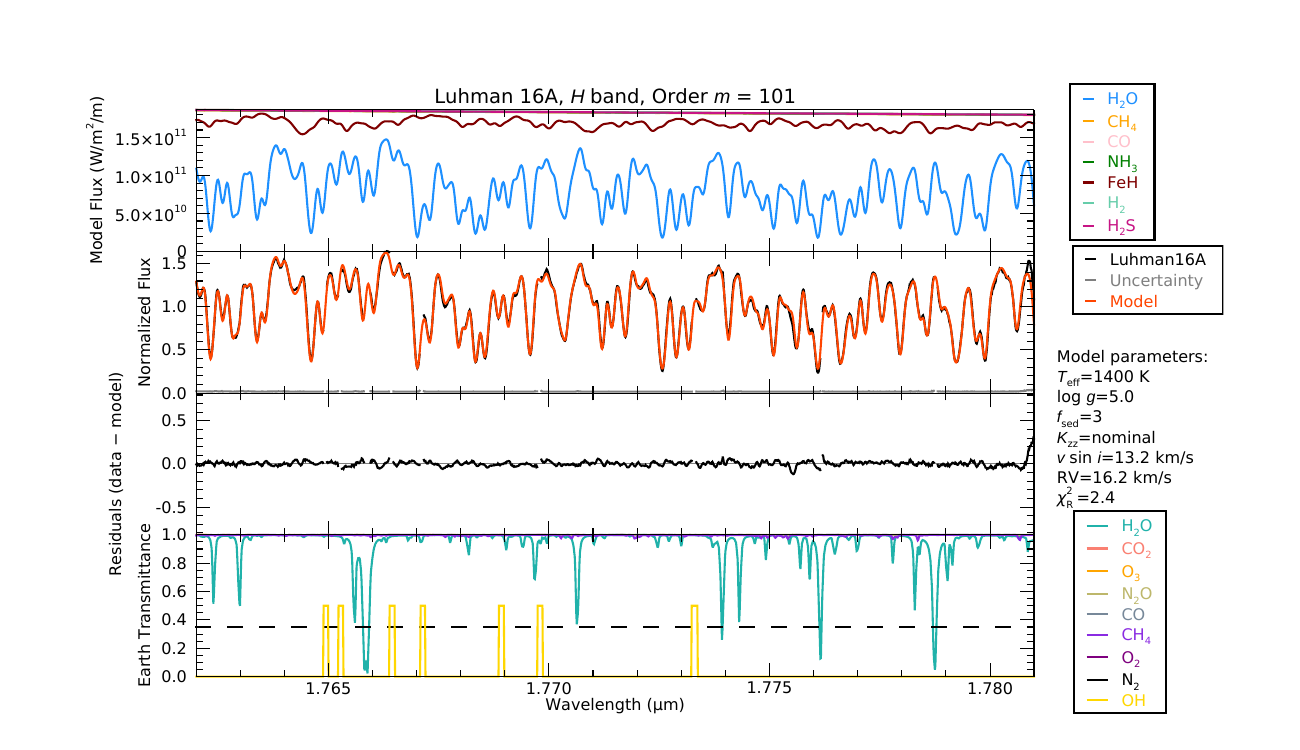}
    \includegraphics[height=0.43\textheight]{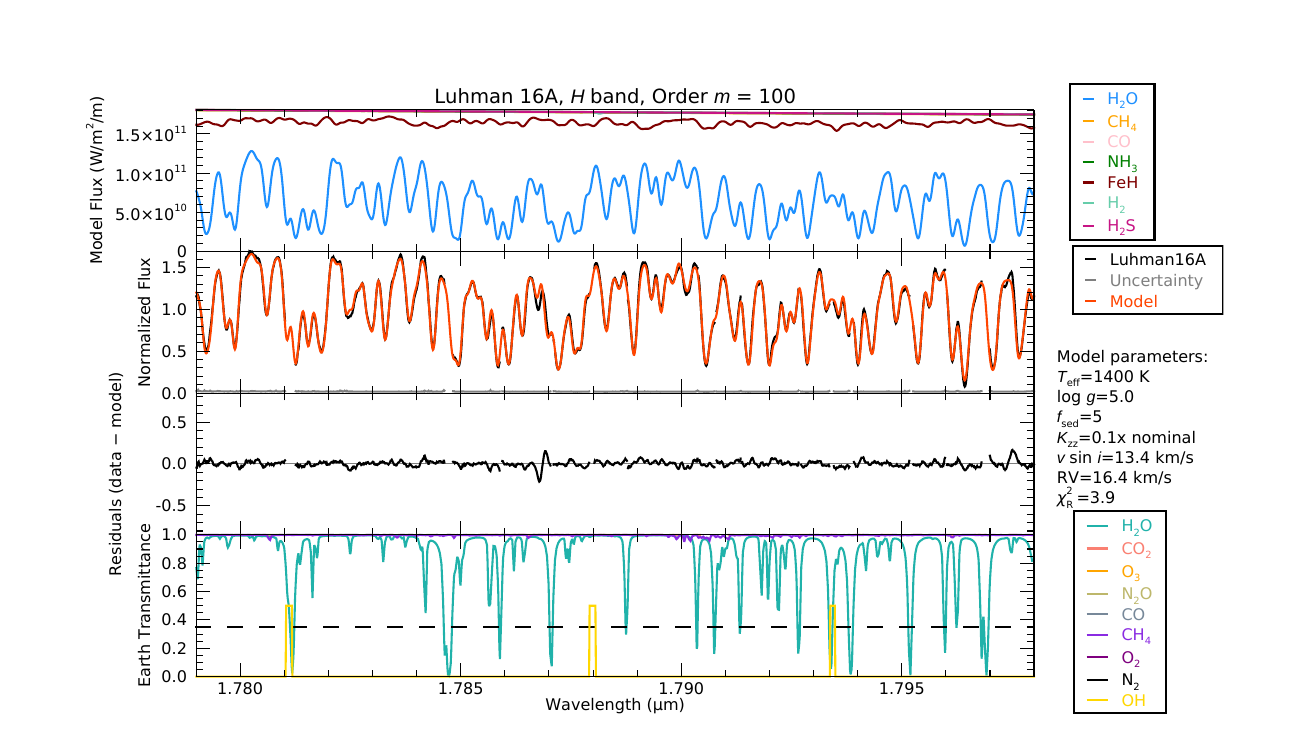}
    \caption{Continued.}
\end{figure*}

\begin{figure*}
    \ContinuedFloat
    \centering
    \includegraphics[height=0.43\textheight]{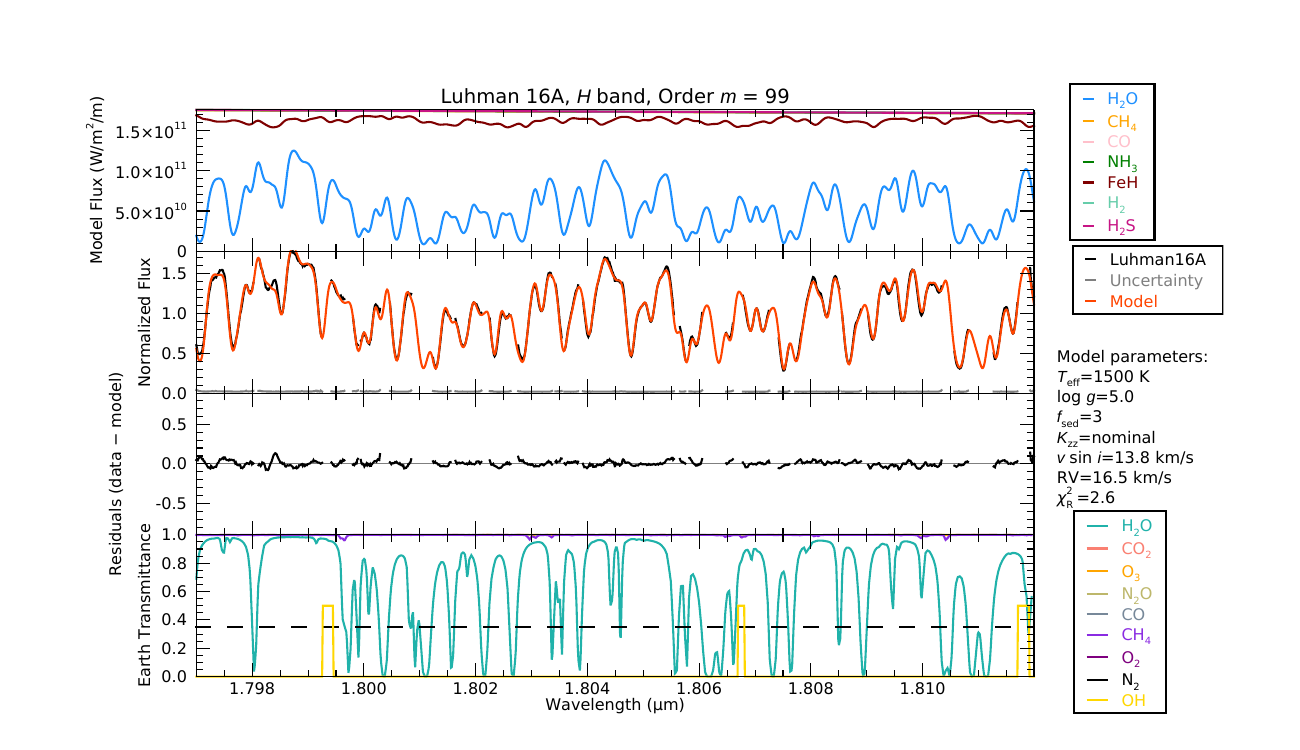}
    \includegraphics[height=0.43\textheight]{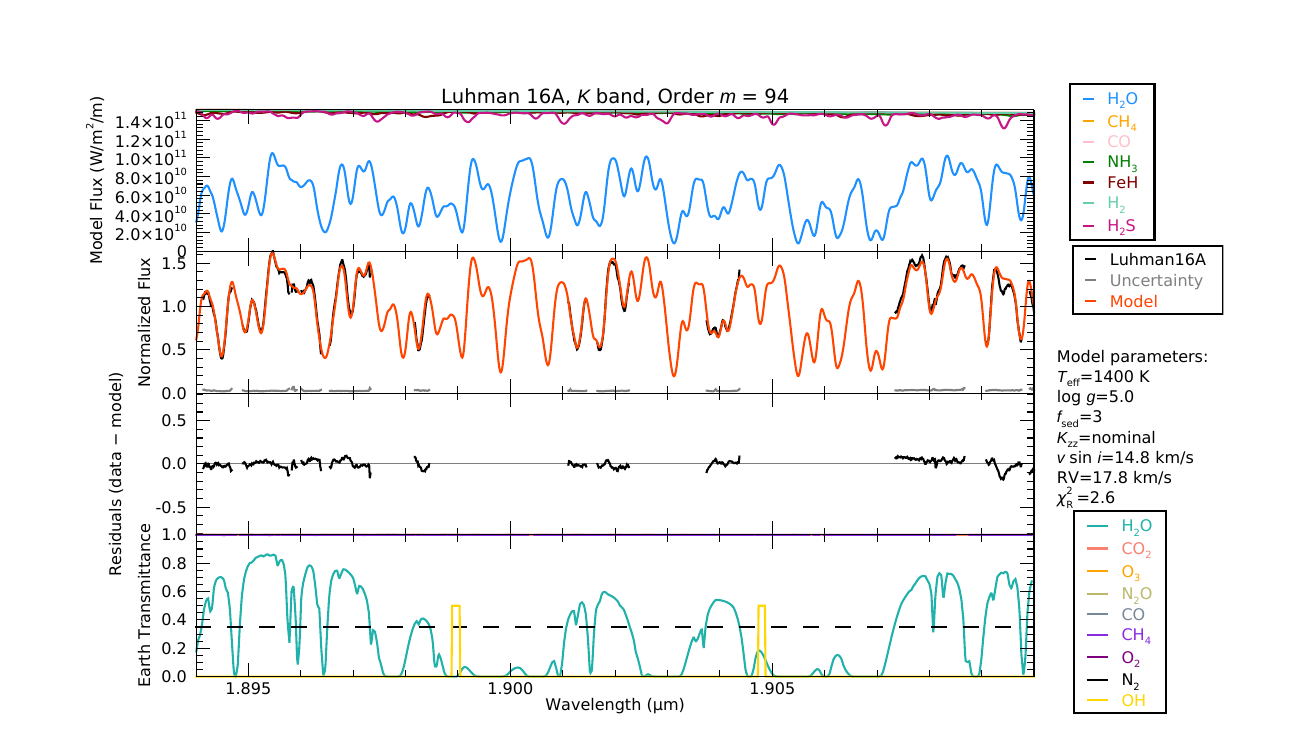}
    \caption{Continued.}
\end{figure*}

\begin{figure*}
    \ContinuedFloat
    \centering
    \includegraphics[height=0.43\textheight]{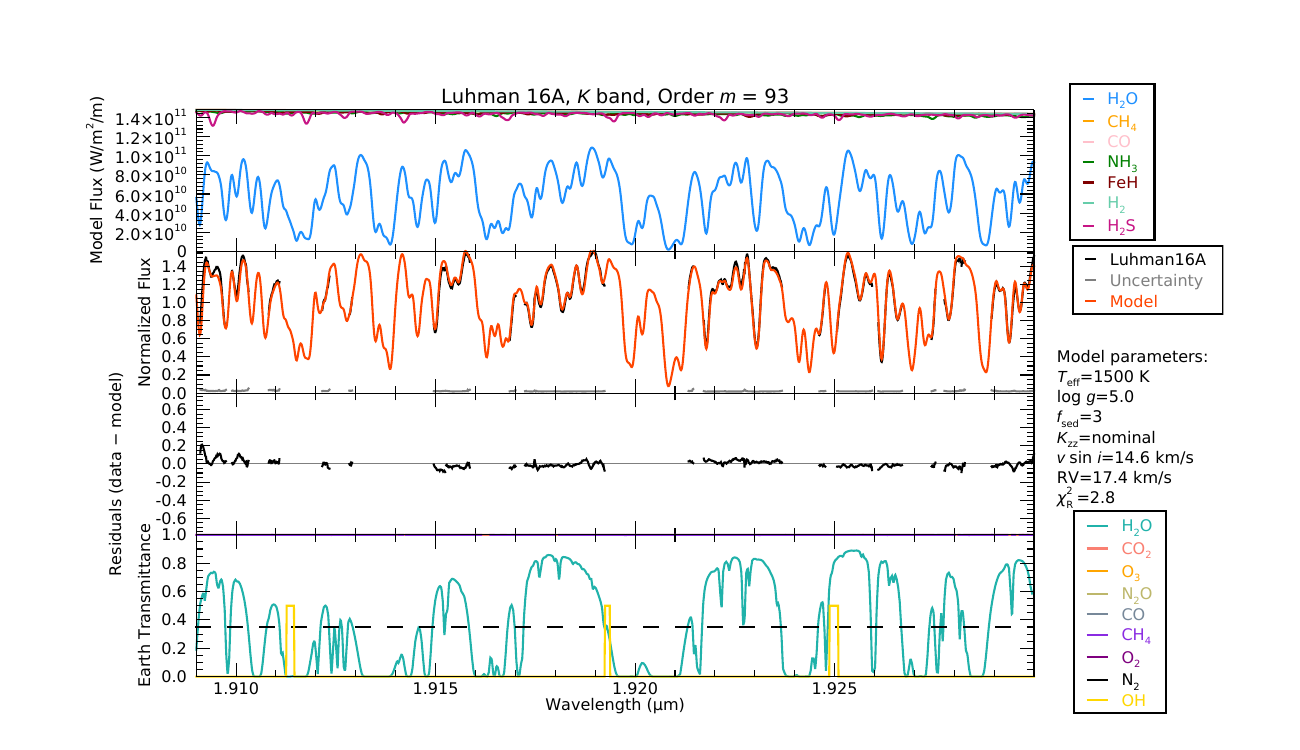}
    \includegraphics[height=0.43\textheight]{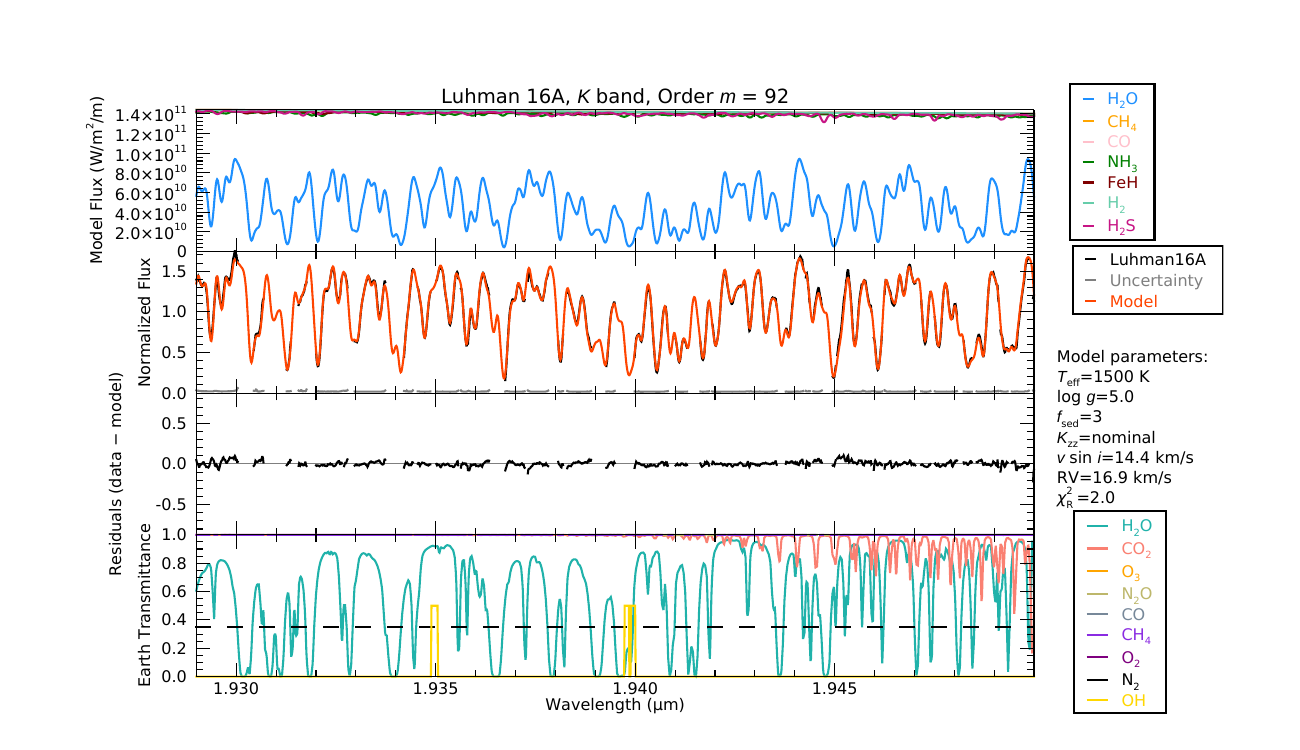}
    \caption{Continued.}
\end{figure*}

\begin{figure*}
    \ContinuedFloat
    \centering
    \includegraphics[height=0.43\textheight]{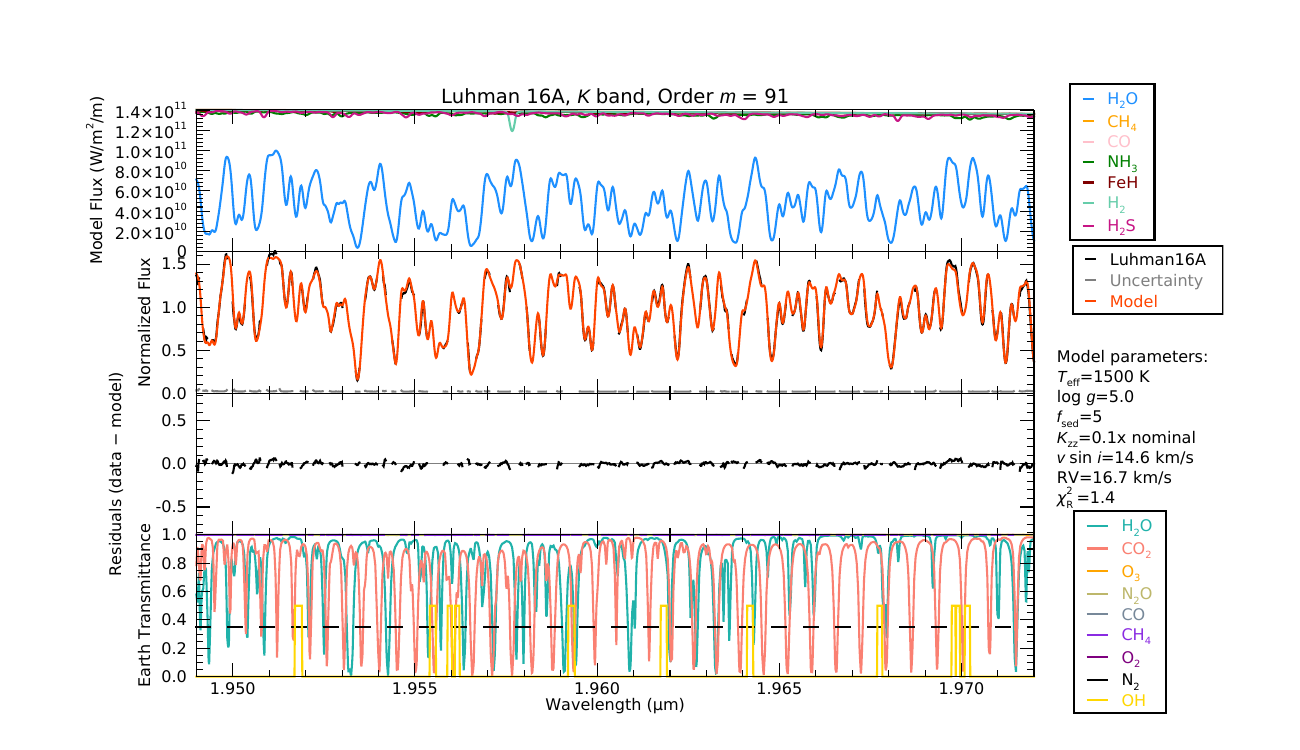}
    \includegraphics[height=0.43\textheight]{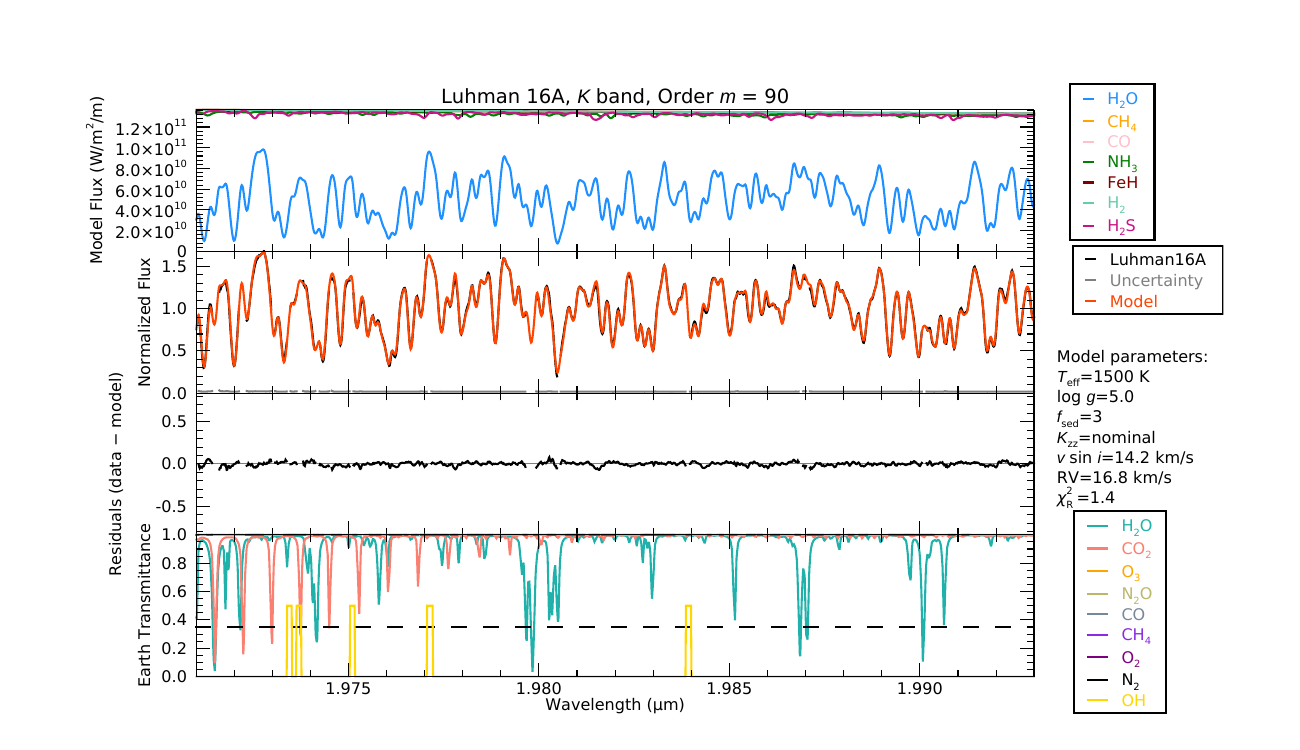}
    \caption{Continued.}
\end{figure*}

\begin{figure*}
    \ContinuedFloat
    \centering
    \includegraphics[height=0.43\textheight]{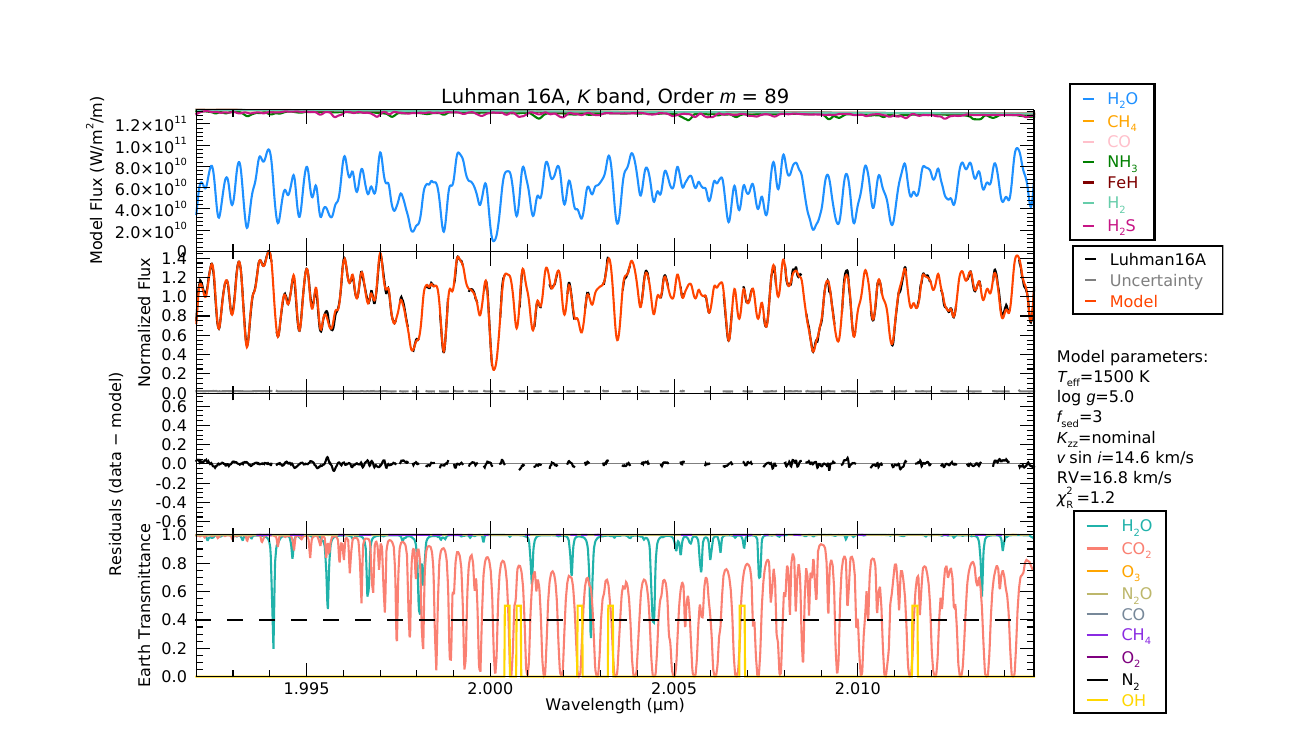}
    \includegraphics[height=0.43\textheight]{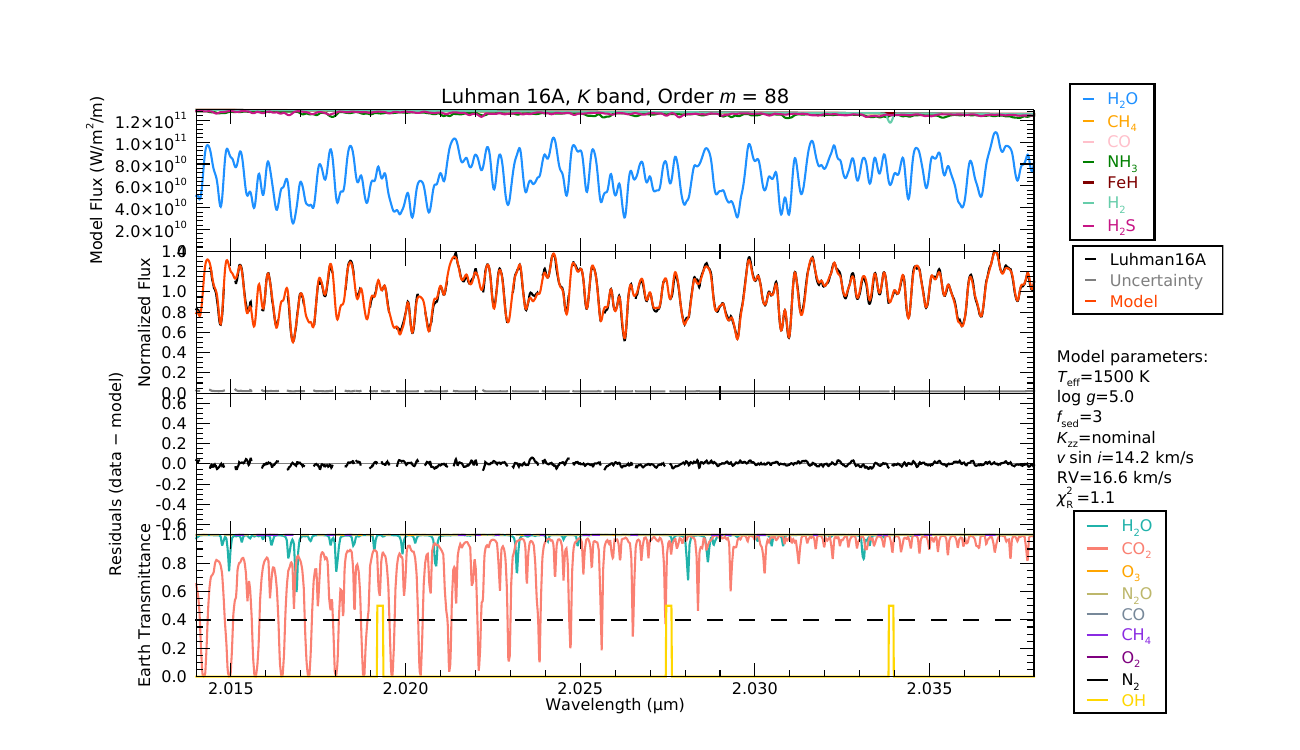}
    \caption{Continued.}
\end{figure*}

\begin{figure*}
    \ContinuedFloat
    \centering
    \includegraphics[height=0.43\textheight]{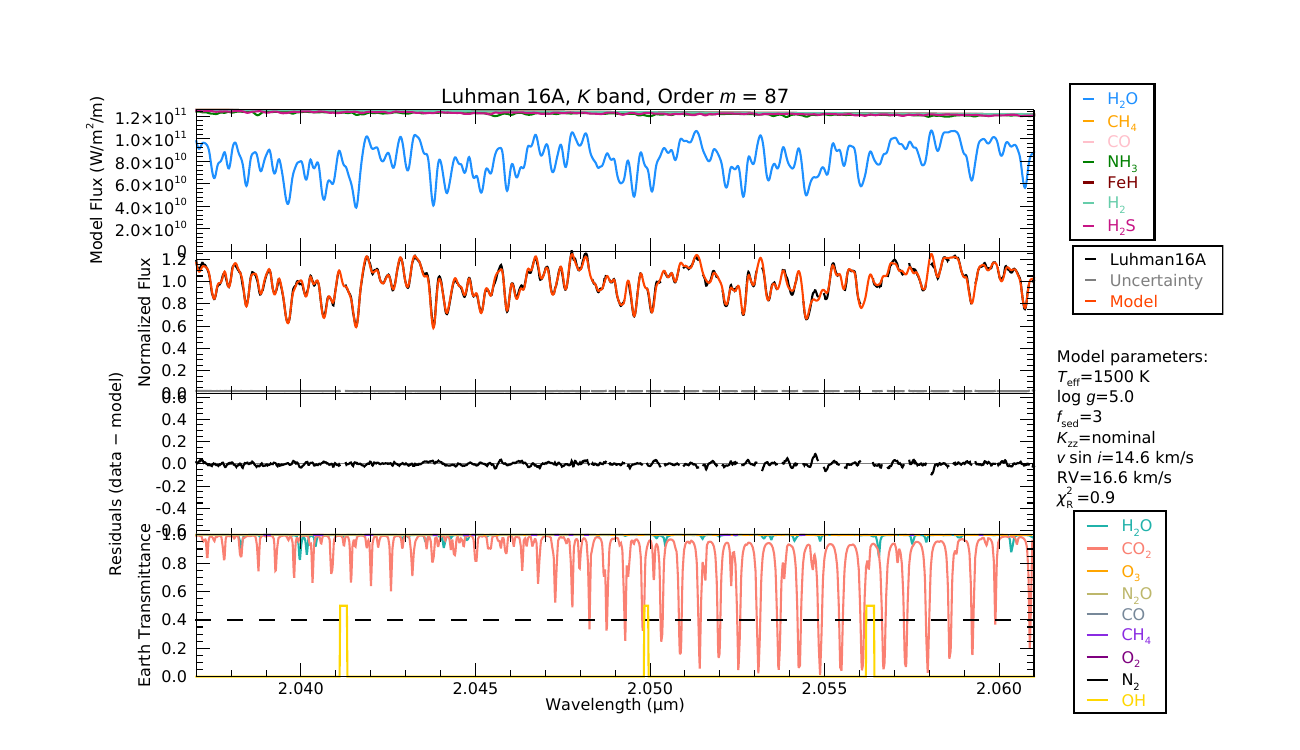}
    \includegraphics[height=0.43\textheight]{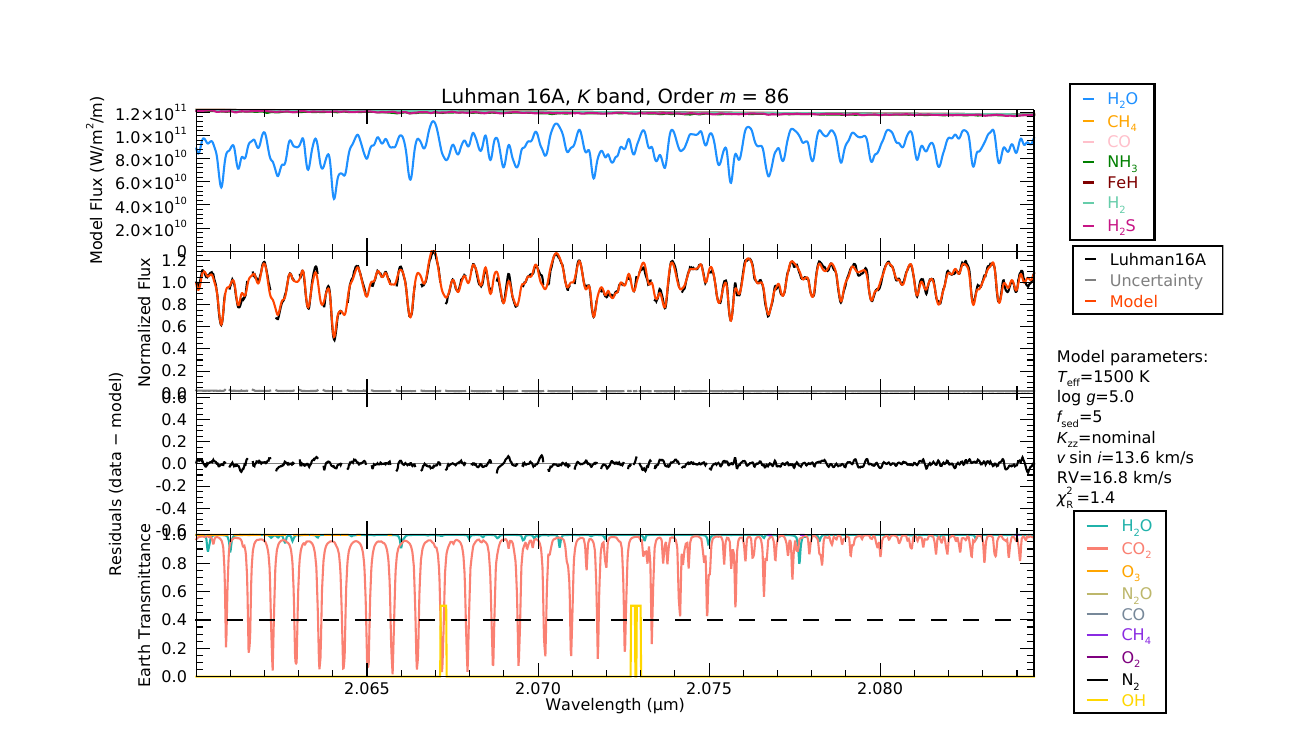}
    \caption{Continued.}
\end{figure*}

\begin{figure*}
    \ContinuedFloat
    \centering
    \includegraphics[height=0.43\textheight]{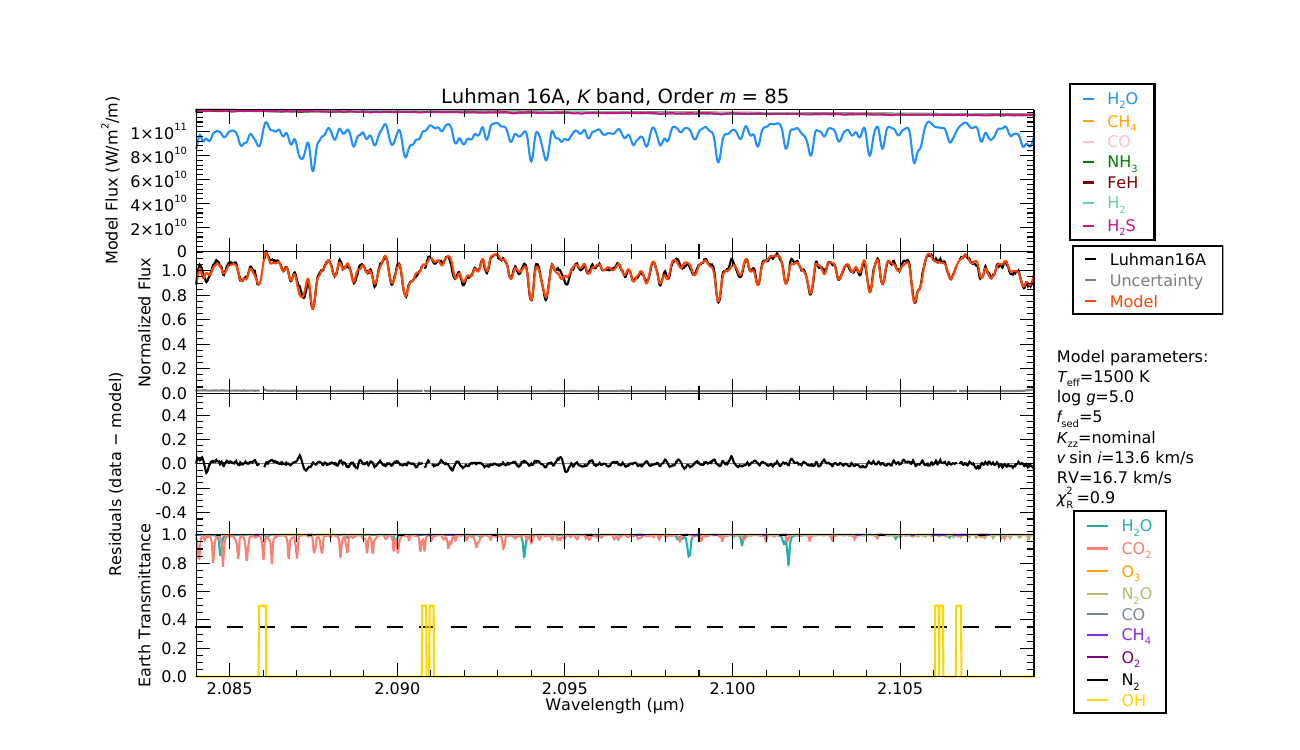}
    \includegraphics[height=0.43\textheight]{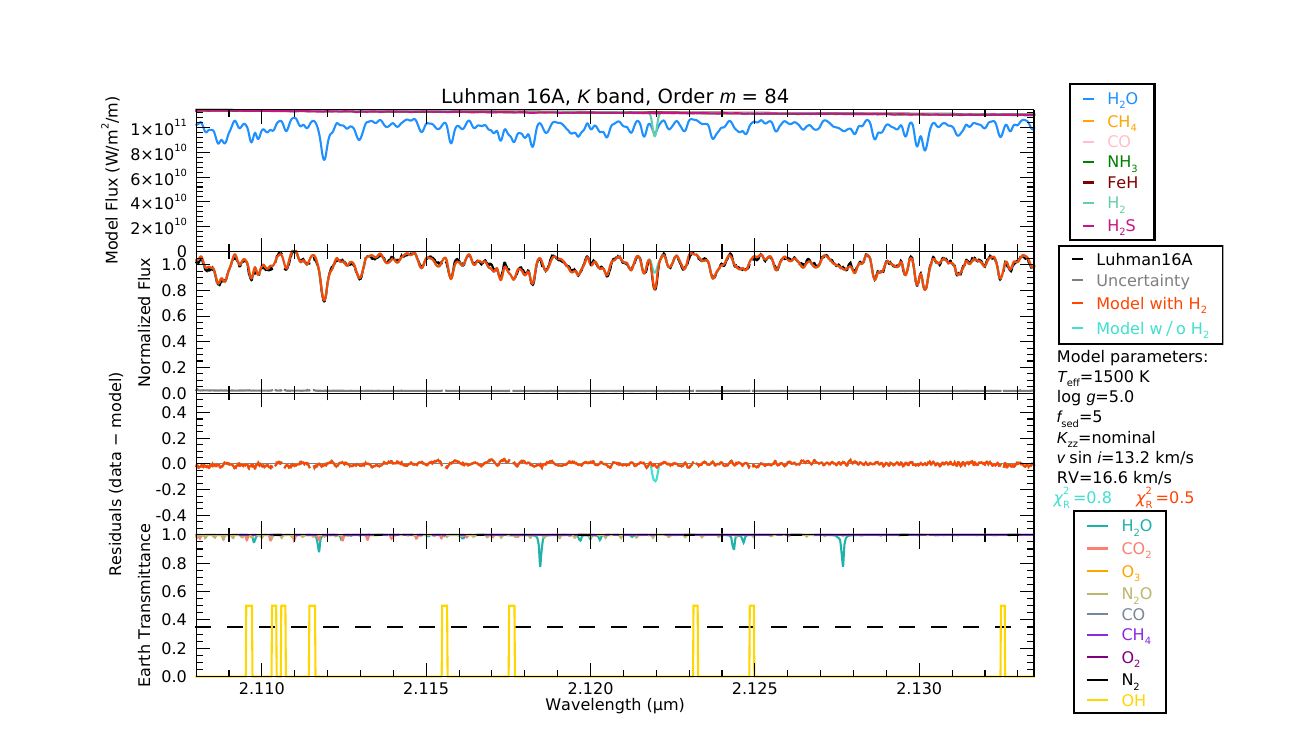}
    \caption{Continued.}
\end{figure*}

\begin{figure*}
    \ContinuedFloat
    \centering
    \includegraphics[height=0.43\textheight]{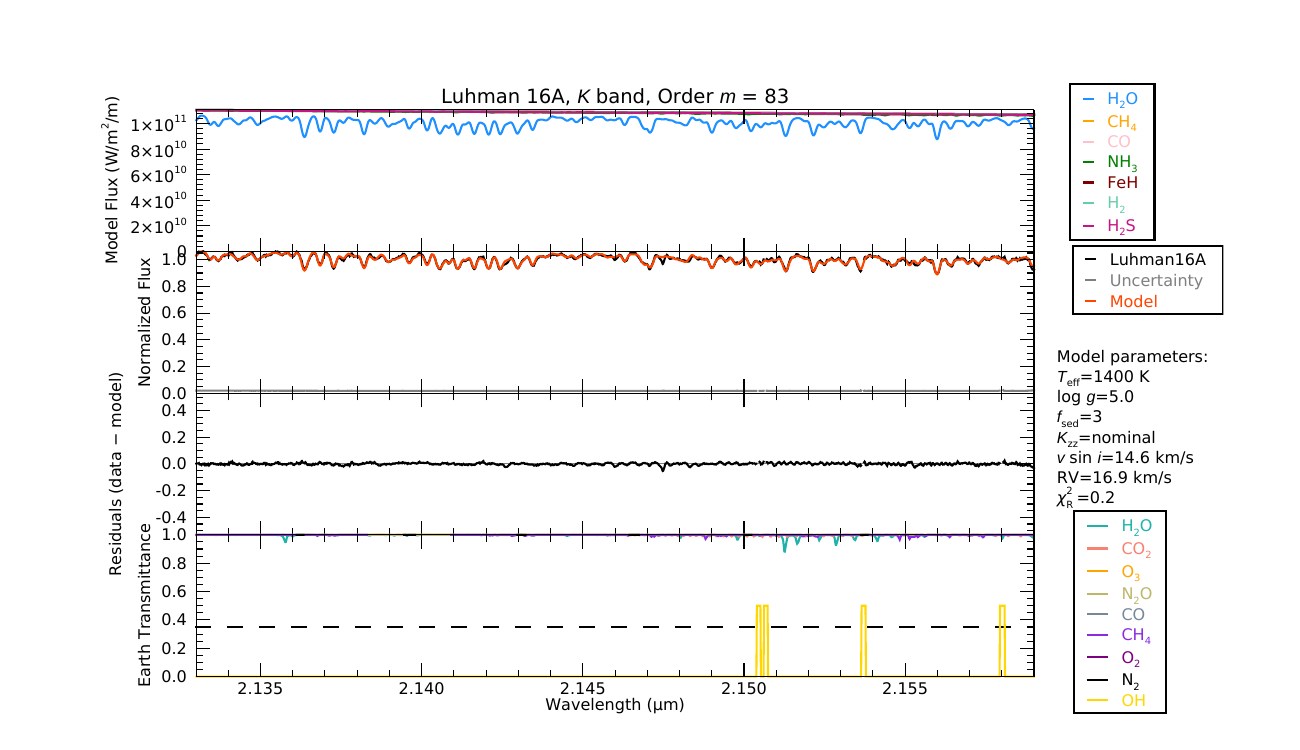}
    \includegraphics[height=0.43\textheight]{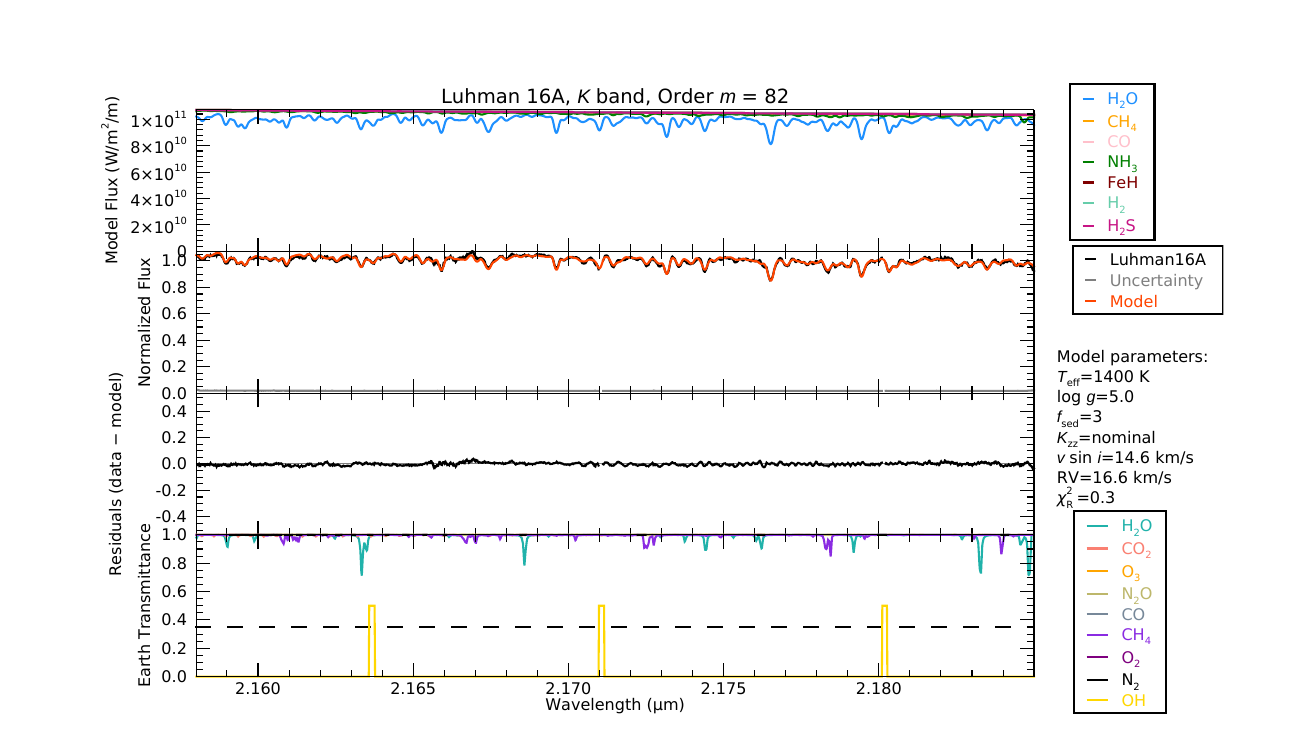} 
    \caption{Continued.}
\end{figure*}

\begin{figure*}
    \ContinuedFloat
    \centering
    \includegraphics[height=0.43\textheight]{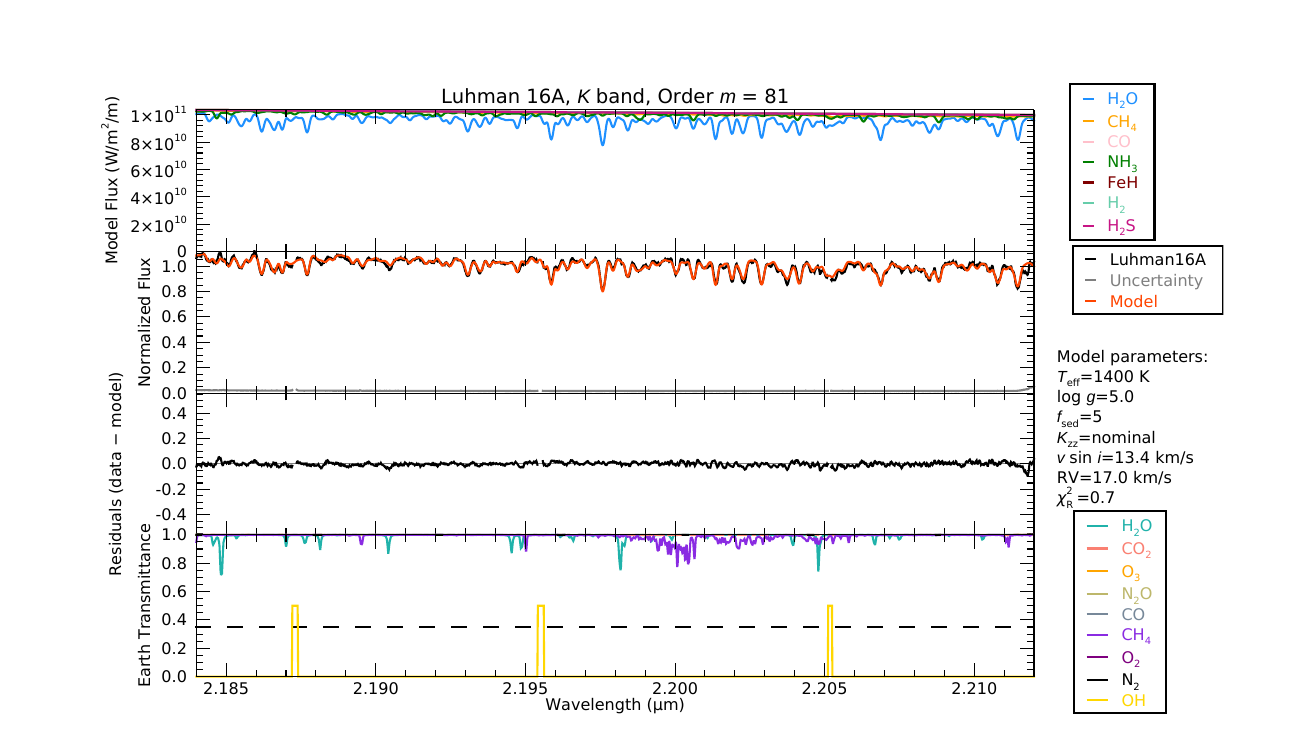}
    \includegraphics[height=0.43\textheight]{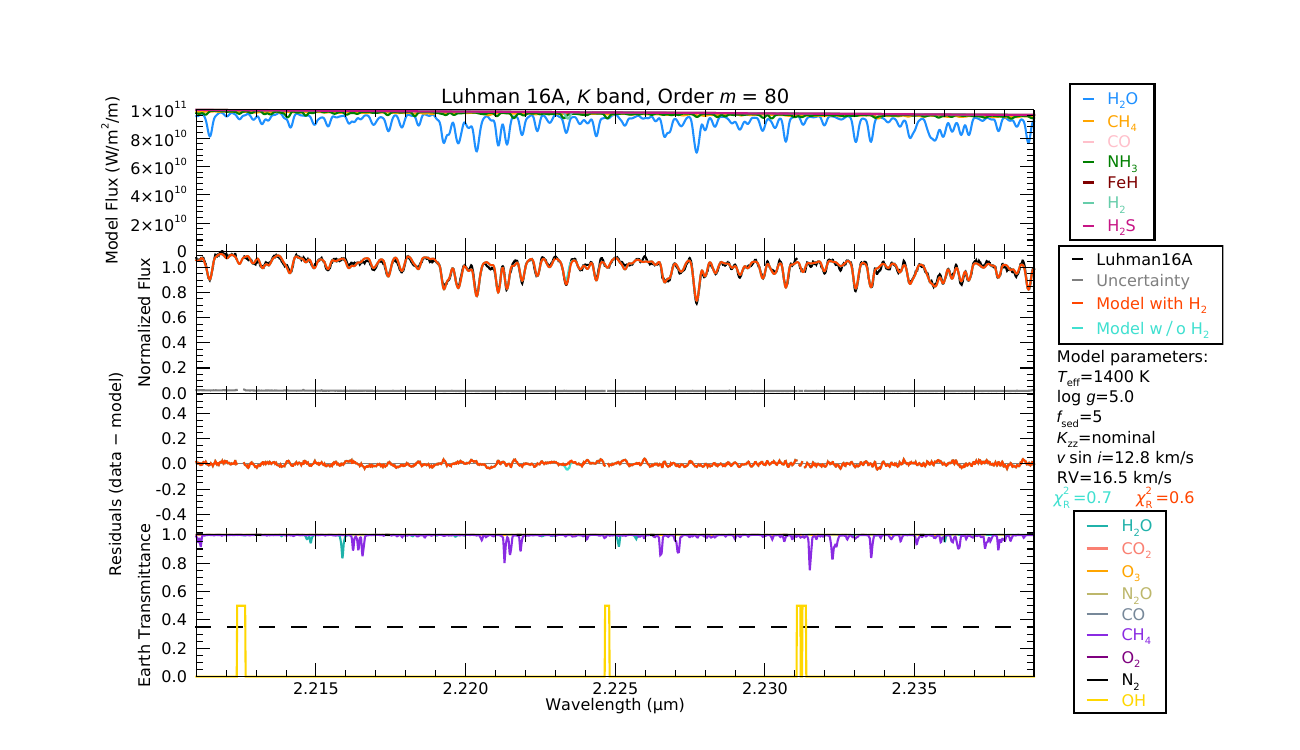}
    \caption{Continued.}
\end{figure*}

\begin{figure*}
    \ContinuedFloat
    \centering
    \includegraphics[height=0.43\textheight]{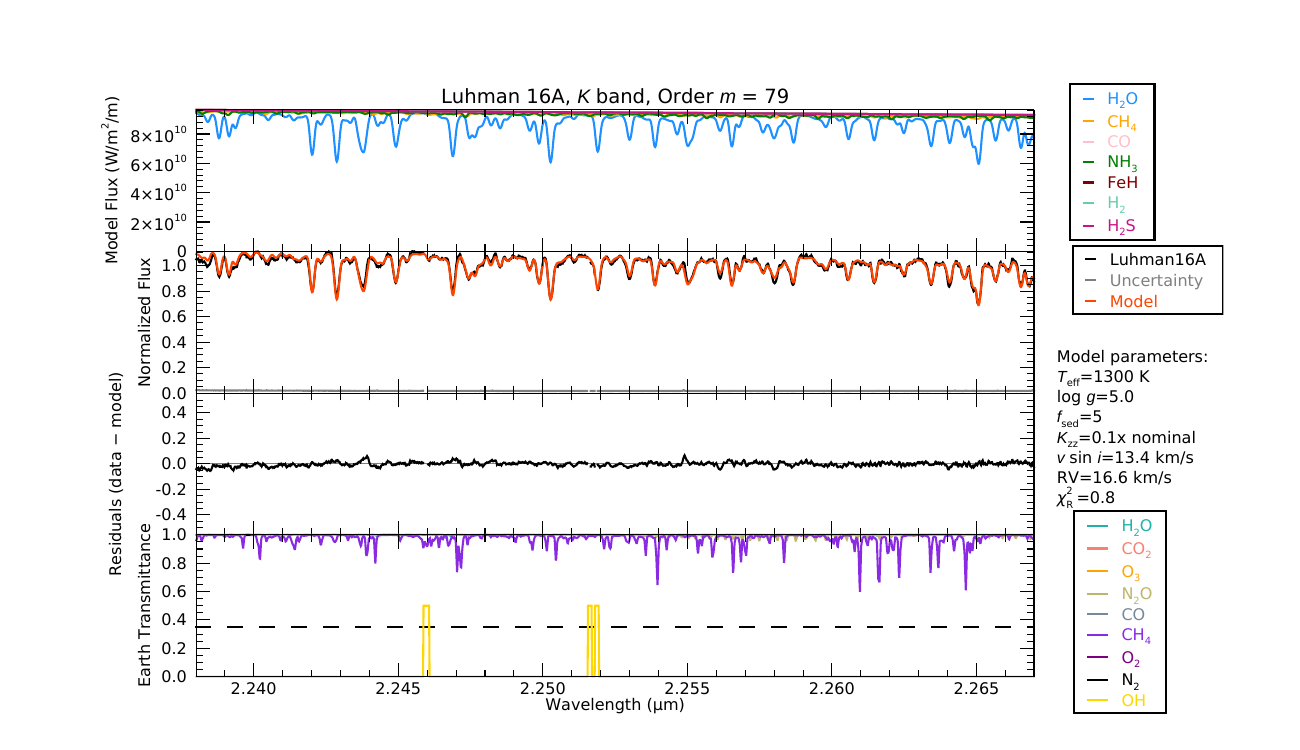}
    \includegraphics[height=0.43\textheight]{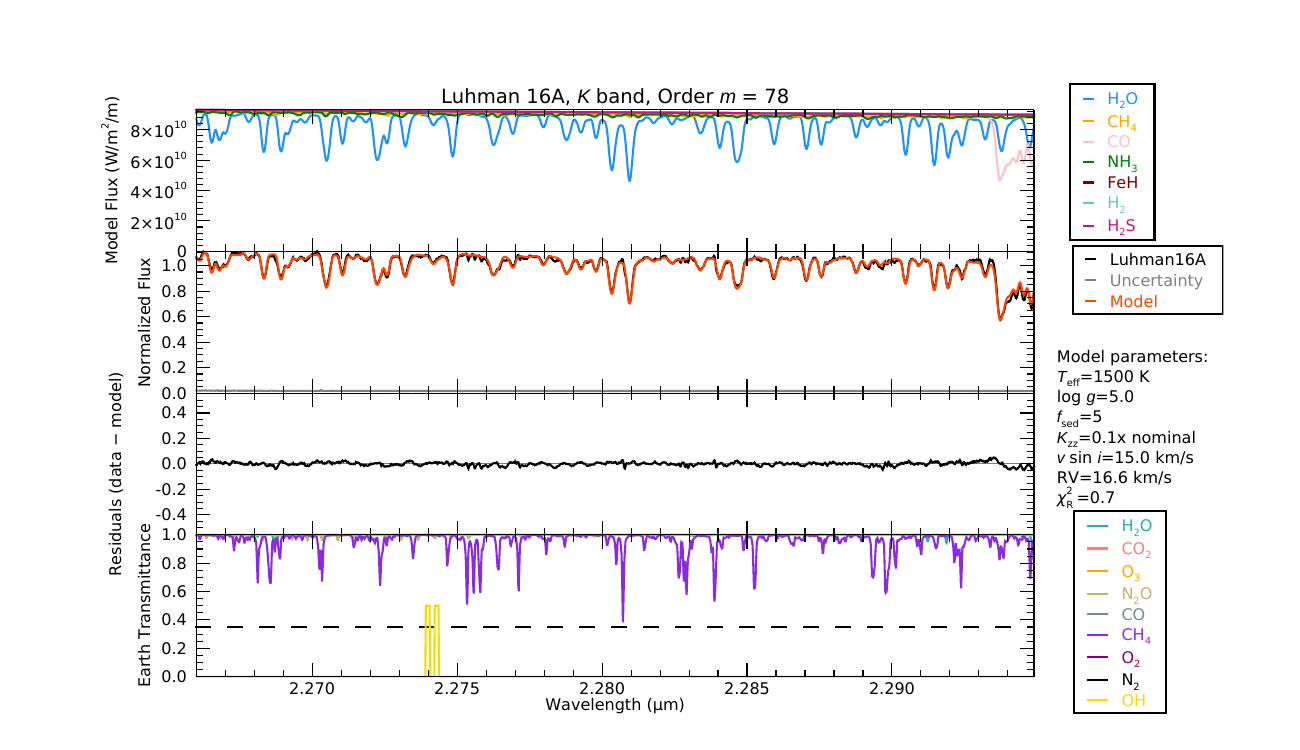}
    \caption{Continued.}
\end{figure*}

\begin{figure*}
    \ContinuedFloat
    \centering
    \includegraphics[height=0.43\textheight]{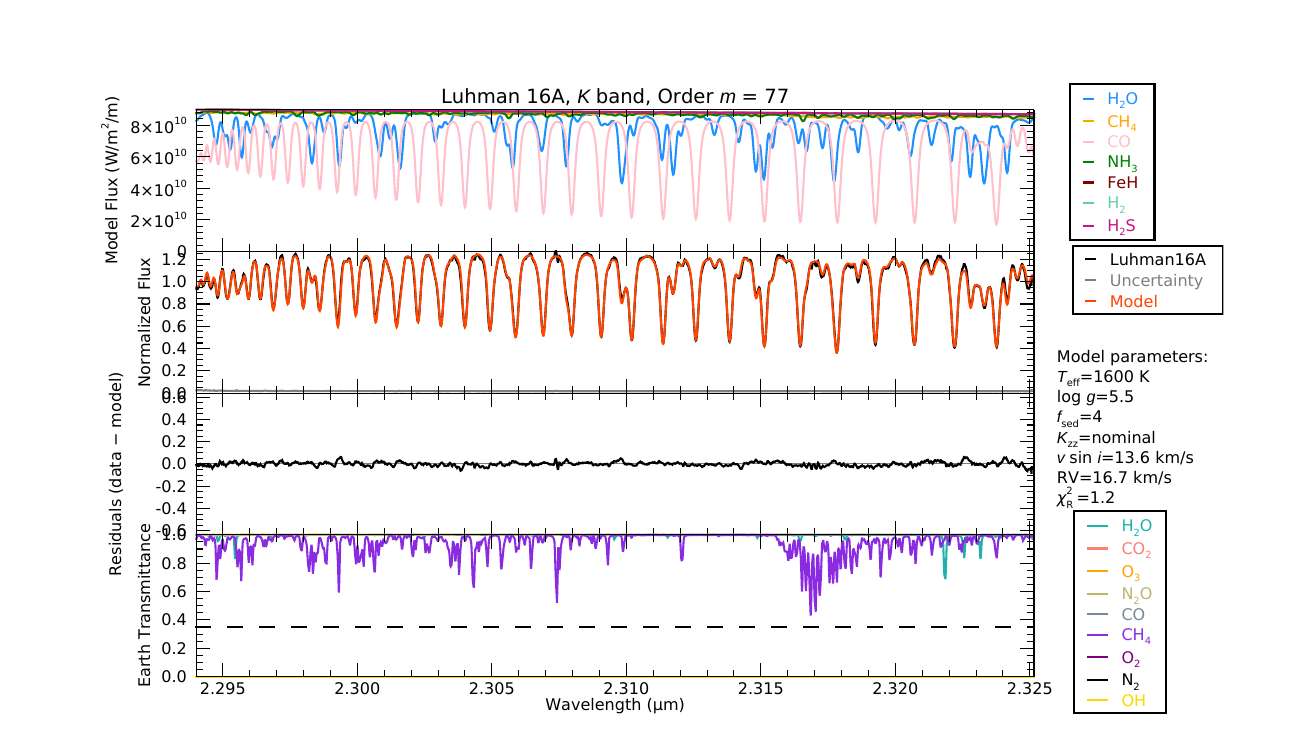}
    \includegraphics[height=0.43\textheight]{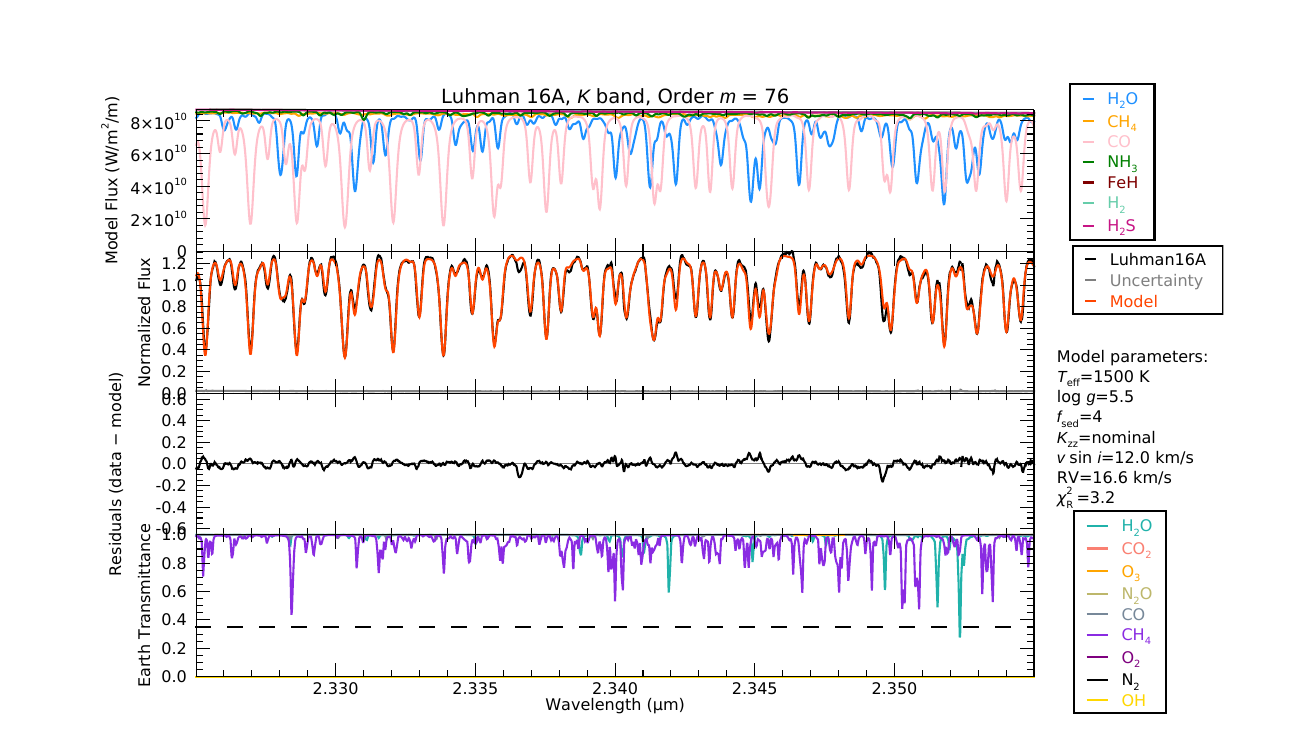}
    \caption{Continued.}
\end{figure*}

\begin{figure*}
    \ContinuedFloat
    \centering
    \includegraphics[height=0.43\textheight]{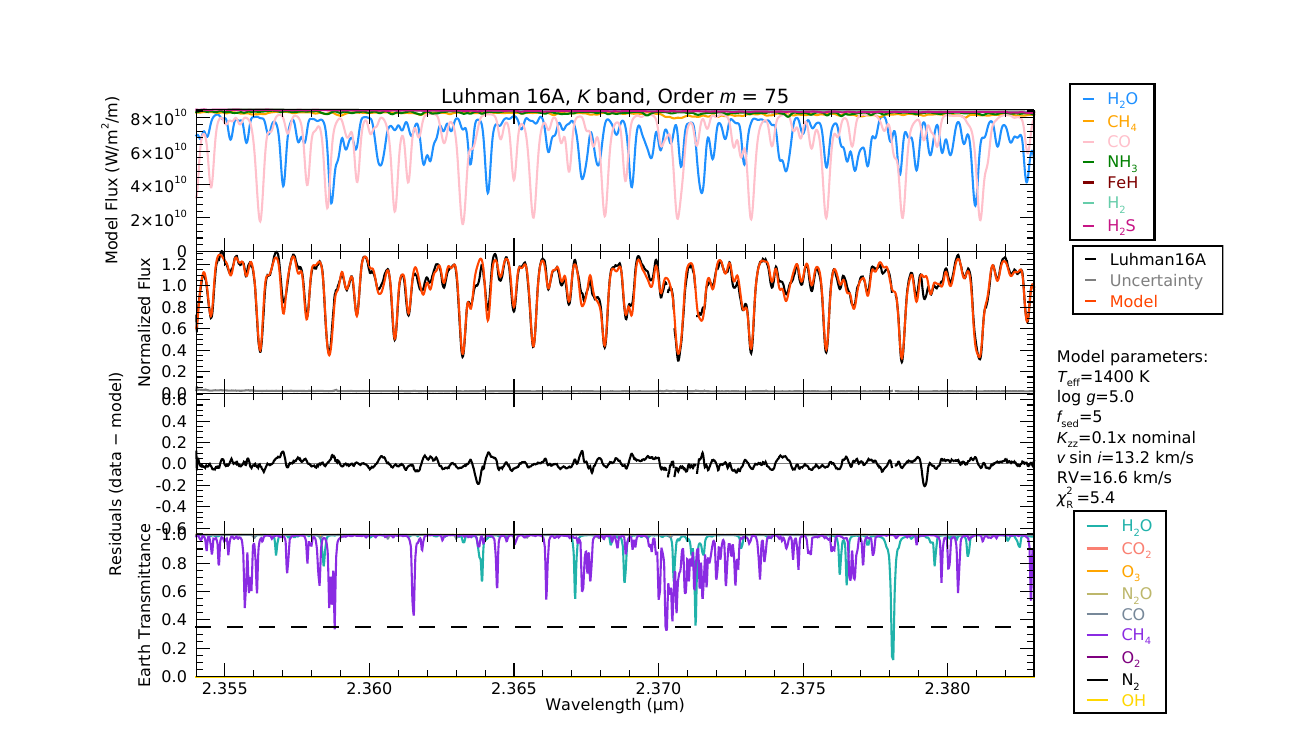}
    \includegraphics[height=0.43\textheight]{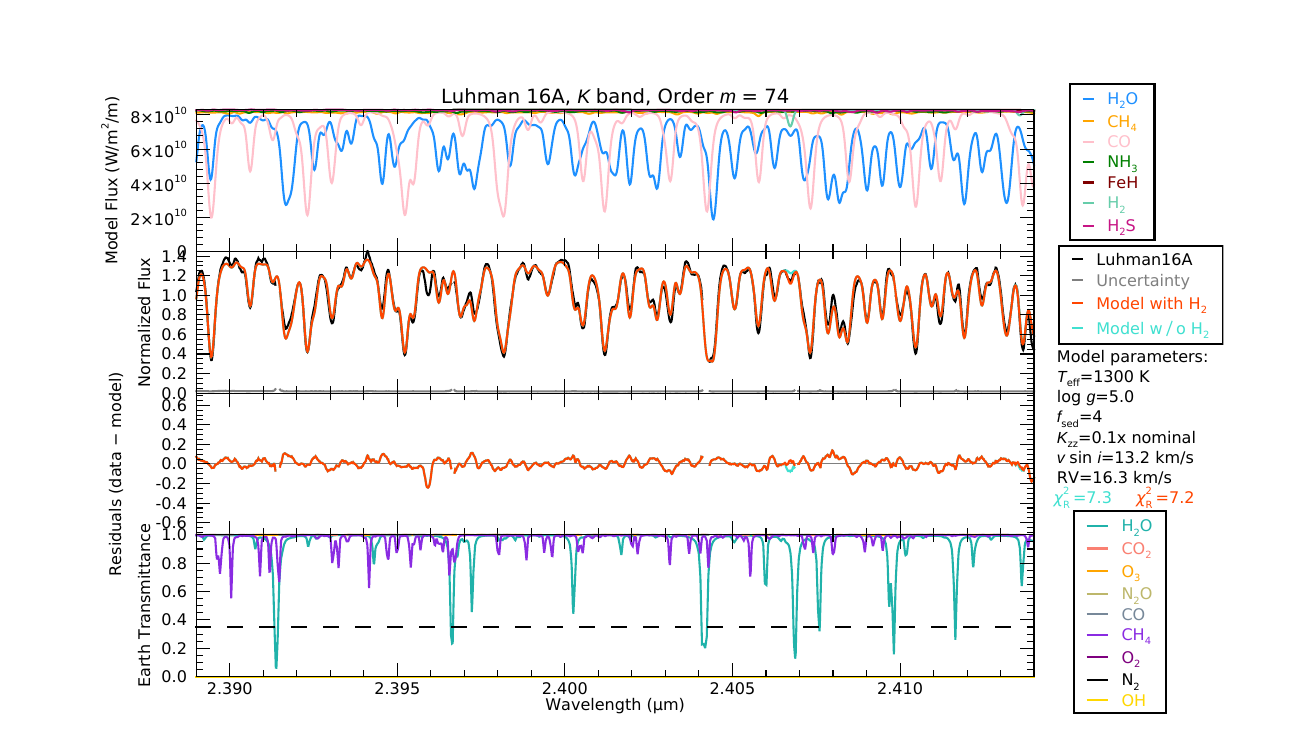}
    \caption{Continued.}
\end{figure*}

\begin{figure*}
    \ContinuedFloat
    \centering
    \includegraphics[height=0.43\textheight]{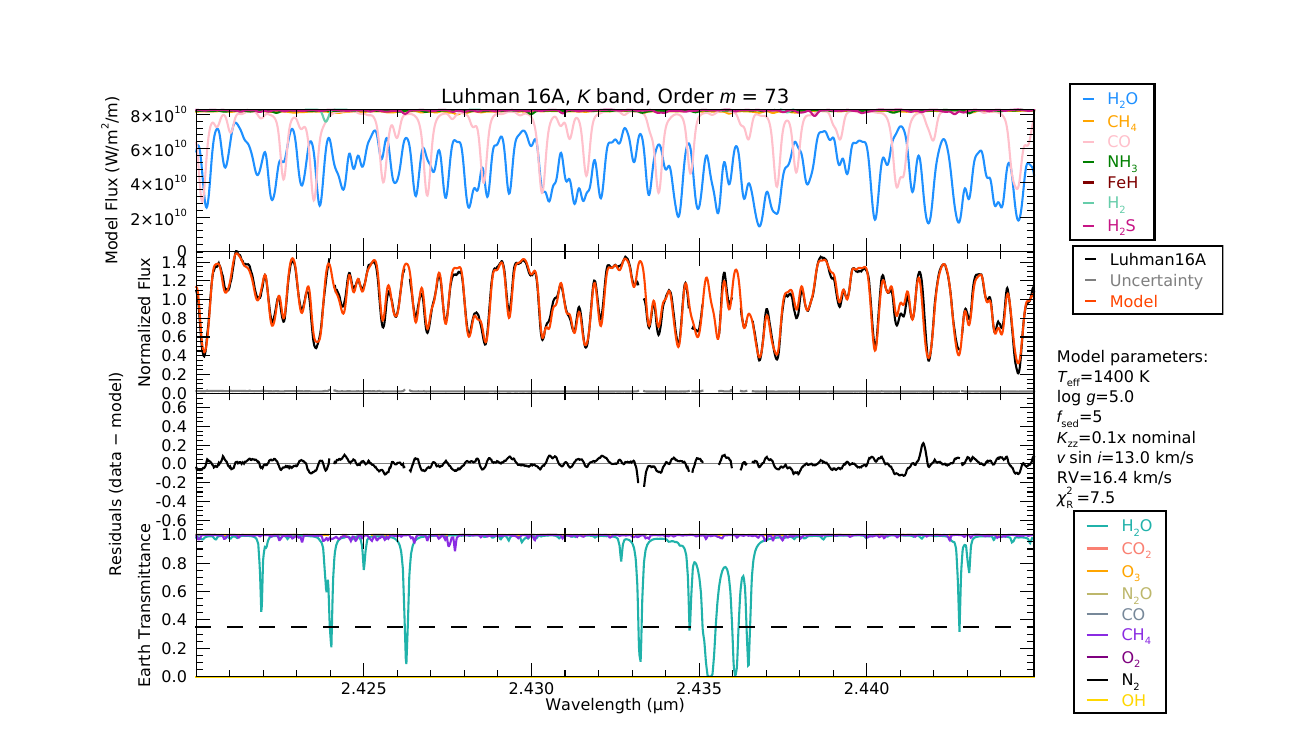}
    \includegraphics[height=0.43\textheight]{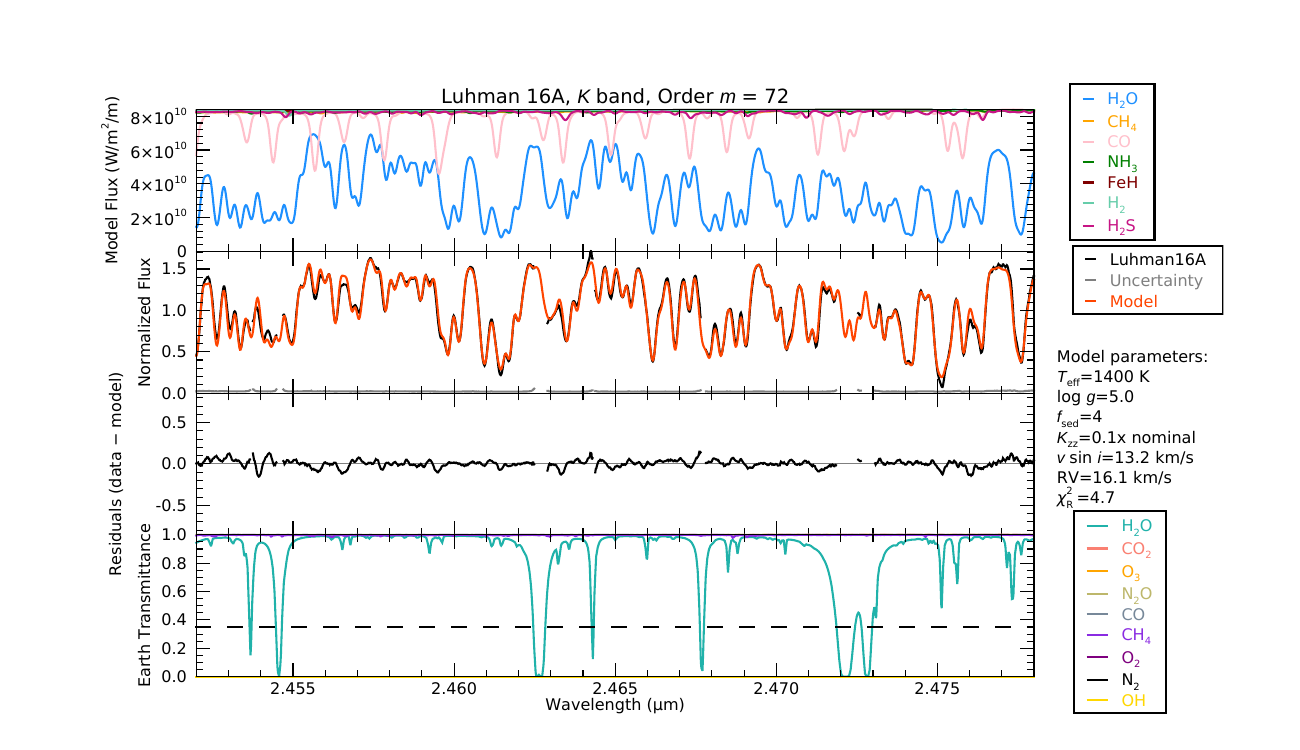}
    \caption{Continued.}
\end{figure*}

\begin{figure*}
    \centering
    \includegraphics[height=0.43\textheight]{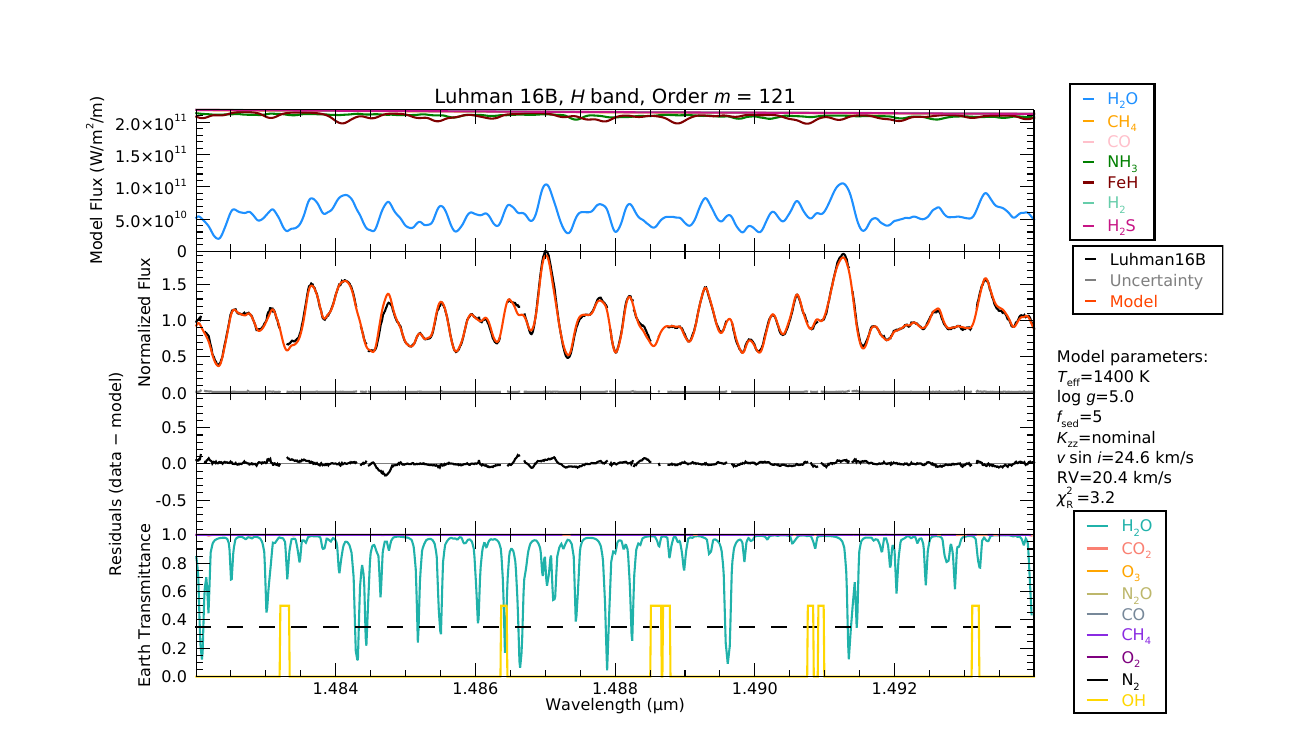}
    \includegraphics[height=0.43\textheight]{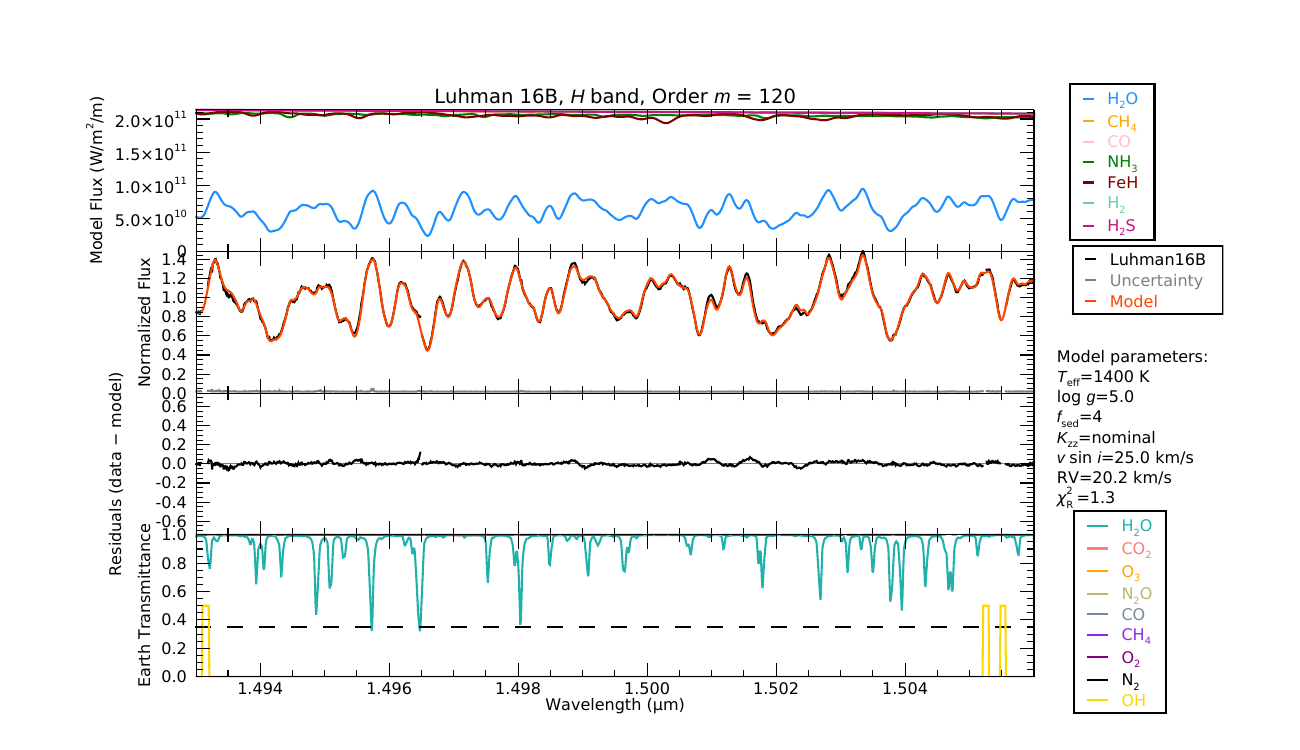}
    \caption{The spectral atlas of Luhman 16B in the same format as \ref{fig:atlasA}.
    }\label{fig:atlasB}
\end{figure*}

\begin{figure*}
    \ContinuedFloat
    \centering
    \includegraphics[height=0.43\textheight]{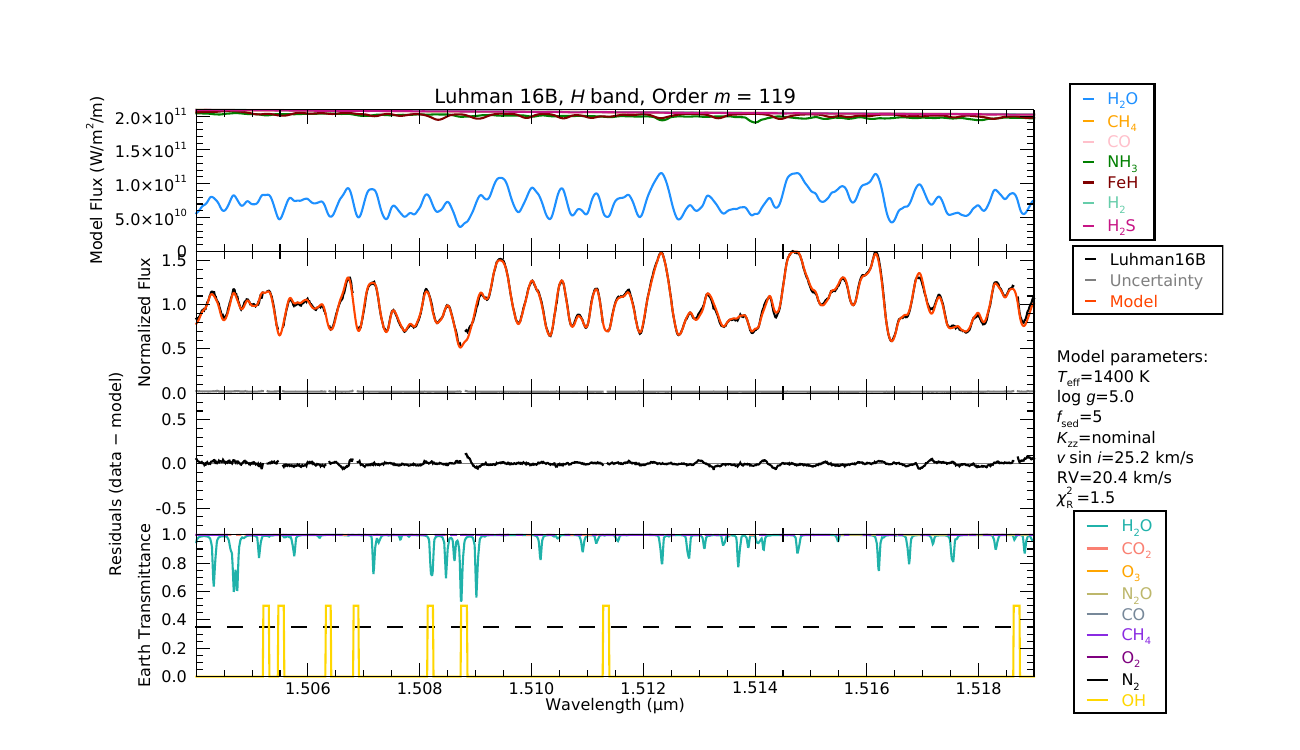}
    \includegraphics[height=0.43\textheight]{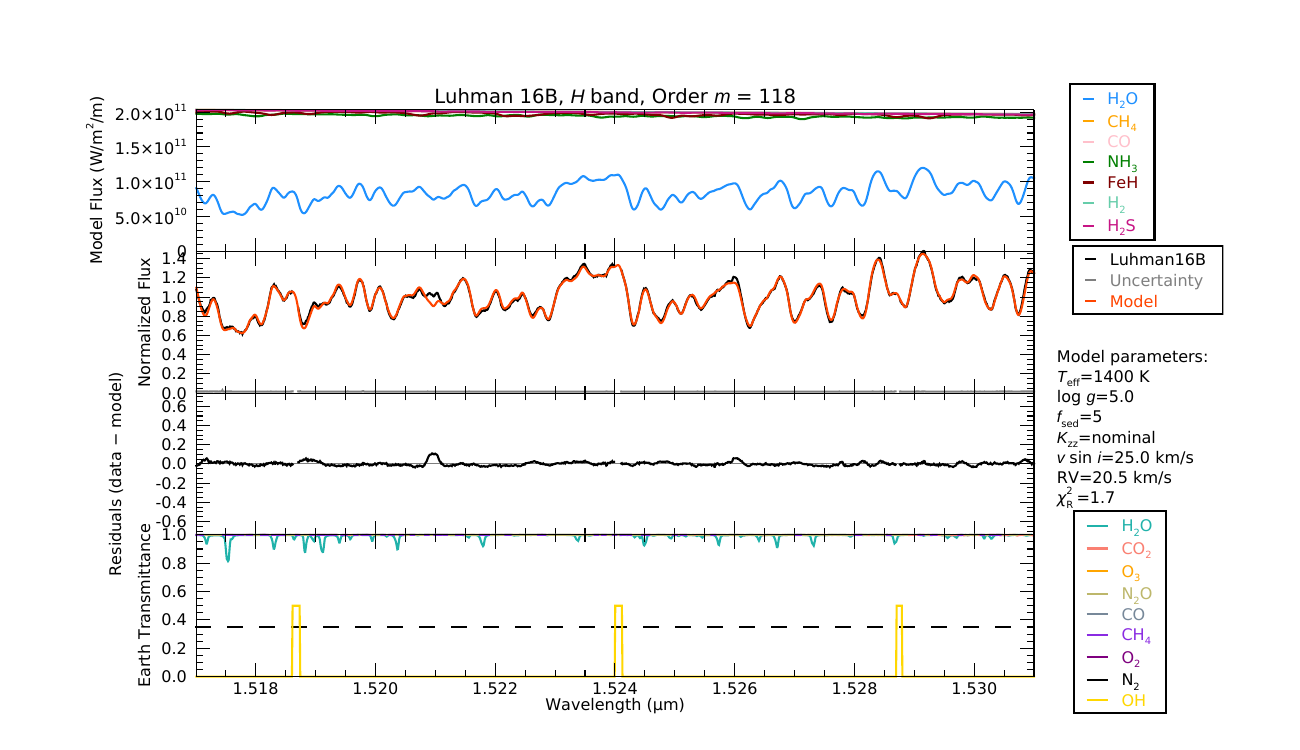}
    \caption{Continued.}
\end{figure*}

\begin{figure*}
    \ContinuedFloat
    \centering
    \includegraphics[height=0.43\textheight]{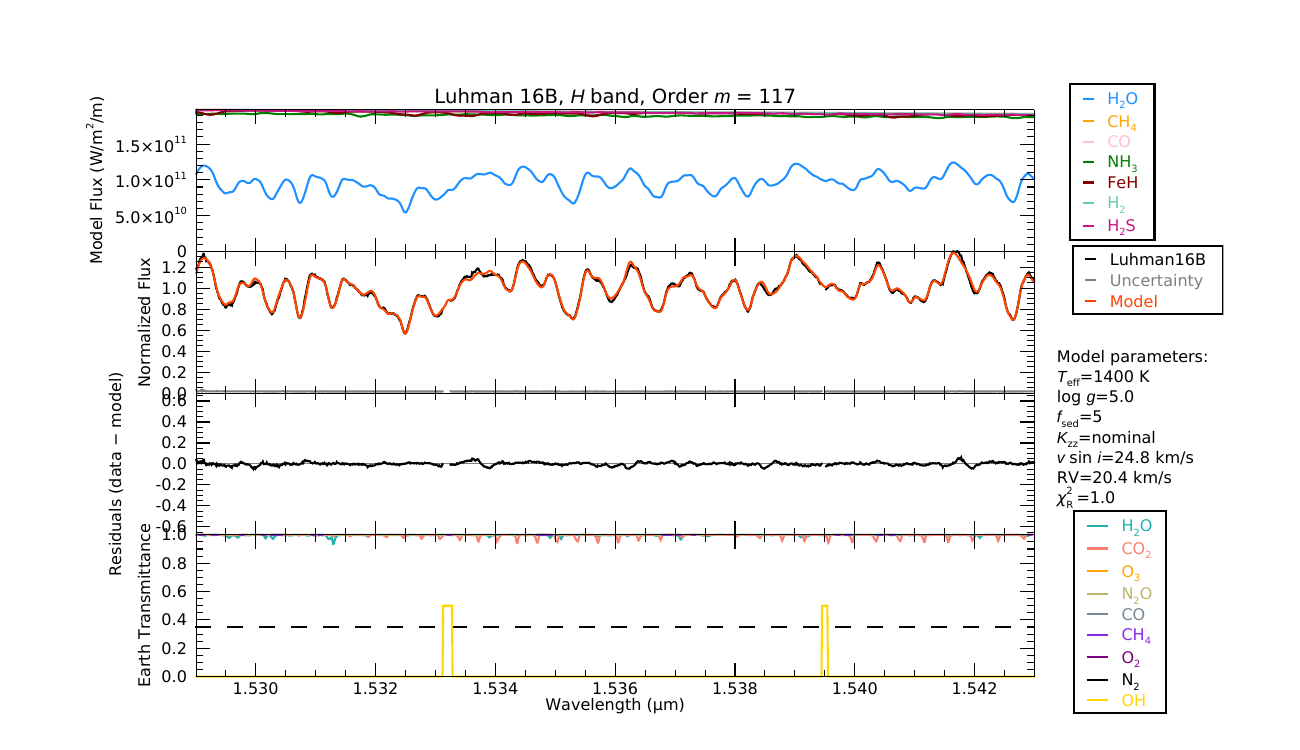}
    \includegraphics[height=0.43\textheight]{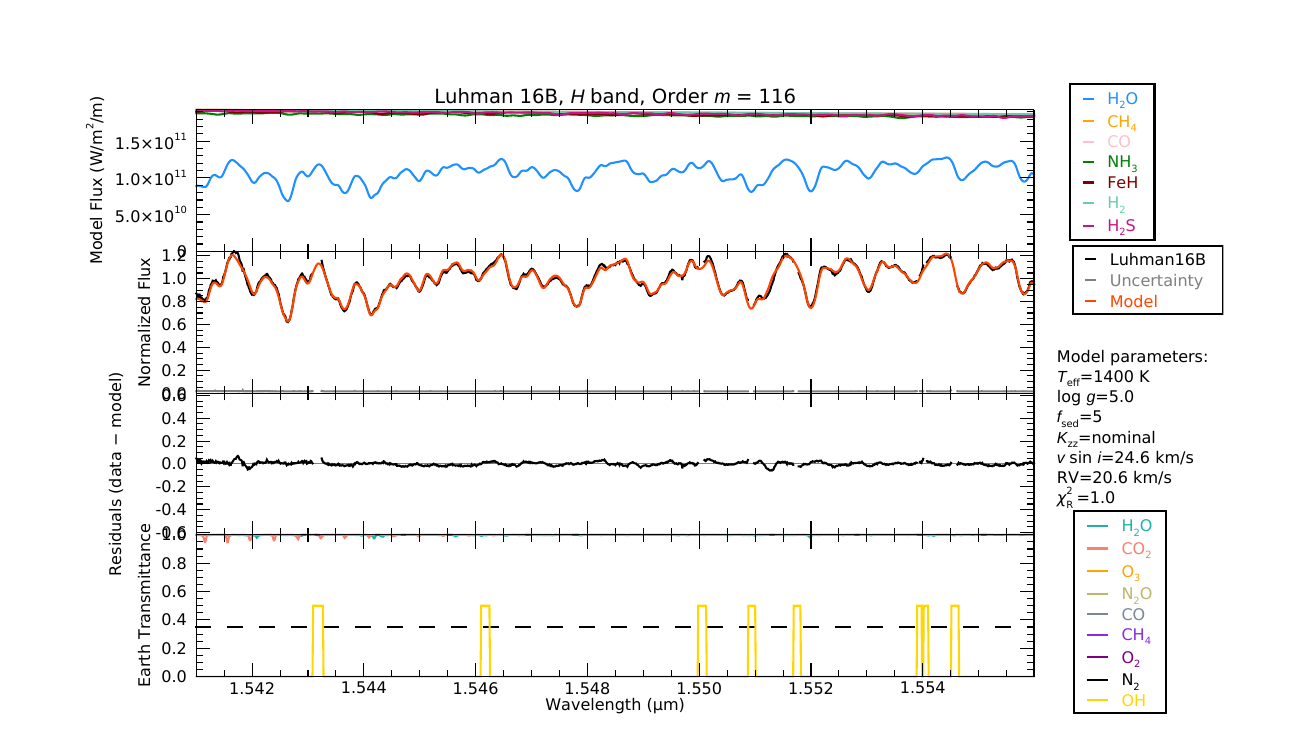}
    \caption{Continued.}
\end{figure*}

\begin{figure*}
    \ContinuedFloat
    \centering
    \includegraphics[height=0.43\textheight]{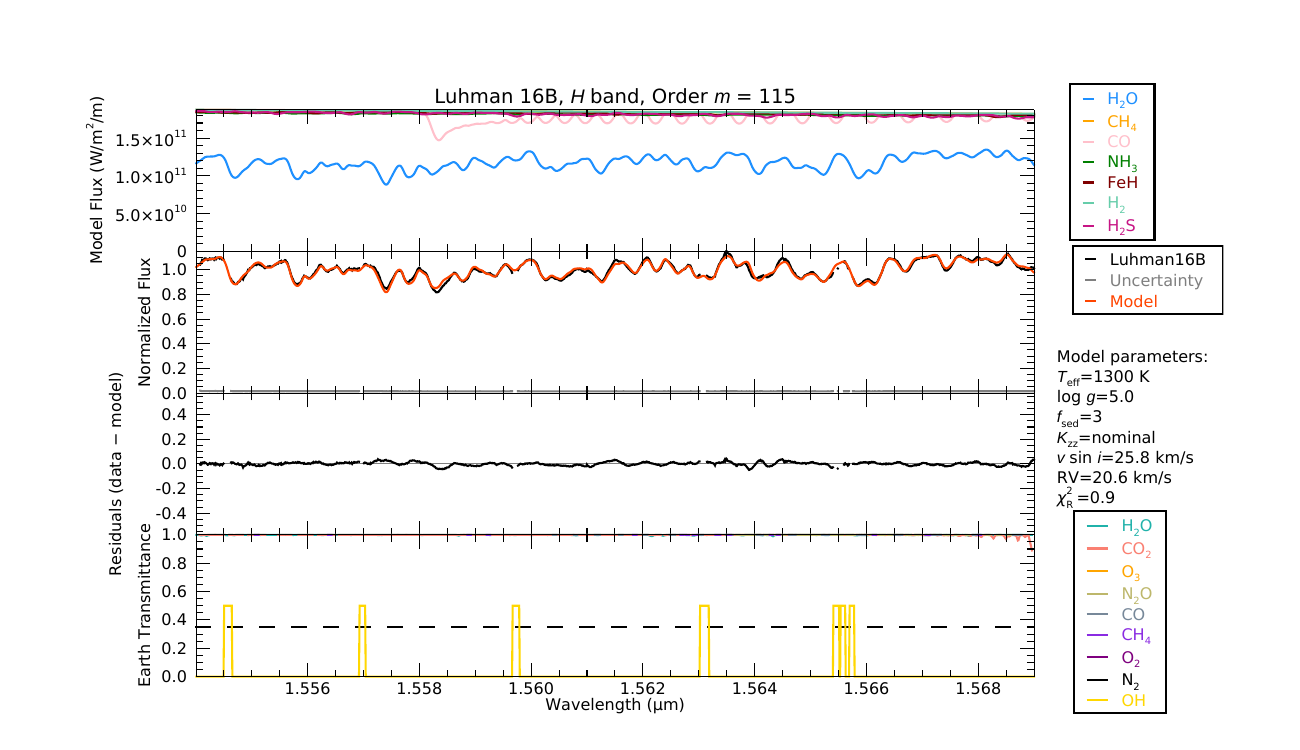}
    \includegraphics[height=0.43\textheight]{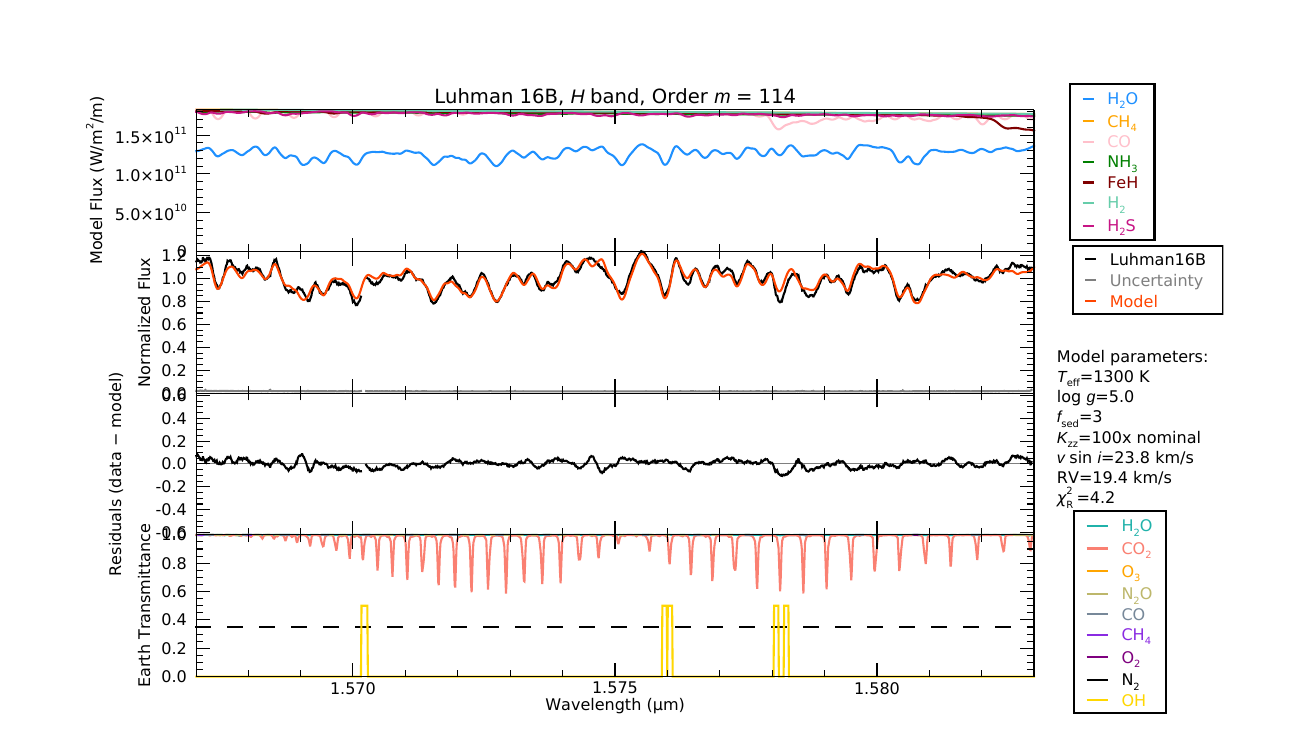}
    \caption{Continued.}
\end{figure*}

\begin{figure*}
    \ContinuedFloat
    \centering
    \includegraphics[height=0.43\textheight]{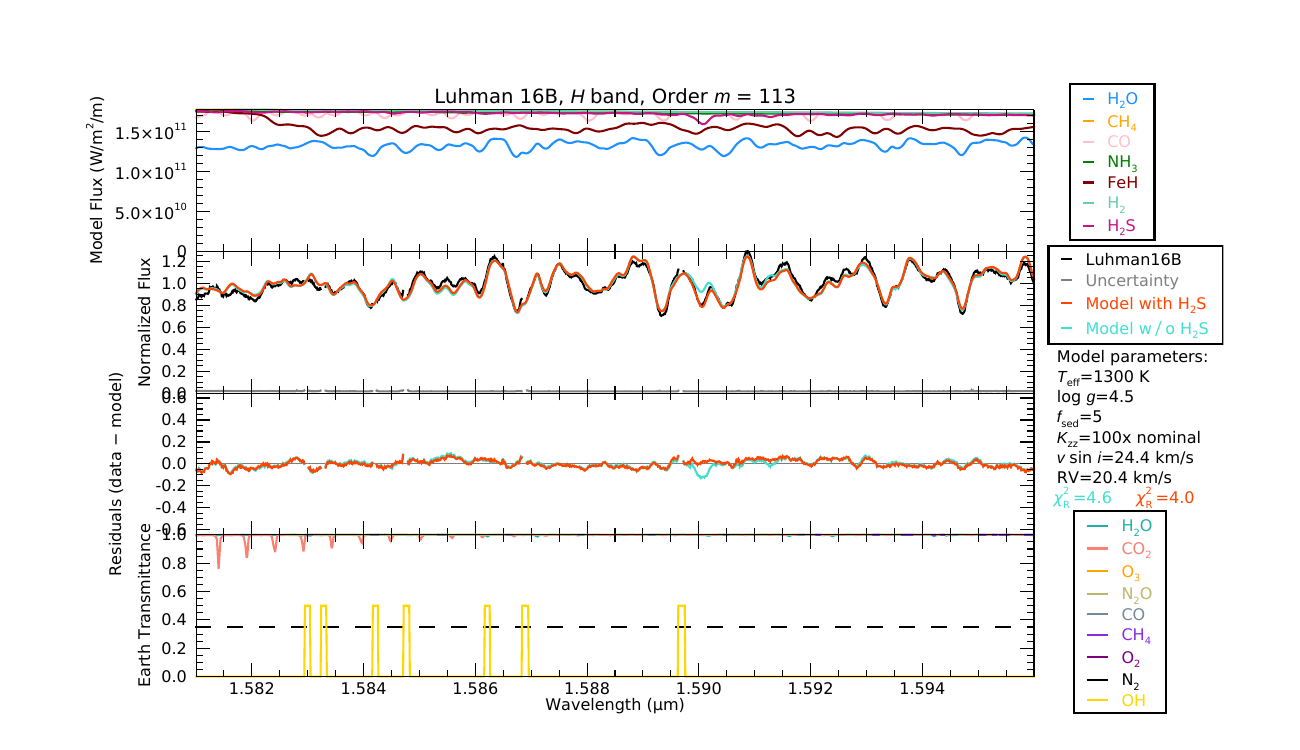}
    \includegraphics[height=0.43\textheight]{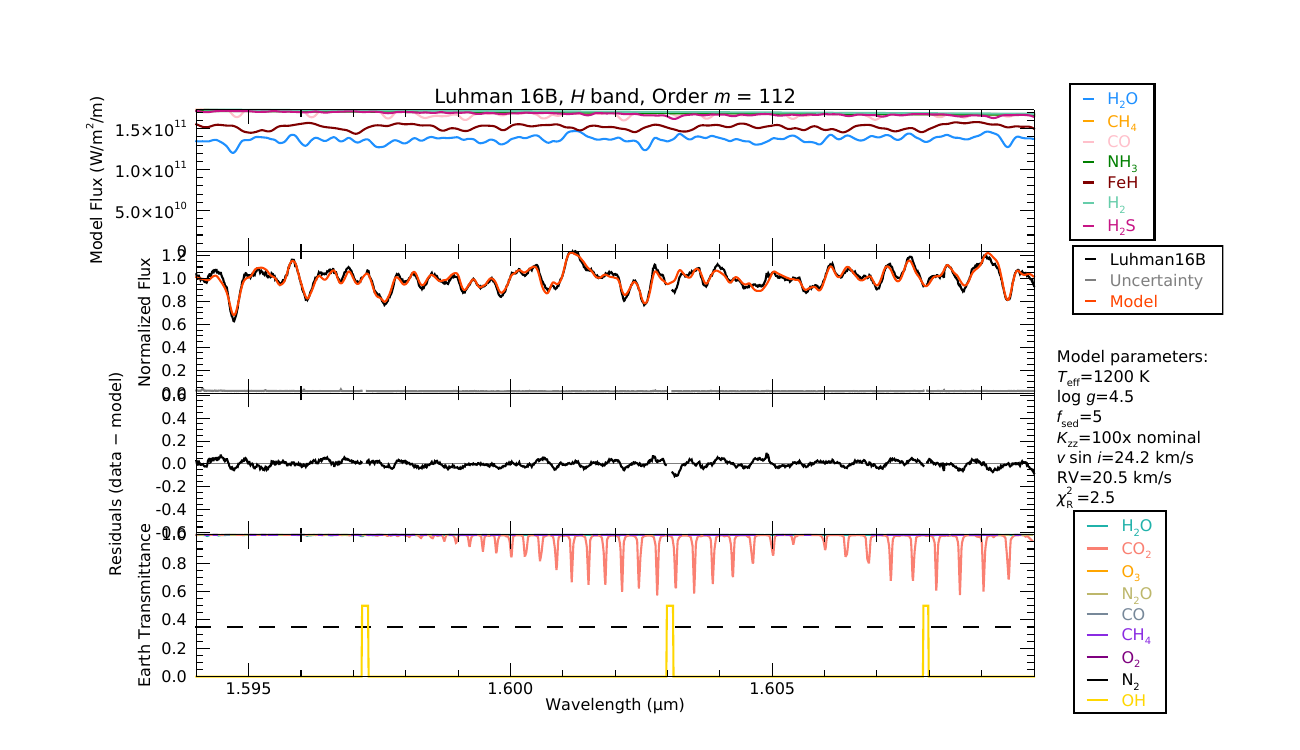}
    \caption{Continued.}
\end{figure*}

\begin{figure*}
    \ContinuedFloat
    \centering
    \includegraphics[height=0.43\textheight]{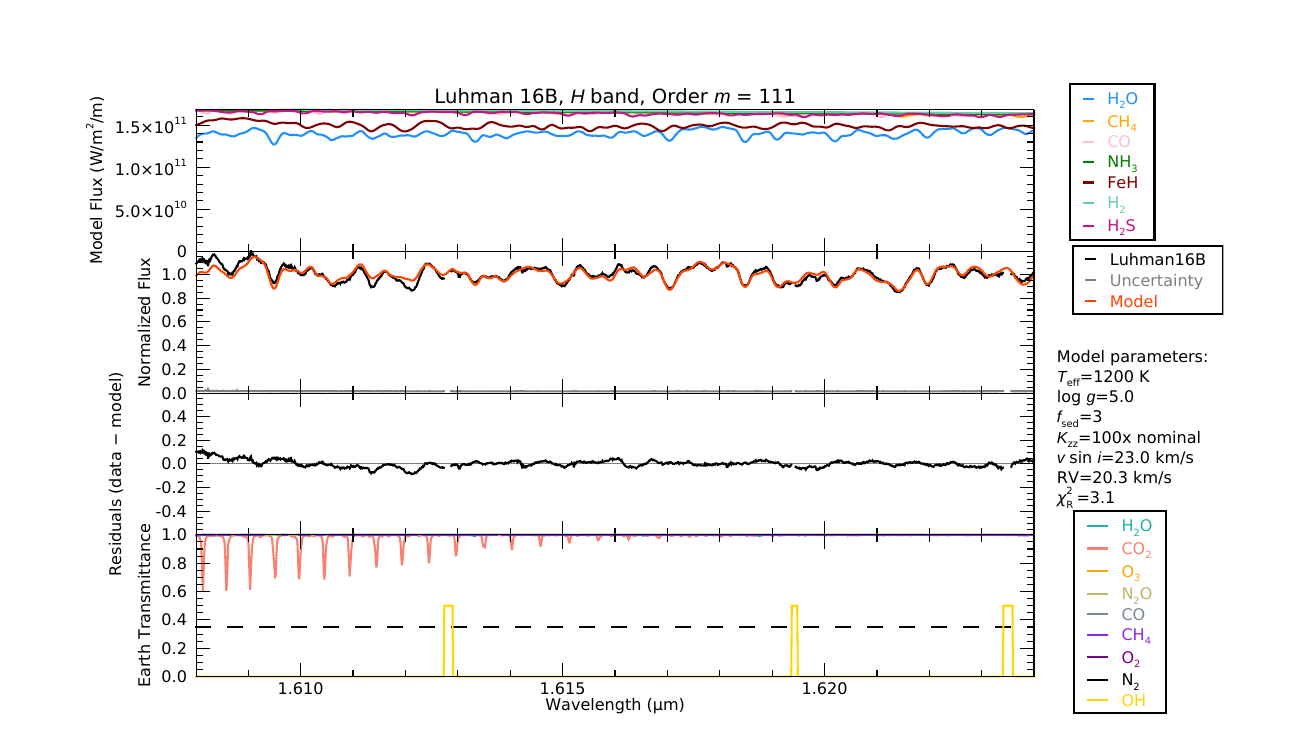}
    \includegraphics[height=0.43\textheight]{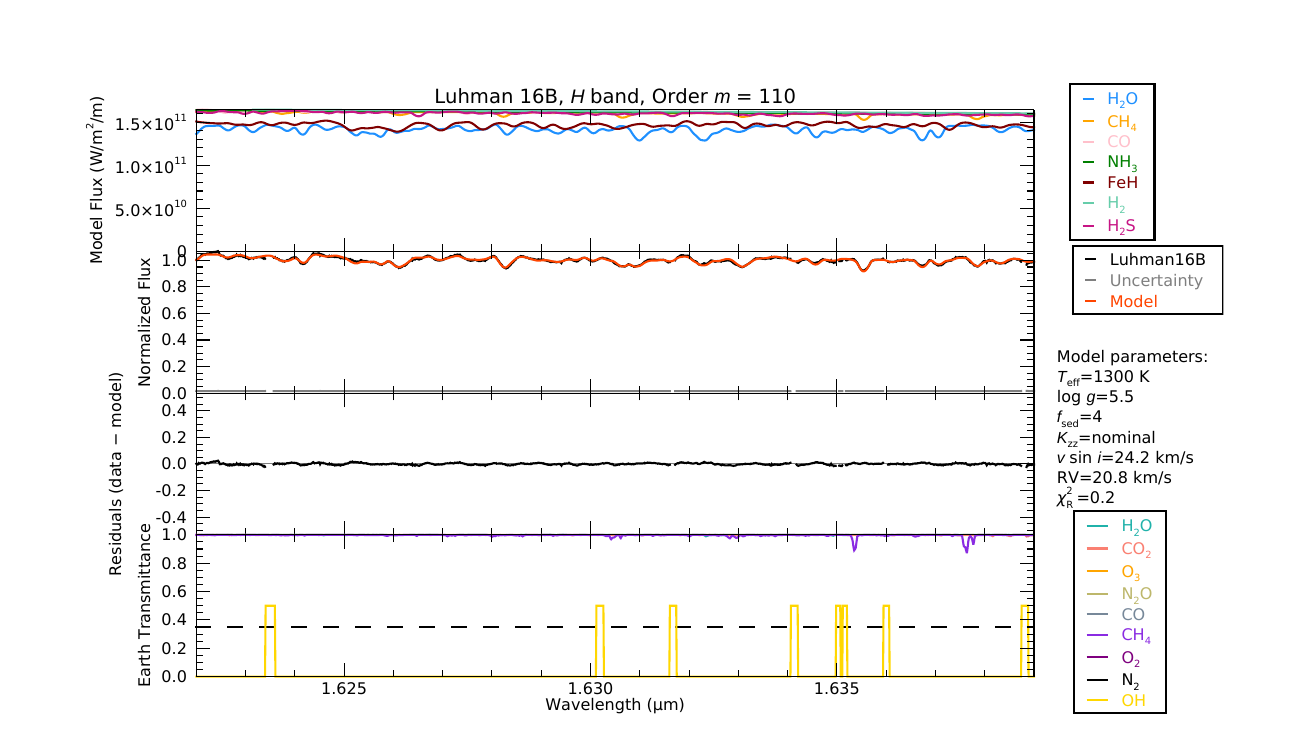}
    \caption{Continued.}
\end{figure*}

\begin{figure*}
    \ContinuedFloat
    \centering
    \includegraphics[height=0.43\textheight]{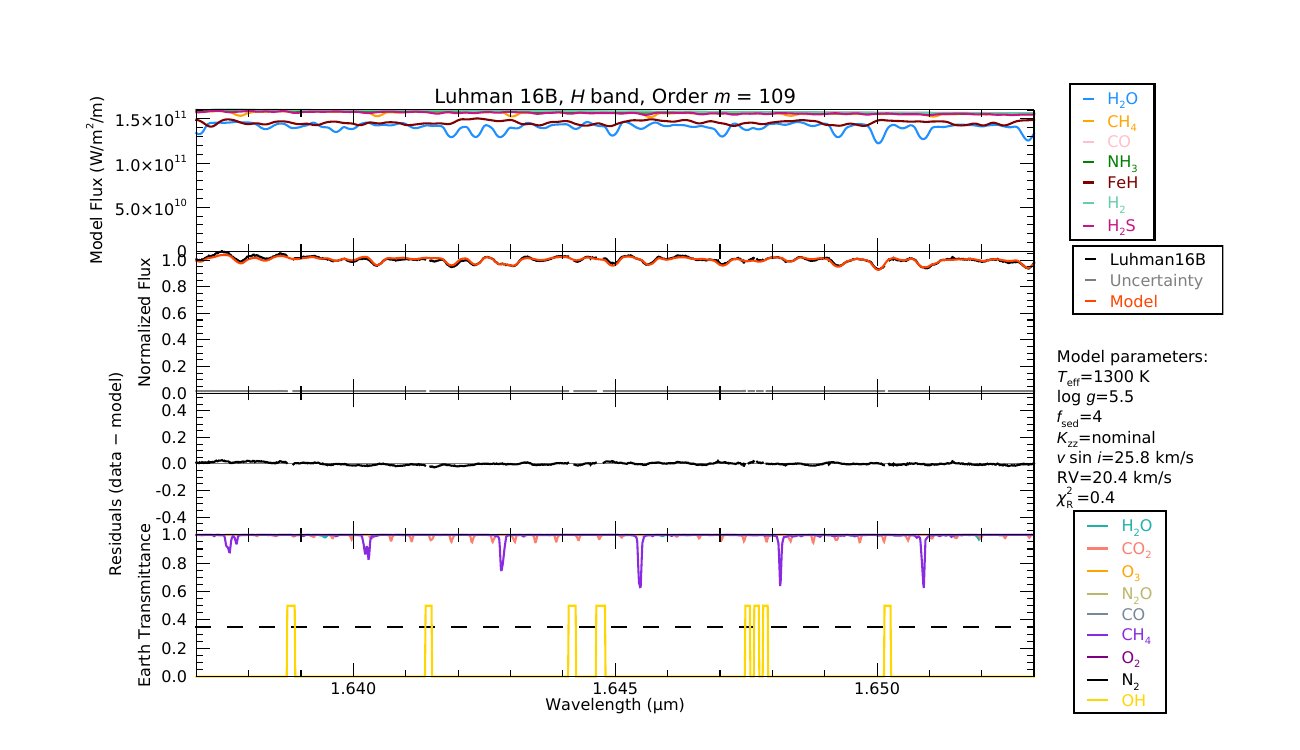}
    \includegraphics[height=0.43\textheight]{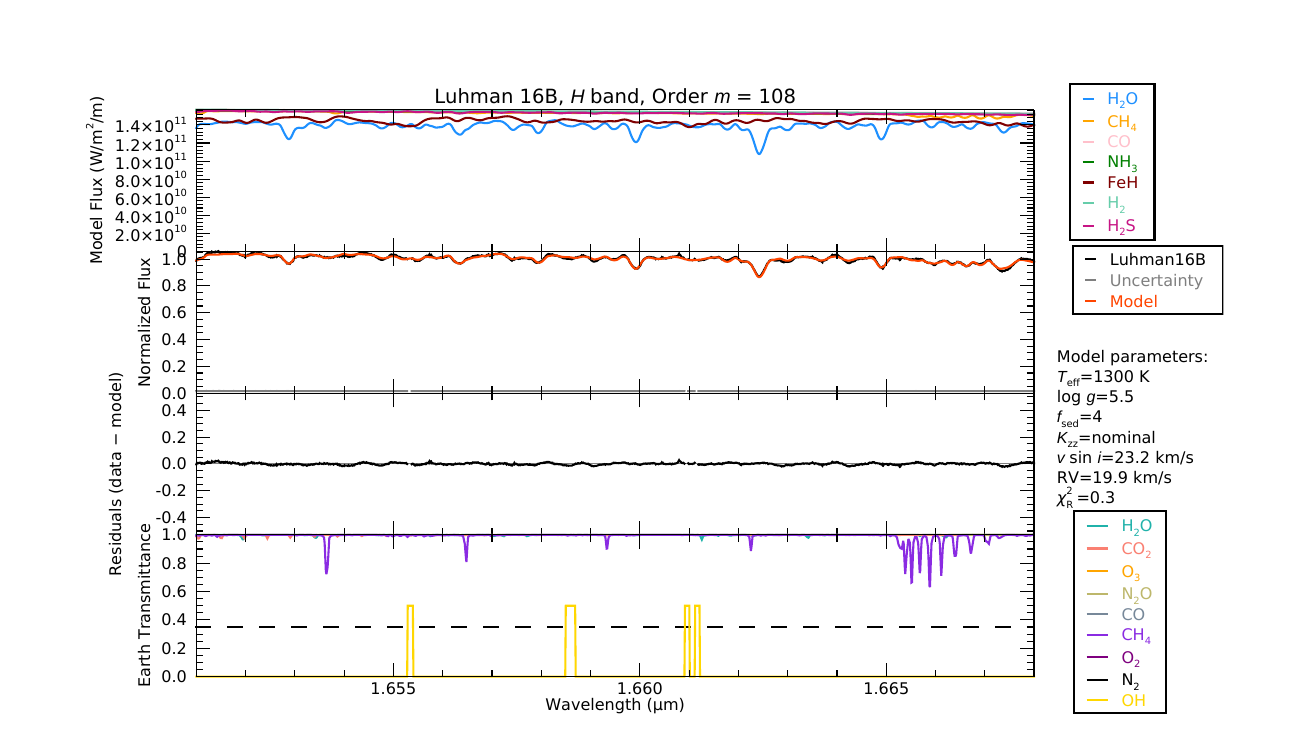}
    \caption{Continued.}
\end{figure*}

\begin{figure*}
    \ContinuedFloat
    \centering
    \includegraphics[height=0.43\textheight]{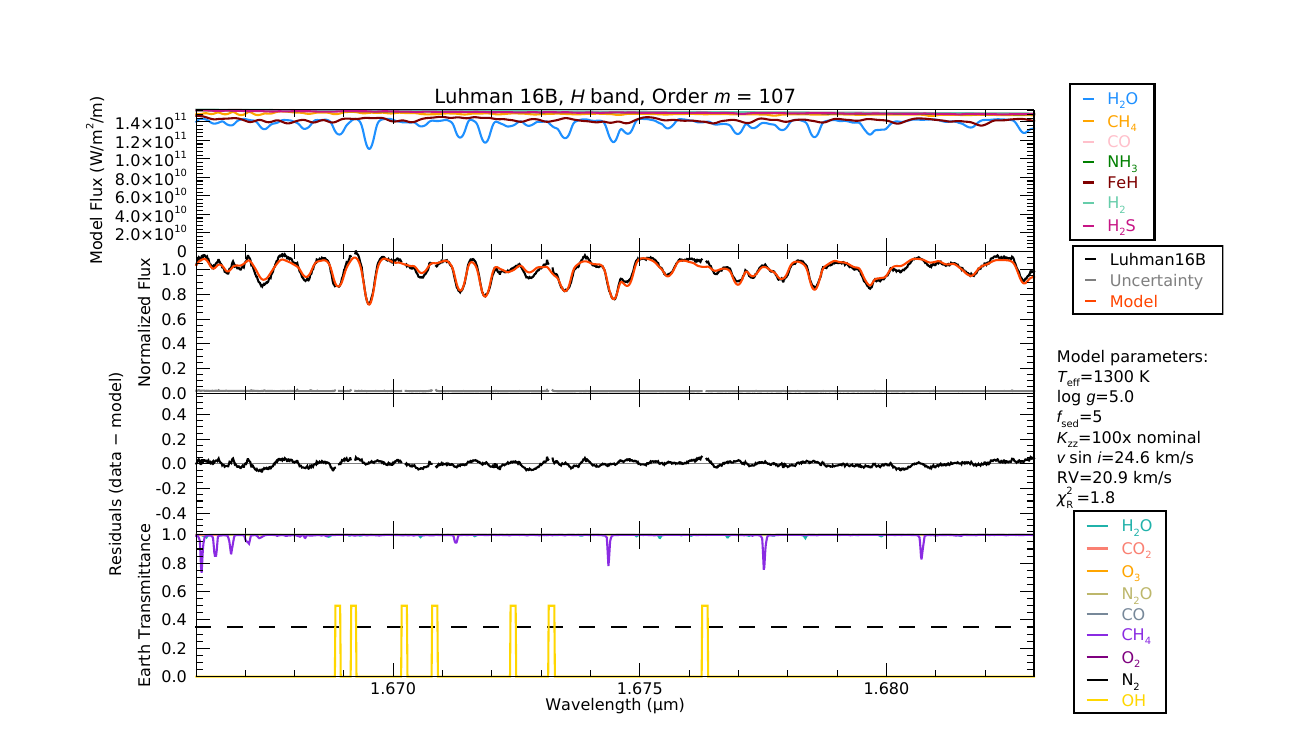}
    \includegraphics[height=0.43\textheight]{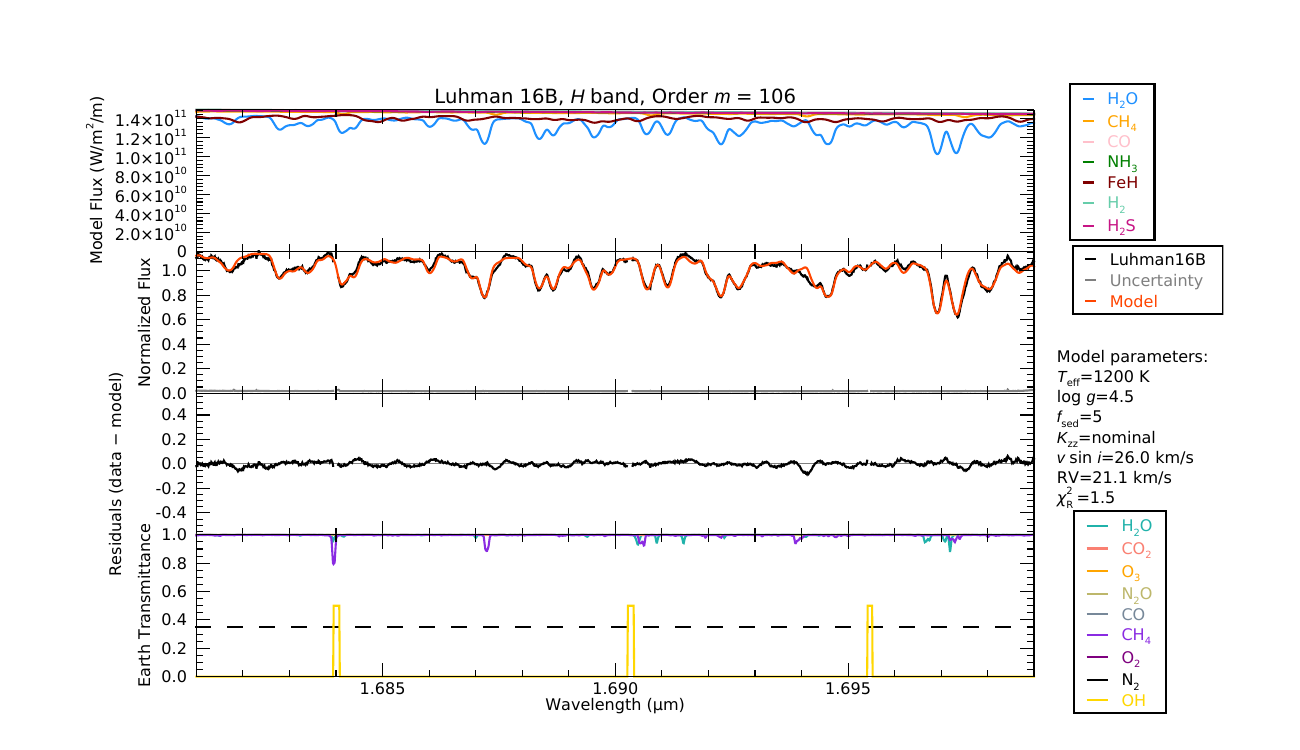}
    \caption{Continued.}
\end{figure*}

\begin{figure*}
    \ContinuedFloat
    \centering
    \includegraphics[height=0.43\textheight]{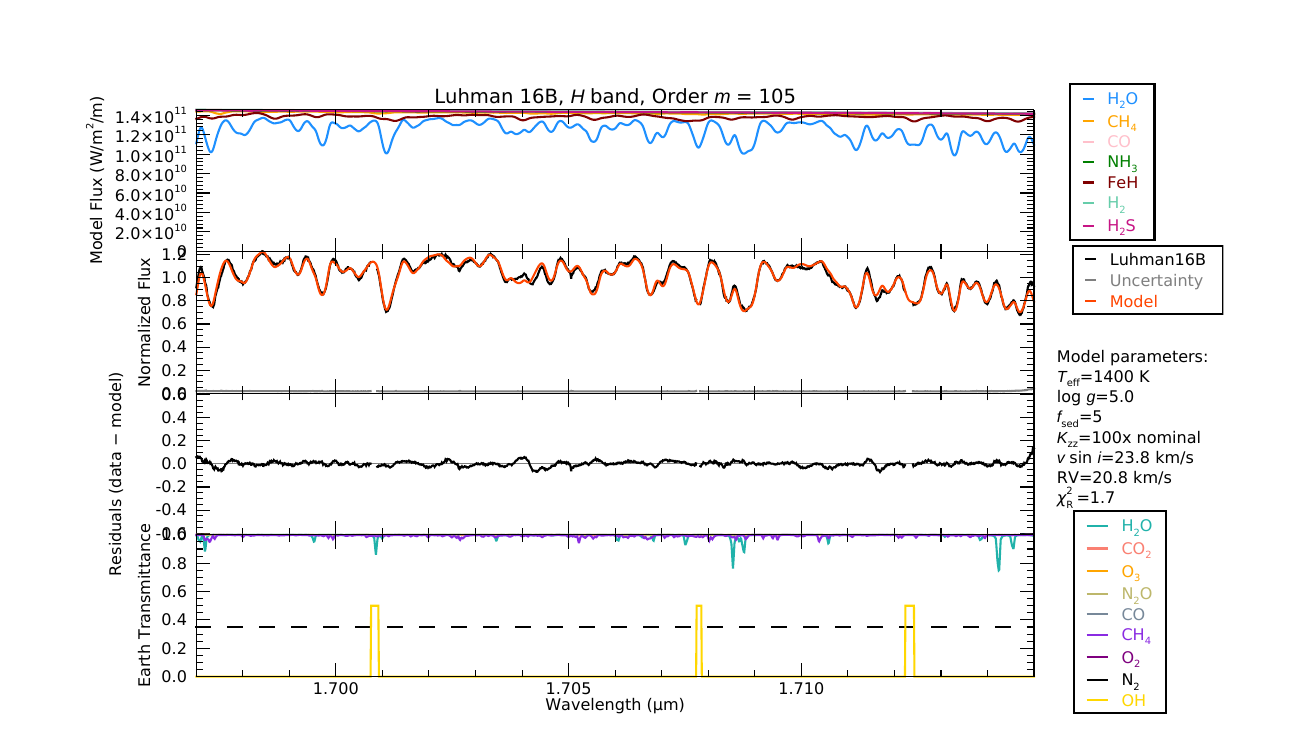}
    \includegraphics[height=0.43\textheight]{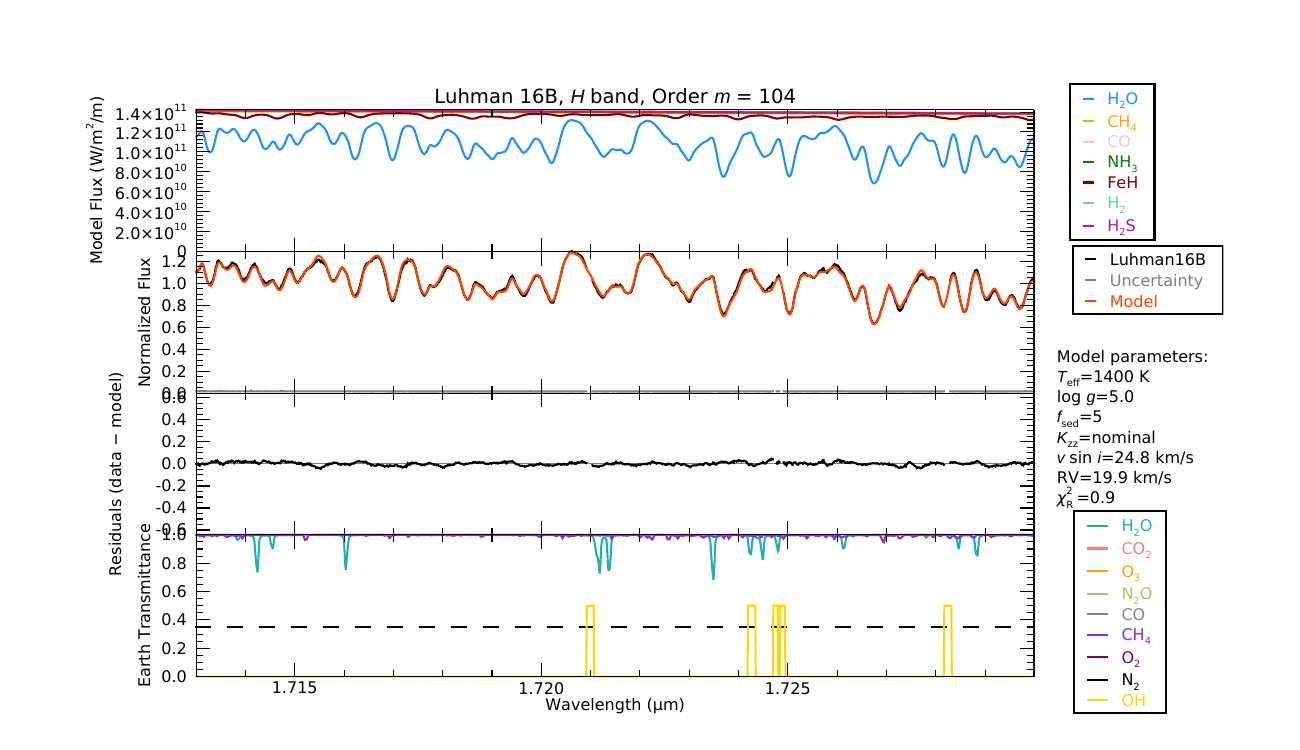}
    \caption{Continued.}
\end{figure*}

\begin{figure*}
    \ContinuedFloat
    \centering
    \includegraphics[height=0.43\textheight]{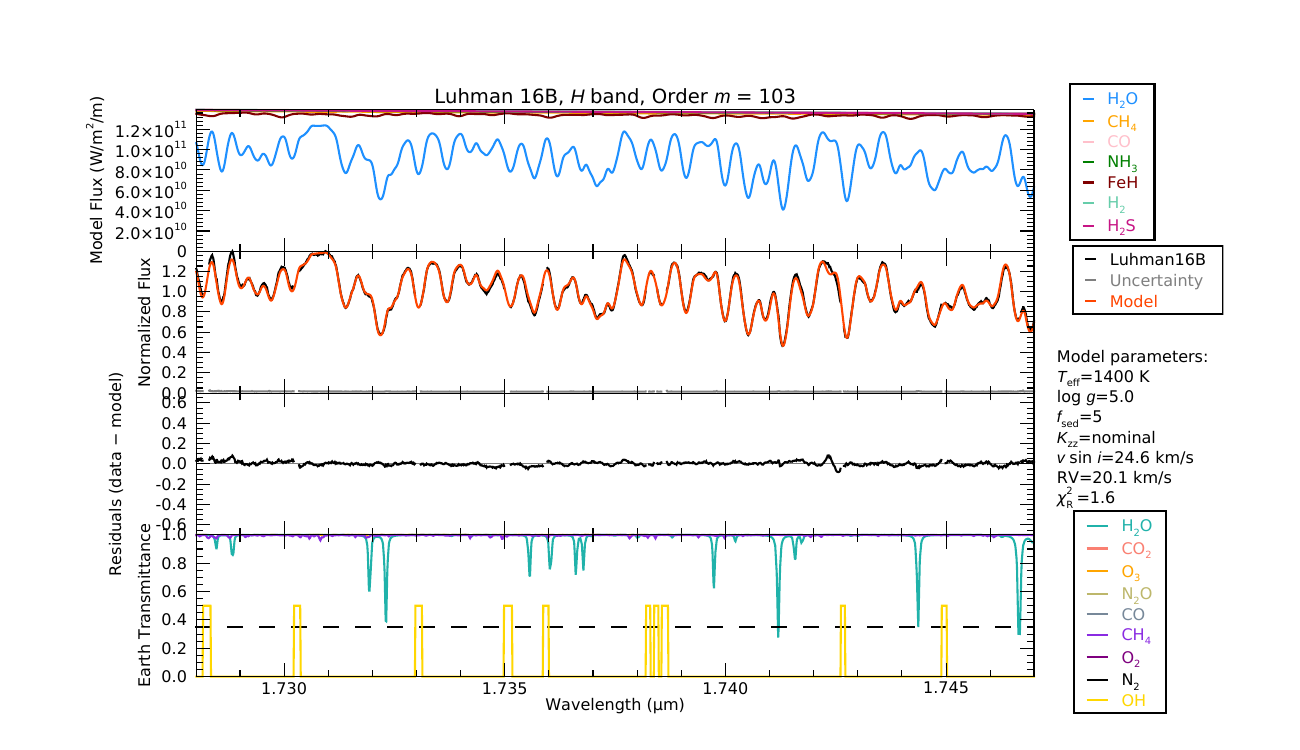}
    \includegraphics[height=0.43\textheight]{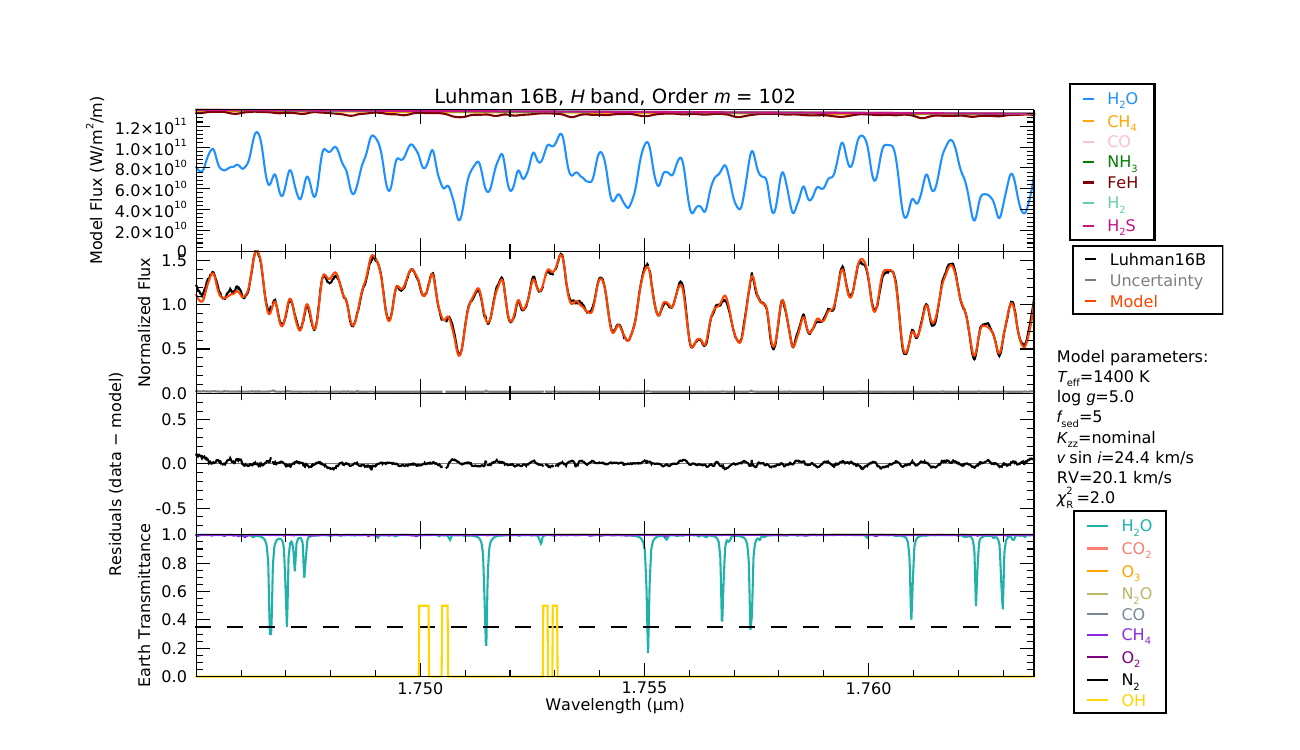}
    \caption{Continued.}
\end{figure*}

\begin{figure*}
    \ContinuedFloat
    \centering
    \includegraphics[height=0.43\textheight]{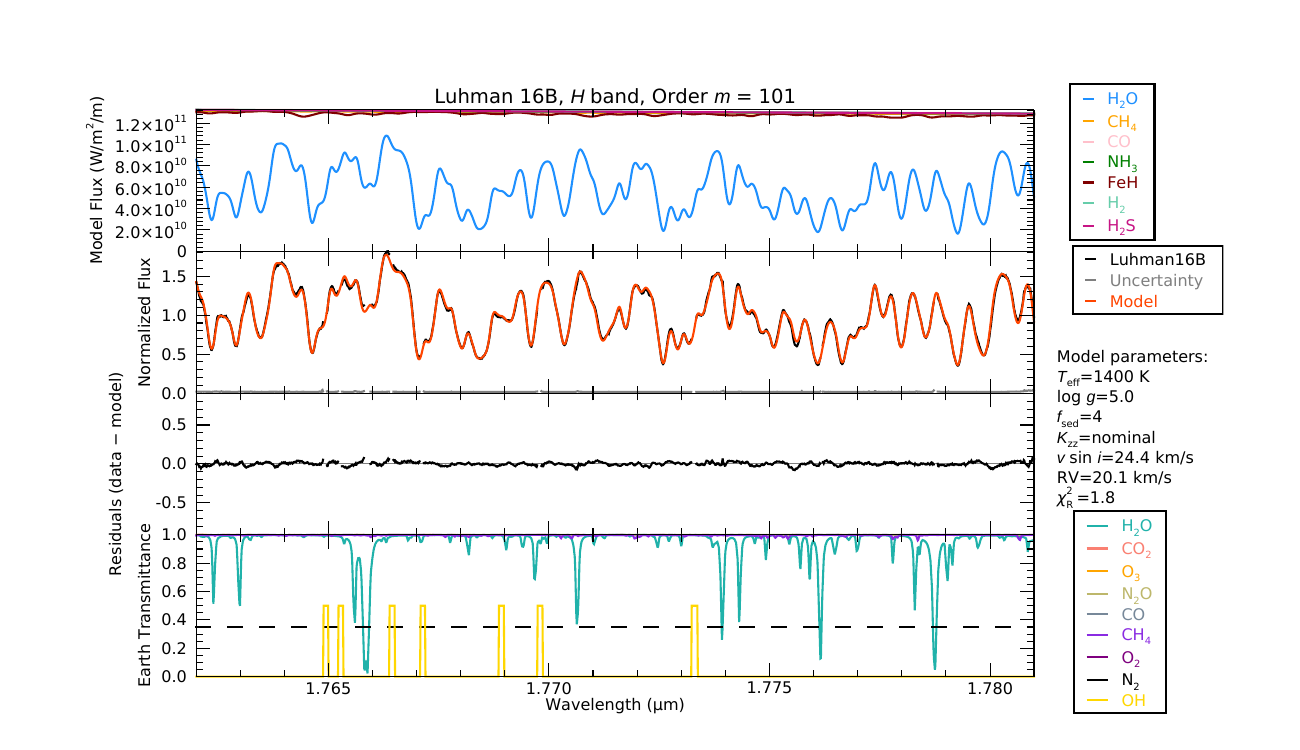}
    \includegraphics[height=0.43\textheight]{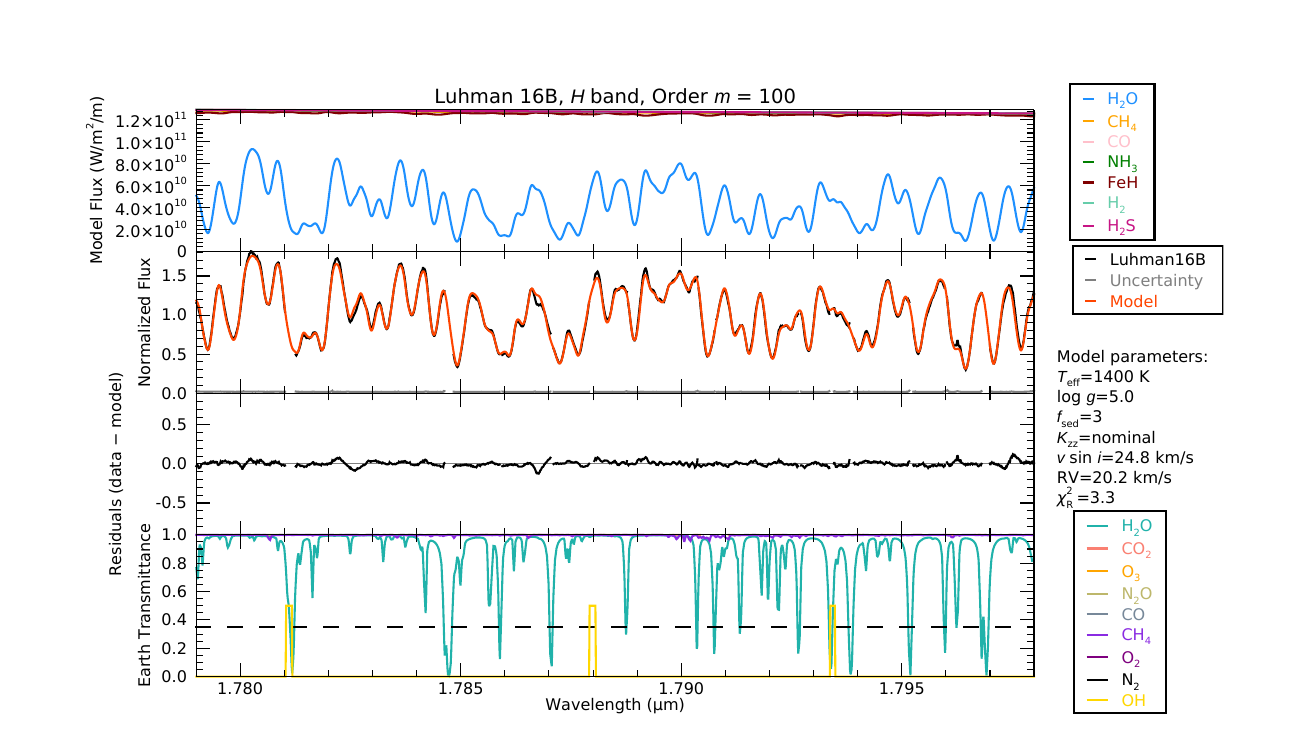}
    \caption{Continued.}
\end{figure*}

\begin{figure*}
    \ContinuedFloat
    \centering
    \includegraphics[height=0.43\textheight]{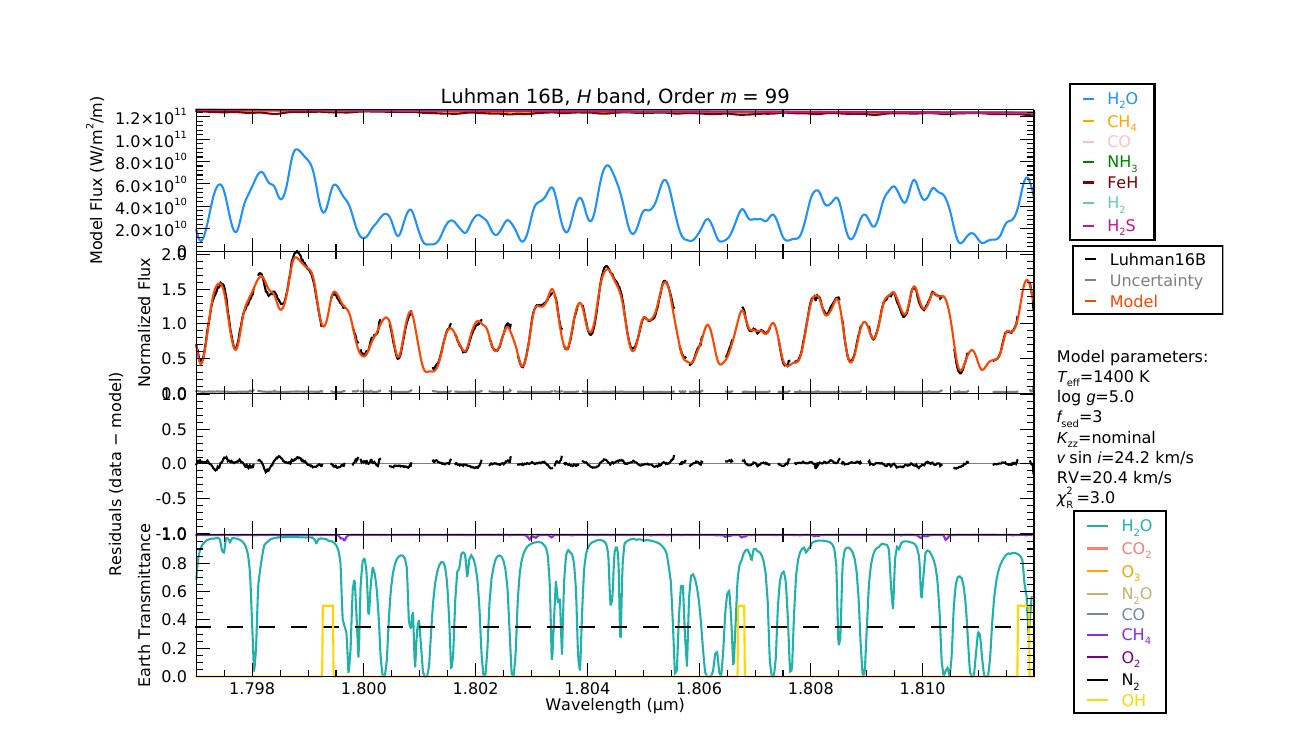}
    \includegraphics[height=0.43\textheight]{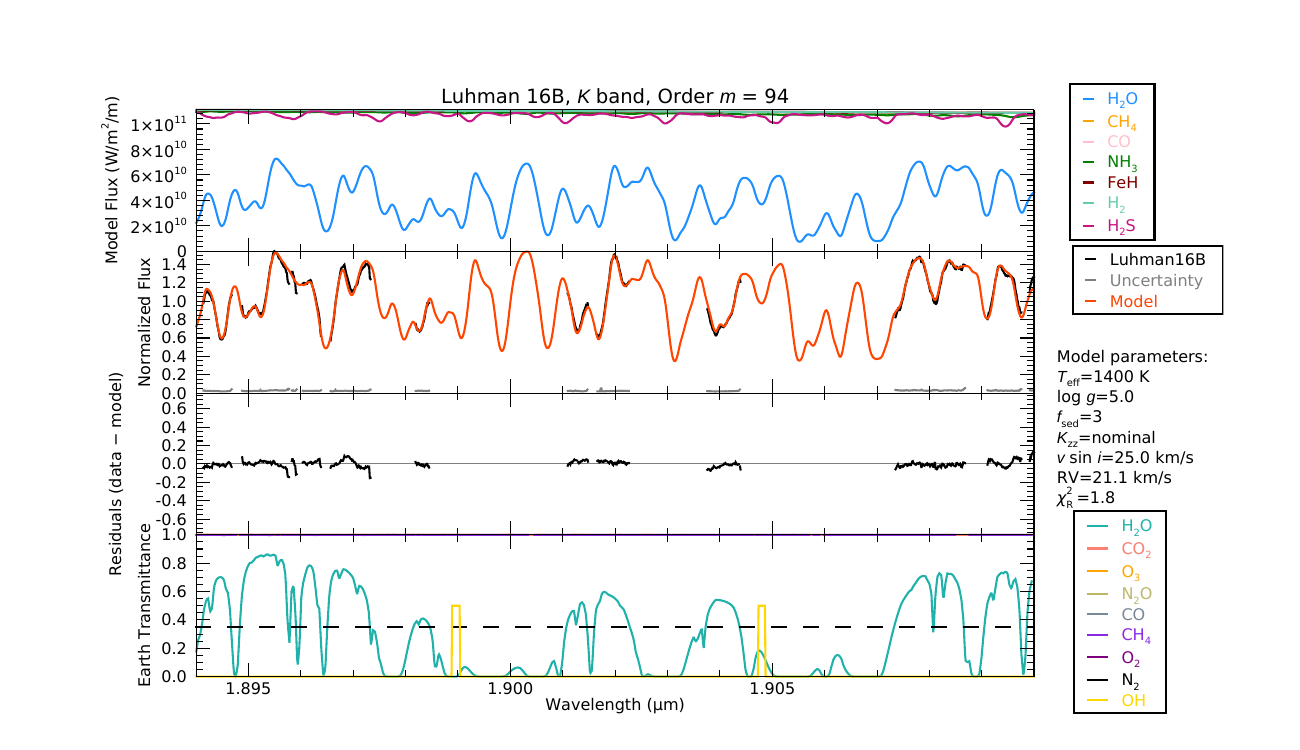}
    \caption{Continued.}
\end{figure*}

\begin{figure*}
    \ContinuedFloat
    \centering
    \includegraphics[height=0.43\textheight]{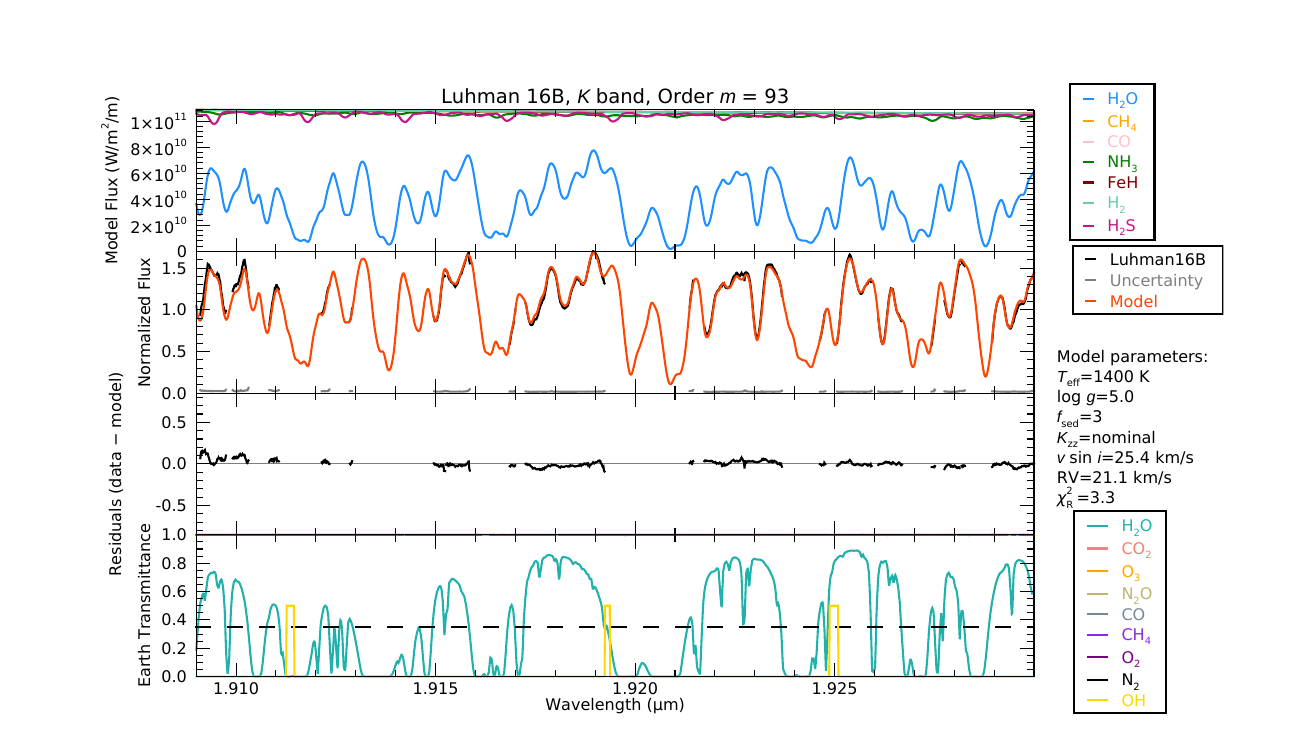}
    \includegraphics[height=0.43\textheight]{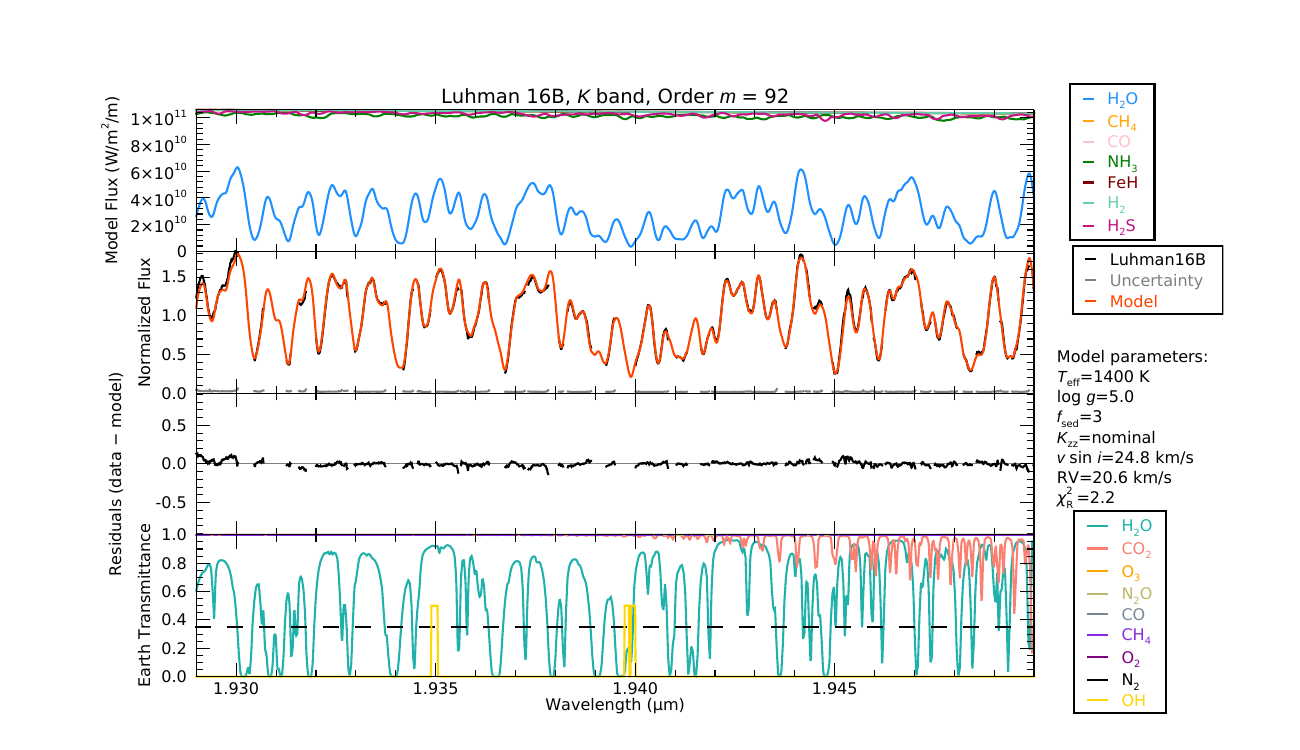}
    \caption{Continued.}
\end{figure*}

\begin{figure*}
    \ContinuedFloat
    \centering
    \includegraphics[height=0.43\textheight]{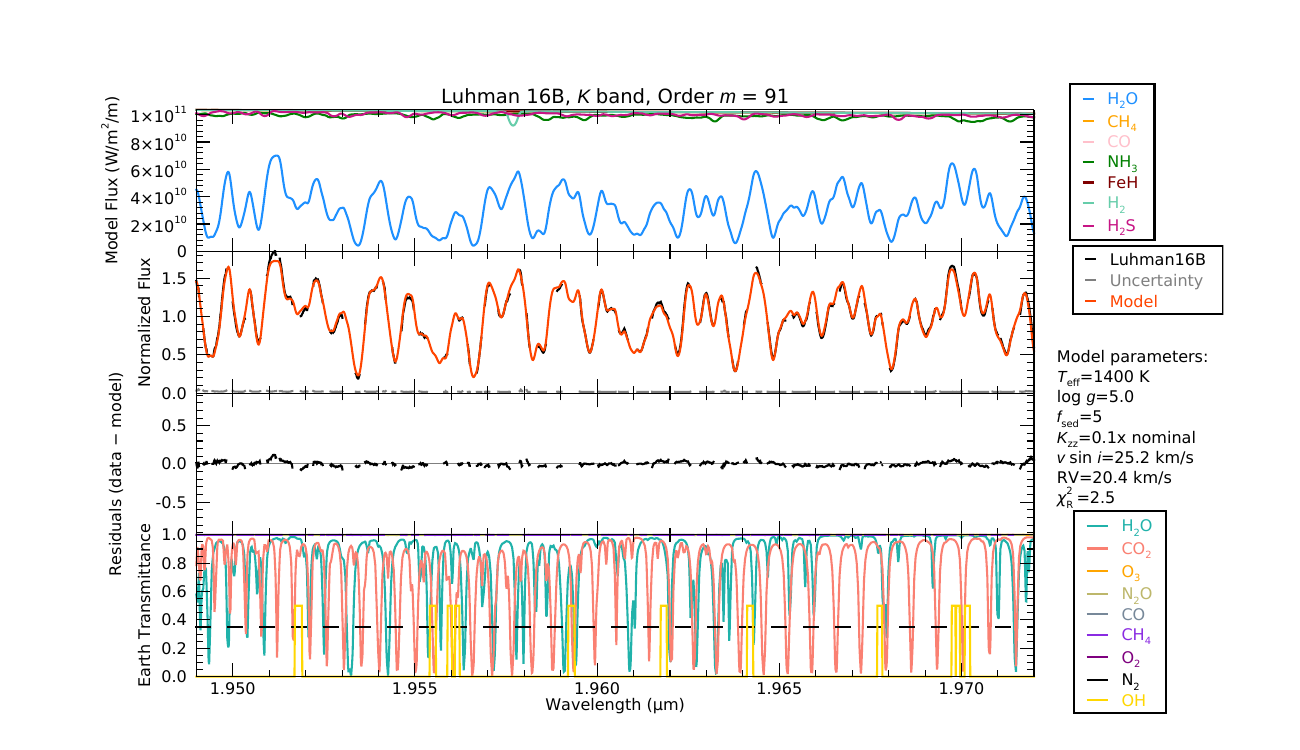}
    \includegraphics[height=0.43\textheight]{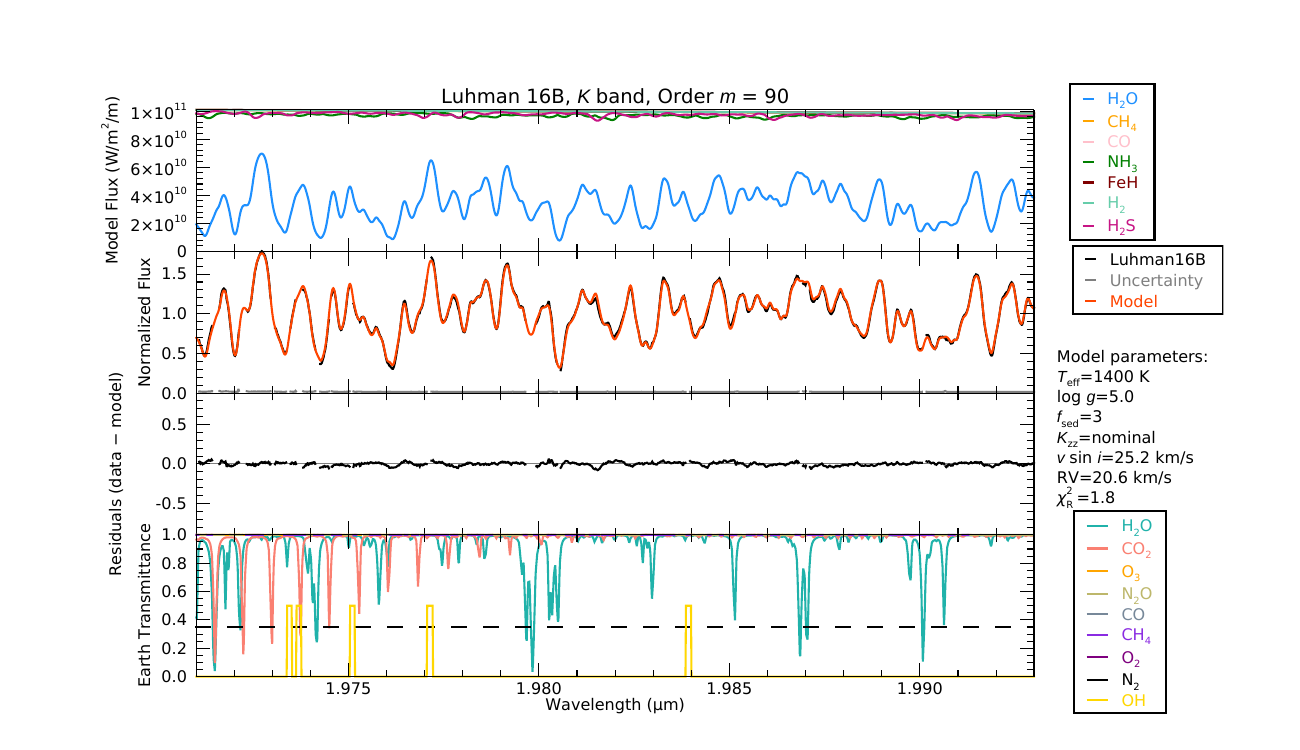}
    \caption{Continued.}
\end{figure*}

\begin{figure*}
    \ContinuedFloat
    \centering
    \includegraphics[height=0.43\textheight]{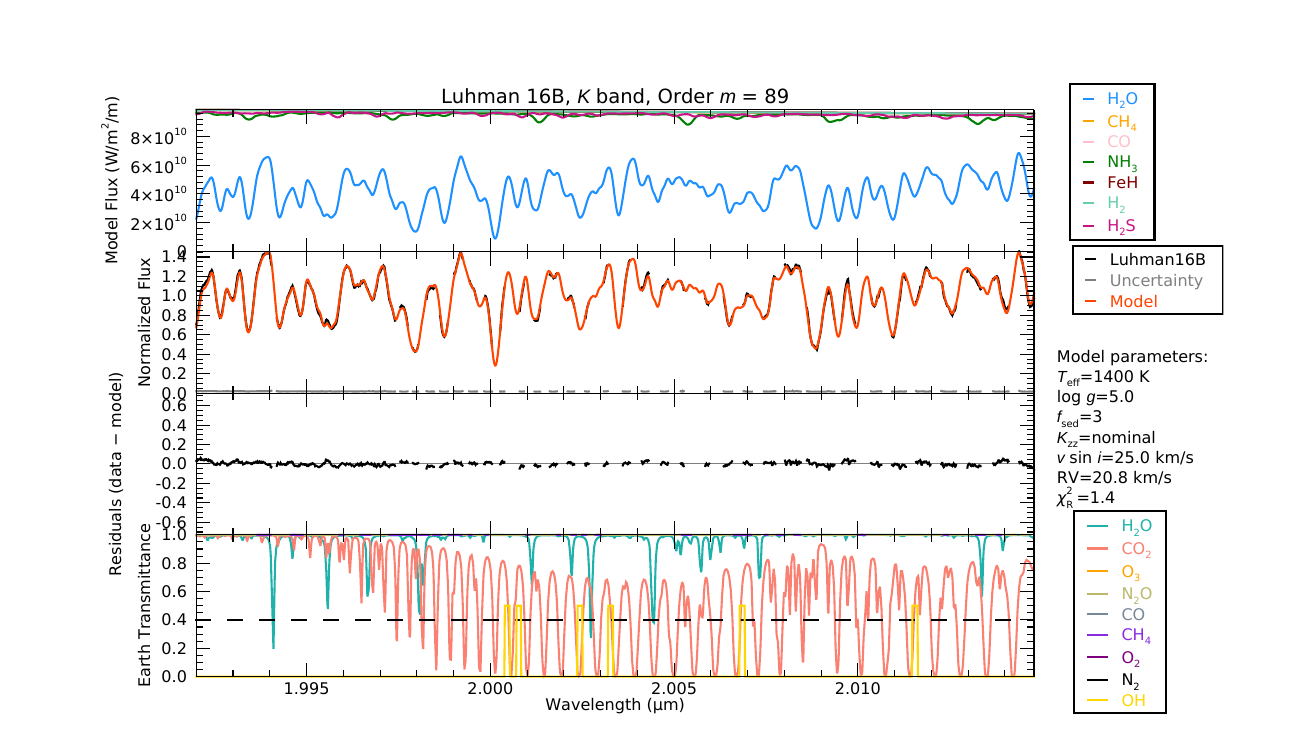}
    \includegraphics[height=0.43\textheight]{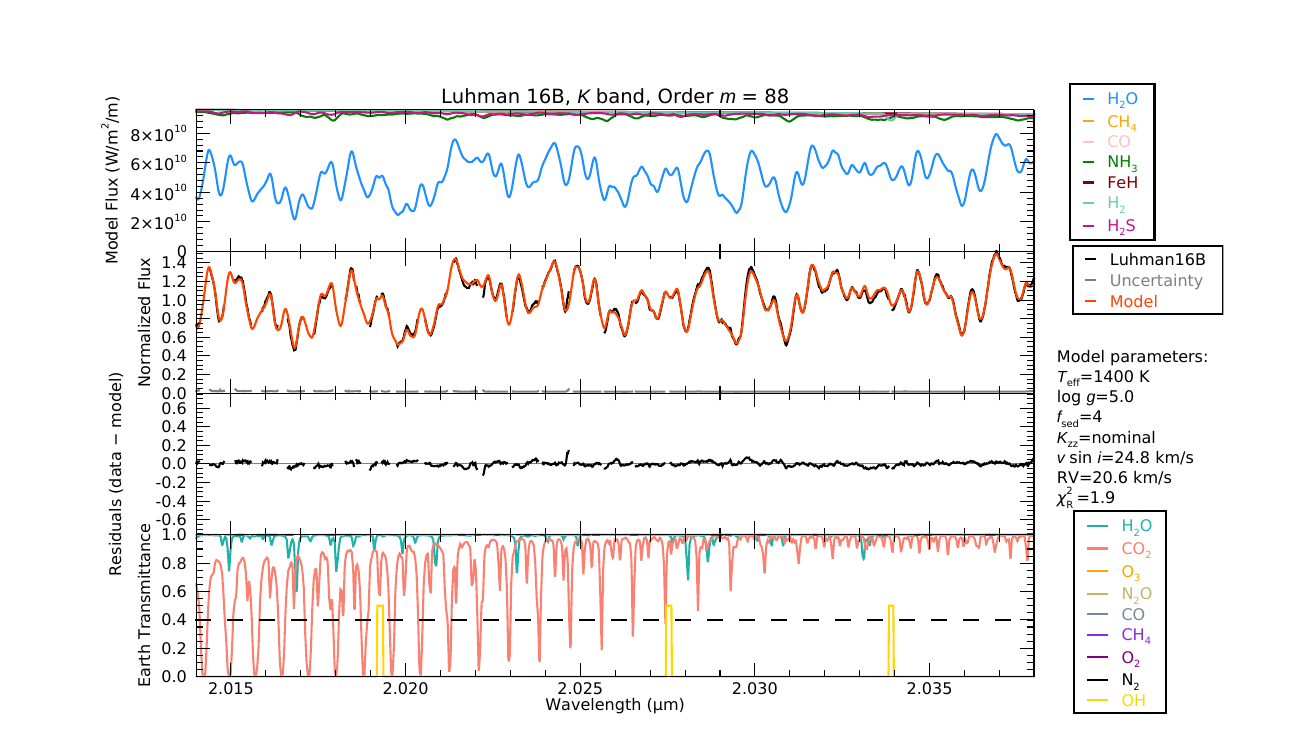}
    \caption{Continued.}
\end{figure*}

\begin{figure*}
    \ContinuedFloat
    \centering
    \includegraphics[height=0.43\textheight]{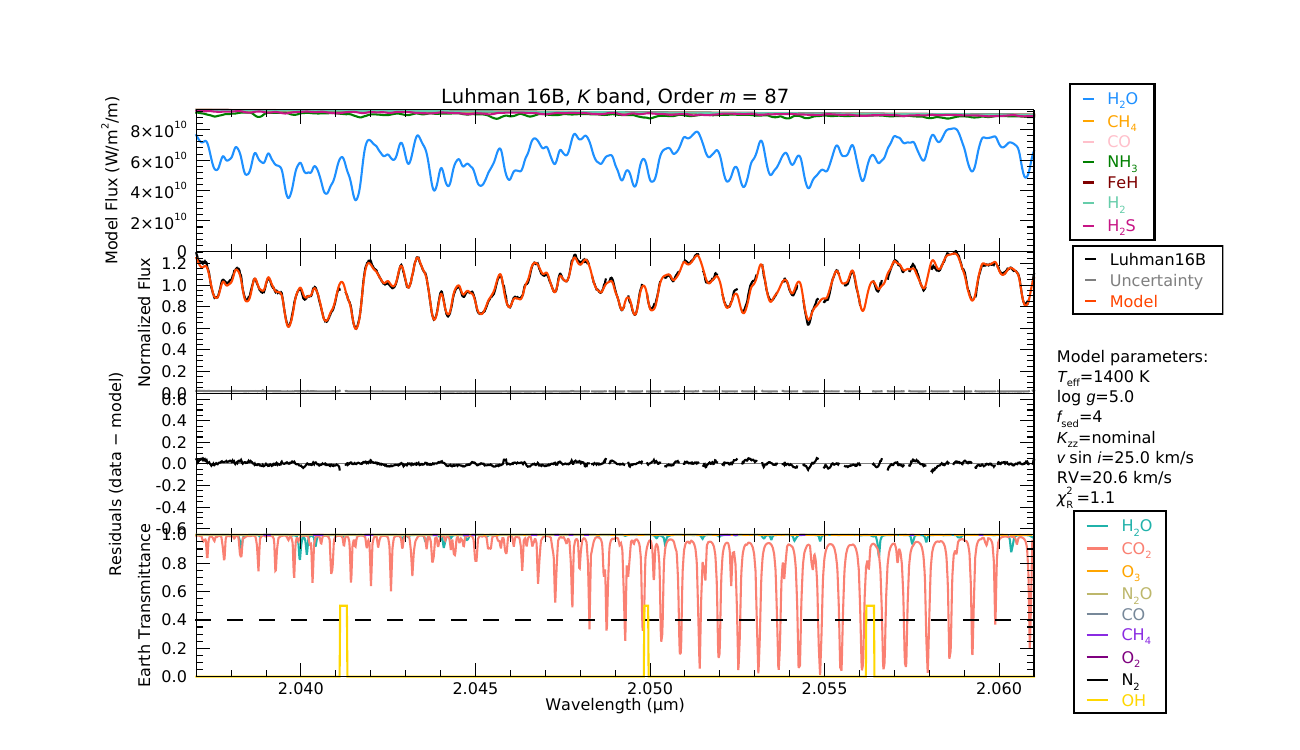}
    \includegraphics[height=0.43\textheight]{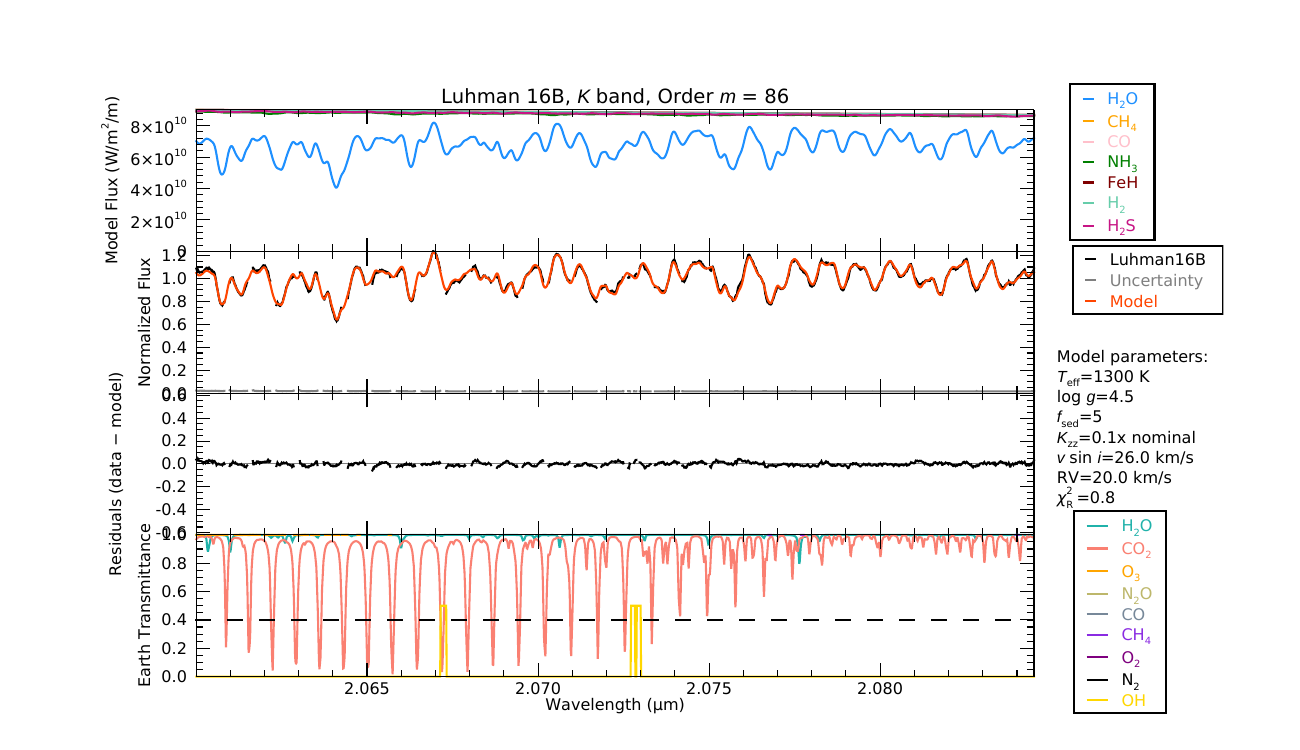}
    \caption{Continued.}
\end{figure*}

\begin{figure*}
    \ContinuedFloat
    \centering
    \includegraphics[height=0.43\textheight]{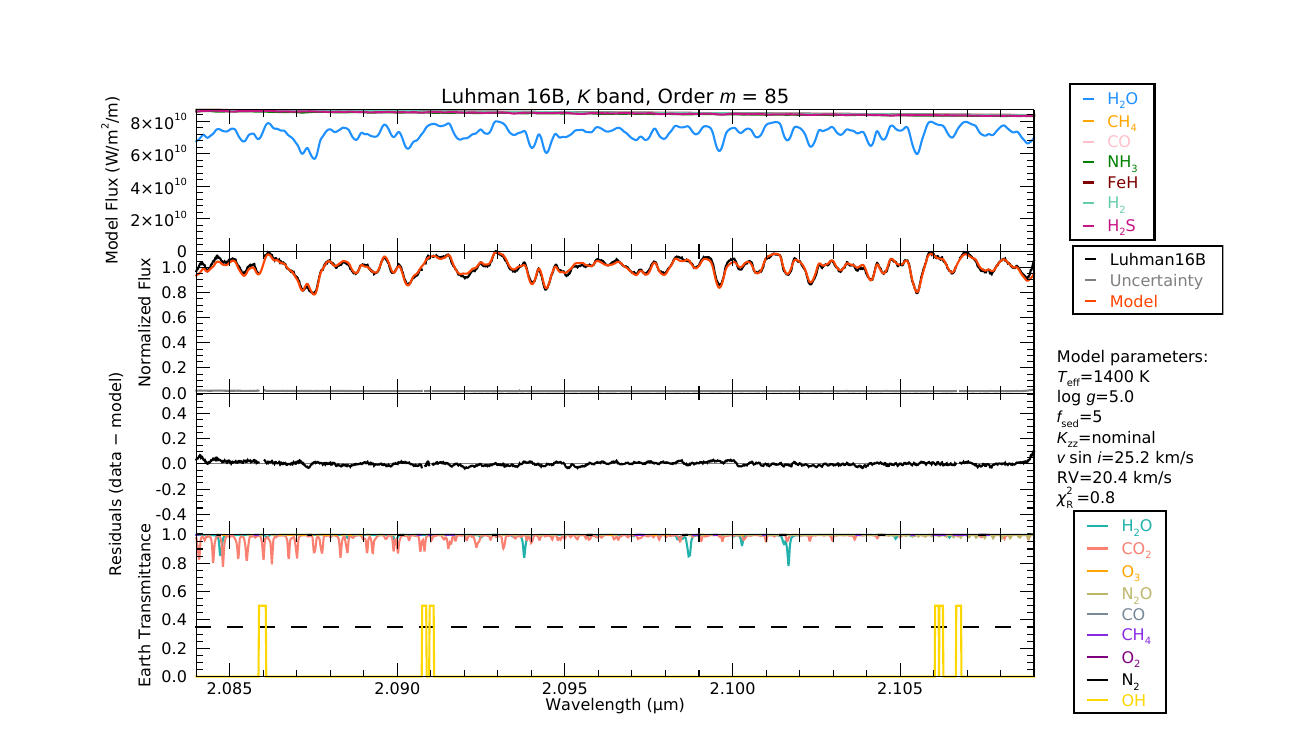}
    \includegraphics[height=0.43\textheight]{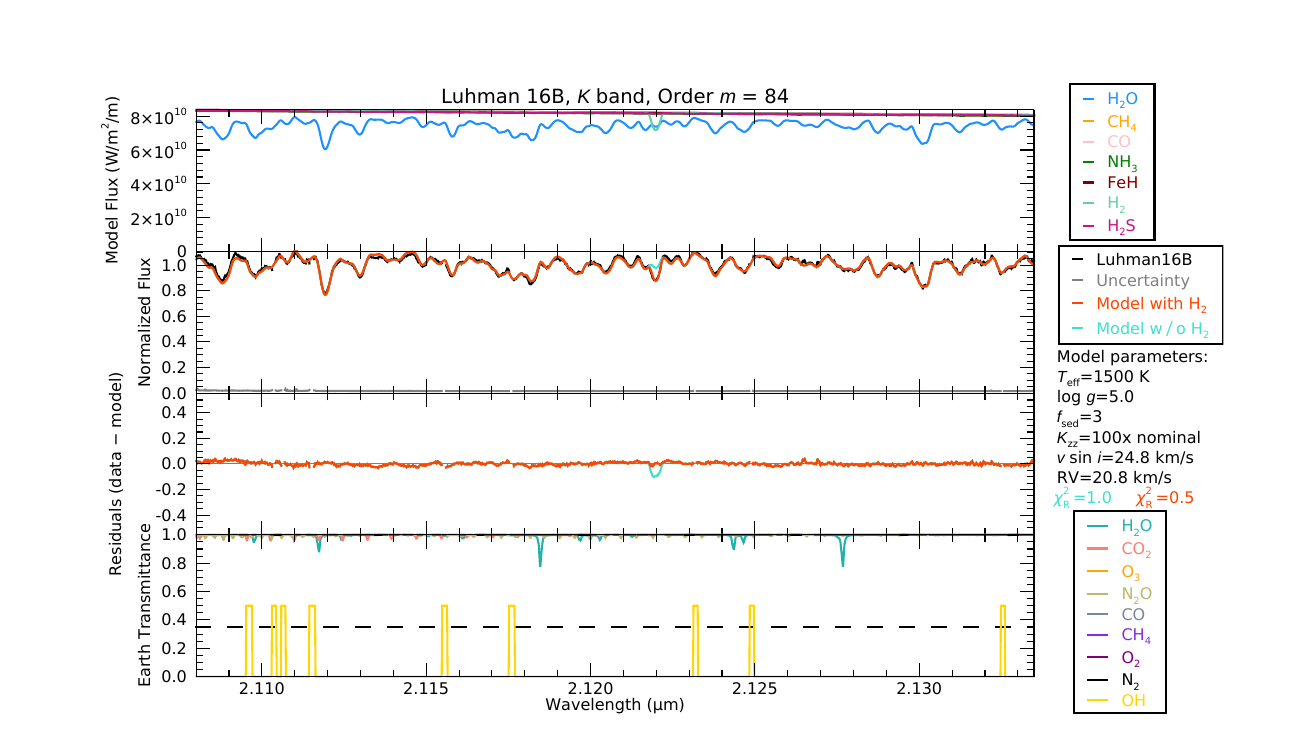}
    \caption{Continued.}
\end{figure*}

\begin{figure*}
    \ContinuedFloat
    \centering
    \includegraphics[height=0.43\textheight]{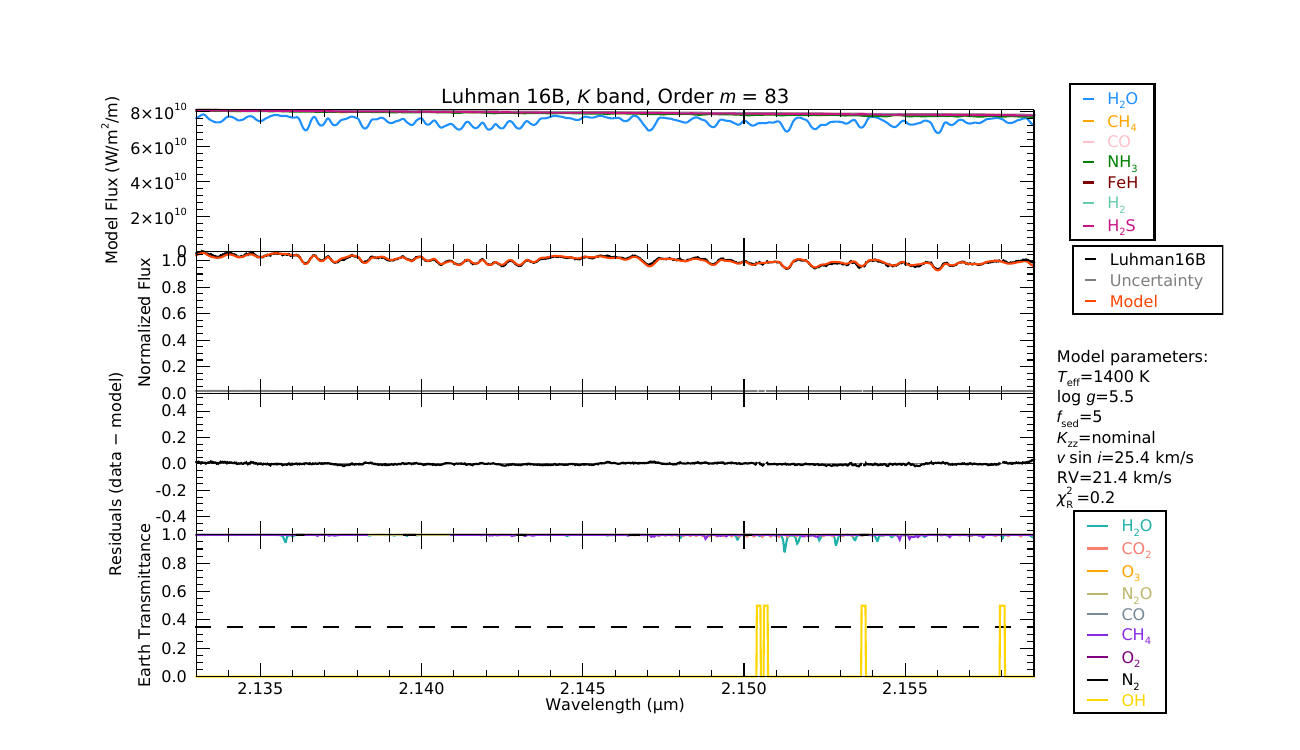}
    \includegraphics[height=0.43\textheight]{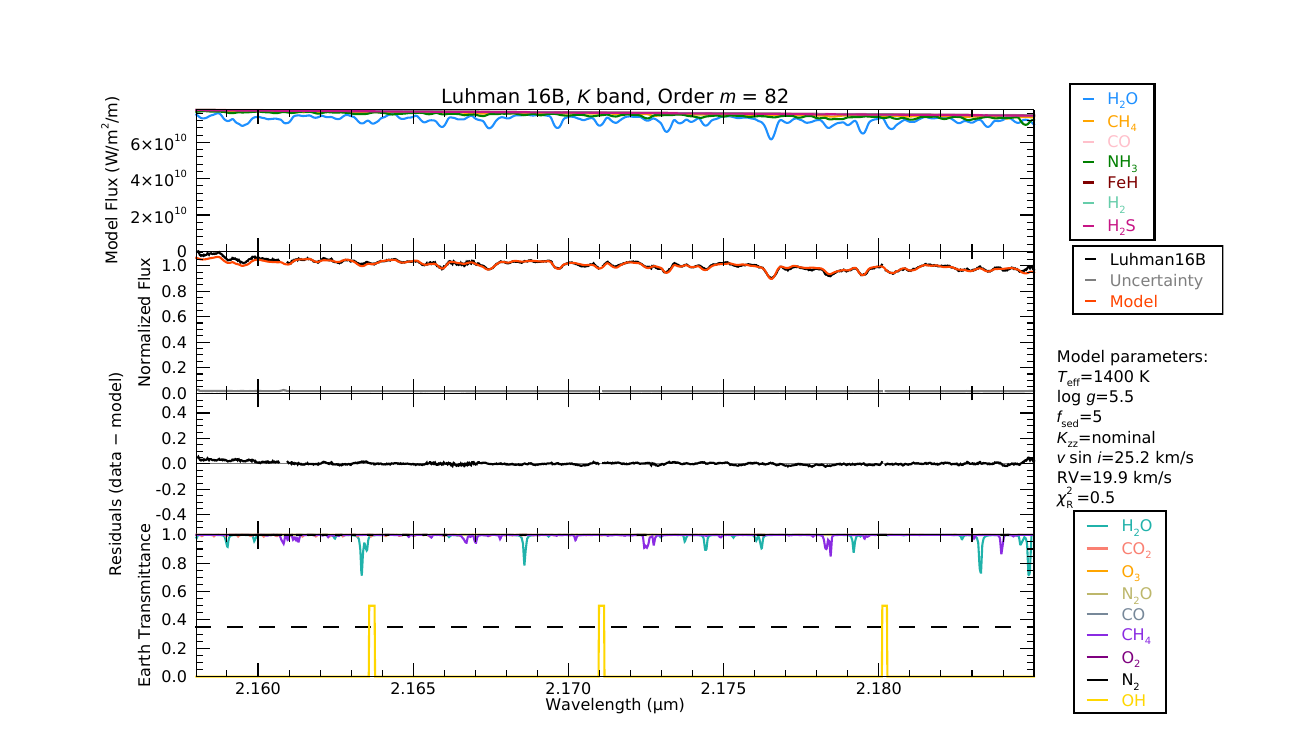} 
    \caption{Continued.}
\end{figure*}

\begin{figure*}
    \ContinuedFloat
    \centering
    \includegraphics[height=0.43\textheight]{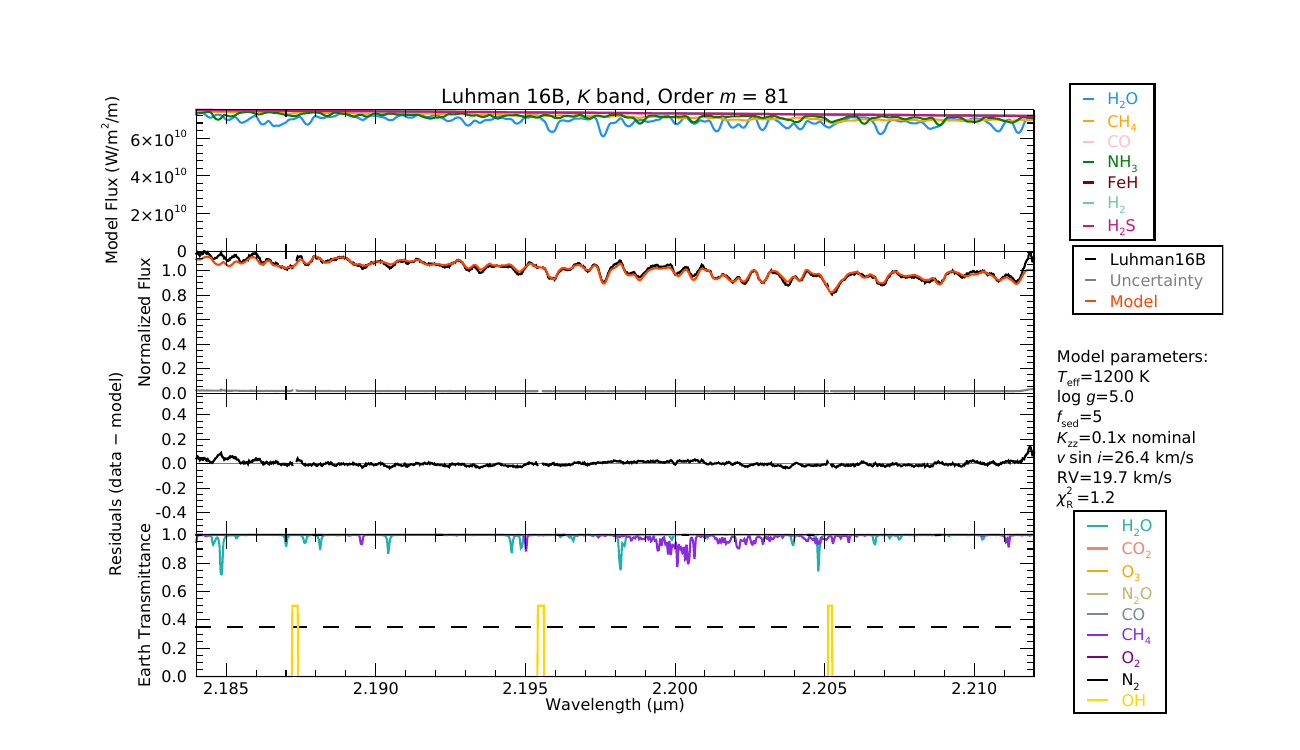}
    \includegraphics[height=0.43\textheight]{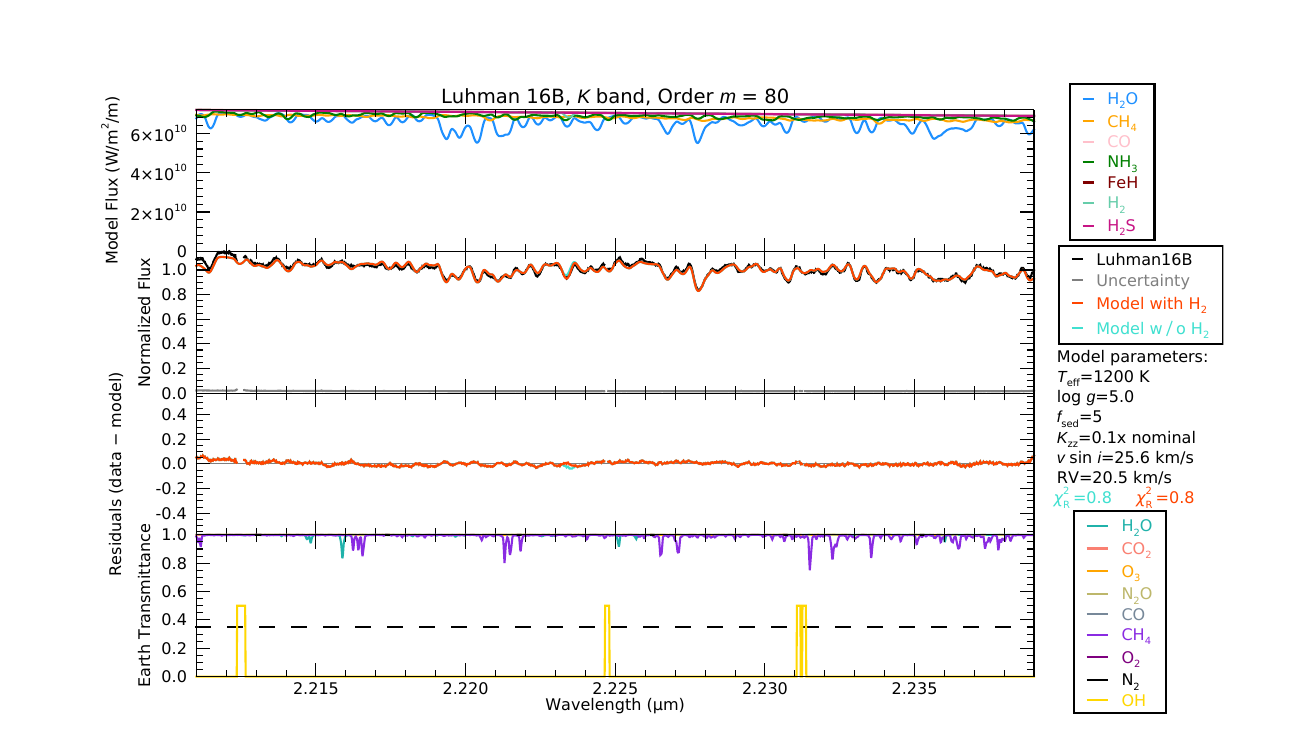}
    \caption{Continued.}
\end{figure*}

\begin{figure*}
    \ContinuedFloat
    \centering
    \includegraphics[height=0.43\textheight]{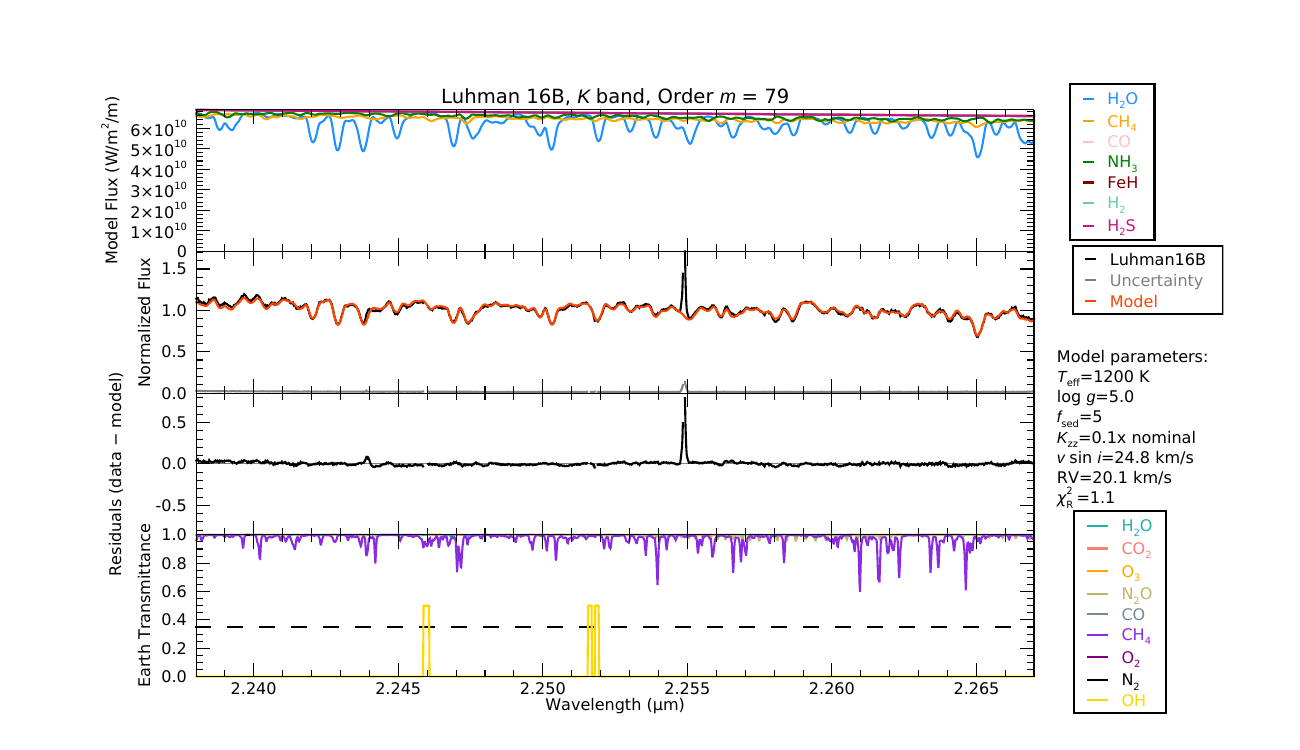}
    \includegraphics[height=0.43\textheight]{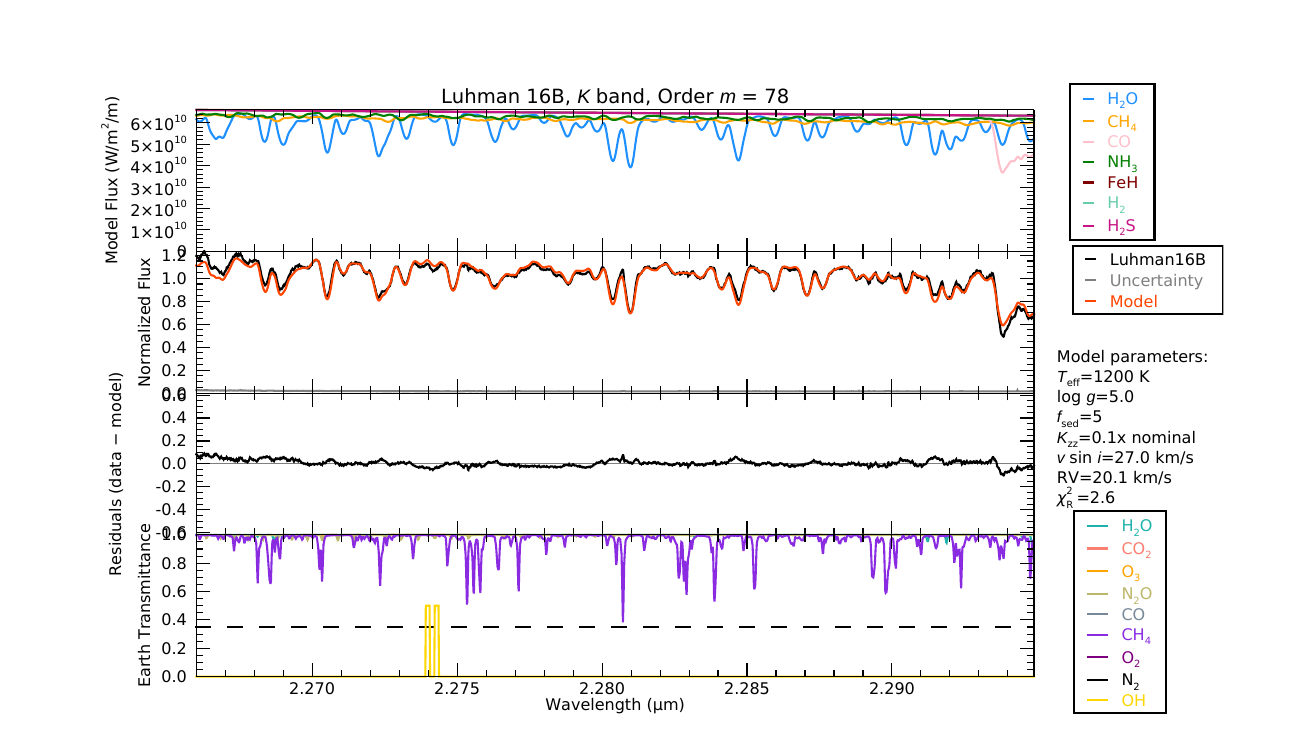}
    \caption{Continued.}
\end{figure*}

\begin{figure*}
    \ContinuedFloat
    \centering
    \includegraphics[height=0.43\textheight]{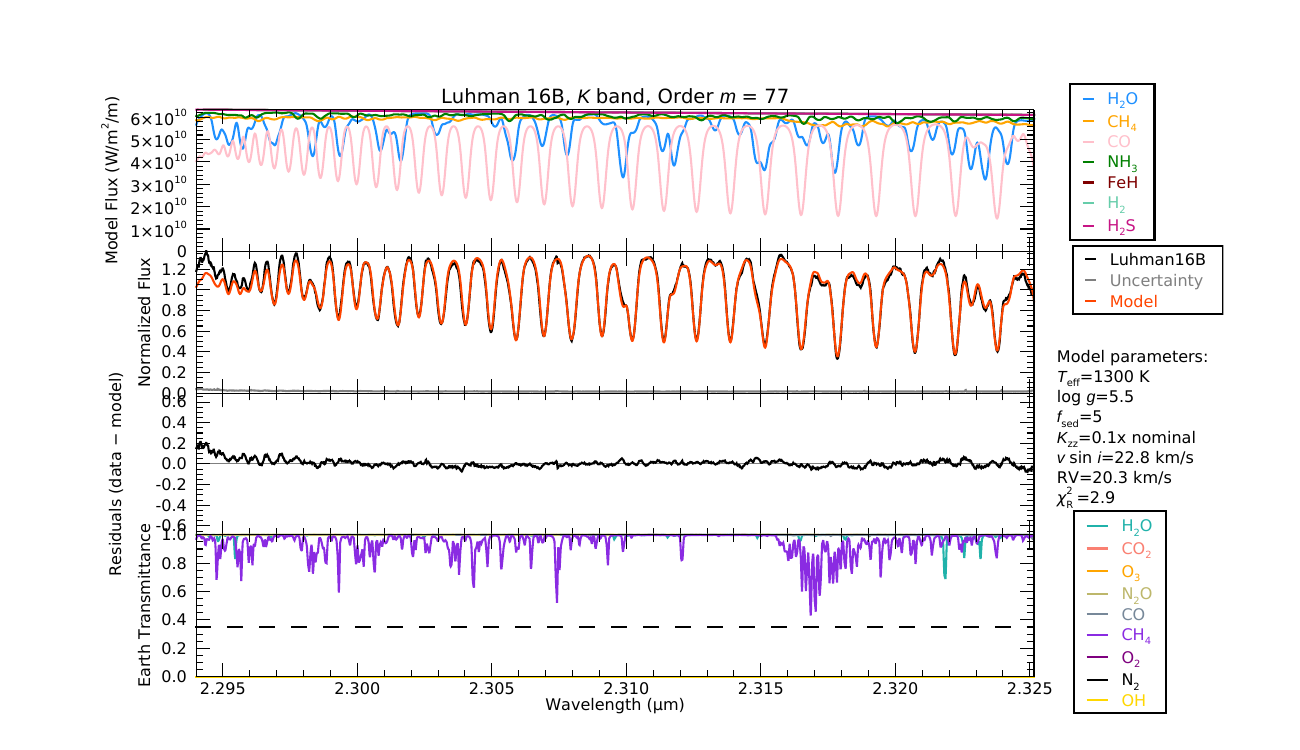}
    \includegraphics[height=0.43\textheight]{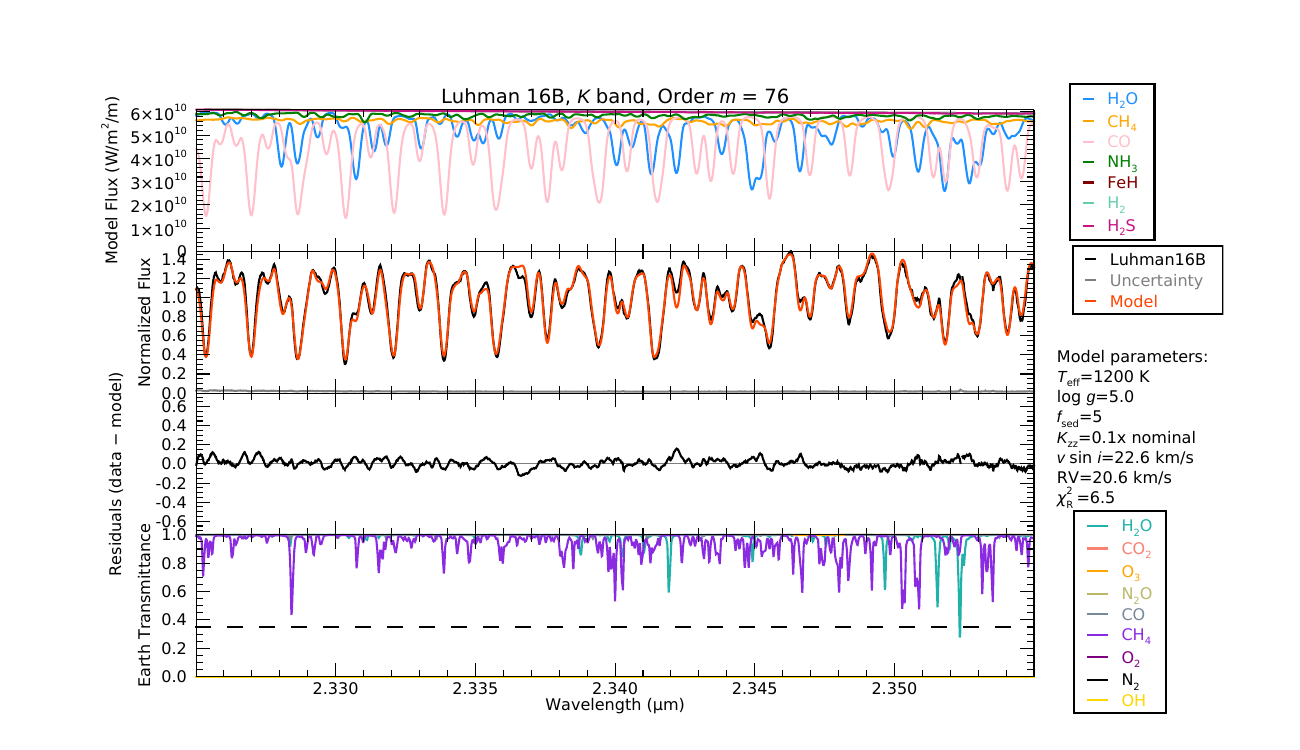}
    \caption{Continued.}
\end{figure*}

\begin{figure*}
    \ContinuedFloat
    \centering
    \includegraphics[height=0.43\textheight]{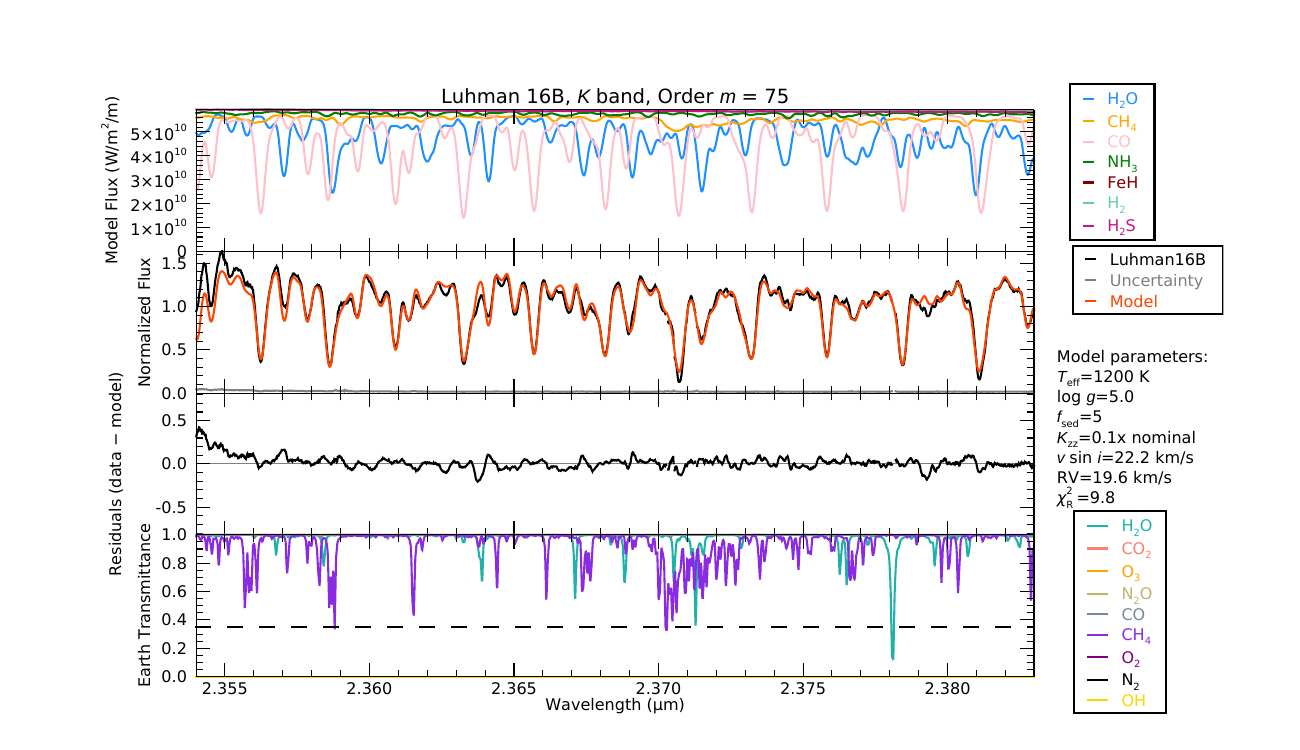}
    \includegraphics[height=0.43\textheight]{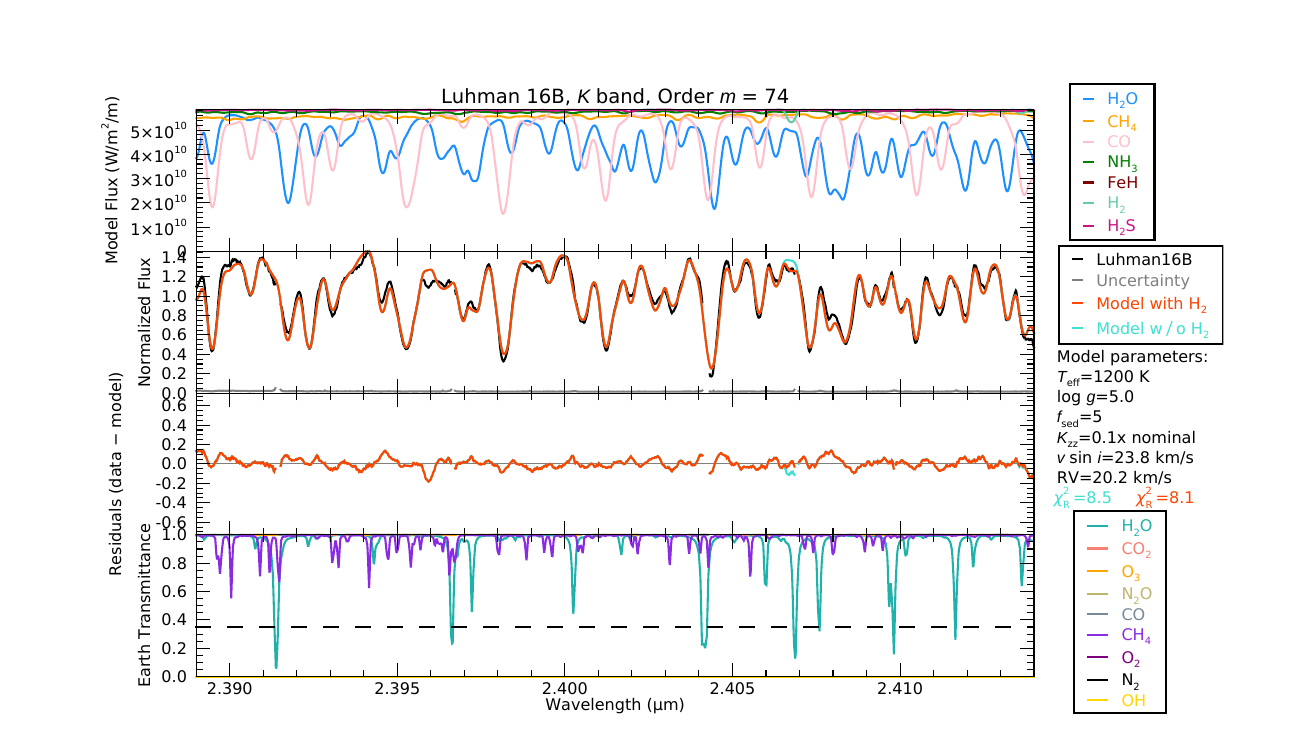}
    \caption{Continued.}
\end{figure*}

\begin{figure*}
    \ContinuedFloat
    \centering
    \includegraphics[height=0.43\textheight]{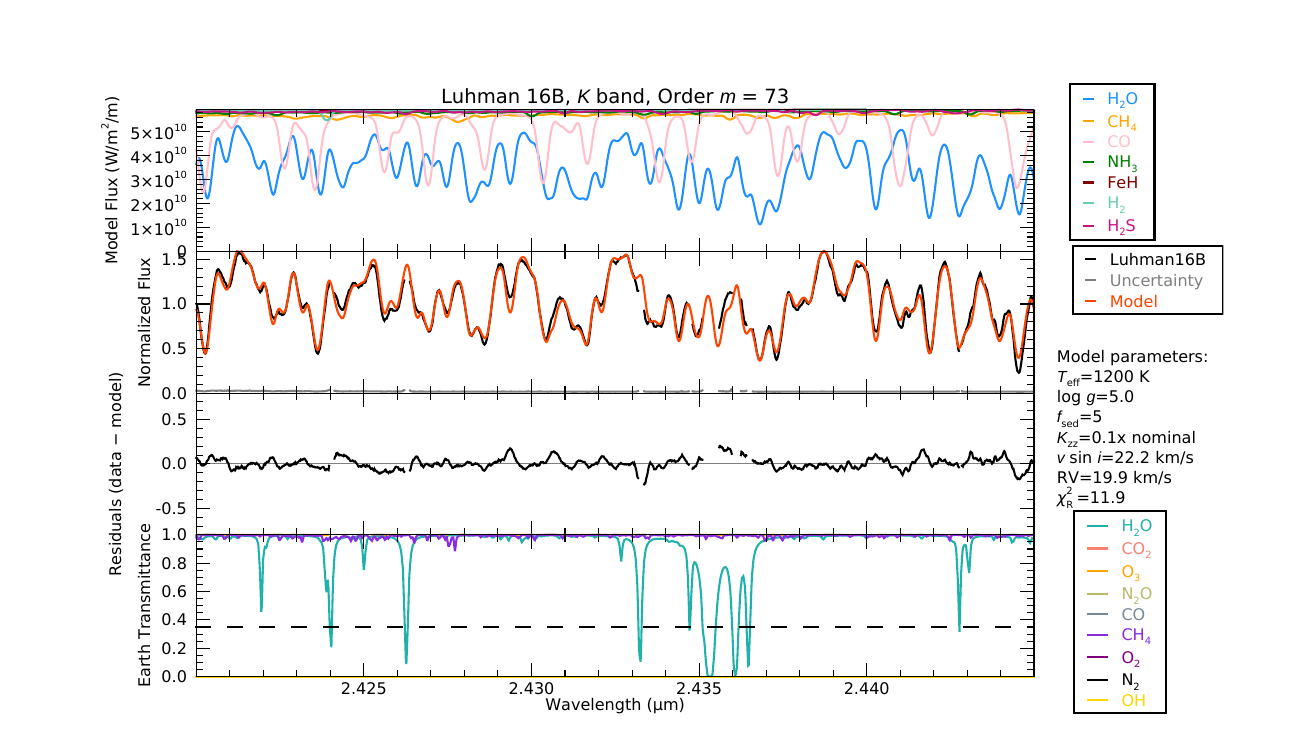}
    \includegraphics[height=0.43\textheight]{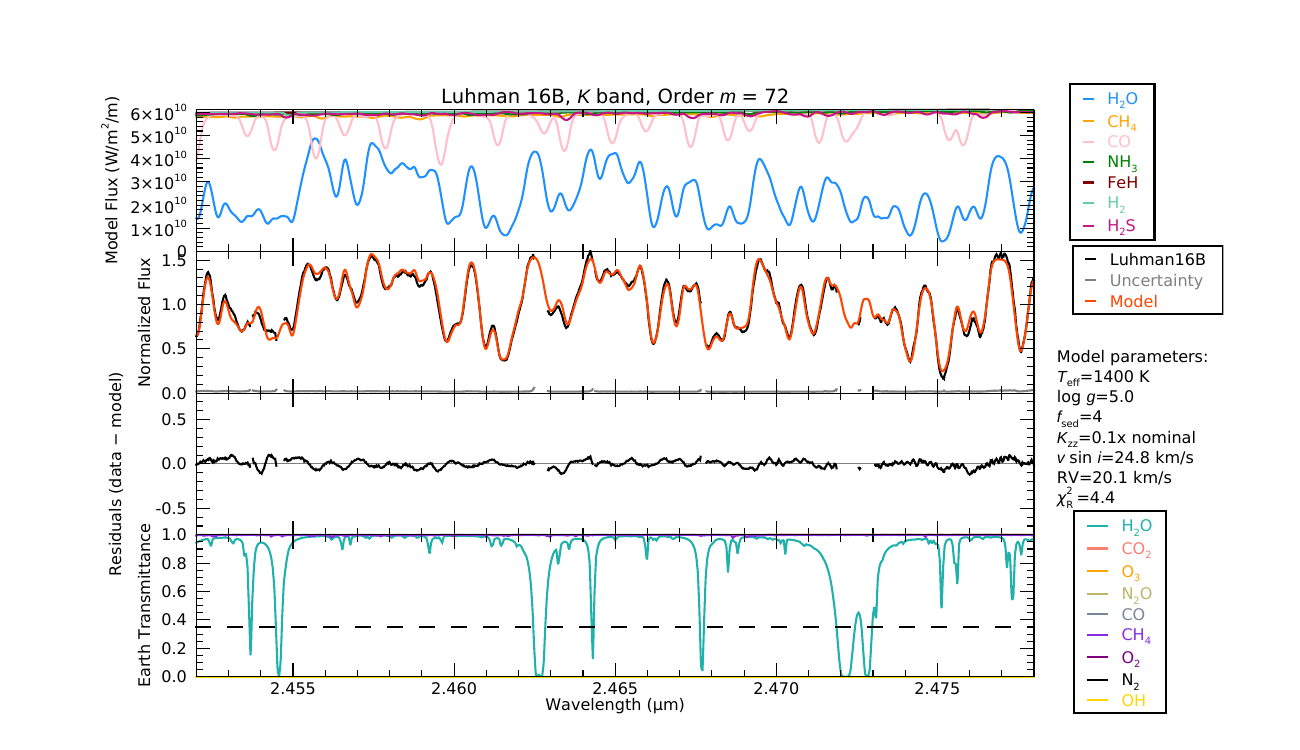}
    \caption{Continued.}
\end{figure*}


\bsp	
\label{lastpage}
\end{document}